# Compositional Metrology of Atom Probe Applied to non-Metallic Materials

Enrico Di Russo

*Dipartimento di Fisica e Astronomia, Università degli Studi di Padova, Via Marzolo 8, 35131 Padova, Italy, Istituto Nazionale di Fisica Nucleare, Laboratori Nazionali di Legnaro, Viale dell'Università 2, 35020 Legnaro, Italy, and CNR-ISMN, Istituto per lo Studio dei Materiali Nanostrutturati, Via Gobetti 101, 40129 Bologna, Italy*

François Vurpillot, Lorenzo Rigutti*

*University of Rouen Normandie, INSA Rouen Normandie, CNRS, Groupe de Physique des Matériaux UMR 6634, 76000 Rouen, France*

* Corresponding author: lorenzo.rigutti@univ-rouen.fr

**Abstract**

Laser-assisted Atom Probe Tomography (LA-APT) has demonstrated a unique potential for the study of the 3D distribution of atomic species in semiconductor materials and devices, and in a growing list of inorganic non-metallic solids. A crucial and often underestimated issue with APT is its accuracy in compositional measurements of non-metallic systems. This work introduces the principles of APT as an experimental technique, recalling the aspects potentially leading to compositional biases and underlining in particular the role of the surface electric field in governing the different physical-chemical phenomena that enable the measurement. It reviews the possible mechanisms of specific losses, as well as the methods for assessing a compositional bias and proposing possible correction methods. Finally, it establishes a state of the art on compositional biases in APT of non-metallic materials, on the basis of which it will be possible to conclude on specific recommendations for best practices, and the perspective of application of APT to new materials.

## List of main acronyms

| | |
|---|---|
| 1D-AP | 1-dimensional Atom Probe |
| aDLD | Advanced DLD |
| APT | Atom Probe Tomography |
| BIF | Best Image Field |
| CSR | Charge State Ratio |
| DE | Detection Efficiency |
| DFT | Density Functional Theory |
| DLD | Delay Line Detector |
| DUV | Deep Ultraviolet |
| EBSD | Electron Back-Scattering Diffraction |
| ECCI | Electron Channeling Contrast Imaging |
| EDS, EDX | Electron Dispersion X-ray Spectroscopy |
| EELS | Electron Energy Loss Spectroscopy |
| ERDA | Elastic Recoil Detection Analysis |
| EUV | Extreme Ultraviolet |
| FIB | Focused Ion Beam |
| FIM | Field Ion Microscopy |
| FlexTAP | Flexible Tomographic Atom Probe (device name) |
| FOV | Field of View |
| FWHM | Full Width at Half Maximum |
| ICF | Image Compression Factor |
| ISO | International Standard Organization |
| LA-APT | Laser-assisted Atom Probe Tomography |
| LaWATAP | Laser-assisted Wide Angle Tomographic Atom Probe (device name) |
| LEAP | Local Electrode Atom Probe (device name) |
| LOD | Limit of Detectability |
| LOF | Length of Flight |
| LPE | Laser Pulse Energy |
| MCP | Multi-Channel Plate |
| MD | Molecular Dynamics |
| OAR | Open Area Ratio |
| PME | Probability of Multiple Events |
| RBS | Rutherford Backscattering Spectroscopy |
| SEM | Scanning Electron Microscopy |
| SIMS | Secondary Ion Mass Spectroscopy |
| (S)TEM | (Scanning) Transmission electron Microscopy |
| STM | Scanning Tunneling Microscopy |
| TKD | Transmission Kikuchi Diffraction |
| TOF | Time of Flight |
| TP-APT | Terahertz-Pulsed Atom Probe Tomography |
| VP-APT | Voltage-Pulsed Atom Probe Tomography |
| ZBEF | Zero-Barrier Electric Field |

# 1. Introduction

Several decades after the introduction of laser-assisted Atom Probe Tomography (LA-APT, also abbreviated as L-APT or La-APT (Kellogg and Tsong, 1980; Cerezo et al., 1985, 1986a, 1989)) and two decades after its crucial update with femtosecond laser pulsing, this technique is today widely used for the nanoscale microscopy and microanalysis of non-metallic systems (Kelly et al., 2007; Lauhon et al., 2009; Gorman et al., 2011a; Giddings et al., 2018), including minerals (Reddy et al., 2020). It has demonstrated a unique potential for the study of the 3D distribution of atomic species in semiconductor materials and devices, and in a growing list of inorganic non-metallic solids. In all these systems, the capabilities of the techniques are being pushed to their limits, in a continuous quest on impurity and interface studied within still shrinking electronic devices (Giddings et al., 2018). Thanks to the recent development of cryo-preparation techniques, its domain of application is still growing and opening to organic and biological systems (McCarroll et al., 2020; Gault et al., 2021; Meng et al., 2022a; Schwarz et al., 2025; Woods, 2025). A crucial and often underestimated issue with APT is its accuracy in compositional measurements of non-metallic phases, including those occurring in the microstructure of metals. Despite APT can detect single ions with a yield which is independent of their mass, it is potentially affected by a set of mechanisms that can significantly – sometimes severely – limit its accuracy in the measurement of chemical composition (Lefebvre-Ulrikson et al., 2016; De Geuser and Gault, 2020).

This expanding context calls for a critical review on compositional metrology in APT applied to non-metallic materials. The timeliness for such a work stems from the experience accumulated by several research groups over two decades of analyses coupled to the new challenges represented by the opening of new domains of application.

### *a. Scope of this review*

The structure of the review (Fig. [1]) includes the following key chapters: (1) This introduction, including a historical overview and the position of APT in the context of composition measurement in microanalysis; (2) Principles of APT as an experimental technique, underlining the aspects that may intervene in compositional biases; (3) A focus on the role of the surface electric field in APT and how it governs the different physical-chemical phenomena that enable the measurement; this section also includes a review of different quantities interpreted as "surface field" and a criticism of the use of the notion of "evaporation field"; (4) A review of mechanisms of loss, i.e. what are the physical or instrumental channels which lead to an error in the measurement of composition; based on these mechanisms, the methods for assessing a compositional bias will be summarized, including possible methods for the correction of ascertained biases; (5) A review of literature data on compositional biases in APT of non-metallic materials; (6) A final chapter will contain the perspectives, specific recommendations for best practices, and the main conclusions of this work. The main advantages and limitations of APT as a microanalytical tool will be clearly set out.

This review develops the problem of compositional biases in APT on the basis of the actual knowledges in the domain of field ion evaporation. It does not exclusively focus on the experimental technique and related methods, but also delves into the fundamental physical processes on which the technique relies. It calls to an expanding readership spanning from theoretical/computational materials scientists interested in surface physics of non-metallic materials to routine APT users that face the everyday problem of the soundness of their APT data. Furthermore, it may be a reference for a broader panel of scientists facing the problem of measuring composition in nanoscale systems and of selecting the most adapted technique and methods.

This article focuses on the application of APT to the measurement of composition of non-metallic inorganic systems. Among these, the most represented are semiconductors and other dielectrics such as oxides, nitrides and carbides. Compositional biases in non-metallic minerals share many common points with the aforementioned systems, along with specific problems related to the variety of impurity distributions. For specific cases concerning minerals, the reader is referred to the comprehensive review by Cappelli et al. (Cappelli et al., 2021). Organic systems are discussed in perspective.

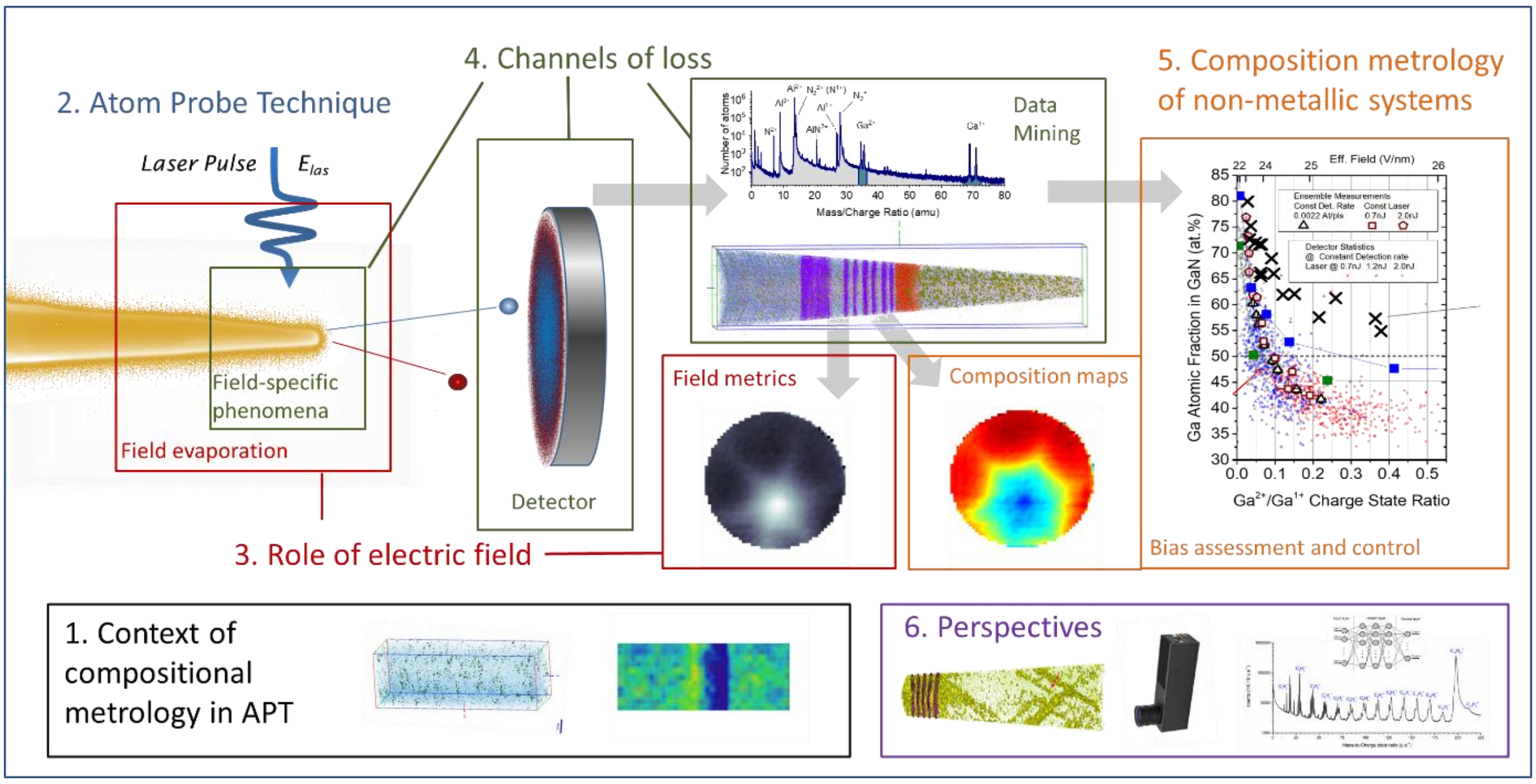

*Figure [1]. The structure of this review at a glance.*

### b. Relevant metrological concepts

Within the above-described scope, we remind here the main metrological concepts and quantities that will be dealt with. The spatial metrology issues and concepts related to the technique have been extensively reviewed in the past (D. J. Larson et al., 2013; Vurpillot and Oberdorfer, 2015; De Geuser and Gault, 2020; Gault et al., 2026) and will be only rapidly reminded here.

*Measurement error*. The error is meant as the measured quantity value minus a reference quantity value. Note that the measurement error may be known or not, depending on the existence of the reference quantity. In APT the reference quantity for the composition is often not known, as APT is in many cases the only technique that can determine it.

*Accuracy.* The accuracy is defined as the closeness of agreement between a measured quantity value and a true quantity value of a measurand. The term 'measurement accuracy' is rather a concept than a quantity and is not given a numerical quantity value. A measurement is said to be more accurate when it offers a smaller measurement error. In composition measurements by APT, the accuracy should thus be meant as the closeness of agreement between the composition measured by APT within a specified region of a dataset and the reference value of the composition, determined either as a standard value or by an independent technique.

*Bias and systematic error.* The systematic error is the component of measurement error that in replicate measurements remains constant or varies in a predictable manner. A bias is the estimate of a systematic measurement error. In APT, these terms are often used as synonyms. A compositional bias refers thus to a systematic error in the measurement of composition.

*Precision.* The precision is the closeness of agreement between indications or measured quantity values obtained by replicate measurements on the same or similar objects under specified conditions. In compositional measurements by APT, the precision can thus refer (i) to repeated analyses performed on similar objects (specimens with similar shape and similar composition) under the same conditions (i.e. within the same instrument and with the same set of experimental parameters: base temperature, laser intensity per pulse, DC voltage, etc.) or (ii) to repeated analyses extracted from different subvolumes of a given dataset from a homogeneous sample, acquired under uniform experimental conditions. In APT, the *compositional precision* is primarily related to the counting error which is introduced by the limitations in the detection efficiency, while the *spatial precision* – often confused with the resolution – can be considered as the closeness of agreement between the positioning of an atom after reconstruction and its effective position within the material (within the lattice if the material is crystalline).

*Spatial resolution.* Spatial resolution is a key quantity in microscopy, meaning the smallest feature (or separation between features) that can be distinguished and measured with specified uncertainty under defined conditions. This quantity has different specific origins in different microscopy techniques. While in optical (photonic or electronic) microscopy it is related to the full width at half maximum (FWHM) of the illuminating beam point spread function (PSF), its definition in APT depends on both the instrument and the feature being quantified (lattice plane, defect, cluster, local order/disorder, etc.). The related term *resolvable feature size* (Gault et al., 2021) is in APT a task-specific measurand: it is the smallest size/spacing for which a chosen measurement procedure (APT experiment + reconstruction + analysis) yields estimates of a feature's properties (e.g., diameter, composition, concentration profile) within acceptable error bounds. This is conceptually analogous to ISO's PSF-based resolution, for APT the equivalent "PSF" is the broadening of reconstructed features and depends on material, microstructure and analysis workflow, so one must report resolution in relation to the feature class (interfaces, clusters, particles) rather than as a universal instrument constant (De Geuser and Gault, 2020; Gault et al., 2026).

*Limit of detection*. The limit of detection (LOD) is the measured quantity value, obtained by a given measurement procedure, for which the probability of falsely claiming the absence of a component in a material is β, given a probability α of falsely claiming its presence (IUPAC recommends $\alpha = \beta = 0.05$). The term "sensitivity" or "analytical sensitivity" is often used as a synonym of the LOD (Reddy et al., 2020; Kelly et al., 2022), but this use is discouraged by ISO, as sensitivity is rather the quotient of the change in measurement indication and the corresponding change in value of a quantity being measured. Another used synonym for this term in radiometry and in APT is the term "detectability" (Currie, 1968; Gault et al., 2012b). In APT, the LOD is related to the ability to detect a given atomic species, requiring both identification of a peak in the mass spectrum and accurate counting of ions. It is first limited by counting statistics, which also include the effect of detector efficiency: for instance, measuring 200 ppm with 50 ppm precision and 100% detection efficiency requires more than $10^5$ detected atoms, which becomes problematic for very dilute species. It is also limited by mass resolution and by the signal-to-background ratio, since a peak must rise above background noise to be distinguished (Currie, 1968). Because mass resolution, background level, and other experimental factors vary between datasets, the limit of detection in APT depends in a complex manner on all these factors and cannot be indicated in a close, simple form.

The main quantities that usually define composition as measured by APT are the following:

*Atomic fraction*. In APT, the atomic fraction $X_i$ of a component *i* is given by the expression

$$X_i = \frac{n_i}{\sum_j n_j} = \frac{n_i}{N_{tot}} \qquad (1)$$

Where $n_i$ is the number of ions, atoms or detected events of the component *i*, while *j* runs over all the components defined for the given analysis, so that the sum in the denominator eventually yields the total number of atoms $N_{tot}$ defined as identifiable signal (i.e. excluding the noise) in the set of detected events. The atomic fraction can be expressed as a percentage, and is sometimes confused with the term *concentration*. Notice that the component *i* can be an element, a charge state of a given element appearing in the mass/charge histogram, an isotope of an element (possibly with a given charge state), a molecular ion species.

*Site fraction*. The site fraction (or *sublattice occupancy*) of element $i$ on a given sublattice (which should be explicitly specified) is

$$x_i = \frac{N_i}{N_{\text{sites}}} \qquad (2)$$

where $N_i$ is the number of atoms of element $i$ on those sites and $N_{\text{sites}}$ is the total number of crystallographic sites in that sublattice. In the context of APT and materials analysis, *site fraction* refers to the fraction of crystallographic sites (or lattice positions) in a crystal structure that are occupied by atoms of a specific element or type, relative to the total number of available sites of that sublattice. This is a dimensionless quantity distinct from atomic fraction, as it considers the structural arrangement of atoms within a crystal. Its use is common in the analysis of oxides or semiconductor compounds in which the lattice structure is known, as for instance in $Al_xGa_{1-x}N$, where $x$ is the III-site fraction of Al within the AlGaN lattice (notice that in this example the V-sites are all occupied by N).

*Isotopic abundance*. The isotopic abundance is the proportion of a specific isotope of an element relative to the total amount of that element in a sample. Denoted as $a(i)$ for isotope $i$, it is given by

$$a(i) = \frac{n(i)}{\sum_k n(k)} \qquad (3)$$

where $n(i)$ is the number of detected ions of isotope $i$ and the sum is over all isotopes $k$ of the element. In APT, it can be defined for a given charge state and compared with the natural isotopic abundance, which can be useful for peak decomposition (i.e. identifying contributions from different elements within a given mass/charge peak, for instance $^{16}O_2^+$ and $^{64}Zn^{2+}$ both occurring at 32 Da) or for assessing specific losses related to detector limitations (Gopon et al., 2022; Ndiaye et al., 2024).

*Relative abundance or relative fraction.* This quantity is the equivalent of a fraction of a component *i* calculated over a specific subset of events and not on the total number of events. It can be useful in order to track spatial dependences of species in the detector or reconstructed space. For instance, the distribution of the relative abundances of the hydrogen species $H^+$, $H_2^+$, $H_3^+$ over the sum of these three components can be tracked in order to disclose their correlation with the equivalent electric field (Rigutti et al., 2021).

*Volume concentration* (or simply, *concentration*). It refers to the amount-of-substance concentration

$$C_i = \frac{n_i}{V} \qquad (4)$$

where $n_i$ is the amount of substance of component *i* and *V* is the volume in which this amount is measured, with SI units mol/m³ or, far more frequently in semiconductor science, $cm^{-3}$. In APT, volume concentration of solute *i* should consider the limited detector efficiency $DE_{MCP}$ (~50-80%, section 4.a), which is applied as a correction factor. Furthermore, the volume V is determined upon reconstruction, with all cautions that this operation requires and with all associated uncertainties.

For other metrologically relevant quantities the reader may refer to online resources ("ISO/IEC Guide 99:2007(en), International vocabulary of metrology — Basic and general concepts and associated terms (VIM)," n.d.). Last but not least, the Charge State Ratio, a paramount quantity for APT field metrology, will be introduced and discussed in section 3.e.

### c. *APT vs other microanalysis techniques*

APT may be compared with similar microanalysis techniques such as (Scanning) Transmission Electron Microscopy ((S)TEM) including Energy Dispersive X-ray Spectroscopy (EDS) and Electron Energy Loss Spectroscopy (EELS), as well as with Secondary Ion Mass Spectrometry ((Nano-)SIMS). Most APT users are well acquainted with SEM based techniques (EDS, Electron Backscattering Scanning Diffraction – EBSD, Transmission Kikuchi Diffraction – TKD, Electron Channeling Contrast Imaging - ECCI) as they are currently applied to APT sample preparation protocols (Saxey et al., 2007; Blum et al., 2016a; Breen et al., 2017; Yang et al., 2025). These techniques are based on different physical principles, instrumental implementation and detection schemes. APT delivers 3D maps of positions and chemical identity of individual atoms with sub-nanometer precision. APT may thus resolve features as small as 1 nm—while simultaneously achieving limits of detection down to parts-per-million (ppm) levels. This combination allows APT to analyze nanoscale features like precipitates, grain boundaries, and segregation layers with quantitative atomic precision, even for light elements such as hydrogen, carbon, and lithium (Gault et al., 2021; Lawitzki et al., 2021).

APT's ability to directly quantify atomic fractions in small volumes—with well-defined statistical uncertainties—makes it uniquely powerful for studying nanoscale phase separation, clustering, and solute distribution in advanced materials. APT also offers extremely high detection limits and isotopic sensitivity, achieving single-ion detection with 50–80% efficiency. It quantifies trace elements (ppm levels), including light elements like hydrogen, carbon, and lithium, which are challenging for EDS or EELS. Time-of-flight mass spectrometry further distinguishes isotopes, supporting tracer and isotope-enriched studies. Finally, APT provides direct number-fraction quantification in small volumes, converting counted ions into atomic fractions with clear statistical uncertainties. This is especially useful for analyzing solutes in nanoscale features like precipitates, Guinier-Preston zones, or segregation at defects, often yielding lower compositional standard deviations than (S)TEM-EDS for comparable volumes.

Table [I] Comparison of microanalysis techniques by some of their typical metrological figures of merit.

| **Technique** | **Resolvable Feature Size** | **Limit of detection** | **Notes** |
|---|---|---|---|
| SEM (EDS, EBSD, TKD, ECCI) | 10 nm – 1 µm | 1% - 0.1% | Surface and bulk analysis; limited sensitivity for trace elements. |
| (S)TEM (EDS, EELS) | 0.1 nm – 10 nm | 1%- 0.1% | High spatial resolution; limited to electron-transparent samples. |
| (Nano-)SIMS | 10 nm – 100 nm | 1 ppm - 1 ppb | Excellent isotopic sensitivity; lower spatial resolution compared to APT. |
| APT | 1 nm – 10 nm | 0.1% - 1 ppm | Near-atomic resolution; quantifies trace elements and isotopes in nanoscale volumes. |

Techniques like (Scanning) Transmission Electron Microscopy ((S)TEM) and Secondary Ion Mass Spectrometry ((Nano-)SIMS) can be considered as complementary to APT (Eswara et al., 2019).

*APT vs (S)TEM-EDS/EELS*. APT delivers 3D atom-by-atom composition mapping, while STEM-EDS/EELS offer 2D nanoscale projections with thickness-integrated signals and potential delocalization, though TEM excels at linking composition with crystallography, defects, strain and phase contrast—areas where APT provides only indirect insights. For quantification, STEM-EDS/EELS face challenges like X-ray absorption, fluorescence, and thickness effects, requiring standards or modeling for accuracy, especially in complex systems (Nilsen and van Helvoort, 2022). APT avoids absorption/fluorescence issues but introduces a number of artifacts (see parts 3 and 4). However, APT uniquely detects and maps dilute solutes and light elements at trace levels within tiny volumes, a task difficult for EDS (De Geuser and Gault, 2020; Lawitzki et al., 2021). In terms of materials systems, STEM-based techniques can address 2D materials, a domain still inaccessible to APT.

*APT vs SIMS*. These techniques differ significantly in resolution, quantification, and application scope. APT delivers sub-nanometer 3D spatial resolution, but only for small, geometrically constrained volumes, while SIMS excels in depth profiling over micrometer-thick films, though with depth resolution of a few nanometers and lateral resolution typically exceeding 50–100 nm (Eswara et al., 2019). For quantification, SIMS is highly matrix- and analyst protocol-dependent, requiring matched standards for accurate concentration measurements. APT, while less affected by classical matrix effects, faces material-specific biases due to field evaporation, often needing empirical corrections or cross-validation for reliable stoichiometry, especially in multi-component systems. The detection limit of APT can be as good as ppm-level, which can be outperformed by SIMS (down to ppb level) as it covers larger surface areas and can thus rely on larger statistics (McPhail, 2006).

Finally, compositional measurements by APT can be compared to and completed with those from other techniques, such as Rutherford Backscattering (RBS), nanoscale and standard X-ray diffraction (XRD), laser-induced breakdown spectroscopy (LIBS), optical, capacitive or thermal spectroscopies.

# 2. Principles and methods of APT

In this part the basic principles and methods in APT will be introduced, with an accent on the subjects relevant to composition metrology.

### *a. Field Evaporation*

The basic theories of field evaporation describe the escape process as thermodynamically determined, thermally activated, with only one escape charge-state involved (Müller, 1941, 1956; Gomer, 1959). The emission and detection rates (measured in ions/s, over the surface of analysis) are then assumed to be proportional to the evaporation rate-constant $K_{EV}$ (measured in $s^{-1}$, on a single atomic site of emission) given by an Arrhenius-type equation:

$$K_{EV} = K_0 exp\left(-\frac{Q(F)}{k_B T}\right) \qquad (5)$$

where $k_B$ is the usual Boltzmann factor, $Q$ is the activation energy for field evaporation, and $K_0$ is the field-evaporation pre-exponential that is also $K_{EV}$ for $Q$=0. Note that experimental measurements validate this expression over large range of evaporation rates (Ernst, 1979; Kellogg, 1984). In the simplest approximation, pre-factors are taken as slowly varying with T and F[1], so that the control of the evaporation rate is made by manipulating the applied voltage (and hence F the surface microscopic field in the electrostatic approximation) and/or the temperature, T of the specimen. Later theories assumed that the atom might initially escape into a charge-state higher than 1, and might then be subject to field-induced post-ionization (see section 3.d). The activation energy, $Q$, is a microscopic energetic parameter that integrates the energy required by an atom to be liberated, expelled, and detected as an ion. In the early ages of APT analysis, it was found that $Q$ was extremely sensitive to the applied microscopic field.

[1] The validity of a constant pre-factor can be debated significantly, since it depends on the variability of the jump frequency over the barrier that is dependant of the strength of the binding interaction of the atom with surface, and of the statistics of vibration frequency that are temperature dependant. Experimentally, Kellogg measured variations of several orders of magnitude for tungsten atoms as a function of the surface field.

For metals, a variation of few percent in the field can change the evaporation rate constant by several orders of magnitude (Tsong, 1971). It was also found that each element has its own magnitude of field required to cause significant field evaporation from the surface. Empirically, $Q$ has been described via two simple properties: the electrostatic local field $F_{EV}$ for which the barrier vanishes or zero barrier evaporation field ZBEF and one (or two) parameter(s) (described below) describing the "field-sensitivity" of $Q$, i.e., how $Q$ depends empirically on F (Müller, 1941, 1956; Tsong, 1978a; Tsong and Müller, 1969; Forbes, 1995; Forbes et al., 1982; Brandon, 1966; D. J. Larson et al., 2013; Miller and Forbes, 2014a). The electrostatic field $F$ (microscopic) applied experimentally is normally always slightly lower than $F_{EV}$.

$F_{EV}$ is one of the most fundamental parameters influencing data analysis. For a given element, some $F_{EV}$ tables are given in literature (Tsong, 1978b) based on either experimental observations or some analytical models. The accuracy of experimental values is widely debated, and is generally reported to be about ±15%. The table is limited to a selection of pure materials, because measurement was made mostly in field ion microscopy mode on pure metals for an evaporation rate value not equal to $K_0$ (much lower). Analytical models gave also values that must be taken carefully due to the roughness of approximations made to calculate them. These tables should therefore be taken with caution. In particular for non-metallic compounds, it is not generally true that the evaporation field is defined for a given species: evaporation fields and activation energy barriers depend not only on the element, but also on the type of bond it builds with its nearest neighbors (Yamaguchi et al., 2009a; Peralta et al., 2013).

The energy barrier is also influenced by the field sensitivity, $S_T$, of the evaporation rate close to the ZBEF. As defined by Brandon (Brandon, 1966), this sensitivity can be defined at the first order mathematically by:

$$S_T = \frac{d\ln(K_{EV})}{d\ln(F)} = -\frac{F}{k_B T}\left(\frac{dQ}{dF}\right)_T \qquad (6)$$

Since AP works close to $F_{EV}$, we can define the quantity $C = \frac{dQ}{d(F/F_{EV})_T}$, so that $C \sim k_B T S_T$ which is an energetic parameter (written in eV) expressing the sensitivity to the field of the field evaporation for a given temperature. Assuming the sensitivity almost constant for a sufficient range of field, a simple assumption takes the activation energy barrier for the evaporation of the considered atom to be given adequately by the linear approximation:

$$Q(F) = C\left(1 - \frac{F}{F_{EV}}\right) \qquad (7)$$

Values of $C$ are experimentally in the range [0.5-3] eV with operating fields close to $F_{EV}$ (Ernst, 1979; Kellogg, 1984; Ashton et al., 2020a; Rousseau et al., 2023; Vurpillot et al., 2024). The higher the $C$ value, the more abrupt is the variation of the evaporation rate with the field. C and $F_{EV}$ are generally higher for materials of high sublimation energy, such as refractory metals (W, Ir, Pt, etc…). The higher these values, the more deterministic the evaporation process becomes. Non-linearities of $Q$ linked to the complexity of the field evaporation mechanism obviously exist (Waugh et al., 1976; Forbes et al., 1982; Wam/z , 1984; Miller and Forbes, 2014a; Ashton et al., 2020a; Vurpillot et al., 2024). However, when they can be considered a small effect in the regime used experimentally (typically $F_{op}$ in the range 0.8 to 0.95 $F_{EV}$).

### b. *Voltage vs Laser Pulsing*

In APT a high voltage $V_{DC}$ (3 ÷ 14 kV) is applied between the specimen and the detection system in order to generate a high electric field. The latter is settled just under the evaporation field $F_{ev}$. During the laser-assisted field evaporation of a tip, positive charged ions $A^{n+}$ are generated and accelerated towards a detector. The initial energy of an ion emitted from a tip surface is equal to the electric potential energy $neV_{DC}$, where $ne$ is the ion electric charge ($e = 1.6 \times 10^{-19}$ C, $n$ is an integer) and $V_{DC}$ *is* the DC voltage applied to the tip. As an ion reaches the detector, its final energy corresponds to the kinetic energy $\frac{1}{2}mv^2$, where $m$ is the ion mass and v = $L/t$ its velocity. $L$ represents the flight path, which as a first approximation corresponds to the distance between the tip and the PSD and $t$ is the time of flight (TOF) interval of the ion (Müller et al., 1968). The mass/charge ratio $m/n$ of ions is derived from the conservation of energy $\frac{1}{2}mv^2 = ne\,V_{DC}$:

$$\frac{m}{n} = 2eV_{DC}\left(\frac{t}{L}\right)^2. \qquad (8)$$

where $t$ is the TOF of ions ($t = L/v$).

Therefore eq. (8) can be used to transform a TOF spectrum into a mass spectrum, where the different ionic species can be recognized. For each ion detected both the impact position on the PSD and the TOF are recorded. In order to measure a TOF, the evaporation should be triggered in time. This occurs by electric or laser pulses, depending on the material which composes the specimen and the atom probe design. Typically, sub-nanosecond pulses are used for conductive materials (i.e. metals). Their effect is lowering the activation barrier $Q$ through the modulation of the surface field $F$ (eq. (Fev4)). Voltage pulsing has been applied to 1D-AP or APT of semiconductors before the introduction of femtosecond lasers, but these analyses are rare, and mostly applied to sufficiently conductive systems (Krishnaswamy et al., 1981). In fact, nanosecond voltage pulses cannot be properly transmitted along a poorly conductive tip. Semiconductors can be analyzed by voltage pulsing, which can be achieved through a conductive coating of the specimen as a final step of preparation (Adineh et al., 2017).

For non-conductive materials, such as semiconductors, oxides and polymers, current practice is using femtosecond laser pulses, which rapidly increase the temperature T of the specimen (eq. (Fev3)) (Gault et al., 2006; Bunton et al., 2007; Cerezo et al., 2007). The mechanisms of laser absorption and energy relaxation are described in several works (Vella, 2013; Kelly et al., 2014; Vella and Houard, 2016; Vella et al., 2018). Significantly, the presence of a strong static field can enhance the optical absorption in dielectrics through the modification of the band structure and the accumulation of positive charges (holes) at the surface (Silaeva et al., 2014). Typical temperature increases are of the order of several tens to one to two hundred K, and have been estimated by exploiting the modifications induced by laser pulses to photoluminescence signals from oxide specimens (Gautam et al., 2025). One of the main drawbacks of laser pulses is that heat relaxation occurs on the timescale of several tens of ns, which translates into the appearance of transients of evaporation probability after the pulse, i.e. into an asymmetric shape of mass/charge peaks (thermal tails on the high mass/charge side of a peak). This phenomenon leads to peak superposition, which generally complicates the analysis of mass/charge spectra and therefore composition assessment (see section 4.b). In order to improve field/temperature for minimal background noise while maintaining evaporation rate, a hybrid pulsing (voltage and laser) has been recently proposed (Larson et al., 2023a).

### c. *Detector in Atom Probe Tomography (APT): Principles, Metrological Limitations, and Performance*

The detector used in Atom Probe Tomography (APT) is arguably the most critical component of the instrument, as many of the technique's metrological limitations are inherently tied to its operation. The final 3D point atom cloud - comprising the reconstructed 3D positions and elemental identities of atoms - is generated by identifying the (X, Y) position and TOF of ion impacts on a time- and position-sensitive detector placed in front of the specimen. Field evaporation of ions, triggered by laser or voltage pulses at repetition rates ranging from 25 to 1000 kHz, produces ions that are accelerated by an applied voltage of 1–20 kV toward the detector (see Fig. [2]). To ensure a detection rate dominated by single-hit events, the evaporation rate is carefully controlled to less than one atom per pulse (typically one evaporation event per 100 pulses, or a detection rate of ~1%). This rate can be practically adjusted from 0.01% to ~10%, with a typical detection throughput of $10^2$–$10^4$ hits per second.

While most events are single hits, a significant fraction involves multi-hit events, defined as the detection of two or more impacts within the detection window following a trigger pulse. Multi-hits can be sequential (with a measurable delay between impacts) or simultaneous (same arrival time but different spatial positions). An ideal detector would capture all events to prevent data loss. However, the physical mechanisms of ion emission often result in correlated, non-random ion emission, leading to frequent species-dependent multi-hits that are closely spaced in time and position. If the detector fails to resolve these events accurately, they may introduce biases into the measurement. A large recent literature on these processes may be found (Meisenkothen et al., 2015; Schiester et al., 2024; Jakob and Thuvander, 2024;Di Russo et al., 2020; De Geuser et al., 2007; Thuvander et al., 2019, 2013; m/z Costa et al., 2012; Peng et al., 2018; Morris et al., 2022; Di Russo et al., 2018a; Cuduvally et al., 2020; Kitaguchi et al., 2014; Larson et al., 2018).

### *i. Detection Principle: Ion-to-Electron Conversion and Amplification*

Detection in APT relies on ion-to-electron conversion using a stack of two microchannel plates (MCPs). Each MCP consists of an array of glass capillaries (or "channels") with high secondary electron emission coefficients, arranged in a hexagonal lattice. A high DC voltage applied across the MCP generates an electron avalanche upon ion impact, providing high gain ($10^4$–$10^6$ electrons per hit for ions in the keV range with masses between 1 and 1000 m/z for a single MCP). MCPs exhibit low intrinsic background noise, typically ~1 count/$cm^2$/s across the detector (Guest, A.J., 1971; Woodhead and Ward, 1977).

The channels, with diameters of 10–20 µm and lengths under 1 mm, contribute to an intrinsic sub-nanometer positioning resolution due to the instrument's magnification ($10^5$–$10^6$). Using two MCPs in a chevron configuration (channels tilted by ~20° to ensure uniform gain) achieves a total gain of $10^7$–$10^8$ electrons per hit. The open area ratio (OAR) of conventional MCPs is ~60%, though recent improvements (e.g., funneled channel inputs) have increased this to 80% (Fehre et al., 2018). Detection efficiency decreases significantly for high-mass elements at low energies (<2–3 keV) (Fraser, 2002). For example, at 1 keV, the detection efficiency for argon ($Ar^+$) and xenon ($Xe^+$) drops by over 30% and 60%, respectively, compared to 5 keV. Special attention is required for large, slow-moving molecules (mass > 1000 Da), as their detection efficiency is severely reduced.

The transit time of the electron burst is less than 1 ns, with a transit time spread of a fraction of a nanosecond. The resulting electronic signals are brief (~ns duration) and detectable (1–100 mV) with standard electronics, exhibiting low sub-nanosecond jitter. MCPs are thus well-suited for ion-to-electron conversion in APT, where TOF typically ranges from 100 to 1000 ns, ensuring precise timing measurements.

### *ii. Event Encoding: Spatio-Temporal Localization*

To achieve optimal metrological performance, the electron burst must be localized in time and space with high precision. This is accomplished by the event-encoding readout system positioned behind the MCPs. Early APT systems experimented with various readout methods, such as Charge-division readouts (e.g., wedge-and-strip detectors in the PoSAP atom probe (Cerezo et al., 1989), 10×10 multianodes in the TAP (Bostel et al., 1989)) or Optical detectors (electron-to-photon conversion via a phosphor screen, as in the oPoSAP (Cerezo et al., 1996) and OTAP (Deconihout et al., 1998)).

Nevertheless, for the past two decades, crossed delay lines (DLDs) have been the standard in modern APT instruments (see Fig. [2]-(c) (Jagutzki et al., 2002a; Kelly et al., 2003; m/z Costa et al., 2005; Monajem et al., 2025). In a DLD, the electron burst is intercepted by two perpendicular propagation lines (one for each axis, X and Y). The generated electronic signals propagate to both ends of each line, and their arrival times (TU, TD, TR, and TL) are used to reconstruct the impact's position and timing. Only four preamplifiers and time-to-digital converters are required, simplifying operation. The electronic pulse from the electron burst has a full-width-at-half-maximum of 3–4 ns. The sum of the propagation times measured at both ends of a delay line (relative to the start signal) equals the line's total propagation time (TL), which ranges from 25 to 100 ns depending on detector size and geometry (e.g., ~0.5 ns/mm for serpentine delay lines on an 80 mm MCP stage, ~1 ns/mm for helical bifilar delay lines). The difference in arrival times determines the (X, Y) position, while the arrival times (corrected for TL) provides sub-100 ps timing accuracy. Sequential multi-hits separated by more than TL are detected unambiguously. Adding a third, tilted delay line or measuring the MCP arrival signal can resolve nearly simultaneous particles if they are sufficiently separated. The sequential multihit decomposition is then limited by the signal separation in time. An intrinsic dead time of ~4 ns (the physical width of the electronic signal) creates a minimum spatiotemporal dead zone of ~8×8 mm² × 4 ns. Advanced systems using high-frequency digital sampling (>1 GS/s) on all four DLD ends can reduce this dead time to 1.5–2 ns, shrinking the dead zone to ~2×2 mm² × 2 ns (Bacchi et al., 2022)(ref.). Nonetheless, multi-hits detected inside this zone are registered as a single event with incorrect position and timing (see section 4.a.ii).

### *iii. Event Encoding: Performance Trade-offs: MCP-DLD Combination*

To conclude, the MCP-DLD system strikes an optimal balance for APT applications, offering:

- High throughput (up to ~50 kHits/s without loss),
- Low intrinsic noise (<0.1% in the best cases),
- High detection efficiency (up to 80%),
- High spatial precision (<50 µm),
- High timing accuracy (<100 ps).

Despite these advantages, limitations persist, particularly for correlated multi-hits and dead zones, which may affect the accuracy of 3D reconstructions, especially for complex materials producing a high amount of correlated evaporation. For this reason, research on detector technology is ongoing and the future developments will be shortly discussed in the perspective section 6.a.i.

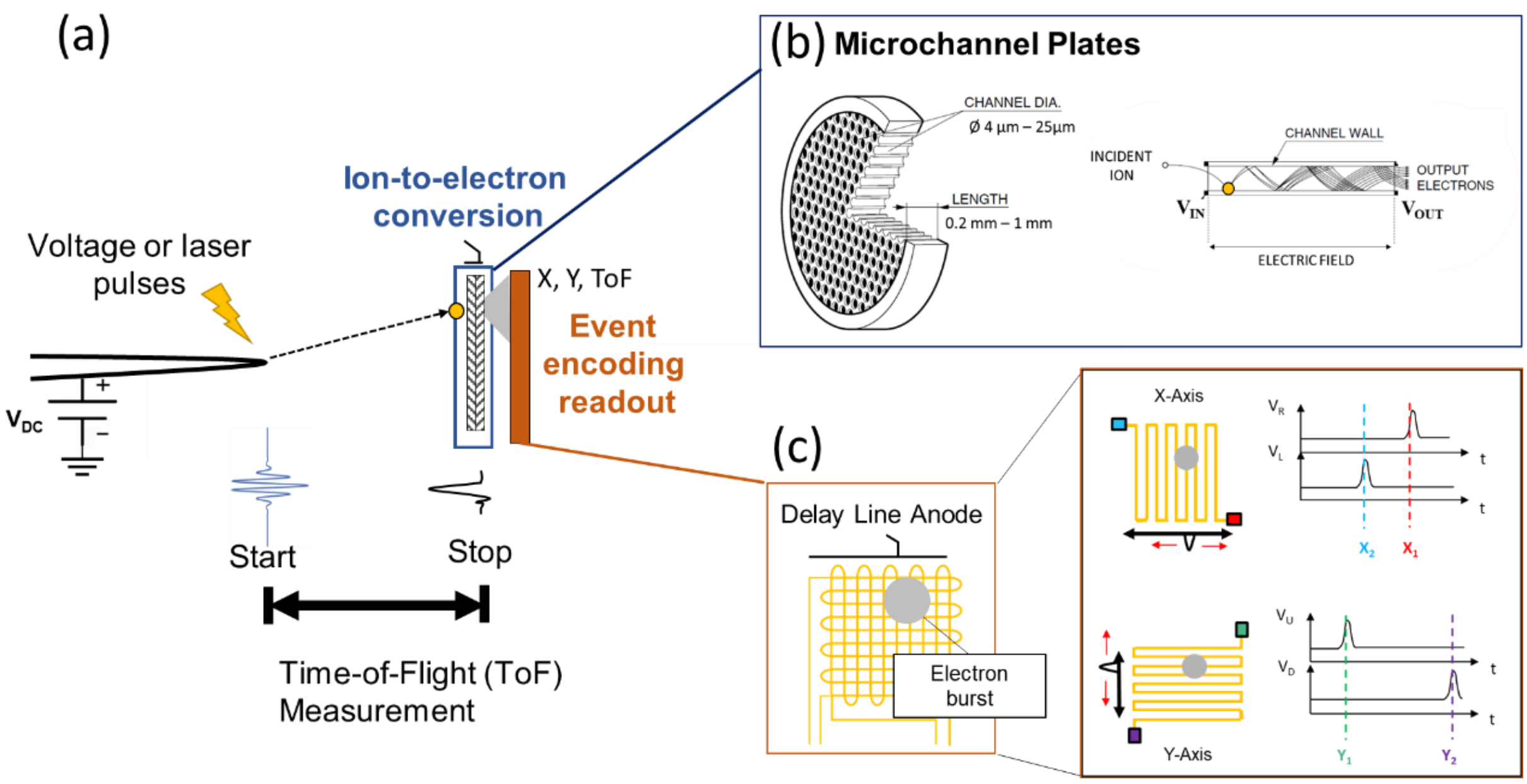

*Figure [2]. (a) Schematic illustration of the APT detection process. The 3D reconstruction of atomic positions and elemental identities relies on measuring both the (X, Y) impact position and the time-of-flight (TOF) of ions using a time- and position-sensitive detector (b) The detector's ion-to-electron conversion system consists of a stack of two microchannel plates (MCPs) arranged in a chevron configuration, with diameters ranging from 40 to 120 mm depending on the atom probe's geometry. Each MCP is biased at ~1 kV, generating an electron avalanche within its tubular channels upon ion impact. This avalanche produces an output charge of $10^3$–$10^6$ electrons, which is directed toward the readout system. The resulting electron burst (~1 ns duration) encodes the event's spatial and temporal information (Source: Hamamatsu) (c) In modern APT instruments, a crossed delay-line detector (DLD) decodes the (X, Y, TOF) coordinates by analyzing electronic pulse signals (VR, VL, VU, VD) collected at the ends of serpentine or helical propagation lines along the X and Y axes. (Image courtesy of C. Bacchi).*

### *d. Mass spectrometry*

Since the development of 1D APs, time-of-flight mass spectrometry based on eq. (8) has been the basis of element-specific detection (Müller et al., 1968). The simple assumptions and basic equations established at that time are still routinely used in modern APs to determine the chemical nature of field evaporated atoms. Both composition measurements and three-dimensional reconstructions rely on mass spectrum quality and processing. The quality of the mass spectrum depends not only on the physics of field evaporation but also on AP design (flight length) and hardware devices eventually used to reduce energy deficits (reflectrons). It also depends on the operator and analysis conditions including voltage, detection rate, sample temperature, pulse fraction, or laser energy. The final result is also related to post-processing carried out by the operator, including calibration, mass resolution optimization, background noise subtraction, peak identification, mass range definition, and peak deconvolutions. These issues have been reported or reviewed in detail in several works (Gault et al., 2012b; Pareige et al., 2016).

In APT, the TOF contribution to composition accuracy is governed by any effect that disrupts the ideal eq. (8) or distorts peak shapes and areas, since composition is derived from integrated peak intensities. Poor calibration or calibration drift can shift peaks and alter ranging of overlapping species, while dispersion in evaporation timing, especially in laser-assisted APT, broadens peaks and increases overlap ambiguity. Variations in flight path and geometry, as well as energy spread arising from field variations, molecular dissociation, or post-ionization, further broaden or skew peaks, complicating integration and masking minor species. Detector timing resolution, jitter, dead-time, saturation, and multi-hit limitations reduce mass resolving power and introduce systematic undercounting, often in a composition-dependent manner during burst or correlated evaporation events. Finally, non-ideal peak shapes, background contributions, imperfect deconvolution, and limits in mass range and dynamic range can bias peak integration, leading directly to errors in quantified composition, particularly for low-abundance elements or closely spaced mass/charge ratios (Pareige et al., 2016). Some of these aspects will be further developed in section 4.b.

### *e. 3D reconstruction*

On modern APT instruments, different geometrical configurations exist. Straight flight path instruments integrate a detector of diameter D at a distance LOF (length of flight distance). D is limited in current technology by MCP diameters (~40 mm, 80 mm or 120 mm, with 80 mm being the most common tradeoff). LOF between the specimen and the detector is in the range 80 mm–400 mm for straight flight path instruments. Some instruments integrate an electrostatic device that enables to collect a larger field of view (FOV) on the specimen. This is the case for reflectron-fitted instrument (Panayi et al., 2006; Tegg et al., 2023; Heller et al., 2024), or more recently with instrument incorporating electrostatic lens. Here, since the projection features, specific to the geometrical and electrostatic parameters of the instrument are taken into account by manufacturers in the reconstruction software, we will focus on the reconstruction process for the former and simple case of the straight flight path atom probe.

The 3D point cloud is built point-by-point by assuming a simple reverse-projection from the detector impact position onto a 'virtual' emitting surface located at the end of an APT specimen. With a specimen radius below 100 nm the projected image of the surface on the detector has a magnification that is in the range of $10^6$. This is the consequence of the highly curved nature of the specimen surface. Ions initially have a near-radially flight path. The trajectory progressively curves towards the detector forming a highly compressed imaged on the surface on the detector. The trajectories of ions are fully determined by the distribution of the electrostatic field, making them independent of the ion's charge, mass, and applied voltage. Final trajectories of ions are very close of a point emitting source of constant position.

The seminal paper of Bas et al. (Bas et al., 1995) based on the recipes from a previous work by Hyde et al. (Hyde et al., 1994) proposed a simple method to 'de-magnify' the detector coordinates ($X_D$,$Y_D$) into the real-space location of each ion on the specimen surface (x,y,z). We may note that the projection relationship used to de-magnify ignores completely the real complexities of the exact trajectories to establish a bijective relationship between a point on the emitting surface and a point on the detector. Some other projections models have been proposed, with the most commonly implemented is a quasi-stereographic point-projection (Wilkes et al., 1974; Smith and Walls, 1978; Cerezo et al., 1999; De Geuser and Gault, 2017). To make simple, the projection is close to a simple angular projection with a compression factor typically between 1.4 and 2 which depends on the specimen geometry, i.e. radius of curvature and shank angle (Hyde et al., 1994; Geiser et al., 2009; B. Gault et al., 2011; Loi et al., 2013). Hence, it varies over the duration of the analysis (Baptiste Gault et al., 2011; Hatzoglou et al., 2023a).

Using the hypothesis of full ionization of surface atom, average and constant detection efficiency, and assuming a conservation of the evaporated volume, an estimation of the depth of analysis is easily produced. The result is a list of 3D coordinates for each ion detected and labelled as an elemental species. This is the basic reconstructed dataset. The accuracy of positioning is linked to simple geometric hypothesis that are used by the reconstruction process. The tip surface is assumed to be mostly spherical, with an estimation of the radius of curvature $R$ (radius of the sphere) that can come from external measurements (e.g. from an electron microscopy image, SEM or TEM), or deduced from the applied voltage V. Indeed, each material is known to have its own evaporation field $F_{EV}$ and electrostatic impose $F=V/k_F R$, with k a field factor (slowly varying with $R$, see equation (10)). Assuming a specimen evaporating at $F_{EV}$, $R$ is simply defined by $R=V/kF_{EV.}$ (e.g. from an electron microscopy image (Geiser et al., 2009)). Despite its simplicity, this method leads to impressive results, allowing for the identification of nanometric or even atomic scale details in the analysis of many materials. Some distortions of the image exist, but do not affect the compositional accuracy of the instrument. Further information on reconstruction methods can be found in dedicated book chapters and references therein (D. J. Larson et al., 2013; Miller and Forbes, 2014b; Vurpillot, 2016).

### f. *Correlation tables and (Saxey) diagrams*

Additional insights into evaporation processes can be obtained by analyzing frequency correlations in multiple detection events. Two data analysis methods are introduced to visualize and interpret these correlations, both introduced in the APT community mainly by D. Saxey (Saxey, 2011). Both of them can be applied for an advanced interpretation of mass spectra and may provide useful information for the assessment of possible compositional errors.

The first, known as the correlation table, helps determine whether the simultaneous detection of two ions in an event occurs more frequently than would be expected under independent evaporation. This involves evaluating all possible ion pairs within each event. For instance, if ions A, B, and C are detected, the resulting pairs are AB, BC, and AC. The strength of the correlation between ions i and j is measured using the significance value $d_{ij}$, given by:

$$d_{ij} = \frac{p_{ij} - e_{ij}}{\sqrt{e_{ij}}} \qquad (9)$$

Here, $p_{ij}$ is the observed number of *ij* pairs in the dataset, and $e_{ij}$ is the expected number of such pairs if no correlation exists, based on the frequency of each ion type. It's important to note that $d_{ij}$ doesn't directly quantify correlation strength - it also reflects the population sizes of the ions involved. Thus, even weak correlations can yield high $d_{ij}$ values if the ions are abundant. For meaningful comparisons between ion pairs or datasets, the ratio $(p_{ij} - e_{ij}) / e_{ij}$ should be used, assuming the $d_{ij}$ value confirms a significant correlation. The $d_{ij}$ values are displayed in a chart where each ion is represented by a row and column as in Fig. [3].

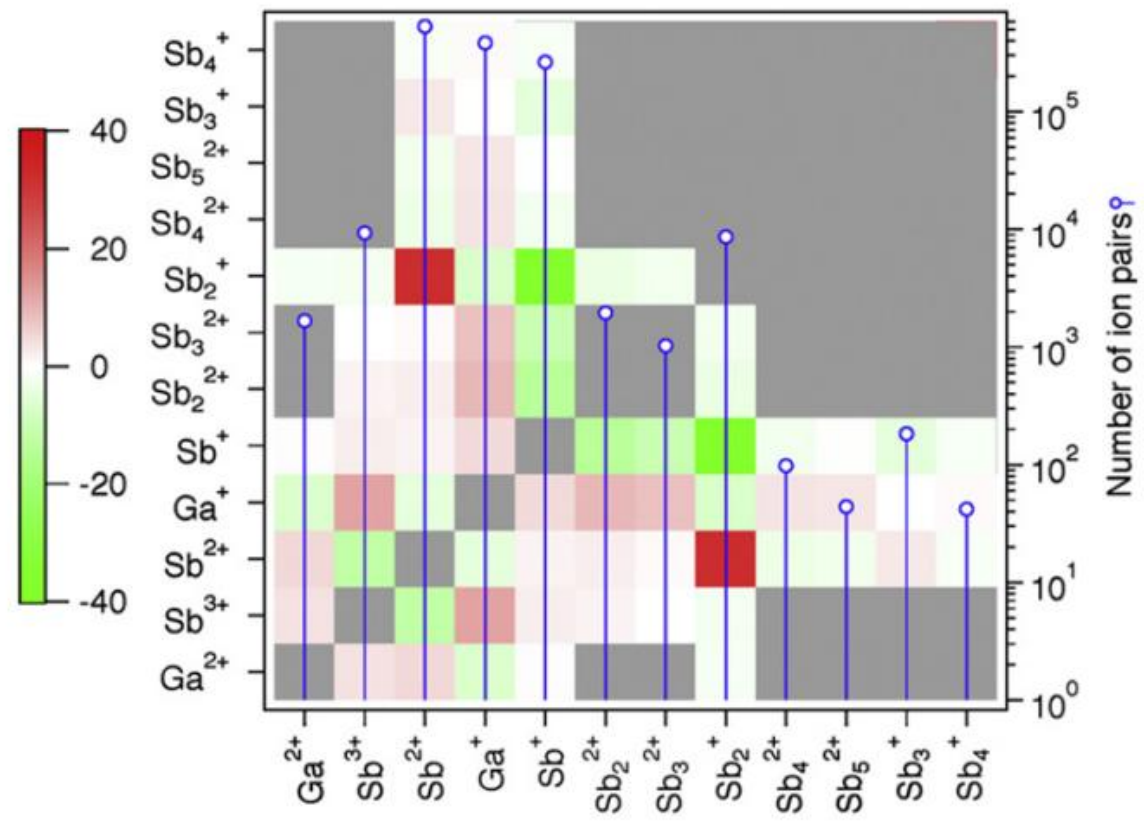

*Figure [3]. Correlation table illustrating the statistical analysis of ion pair combinations detected within the same multiple event in the analysis of GaSb under high field conditions (19.75 V/nm). Strong colors highlight significant correlations (in red) and anti-correlations (in green). The color intensity corresponds to the $d_{ij}$ values described in eq. (9), while grey entries correspond to a statistically insignificant set of pairs. As the order of ions in each pair is not considered, these tables are symmetric. Additionally, the vertical axes display the total number of ion pairs that each ion species is involved in. This tables indicates a strong correlation of $Sb_2^{+}$-$Sb^{2+}$ pairs and a strong anticorrelation of $Sb_2^{+}$-$Sb^{+}$ pairs (Reproduced with permission from (Müller et al., 2011), Elsevier).*

The second method for analyzing correlations in multiple events is the correlation histogram (Saxey, 2011), originally introduced in molecular physics for the covariance analysis of multiphoton ionization and dissociation experiments of gas jets (Frasinski et al., 1989). This plot resembles the correlation table but instead of identifying ions by detector range, it displays their mass/charge ratios on a two-dimensional histogram. Each entry represents one ion pair, with the $x$ and $y$ axes corresponding to the mass/charge ratios of the two ions. Since the pair AB is equivalent to BA, each is counted twice, making the histogram symmetric along the diagonal. Figure [4]-(a) shows a typical correlation histogram for SiC, where the redundant part below the diagonal has been partly masked. The interpretation of the main features visible in the histogram is assisted by the scheme of Fig. [4]-(b).

Peaks in the histogram correspond to ion pairs and are often followed by tails extending in various directions. The most prominent tails have a positive slope and align with the diagonal. When the same data is plotted using corrected time-of-flight (adjusted for voltage variation and flight path differences), these tails appear as straight lines with unit slope. This corresponds to correlated evaporation shortly after the laser pulse, during the cooling phase, with identical delays. Other tails in the correlation histogram appear horizontal or vertical. These represent cases where one ion evaporates at the time of the laser pulse, while the second ion evaporates later without any time correlation.

Additionally, tracks with a negative slope can be seen. These tracks usually indicate the presence of frequent molecular dissociation events during flight (Blum et al., 2016b; Saxey, 2011). Dissociations involve metastable ions. Some of these ions are known to dissociate in specific directions with respect to the electric field (Tsong and Cole, 1987; Zanuttini et al., 2017a), and it has been assessed that evaporated ions should be in an excited state in order to dissociate (Zanuttini et al., 2018). Such dissociation affects the time of flight of the resulting ions due to their change in mass/charge ratio as they traverse the accelerating field (Tsong and Liou, 1985). More details about molecular dissociation will be given in section 4.c.iv. These features are found in straight flight-path instruments, while do not apply to instruments with a reflectron, in which trajectories are strongly modified and focused in time (Di Russo and et al., 2020).

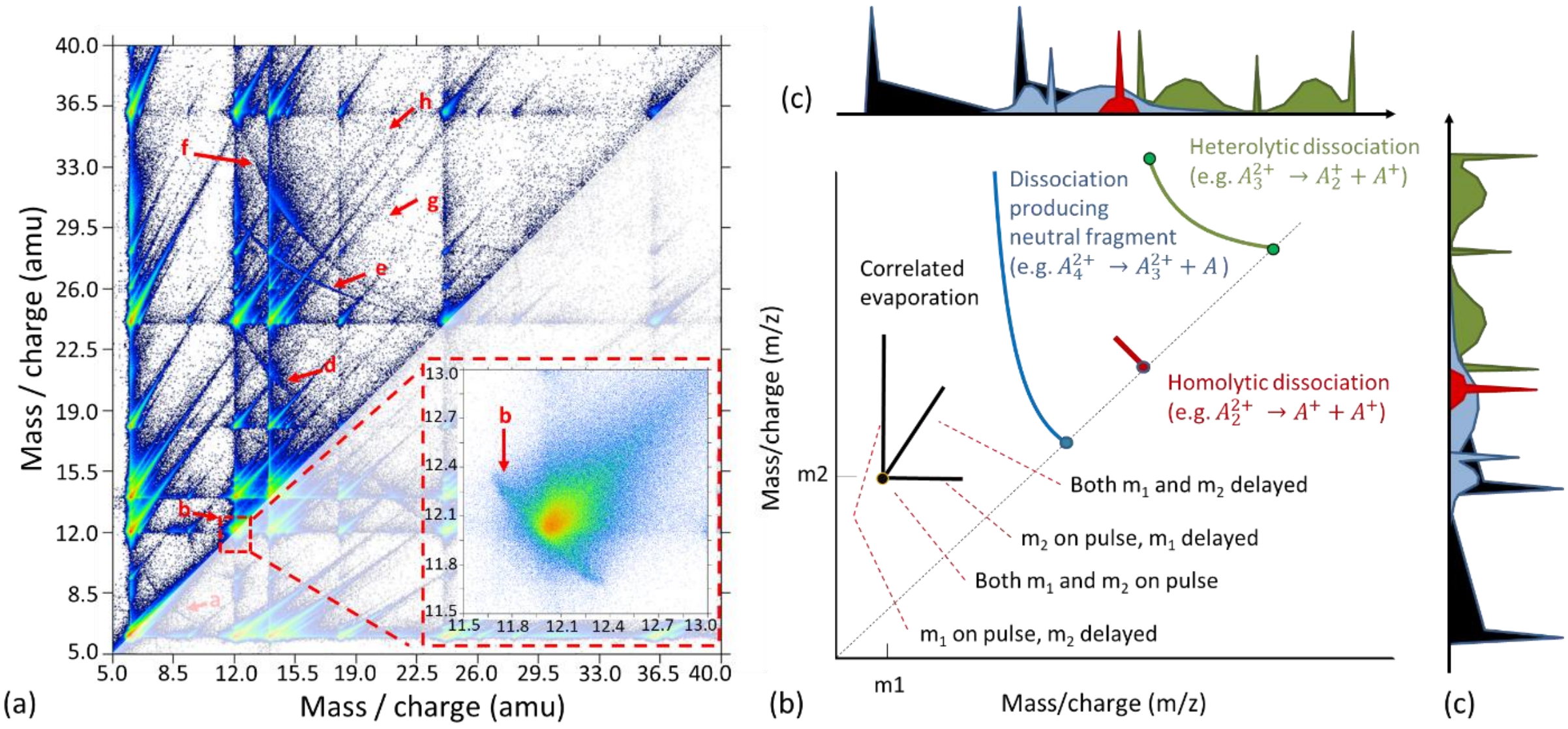

*Figure [4]. Use of correlation histograms for advanced interpretation of mass spectra and insight in field evaporation mechanisms. (a) The correlation histogram extracted from an analysis of SiC. The red letters and arrows indicate several dissociation tracks. The inset is a zoom in the neighborhood of m/z=12 and shows the detail of the peak related to the detection of $C^{+}$-$C^{+}$ pairs; here, the diagonal track marked with a red "b" is due to the homolytic dissociation $C_2^{2+} \rightarrow C^{+} + C^{+}$. (b) Schematics indicating the different possible interpretation of features (dots refer to correlation peaks, lines to tracks) appearing in the correlation histogram (Adapted with permission from* (Ndiaye et al., 2023a)*, American Chemical Society). (c) Schematic representation of how the different processed highlighted in the correlation diagram would appear within the mass spectrum containing multiple events only.*

The interest of correlation histograms in APT compositional metrology is twofold. On one hand, they allow for the identification of ions that would be typically out of the ranges defined for standard mass spectrometry. Despite most of the reported studies assessed that the impact of dissociation products on the composition measurement is moderate in dielectrics (Santhanagopalan et al., 2015; Di Russo et al., 2018a; Cuduvally et al., 2020; Ndiaye et al., 2023a), the situation could be different in the expanding domain of soft materials and organic compounds. On the other hand, correlation histograms (as well as correlation tables) allow for the direct visualization of dissociations and ion correlations. They constitute thus a very important tool for the investigation of the physico-chemical processes occurring at the specimen surface or during flight. This can in turn be useful for the interpretation of compositional biases (see sections 4.c.iv and 5.a.iv).

# 3. The role of surface field in APT

The driving force of almost all physical mechanisms involved in APT is the strong electrostatic field ***F*** existing from the specimen surface to the micron scale around it. This field is generated by the applied voltage V to the pointed needle of interest with reference to surrounding electrodes existing generally held to ground voltage. These electrodes can be placed a few centimeters ahead (in simple atom probe designs such as the LaWATAP), or more locally, if an extraction microelectrode is used (case of the local electrode atom probe). Different fields can be defined and used in APT, as reported in Tab. [II], corresponding to different mechanisms or phenomena. The electrostatic field distribution is at the source of atomic effects such as field polarization, induced field driven diffusion, field ionization effects, field evaporation effect, field dissociation effect, and field projection. The surface field and the local field in the micron scale distance to the surface is of interest to understand all these processes. In this review, we will address only electrostatic effect, which is valid generally in Laser Assisted Atom Probe Tomography (The intrinsic electromagnetic oscillation is neglected in LA-APT) and in Voltage Pulsed Atom Probe Tomography (VP-APT) for metals or good electronic conductors. However, we may note that more advanced treatment should be necessary to understand Terahertz Pulse driven Atom Probe (TP-APT) that was recently developed, where electric field oscillation with sub picosecond timescale is used to trigger field evaporation. TP-APT is therefore out of the scope of this review.

Table [II]. A set of differently defined electric fields used in APT and field ion emission theory and experiments. Ref. a = (Müller, 1941, 1956; Tsong, 1978b; Miller and Forbes, 2014a); Ref. b = (Ernst, 1979; Kingham, 1982; Mancini et al., 2014a; Tegg et al., 2024); Ref c = (Müller, 1960; Tsong, 1978b, 2005); Ref. d =(Gomer, 1994; Rigutti et al., 2017) ; Ref. e =(Suchorski et al., 1995) ;

| **Type of field** | **Typical Notation** | **Definition / description** | **Refs.** |
|---|---|---|---|
| Evaporation field, escape field | $F_e$, $F_{ev}$, $F_{EV}$, ZBEF | Field value for which the energy barrier for the transition of an atom from a bound state to an ionized state vanishes | a |
| Equivalent field, effective field | $F_{eq}$, $F_{eff}$ | Electric field calculated through charge sate ratio (CSR) statistics. Within the limits of the post-ionization models, it yields an approximation of the microscopic field $F_\mu$. | b |
| Best Image field (FIM) | BIF | Field at which the images formed in FIM have the best atomistic contrast and spatial resolution. The value of the BIF is primarily related to the ionization energy of the image gas used and therefore has a given value for each image gas. | c |
| Average, nanoscale field | $F_a$, $F_s$ | Average value of the electric field on the surface of a field emitter. Determines the stress state at few layers from the surface and in the sample bulk. | d |
| Microscopic field | $F_\mu$ | Field at a given point of the surface of a field emitter. Varies following the atomic roughness and the crystal symmetry. In FIM images recorded at the BIF, the BIF yields an approximate measure of the microscopic field in correspondence of the imaged atoms. | e |

### a. *Field distribution from analytical to atomistic distribution*

The field vector $\boldsymbol{F}(\boldsymbol{r})$ can be derived from the electric potential distribution $\boldsymbol{F}(\boldsymbol{r})=-\nabla V(\boldsymbol{r})$. $\boldsymbol{r}(x,y,z)$ is here the position vector describing the space. $\boldsymbol{F}$ can also be derived from the distribution of charge existing on the surfaces of the electrodes surrounding the ion considering that any surface $dS$ with a surface charge $\sigma$. The APT specimen, a sharply pointed needle, in vacuum, is polarized to a DC voltage $V$, with $V$ in the kV regime. The surface electric field generated on APT specimen is about 10-50 V/nm, which means surface charge of about 1 electron charge per nm². The end apex of the sample is roughly described by a hemispherical cap of radius R. (R in the range 10-100 nm). The specimen is situated at a close distance of grounded electrodes. LEAP instrument uses conical extraction electrodes with aperture of about 50 microns placed at 50 microns ahead. The 3D distribution of the electric field in this geometry is a classical problem that can be answered by solving the Laplace equation (electrostatic approximation) using a known geometry. Note that in DC voltage, the approximation of the specimen as a sufficiently good electrically conductive medium is fairly sufficient. Indeed, as we will show, the amount of total charge existing at the polarized specimen apex is about $10^4$ electrons for the application of the voltage V. It means any variation of surface electric field in the sub-second range is equivalent to current in the sub-fA regime, that can be easily drained in the micron scale apex, with limited voltage drop and internal field. Anyway, over the surface and in the free space in front of the specimen the field intensities and gradients are strong.

Conditions of constant voltage are used on the specimen surface. For certain smooth, symmetrical emitter shapes, the problem is easy to solve analytically. A number of mathematical models of emitter shape have been used and some have an exact solution to Laplace's equation for the electrostatic potential. The simplest approach imposes a specimen shape very close to a surface of constant curvature radius R, at least close to the specimen axis. For a polarized sphere $F$ is simply $V/R$. So it is tempting to adapt this expression for a real APT geometry. To a first approximation, the electric field magnitude F was found to follow the classical expression

$$F = \frac{V}{k_F R} \qquad (10)$$

where $k_F$ a dimensionless semi-constant, i.e. that is slowly varying as the specimen is field evaporated. $k_F$ is in the range of 2–10, and depends primarily on the specimen's shank angle and end radius (Vurpillot and Oberdorfer, 2015). For a smooth emitter, the hemispherical approximation ±45° about the specimen's main axis is a good approximation of the morphology of the specimen's tip. However, when approaching the outer edge of the end apex, the electric field decreases, the steady state radius of curvature tends to decrease, which makes the specimen surface more elliptical. Nevertheless, in atom probe, ±45° is approximately the maximal field of view of most of instrument. The spherical approximation is used for field evaluation, and standard process of reconstruction of the dataset. For a homogeneous material, analyzed in atom probe under a limited variation of voltage, $k_F$ is considered as a constant, and the variation of field (lateral to the surface and in the first nanometers above the surface) are neglected. It is also considered that the voltage penetration in the material is extremely abrupt. A first order evaluation of the local electric field at the tip surface is to consider locally the surface curvature ($1/R_C$), which is defined as the average of the principal curvatures $1/R_1$ and $1/R_2$ for an arbitrary shape surface. Field is thus inversely proportional to $R_C$ ($R=R_C$ for a sphere) in a first approach. An evaluation of the surface geometry and its local curvature is a first indicator of the field distribution existing on the surface. This approach was used for instance to study morphology variation of heterogeneous materials under field evaporation and related distortion in image projection of ions (Figure [5]) (N. Rolland et al., 2015a; Vurpillot et al., 2016; Beinke et al., 2016).

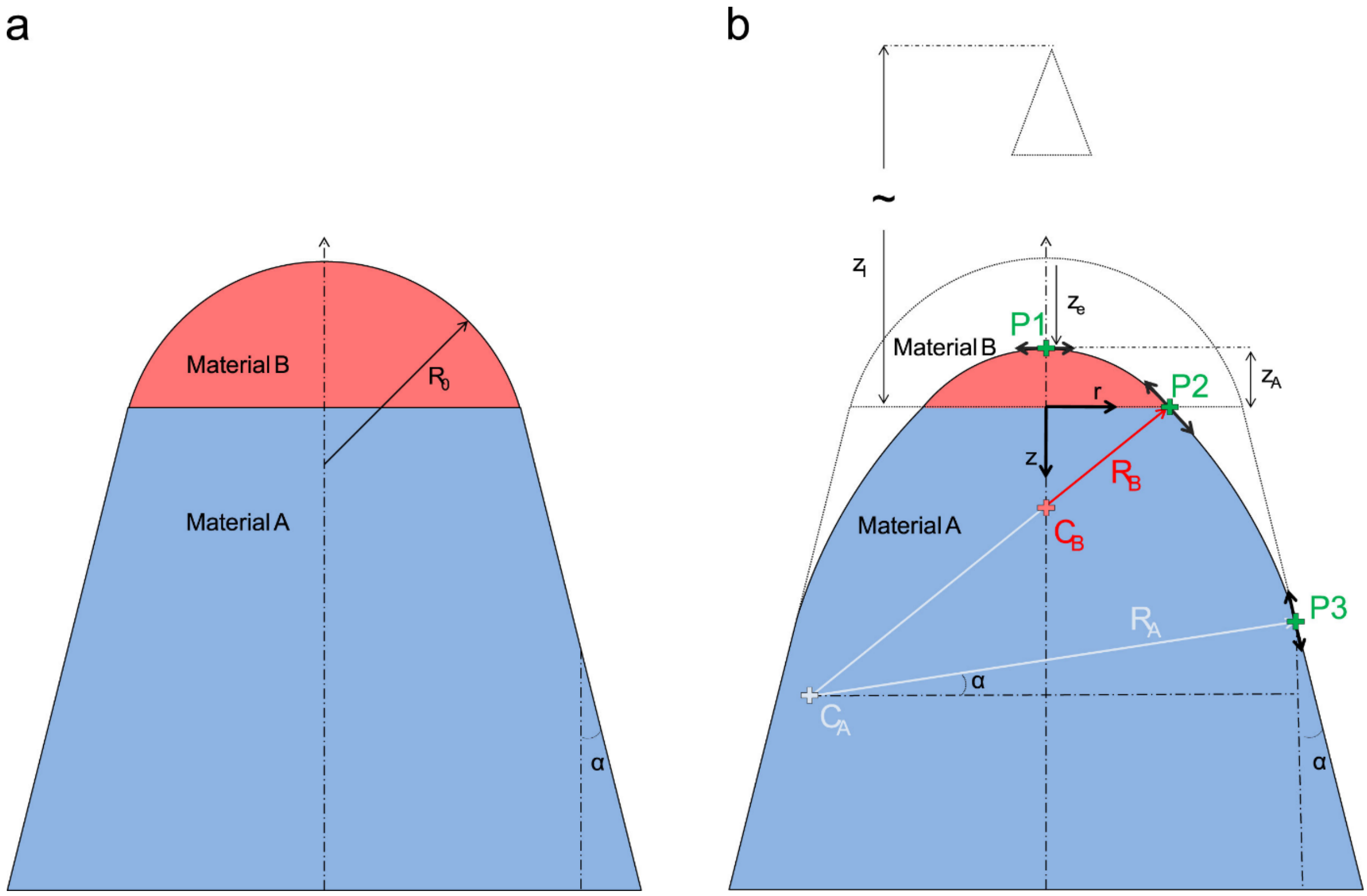

*Figure [5]. (a) Geometric modeling of a specimen geometry composed of two phases having evaporation field different by 20%. (b) Modeling of the tip shape at the evaporation depth $z_e$ in the case of a high field material B (in red) seated on a low field material A (in blue). (Adapted with permission from* (N. Rolland et al., 2015a), *Elsevier).*

Anyway, when differences in specimen surface becomes too strong (about 20% is curvature), the validity of a constant $k_F$ is not sufficient, and non-linear effects are observed. In addition, the notion of curvature is an ill-posed problem. Classical electrostatic textbook formulates that the relationship between the local curvature and the amplitude of the field is not straightforward, since the buildup of charge on the surface giving rise to the electric field is a long-distance process that cannot be solved only by local inspection of the morphology. This over-simplification is the main cause of misinterpretation and errors in the final dataset analysis. In addition, curvature can be defined at different scale, for macroscopic, micro or mesoscopic, but also atomic. A single atom on a flat surface can be considered to produce a sub-nanometer roughness. Ab-initio calculation shows geometric curvature of material surfaces at this scale are not smeared out and induces local enhancement of field close to the geometric electrostatic description (Silaeva et al., 2013) . To resume, the control of the electric field from the atomic scale to the mesoscale is at the roots of metrological performances and experimental, empirical or computational means are necessary to evaluate dynamically the field distribution over the specimen surface.

Assuming physical and geometrical properties of the specimen in the configuration of an atom probe enables a direct modelling of electrostatic field distribution over the specimen. In older models applied to an arbitrary emitter shape, determining this field was difficult because of the nine or ten orders of magnitude variation in length scale going from the specimen (defined at the nanometer scale) to the surrounding electrodes (defined on the microscale) to the detector (defined at nearly the meter scale). Currently, the electrostatic field is often calculated using finite element methods (FEM) when the space between electrodes is meshed so that electrostatic voltage can be deduced everywhere or boundary element methods (BEM), where the distribution of surface charge on electrodes can be calculated in 3D without the need to mesh the 3D space entirely (Figure [6]-(a-d) (Loi et al., 2013; Vurpillot et al., 2013; Oberdorfer et al., 2013; Vurpillot and Oberdorfer, 2015; Nicolas Rolland et al., 2015; Rousseau et al., 2020; Fletcher et al., 2022; Hatzoglou et al., 2023b; Lüken et al., 2024). Using this approach, it is possible to use a fine mesh to define the specimen (sub-nm) and a much coarser mesh for electrode of macroscopic dimensions. Both models (Laplace computation or BEM computation) are equivalent, since they both assume that the density of charges are located on the surface with a penetration depth smaller than the size of a single atom (Figure [6]). Models can describe mesoscopic details in continuum approaches (Fletcher et al., 2022; Hatzoglou et al., 2023a), or atomistic details (Vurpillot and Oberdorfer, 2015). This last approach requires more computational resources due to the large number of points to be defined in an atomically rough model Figure [6]-(f). The field distribution determination on realistic modelled specimen enable to explore surface effects (Katnagallu et al., 2018; Vurpillot et al., 2018), of dynamic effects during the flight of ions from the surface to the detector (Figure [6]-(d)) (Blum et al., 2016b; E. Di Russo et al., 2020; Gault et al., 2016; Rousseau et al., 2020).

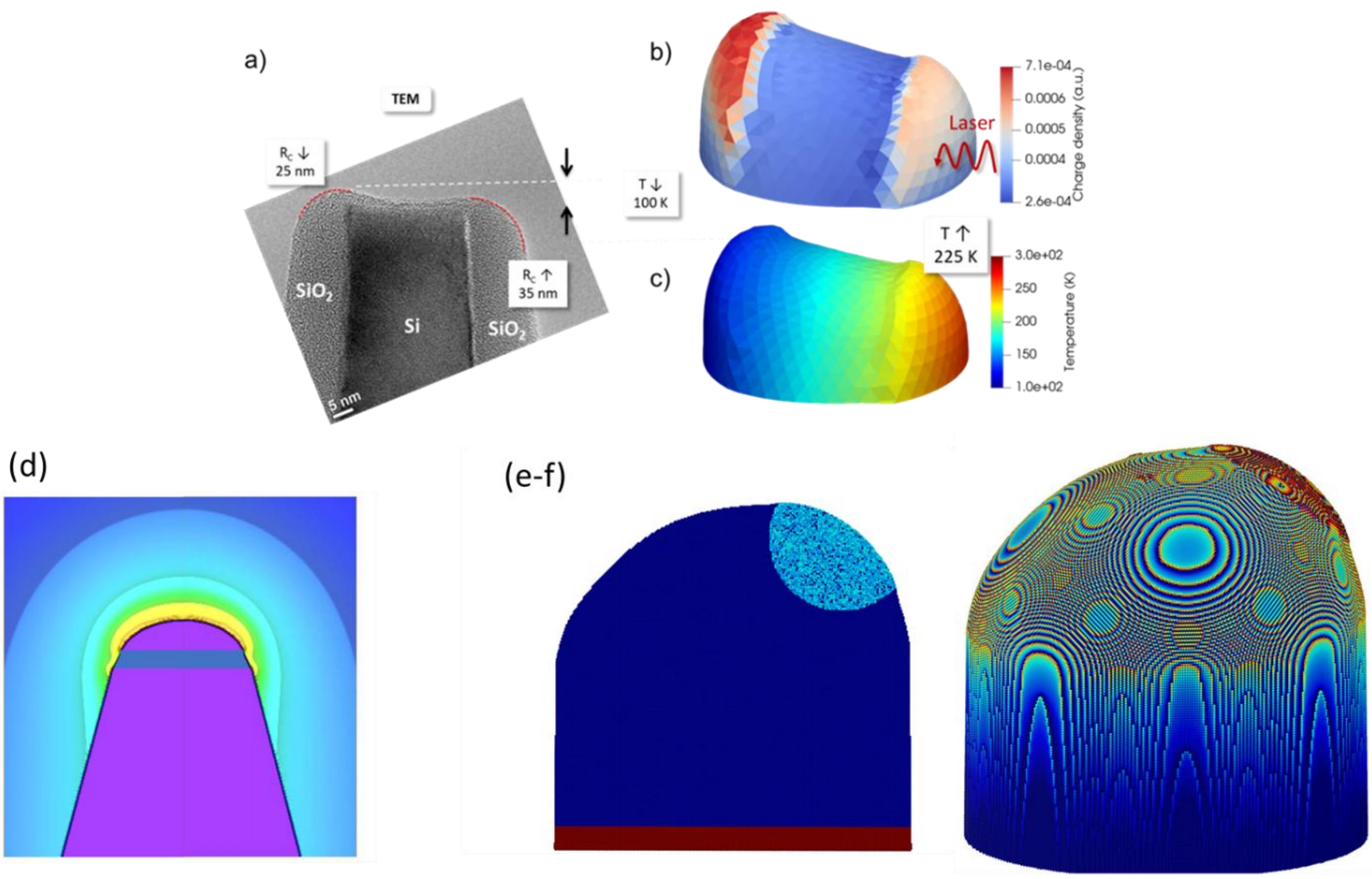

*Figure [6]. (a) TEM image of a specimen composed of heterogeneous materials with a large difference in evaporation fields ($SiO_2$ and Si), analysed by APT . Note the complexity of local curvatures of the surface (negative curvature which not exist considering equation (10). Simulation of the field evaporation of this specimen, and representation of (b) surface charge and electric field (c) local temperature under laser pulsing showing the interplay between geometry, temperature and local field for this structure (Adapted with permission from (Hatzoglou et al., 2023b), IOP). (d) Simulation of the field evaporation of a heterogeneous material with representation of the electric field in a realistic Atom Probe geometry. This field distribution can be used to study the trajectories of ions or cationic molecule from the surface to the detector (Adapted with permission from (Vurpillot et al., 2013), Elsevier). (e-f) Modelling field evaporation of a heterogeneous material composed of 2 crystallographic structures. The internal and surface structure can be observed. Note the strong atomic variation of the local electric field induced by local roughness at the atomic scale.*

### *b. Field polarization*

In the presence of the strong electric field, any atom or molecule reaching the apex region are submitted to polarization effect. Under this electric field, a neutral particle is transformed into a dipole interacting with the strong variation of field existing in the first 100 nm to the surface. Due to polarization forces, these particles migrate towards the highest field region at the apex and accommodate to the temperature of the specimen. The end apex field distribution acts as a potential well with an energetic depth depending on the intrinsic polarizability of the particle. The same forces may act on any neutral particle emitted from the surface. The volume isotropic polarizability α of most of atoms and molecules may be found in literature (in Å³) and are generally around ~1 Å³ for single atom or diatomic molecule. Under the surface electric field intensity (10-50 V/nm), final binding energy $E_b$ is in the range 0.5-10 eV ($E_b \sim \frac{\alpha F^2}{2} / 4\pi\epsilon_0$).

The effect of the electric field on gas atoms or molecules near a field emitter has been studied in the context of field-ion microscopy (FIM) (Miller and Smith, 1981). FIM is a predecessor of APT, utilizing similar specimen geometries and electric field intensities. In FIM, mono-atomic gases (e.g. He, Ne) are typically used to produce a highly-magnified image of the surface of an emitter via field ionization near the specimen surface. Eventually, the atoms cross a region where the electric field is sufficient to cause their ionization and projection towards a detector screen. The polarization effects act as a local well, increasing the local concentration of atoms. Maximal enhancement of this concentration to $10^4$ was evaluated. In a first approach, the local pressure at the apex may be described using Maxwell Boltzmann distribution of velocities as

$$P = P_0 e^{\frac{\alpha F^2}{2kT} / 4\pi\epsilon_0} \qquad (11)$$

In FIM, this effect enables a strong enhancement of the ionization probability (described in the next section), close to high field region as it participates to the increase of the image gas supply function.

The same effect was recently used to elucidate local oxidation of Fe in an Environmental Atom Probe. The specimen (pristine Fe single crystal) is simultaneously exposed to low pressures of pure oxygen gas, on the order of $10^{-7}$ mbar, while applying intense electric fields on their surface of several tens of volts per nanometer. The local composition of the different surface structures is probed directly and in real time and successfully compared with first principles-based models. Rough Fe{244} and Fe{112} facets were found more reactive toward oxygen than compact Fe{024} and Fe{011} facets. Ultimately, externally applied electric fields can be used to dynamically exploit reaction dynamics at the nanoscale towards desired products in a catalytic reaction (Lambeets et al., 2025).

Field polarization is also an important phenomenon limiting the probability of emitting desorbed atoms of molecules. Indeed, if atoms are emitted from the specimen surface, under the application of laser pulses, the kinetic energy of these atoms may not exceed the induced surface thermal energy (<1000K). As a result, thermally desorbed energy would never exceed 100 meV. This energy is much smaller than the binding energy induced by the polarization (>1 eV in practice). We may note that the spatial extension of the high electric field region may achieve several nanometers in thickness, so that a displacement of the emitted particle is however possible inside the potential well induced by the gradient of field (Gault et al., 2016).

### *c. Field ionization*

Field ionization is a quantum mechanical mechanism that occurs when atoms are subjected to sufficiently strong external electric fields. In the absence of a field, an electron in a bound atomic state is confined by the Coulomb potential of the nucleus. The application of an intense electric field, however, modifies this potential by introducing a linear term that effectively lowers and narrows the barrier on the vacuum side. As a result, the electron wavefunction acquires a finite probability to tunnel through the barrier, a process formally described by quantum tunneling theory and closely related to the Fowler–Nordheim formalism originally developed for field electron emission (equation (12)). (Fowler and Nordheim, 1928). As a result, the theoretical treatment is very similar to the calculation of electron tunneling on a field emitter surface.

When the local field strength reaches several tens of volts per nm, as is the case at the apex of sharply pointed needle, the tunneling probability becomes appreciable and field ionization takes place. This process underpins the operation of field ion microscopy, where imaging gas atoms (e.g., He, Ne) are field-ionized in the vicinity of the tip apex and the resulting ions are projected onto a detector, enabling imaging with near-atomic resolution (Müller and Bahadur, 1956; Forbes, 1995; Gomer, 1959).

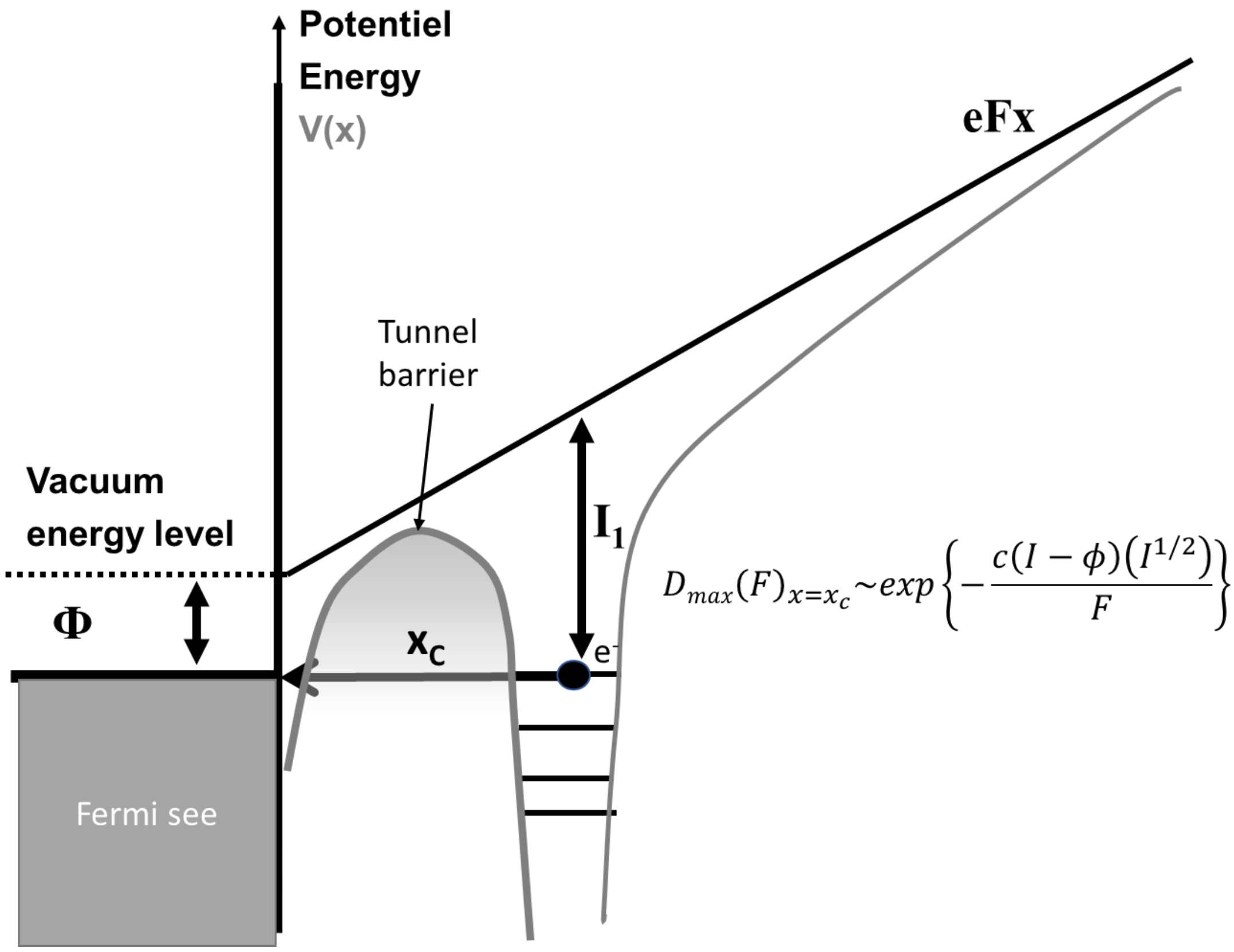

*Figure [7]. Schematic potential energy diagram for field ionization of an atom of ionization energy $I_1$ close to a clean metallic surface defined by Φ the work function of electron (the energy difference between the Fermi level and the vacuum level) situated at the optimum tunneling conditions. A clear relationship between the maximum tunneling rate and the field F is found, with $x_c$ being the critical distance for tunneling.*

Potential energy diagrams (V(x)) for field ionization, in close proximity to a material surface, are simplified in Figure [7]. The barrier tunneling rate D(V,E) for an electron in a bound atomic state close to the surface is generally calculated using the Wentzel-Kramer-Brillouin (WKB) approximation, following

$$D(E,V) \propto \exp\left\{-\sqrt{\frac{32\pi^2 m_e}{h^2}}\int_{barrier}\sqrt{V(x)-E}\,dx\right\} \qquad (12)$$

With $m_e$ the electron mass, and E the energic level of the electronic wavefunction. V(x) is the sum of electrostatic potentials induced by the material surface, the electric field, and the coulomb potential of the atom nucleus.

For a given surface, the transparency can be calculated as a function of physical constant of the selected atom (essentially the ionization energy I), the work function of electron in the material (Φ), and a refined expression of the shape of potential energy close to the surface (V(**r**) or V(x), in a one-dimensional form). In the simplest approach a triangular shape of the barrier may be assumed giving a transparency with following

$$D(E,V) \propto \exp\left\{-\frac{b}{F}\right\} \qquad (13)$$

With *b*, a constant related to I and Φ. It reveals that there is a strong dependency of the ionization rate over the surface field. We may note that the field is generally higher at close distance (more precisely, around a given critical distance $x_c$) from surface roughness, so that ionization can be used as a local probe of the surface field. The ionization probability thus acts as a local probe of the surface field, analogously to the tunneling current in STM probing the local density of states.

From τ=1/D, the probability of ionization P(t) for a single atom in a time interval t may be evaluated given by

$$P(t) = 1 - \exp\left\{-\frac{t}{\tau}\right\} \qquad (14)$$

if τ remains constant. Generally, τ varies as the particle changes position and velocity close to the surface. Equation (14) must be replaced by

$$P(t) = 1 - \exp\left\{-\int_x \frac{dx\prime}{v(x\prime)\tau(x\prime)}\right\} \qquad (15)$$

where v(x)=dx/dt the velocity of the particle

The probability of ionization is therefore not only dependent of the quantum ionization rate, but also from the time of residence of the particle in the high field region. We may note that it was demonstrated by ab-initio calculations that the probability was also dependent of the electronic density of state of the surface. Using the Tersoff-Hamann approximation known from scanning tunneling microscopy it is possible to study the tunneling of gas atoms hovering above the surface in the presence of a very strong electrostatic field. It was demonstrated that the tunneling rate could vary significantly at the atomic scale depending on the nature of the atom of the surface. This difference is linked not only to the work function of electrons but also to the electronic density of states of the surface (Bhatt et al., 2023).

### *d. Post-ionization*

The process called post-ionization can be understood as the sequential ionization of an ionized atom, as described in the previous section, beyond the escape charge state $n=n_i$ (Fig. [7]). A field-evaporated or field-ionized atom can lose one or more extra electrons provided the tunneling probabilities of these electrons are sufficiently high. It finally acquires the arrival charge state $n=n_f$. This process occurs therefore after an ion has already been created and is being accelerated away from the surface by the intense field. Post-ionization processes occur nevertheless extremely close to the surface. Microscopically, post-ionization has the same features of ionization: it is field-assisted electron tunneling from the ion to the sample surface. The role of the external field is lowering and thinning down the potential barrier, so that electrons can tunnel out with appreciable probability. The theory of post-ionization has been developed by Ernst and Jentsch for $Rh^+$ (Ernst and Jentsch, 1981) and later generalized by Kingham (Kingham, 1982) based on an analytical approximation of a WKB Hamiltonian. This probability is largest very close to the surface, where the local field is of the order of tens of V/nm, at a specific critical distance $x_c^n$. Most importantly in APT metrology, it states that sequential charge states $M^+$, $M^{2+}$, $M^{3+}$ etc. have increasing field threshold for tunneling probability, as well as increasing values of critical ionization distances $x_c^n$ depending on the ionization state n. This mirrors the fact that ionization energies increase with ionization state. The relative abundance of different ion species becomes therefore a function of the electric field and may be used to determine or, at least, estimate it. This finding lies at the foundation of the charge state abundance or ratio metrics.

### *e. Charge state ratio metrics (Kingham curves)*

As mentioned above, the phenomenon of post-ionization translates into a monotonous increase of the abundances of higher charge states from a field emitter with increasing field intensity. This behavior is illustrated in Fig. [8]-(a): the abundance of the $Ga^{2+}$ ion increases by almost one order of magnitude when the $V_{DC}$ bias is increased from 4.5 kV to 6.5 kV during a series of parametric analyses of a GaN specimen. As the Kingham post-ionization theory clearly identifies a univocal relationship between the abundance of charge states and the field, it becomes possible to exploit the former in order to estimate the field at the surface of a field emitter. An example of this is reported in Fig. [8]-(b), where the charge state abundances of W ions are reported as a function of the field intensity (Kingham, 1982; Tegg et al., 2024). For practical purposes it is common to introduce the quantity known as Charge State Ratio

$$CSR\left(A^{(n+1)+}/A^{n+}\right) = \frac{N(A^{(n+1)+})}{N(A^{n+})}, \qquad (16)$$

Where $N(A^{i+})$ is the number of ions of charge state $i+$ counted within a given region of the dataset. If only two charge states are present in the dataset, the shorter expression $CSR(A)$ can be adopted without ambiguity. The expressions relating CSR and field $F$ in the theory by Kingham are the result of a full quantum mechanical calculations, and their application to common situations in APT is not straightforward. In order to circumvent this problem, Tegg and co-authors have recently provided a set of simplified expressions of this form:

$$F_{eq} = a\left(1 - \frac{b}{\left[CSR\left(A^{(n+1)+}/A^{n+}\right)\right]^{0.3} + b + 0.256}\right), \qquad (17)$$

Where a and b are parameters listed in the reference (Tegg et al., 2024). The $F_{eq}$(CSR) function (also commonly called the *equivalent field* or the *effective field*) is given for three combinations of charge states of W in Fig. [8]-(c), which also visualizes the excellent approximation of the analytic expression (17) to the relations directly issued from the more complex expressions given by Kingham. The CSRs provide thus a commonly used metrics for the estimation of the microscopic field $F_\mu$ at the surface of a field emitter, as $F_{eq} \sim F_\mu$. The main advantages of it are the following;

*(i) Instrument and sample independence.* The CSR is issued by a quantum model in which the only experimental parameter is the surface field $F$. In other words, it is independent of temperature and also of the used instrument. Notice that other experimental parameters, such as the laser energy per pulse or the DC voltage, are instrument and/or sample dependent, i.e. a given set of their values does not yield the same effect in different instruments and for different samples.

*(ii) Material independence.* As the phenomenon of post-ionization occurs in vacuum, the CSR is in principle independent of the material from which a given element is evaporated. However, this statement is only approximately valid. A given element may indeed be found in different oxidation/reduction states within different materials, which could translate into a significant abundance of higher charge states at fields for which post-ionization is not predicted to produce them. As an example, $Al^{2+}$ is not supposed to be post-ionized to $Al^{3+}$ at the typical fields at which AlGaN is evaporated, yet $Al^{3+}$ is found in the mass spectrum (Rigutti et al., 2016a).

*(iii) A global or microscopic quantity.* The possibility of representing the CSR not only as a *global quantity*, but to define it within an arbitrary region of a given APT dataset is perhaps the main advantage of this metrics. An example of the detector space distribution of the CSR(Ga) from the analysis of GaN is reported in Fig. [8]-(d), while the CSR(Ga) distribution within a thin slice extracted from the 3D reconstructed volume of a GaN/AlGaN heterostructure is reported in Fig. [8]-(e). In the first example the CSR distribution provides information about the inhomogeneity of the surface field and its correlation with the specimen crystallography (Di Russo et al., 2018a); in the second example, the alternation of layers with higher and lower fields correlates with the presence of layers of different composition and, consequently, different binding energy; in this same example, the gradual decrease of the field from the beginning to the end of the analysis is linked to the decrease of the evaporation rate necessary to keep the detection rate constant while the imaged surface increases (Di Russo et al., 2018b). These two examples illustrate how the CSR can be displayed as a *microscopic quantity*, which can yield important information about the details of the APT analysis.

The CSR has become in the last two decades an extremely important parameter for the assessment of the surface field intensity, and its use is nowadays current in most APT works studying the fundamental aspects of field evaporation of new systems and/or their compositional metrology. The application of the CSR to specific problems will be illustrated in the case studies of part 5.

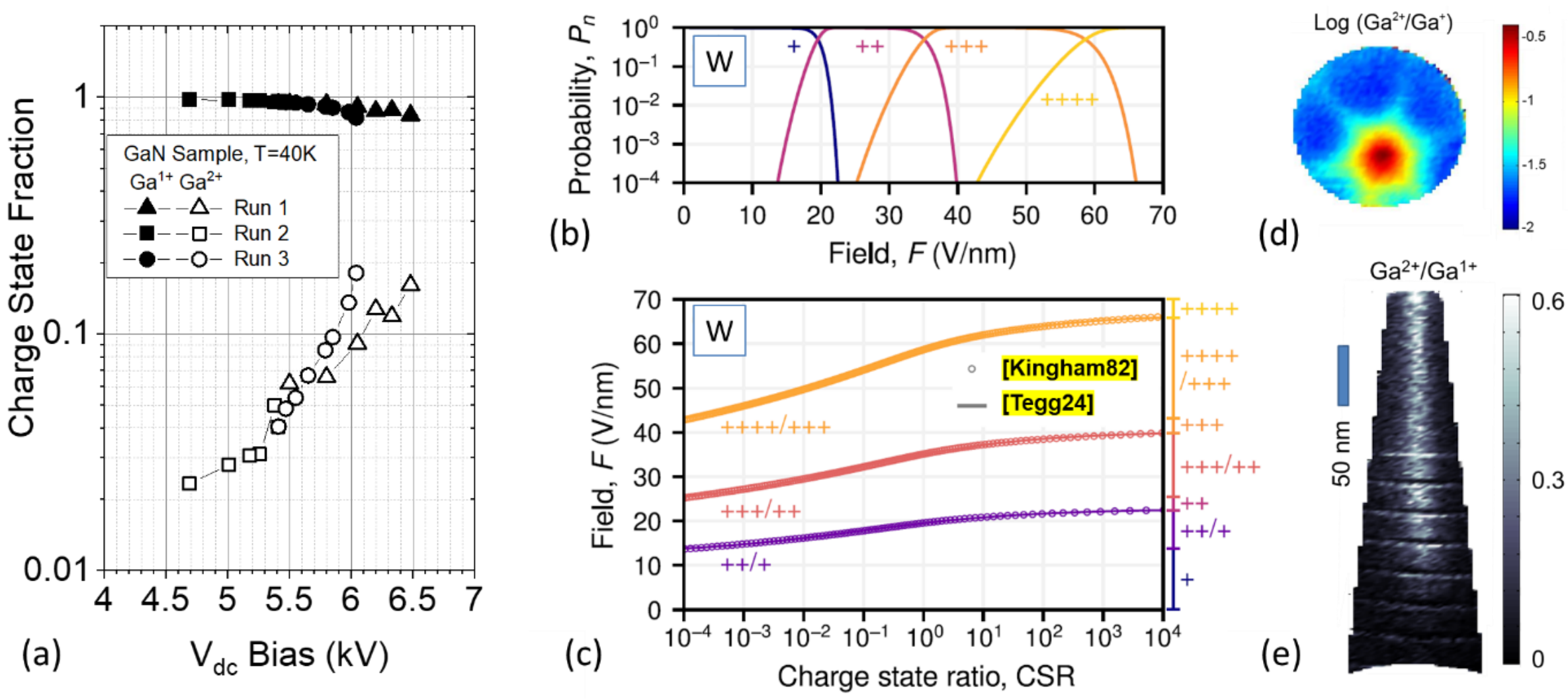

*Figure [8]. Relevance of charge state ratio (CSR) metrics in APT. (a) Example of experimental dependence of the fractions of $Ga^{2+}$ and $Ga^{+}$ on the applied voltage during several analyses of a GaN field emission specimen. (b) Field dependence of the probability of finding a field-evaporated atom in a given charge state according to the Kingham post-ionization model applied to tungsten. (c) Relationship between the CSR($W^{(n+1)+}/W^{n+}$) and the surface electric field as calculated by Kingham (Kingham, 1982) and by (Tegg et al., 2024) (Adapted from (Tegg et al., 2024), Oxford Academic). (d) Two-dimensional detector space map of the CSR($Ga^{2+}/Ga^{+}$) issued from the analysis of a GaN field emitter. (e) Two-dimensional map of the CSR($Ga^{2+}/Ga^{+}$) issued from the analysis of a GaN/AlGaN heterostructure realized within a thin slice extracted from the 3D reconstructed space.*

### *f. Field-induced stress*

The average surface electric field $F_s$ introduces a tensile stress state in the specimen. Due to the accumulation of a high density of free holes on the apex surface, the field does not penetrate deep into matter but "pulls" on the system, generating a Maxwell stress on the emitter surface (Gomer, 1994):

$$\sigma_{apex} = \frac{1}{2}\varepsilon_0 F_s^2 \qquad (18)$$

Which is independent of the analyzed material. Significant stress intensities can be achieved during APT analyses, as reported in Fig. [9]. Approximating the apex as a hemisphere and integrating the forces $d\boldsymbol{T}$ that develop on a surface element $dS$, as illustrated in the inset of Fig. [9], the tension $\boldsymbol{T}$ along the field emitter axis is

$$\boldsymbol{T} = \pi R_{apex}^2 \sigma_{apex} \hat{\boldsymbol{n}}_{<\boldsymbol{axis}>} \qquad (19)$$

where $\hat{\boldsymbol{n}}_{<\boldsymbol{axis}>}$ is a unit vector pointing along the field emitter axis. The stress at the apex may be assumed as hydrostatic, as the Maxwell forces are directed approximately along the needle radius. At a distance from the apex approximately equal to several $R_{apex}$, the stress becomes uniaxial, with the only non-negligible component

$$\sigma_{zz}(z) = |\boldsymbol{T}|/S(z), \qquad (20)$$

where

$$S(z) = \pi \cdot \left( R_{apex} + \frac{\left(R_{base} - R_{apex}\right) \cdot z}{L_{tip}} \right)^2 \qquad (21)$$

is the area of the axial cross section at the axial coordinate $z$, $R_{base}$ is the specimen radius at its base and $L_{tip}$ is the specimen length (inset of Fig. [9]) (Rigutti et al., 2017). An upper limit for the intensity of stress may be estimated for an ongoing experiment though the CSR metrics. So far, a measurement of stress intensity cannot be achieved within typical APT setups. Photonic Atom Probe measurements, exploiting the known dependence of photoluminescence lines as a function of stress could assess stress levels up to 9 GPa in diamond (Rigutti et al., 2017) and around 1.25 GPa in ZnO (Dalapati et al., 2021). Mechanical stress may produce plastic relaxation, which may turn into specimen fracture (some specific values of yield strength are reported in Fig. [9] for reference). This mechanism is particularly critical during the analysis of multilayers with strongly differing evaporation fields (N. Rolland et al., 2015b). Due to the complexity of the stress measurement, the role of mechanical stress on composition measurements in APT has not been studied yet, but it is known that stress gradients may promote formation, diffusion and migration of point defects (Aziz, 2001; Connétable and Maugis, 2020) as well as dislocation motion (Chakravarthy and Curtin, 2011) in bulk materials.

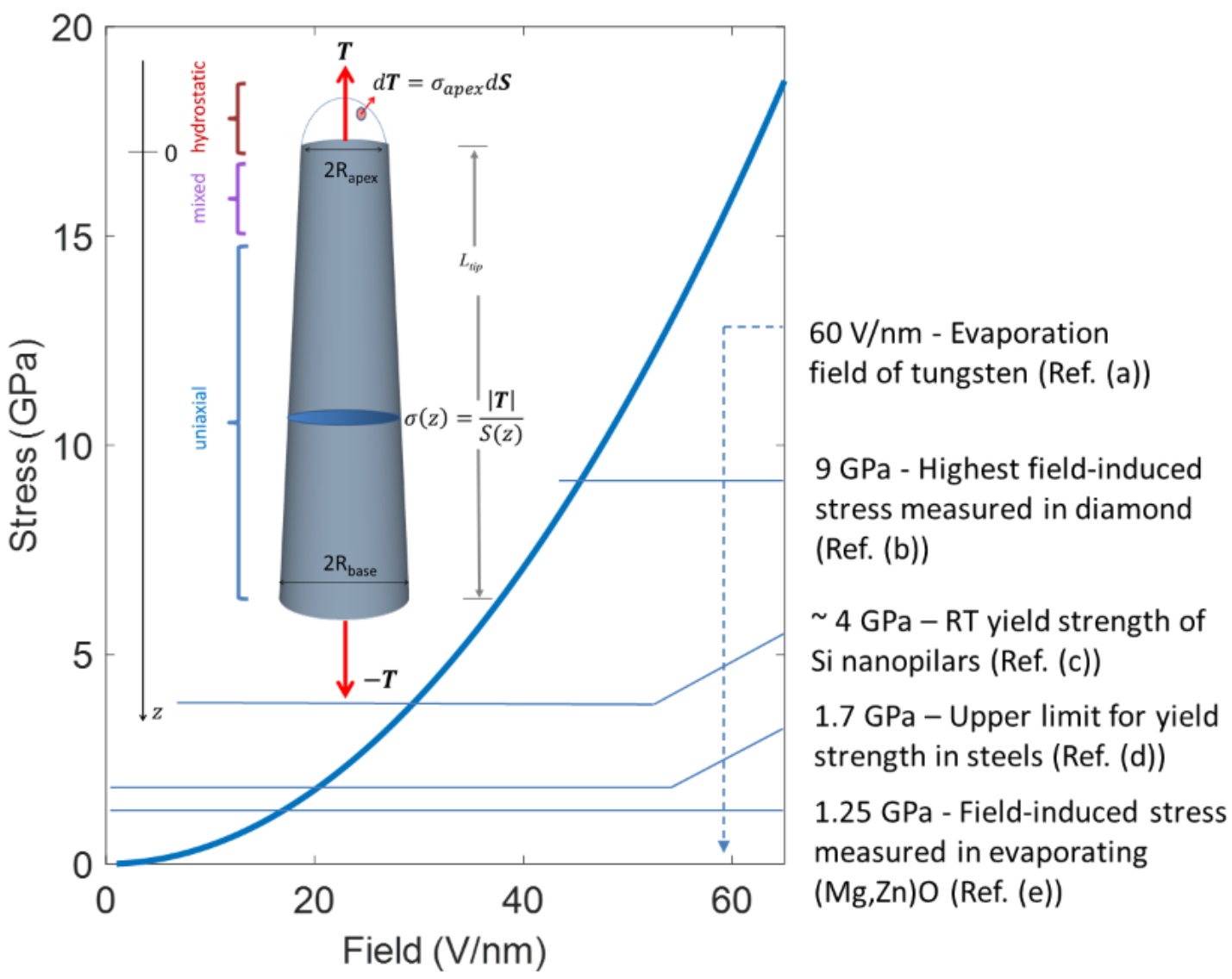

*Figure [9]. The hydrostatic tensile stress generated at the apex of a field emitter as a function of the electric field. Reference values of stress and field are reported from the literature in the diagram: ref. (a) = (Ono et al., 2005); ref. (b) = (Rigutti et al., 2017), ref.(c) = (Chen et al., 2020), ref.(d) = (Pavlina and Van Tyne, 2008), ref. (e) = (Dalapati et al., 2021). In the inset, schematics illustrating the spatial evolution of the stress field within the specimen.*

### g. *Field-induced migration*

Some elements are quite prone to movements on the nanoscale under the effect of electric or strain fields (Tsong and Kellogg, 1975). *Electromigration* – a thermally activated transport mechanism where an applied electric field reduces the activation energy barrier, enables directional ion movement driven by thermal energy. This effect allows ions with lower migration energy barriers to move efficiently, even against concentration gradients, and typically occurs faster than diffusion. The migration, also known as *directional walk,* has usually specific directions, towards one or more crystal poles, and primarily concerns solute species in both metals and semiconductors. In metals this process has been studied for solute species such as P, Si, Mn, B, C and N (Kobayashi et al., 2011; Gault et al., 2012a). The proneness to directional walk is generally increasing with increasing elemental evaporation field and decreasing atomic radius (Gault et al., 2012a). In semiconductors, an example of directional walk is illustrated in Fig. [10], showing that for low (actually: less homogeneous) field conditions the distribution of B in Si is apparently concentrated close to the axis of the specimen (Guerguis et al., 2024). These solute species are accumulated in one subregion of the surface. On one hand, this induces a degradation in the associated spatial precision and a bias in the calculation of concentrations. On the other hand, migrating species may evaporate with a strong degree of correlation (section 4.c.iii), which can lead to specific losses by pile-up effects (section 4.b.ii). In battery materials, for instance, the in situ delithiation process can be understood as an electromigration process (Greiwe et al., 2014; Kim et al., 2022). Notably, at the low temperatures used in APT analyses (typically 30–80 K), the electrostatic field still facilitates ion migration. To mitigate unwanted effects, effective shielding—such as coating the specimen with a conductive material—can be employed (Kim et al., 2022).

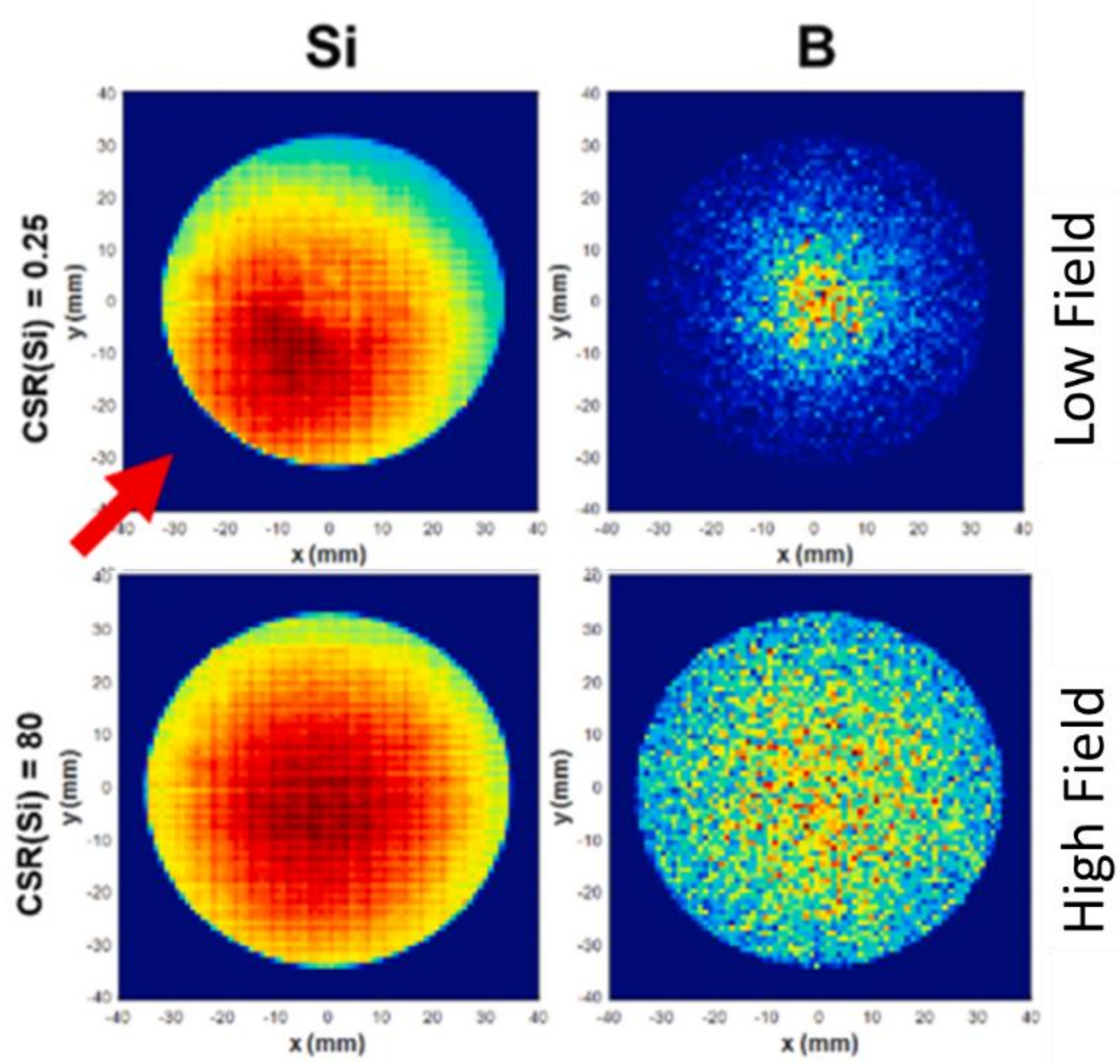

*<span>Figure [10]</span>. Normalized detector density maps of Si and B from the APT analysis of Si:B at two different field conditions according to the CSR(Si) metrics. The red arrow indicates the laser direction. The analysis conducted at low field clearly point out that B is subjected to field-driven migration towards the axis of the specimen. (Adapted with permission from (Guerguis et al., 2024), Elsevier)*

## 4. Mechanisms of loss

In APT there are different mechanisms leading to losses. These can be classified into three categories, which are often interdependent, i.e. detector losses, losses related to difficulties in the interpretation of mass spectrum, and physical mechanisms, related to the evaporation process itself. It is crucial that the operator applies empirical methods allowing for the assessment of a loss. In some cases, the assessment and sufficient complementary information may allow for the correction of the error in the composition measurement. Table [III] reports the main mechanisms of loss and lists some of the systems impacted by the different mechanisms.

Table [III]. Synoptic view of the main channel of loss or error in composition measurement in APT, the conditions at which they typically occur, and examples of the systems impacted, which are detailed in part 5.

| Category | Mechanism | Conditions of occurrence | Examples of systems affected |
|---|---|---|---|
| **Detector Performance** | Intrinsic detection efficiency – Open Area Ratio | All; instrument-dependent | All |
| | Pile-up – dead or dazzled zone | Large fraction of multiple detection events, high evaporation rates | See entries for correlated evaporation and/or molecular dissociation |
| **Data Analysis** | Peak Overlap | Composition-dependent | $^{64}Zn^{2+}$,$O_2^+$ in ZnO; $^{14}N$, $^{28}Si$ in Si, N containing systems (e.g. TiSiN, GaN:Si); |
| | Peak Ambiguity | Composition-dependent; Molecular ions at low field. | As, P clusters in III-V semiconductors; SiGeC; |
| | Thermal tails | Poor thermal management due to low conductivity or conductance; High Laser Intensity | GaAs; $SrTiO_3$ |
| | Background noise | Instrument-dependent (DUV, EUV); High-Field | $SrTiO_3$ |
| **Physical Channel** | Neutral Evaporation | Systems ejecting neutral species directly or upon molecular dissociation; Low-Field | As in III-As, N in nitrides, O in oxides |
| | Preferential Evaporation | Systems with chemical species with different ZBEF; High-Field, Low Field-Fraction, Low Laser Intensity | Ga in GaAs, GaN, AlGaN, InGaN; In in InN; Zn in (Mg,Zn)O; |
| | Correlated Evaporation | Chemical bond-dependent; Possibly related to retention and migration; Feeding Pile-up | GaSb; GaAs; GaN; TiN; $Fe_nO_m$; $ErMnO_3$; SiC; B in Si:B and SiGe:B; |
| | Molecular Dissociation | Sample-dependent (Low-field) | InP, GaN, Al(Ga)N; $Ga_2O_3$; SiC; |

### *a. Detector losses*

The average system detection efficiency DE is a key parameter to obtain accurate reconstruction of atom coordinates, but also to get reliable measurement of composition, in particular if the efficiency is not homogeneous over the mass range or the energy range of impinging ions. It is therefore important to develop ways to measure efficiently this parameter as a function of the system of interest and conditions of analysis. When focusing on the detection system, we need to evaluate the contribution of the amplification system (the MCPs), and the contribution of the Time and position encoding system (the delay-line anode) (Sijbrandij et al., 1996; Deconihout et al., 2002; Prosa et al., 2014).

#### *i. Multi-channel plate performance*

Microchannel plates (MCPs) are electron multiplier devices that can detect charged particles and fast neutral particles, among other high-energy particles. The MCP consists of micropores that form a honeycomb structure, as shown in Figure [2]. In the event that an incoming particle generates secondary electrons at the input surface of the MCP and these electrons enter the pores under the electric field, then avalanche multiplication of the electron bunch provides a pulsed current. The first step toward generating secondary electrons in the pore is important. In APT, the ratio of the number of detected ions to the number of effective signals that reach the delay-line anode, that is, the MCP detection efficiency ($DE_{MCP}$), is therefore critical.

$DE_{MCP}$ varies with the incident energy, mass, incident angle of detection, and charge state of an ion, and several experimental studies of detection efficiency in an MCP operated in the counting mode have been reported (Fraser, 2002). Recent MCPs have generally a large angle of channels with respect to the face entrance of MCPs (~20°), to avoid variation of gain with angle of incidence of striking ion. $DE_{MCP}$ is therefore considered almost homogeneous over MCPs surface. The main factor affecting $DE_{MCP}$ is the ion of interest velocity. To summarize these previous investigations, the detection efficiency generally increases with the ion impact energy above the threshold for secondary electron emission, and the detection efficiency varies with the ion species. Above 1-2 keV for low mass element (high velocity) and above 3–6 keV for high mass element (low velocity), the detection efficiency reaches a plateau roughly proportional to the open-area ratio (OAR) of the MCP. The fact that the OAR limits the detection efficiency indicates that particles are poorly detected if they do not enter an MCP pore. The OARs of conventional MCPs (hereafter MCPs) are typically 50–65%. Note that this value was confirmed experimentally in APT measurement using the lattice of tungsten atoms as a reference (Deconihout et al., 2002).

In the early 2000s, several attempts to exceed the OAR limit in APT have been made by applying a repelling electric field to reflect secondary electrons back to the MCP surface (Sijbrandij et al., 1996; Deconihout et al., 2002); however, this solution generally increases the occurrence of noise signals due to the emission of secondary electrons by the mesh used to generate the field region. One solution was to increase the open-area ratio of the MCP. An efficient way to accomplish this goal was to narrow the walls of the pores. However, a certain wall thickness is needed to maintain mechanical durability. That is, the OAR has a practical upper limit, usually 80%, and the OAR never approaches 100%. An alternative solution is to expand the entrance of the pores, leaving the inner walls untouched. Detection efficiency up to 90% were reported with some prototypes (Matoba et al., 2011). Commercial MCPs are certified to 80% detection efficiency. Using a protocol of encapsulating a defined volume of a known material, an overall detection efficiency measurement was performed. In Nickel/Chromium multilayer samples (NIST SRM 2135c), DE was found to reach 77±2 % close to the manufacturer value (Prosa et al., 2014).

### *ii. Dead or dazzled zone*

As soon as ions are detected one by one, the detection efficiency DE is equal to $DE_{MCP}$. However, difficulties arise when more than one ion is evaporated and detected from the probed area on a single evaporation pulse. Incident ion hits on the walls of MCP channels generate a short electronic (~ns) pulse from the output. We may note that at high gain, the recovery time of a MCP channel is relatively long (up to 100 µs (Beavis et al., 1989; Westman et al., 1997)), so that the amplitude of electronic pulse is lower for successive impacts at the same position, but this effect generally affects MCP gain at a high flux, much higher than usually used detection rate in APT. The main cause of DE loss, is failure of the position-time encoding system to detect some impacts. Indeed, when two ions reach the detector closely separated in time and in space, one of these ions can be lost due to a masking effect. This phenomenon is named ion pile-up (Figure [11]-(a), (Tsong et al., 1978; Cerezo et al., 1984; Rolander and Andrén, 1989)) and was known to affect atom probe measurement very early in the development of the instrument. It affects in particular correlated field evaporation events and dissociation events, which are the main cause of multi-hit events (or multiple events). We may recall, that even working at low detection rate (<<1 ion/pulse in average), the probability of multiple event (PME), due to intrinsic physical mechanism of emission (De Geuser et al., 2007; Cojocaru-Mirédin et al., 2024), can be large. In some non-metallic alloy the PME can reach 100% of the detection events. In addition, the probability that these multiple events arise from close location is generally high.

Ion pile-up and relative DE loss is linked to capability of the position-time encoding system (namely the DLD) to detect close impacts, in latest atom probe designs. A typical DLD system has an electronic dead-time that is between 1.5 and 20 ns depending on the acquisition electronic system (Da Costa et al., 2005; Gribb et al., 2002; Jagutzki et al., 2002a; m/z Costa et al., 2012; Prosa and Oltman, 2022). During this dead-time, multiple ion impacts on the detector cannot be differentiated from one another. However, if the ions in the multi-hit event arrive sufficiently far apart on the detector, the positions of individual ions may be resolved even if they arrive simultaneously. In DLD this is possible through the time-to-position conversion, 1 ns corresponding to about ~1 mm on large DLD. This defines a 3D dead region in (X,Y,time) space, The physical dead-zone is thus a few millimeters (5 to 20 mm) wide around each ion impact site and span over a few nanoseconds after the initial hit. When plotting the time and space distribution for every second ion related to each first ion hit in multiple events, a depleted region appears; where TOF differences and relative positions do not exceed, respectively, the dimensions of the dead zone around the first point of impact. It was highlighted by Peng et al. (Peng et al., 2018) and Meisenkothen et al. (Meisenkothen et al., 2015) with the analysis of materials producing high PME Figure [11]-(b,c). We may note that the exact shape of the dead region is neither homogeneous nor isotropic, because of the complexity of the encoding process as it was demonstrated by Bacchi et al. (Bacchi et al., 2019).

Depending on the PME and on the complexity of the mass spectra, the selective losses and the exact estimation of the DE can be complex to determine. Some recent corrections procedures, assuming a model of multiple events, were proposed in some cases, but it remains difficult to generalize (Ndiaye et al., 2024; Prosa and Oltman, 2022). However, we may note that some additional information can be used to reduce or compensate the detection losses induced by multiple events. In 1978, Tsong et al. noted that the amplitudes of detection signal vary in height as a function of the number of hits (Tsong et al., 1978). The MCPs are indeed used in a saturation regime, which means that every single-ion impact induces almost the same signal amplitude. However, during a multiple impact when the electrical signals of the ions are superimposed on the MCP, their amplitude is added together. In 1987, using an imaging atom probe, Kellogg demonstrated the good linearity between the signal amplitude and the number of spots induced by field evaporation (Kellogg, 1987). As a result, on modern DLD atom probes, the dead region is indeed not completely "dead", if amplitude of the signal is measured. This region should be rather referred to as the "dazzled region" because the metaphor indicates a reversible state and the region is not completely "blind" with respect to the generation of a signal issued by a second impact. The comparison of the amplitude histograms of single events in Figure [11]-(d,e) shows that the histograms relative to carbon in a SiC APT analysis which appear to be composed of successive convolutions of a histogram like the one produced by silicon. The multiple peaks appearing in the carbon amplitude histogram correspond thus to the trace of double (triple, etc.) impacts. This phenomenon has been consistently interpreted due to the specific tendency of C to correlate evaporation in carbides, yielding a higher fraction of multiple evaporation and detection events for this element, which is consequently more exposed to measurement errors due to the limitations in the detector performances. The observed trend of increase of C loss with increasing depth of analysis can then be explained by the decrease in geometric magnification, making it more and more likely for correlated evaporated ions to fall within the so-called "dazzled region" of the detector (Figure [11]-(f), (Ndiaye et al., 2023a)).

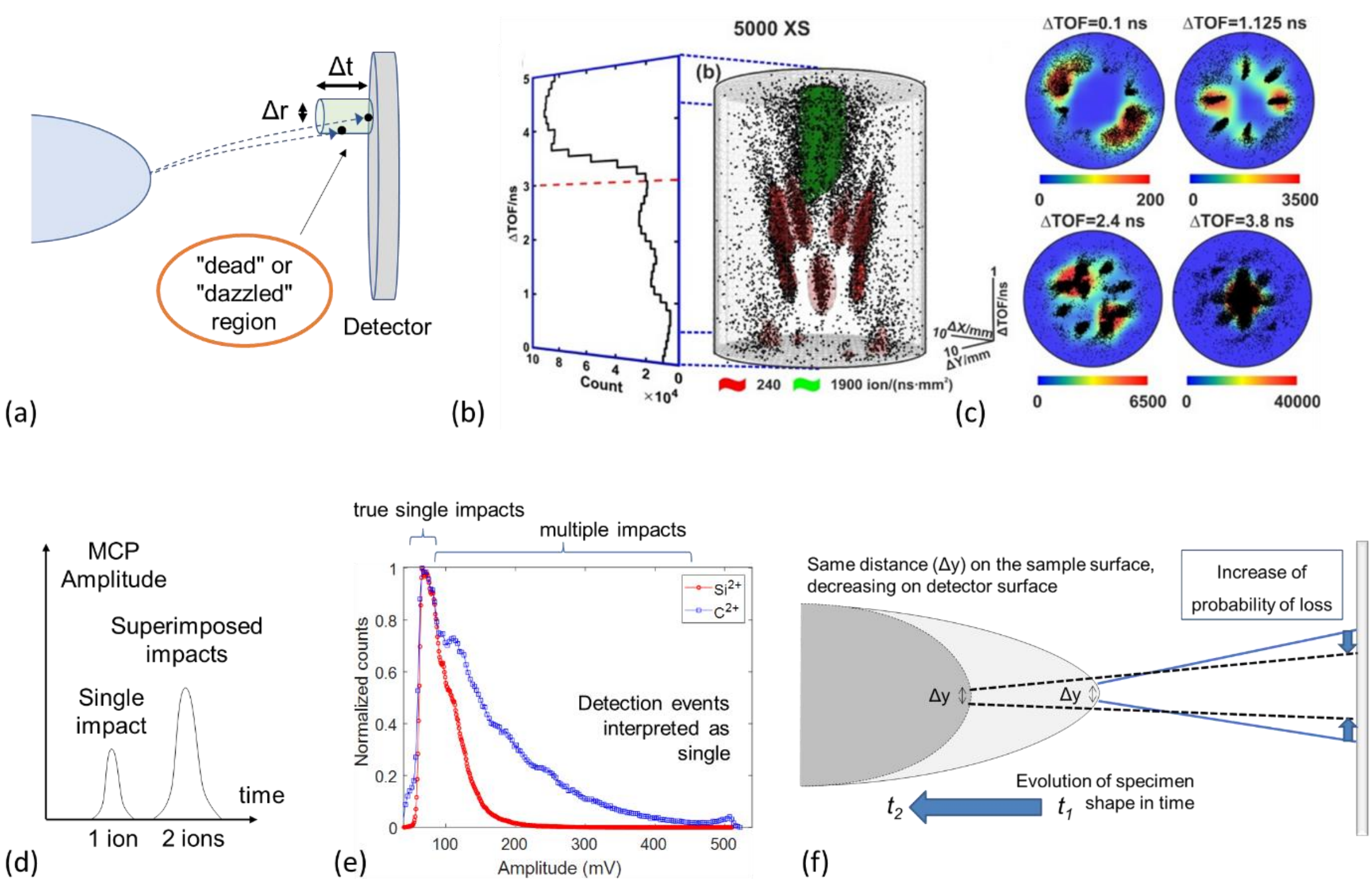

*Figure [11]. Detector losses in APT. (a) Schematic representation of the so-called "dead" or "dazzled" region of a detector, approximately corresponding to a volume within a radius Δr and a time Δt from the first impinging ion issued from a multiple evaporation event; (b, left) Dead zone characterization for a LEAP 5000 XS: amount of ion pairs detected in multiple events as a function of the TOF difference between the two constituent ions and (b, right) 3D time and spatial distribution of the second ion in the ion pair with respect to the first ion; selected iso-density surfaces are marked in red and green; (c) serial top views of density maps over thin slices at different positions of (b,right) (Reproduced with permission from* (Peng et al., 2018)*, Elsevier). (d) Sketch of the dependence of MCP amplitudes on the number of impacts at neighboring distance (i.e. within the dazzled region); (e) Histograms of the amplitudes of $Si^{2+}$ and $C^{2+}$ detection events interpreted as single impacts in the analysis of SiC: the histogram for $C^{2+}$ shows a higher presence of single detection events whit large amplitudes, which implies the presence of a significant number of multiple impacts erroneously interpreted as singles. (f) Dynamical effects influencing detector losses. At increasing time points of the evolution of the specimen, correlated ions of the same mass/charge evaporating from sites at the same distance on the specimen surface impinge with decreasing distance on the detector. This may translate into an increased probability of detector loss via the pile-up effect.*

### *iii. Detection efficiencies*

An evaporated atom has a certain probability of detection, which is related on one hand to the intrinsic efficiency of the detector, and on the other hand to possible physical mechanisms limiting its probability of being detected. For this reason it is useful to introduce some classification in what is referred to as "detection efficiency" (Rigutti et al., 2016a; Di Russo et al., 2018b).

*Efficiency of the detector system*. (section 4.a.i.) The intrinsic efficiency of the detector $DE_{MCP}$ (sometimes indicated as $\eta_D$) corresponds approximately to the OAR. A further reduction in the intrinsic efficiency must be taken into account for the instruments including a reflectron. It can be considered as independent of the incident ionic species (Da Costa, 2016). Typically $DE_{MCP}$ = 60-80% depending on the detector type and generation.

*Specific detection efficiency***.** It is convenient to define a specific detection efficiency $DE_i$ ($\eta_i$), i.e. relative to a chemical species i (atomic ion, molecular ion, element, isotope,...) that can be identified by the analysis of the spectrum of mass:

$$DE_i = \alpha_i DE_{MCP} \quad (22)$$

where $\alpha_i$ corresponds to the fraction of detectable ions, i.e. those which do not evaporate in a loss channel such as preferential evaporation, neutral evaporation or affected by molecular dissociation mechanisms after evaporation (which will be discussed in the following sections). This detection efficiency is not only a characteristic of each ionic species, but can also be considered as a function of the electric field, and of the starting position of the ion on the tip. The specific efficiency is therefore also a local, microscopic efficiency.

*Reconstruction, or global detection efficiency.* It is also possible to define a detection efficiency $DE_{rec}$ ($\eta_{rec}$), related to the reconstruction (i.e. to the algorithm used in order to retrieve volume information from the evaporation sequence of a sufficiently large specimen). This parameter is a constant that the user determines when reconstructing a certain volume. The chosen value is the one that makes it possible to obtain the best spatial fidelity of the 3D reconstruction (section 2.e). Its physical meaningfulness is conditioned by the knowledge of the specific atomic volumes within the analyzed compound. For a structure with known thickness layers the global detection efficiency is that which makes it possible to recover these thicknesses in the 3D reconstruction. It is possible, however, that for particularly deep volumes a constant value of $DE_{rec}$ is not sufficient to ensure the spatial fidelity of the entire reconstruction. In this case, in the absence of a method for determining the detection efficiency as a function of the analysis depth, it is appropriate to carry out the reconstruction of separated layers, each layer adopting a different value of $DE_{rec}$. An estimation of measurement of this detection efficiency was performed in various alloys and compounds by Diercks and Gorman (Diercks and Gorman, 2018), by using correlated electron microscopy measurement of APT specimen before and after APT measurements (Fig. [12]).

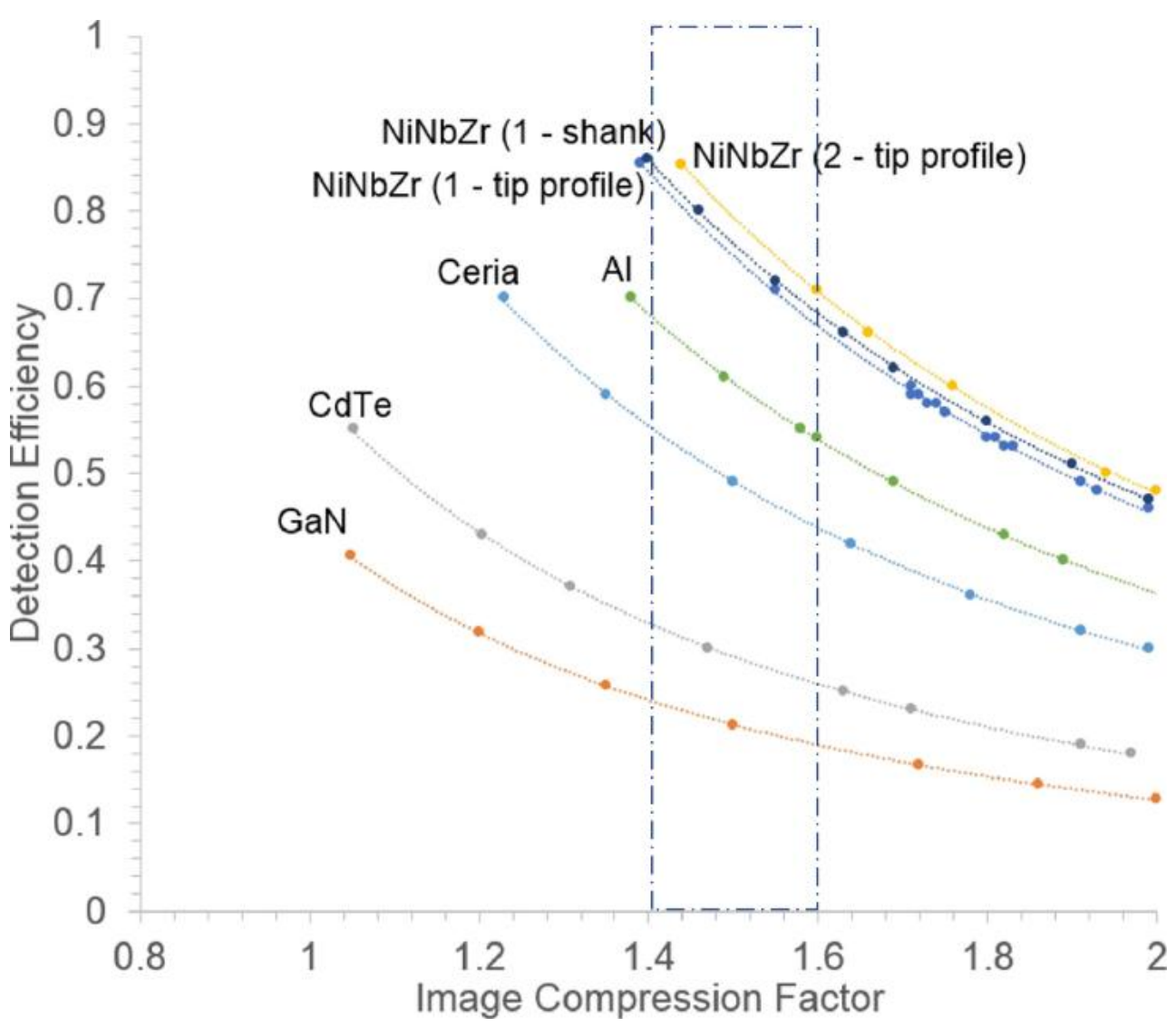

*Figure [12]. Curves of the detection efficiency and image compression factor values for GaN, CdTe, Ceria, and some metallic specimen (Al, NiNbZr alloy) that produce a self-consistent reconstruction. Assuming common values of ICF in the range 1.4-1.6, the detection efficiency in the analysis of GaN, CdTe and Ceria is found much lower than metallic samples, indicating a strong channel of undetected emission (reproduced with permission from (Diercks and Gorman, 2018), Elsevier).*

### b. *Mass peak interpretation*

After detection and data collection, attention must be paid to the interpretation of mass spectra, which should be as accurate as possible. This is not always straightforward in mass spectrometry and in TOF spectrometry in particular, as ambiguities may affect mass spectra, depending on the system under study.

#### i. *Mass peak superposition*

*Mass peak superposition* is a current problem in APT. Some examples of this are given in Fig. [13]-(a,c,d,e). In part (a) the typical mass spectrum of ZnO is affected by the superposition of the $O_2^+$ dication and of the $^{64}Zn^{2+}$ atomic ion at 32 m/*z*. In this case it is quite straightforward to separate the amount of O and the amount of Zn within the peak, under the hypothesis that the isotopic abundances of $Zn^{2+}$ correspond with the natural abundances, and that no other O-containing species are present. A slightly more complex case is presented in part (c), showing a close up of a mass spectrum issued from the analysis of Cu-duped ZnS. The family of $Zn^+$ isotopes can be quite well resolved. However, these peaks are superimposed to $ZnH^+$ peaks, which count for about 1/20 of the total amount of Zn and which are superimposed to the $^{67}Zn^+$ and $^{68}Zn^+$. Furthermore, the detection of Cu is complicated by two factors: first, the $^{63}Cu^+$ peak is quite clearly resolvable, but it is affected by a high background originating from the low-mass tail of $^{64}Zn^+$. A case in which hydride molecules are in even larger amount is presented in part (d) relative to the analysis of a ZrH phase. In this close up, the $Zr^{2+}$ peaks are superimposed to the $ZrH_n^{2+}$ molecular ions. The peak decomposition is shown in the inset, and has been obtained under the hypothesis that $ZrH^{2+}$ ions are present only (Diagne et al., 2025). Peak decomposition or, with a deprecated term, *deconvolution* methods are not within the scope of this review (Johnson et al., 2013; Coakley and Sanford, 2022; Li et al., 2026). They can provide quite accurate composition measurements, but it must be kept in mind that the identity of individual detection events cannot in general be provided (for exceptions and for the statistical interpretation of individual detection events, see for instance (London et al., 2017) and (London, 2019)). This, in turn, complicates and questions the possibility of performing further statistical analyses on the events contained in superimposed mass/charge peaks. Strategies for overcoming these problems are quite demanding and imply either the synthesis of materials using isotopic substitution (D. L.J. Engberg et al., 2018) or the implementation of the instrumentation. The close-up of the spectrum in Fig.[13]-(e) shows the possibility of resolving $^{28}Si^{+}$ and $^{14}N_2^{+}$ at 20 m/z in the analysis of SiN. This spectrum has been obtained by exploiting the capabilities of an instrument featuring a dual Einzel lens configuration and an elongated flight path of about 1m. Furthermore, the data have been selected from a reduced zone of the detector, which increases the mass separating power up to $\Delta m/m$~1400 but also reduces the field of view (Chae et al., 2025).

### *ii. Mass peak ambiguities*

*Mass peak ambiguities* occur when the identity of a peak presents some uncertainty between two or more ion species. This situation may involve a mass peak superposition or not, and is often related to the evaporation of molecular ions or clusters. An example of this is provided in Fig. [13]-(a), where the peak at 16 m/z can be attributed to $O^{+}$, $O_2^{2+}$ or to a mixture of both. A similar situation applies to the peak at 14 m/z in N-containing systems. Currently, arguments based on the high ionization energy for the process $O_2^{+} \rightarrow O_2^{2+}+e^{-}$ tend to suggest that the peak is actually composed of atomic $O^{+}$ (or $N^{+}$, respectively) only (Jaroń-Becker et al., 2004). This sort of ambiguities may be present also in spectra with a large number of relatively complex molecular ions. III-V materials such as arsenides (Fig. [13]-(b)) or phosphides produce a large number of molecular clusters $As_n^{k+}$ or $P_n^{k+}$. In these cases, ambiguities may occur concerning mass peaks potentially formed by ions with the same mass/charge ratio but with different masses. In all these cases, the multiplicity of the atom within the molecular ion is a factor directly influencing the calculation of the composition (Di Russo et al., 2017; E. Di Russo et al., 2020).

The occurrence of molecular ions, usually enhanced in low-field conditions, is a common phenomenon in field evaporation of dielectrics and often complicates the interpretation of mass/charge spectra. If they are not producing mass peak superposition or other ambiguities, and if they are not associated to more complex phenomena such as molecular dissociation and correlated evaporation (see sections 4.c.iv and 4.c.iii, respectively), their presence does not prevent the user from the determination of a correct composition, provided the multiplicity of each atomic species is correctly accounted for. However, molecular ions are generally a hint about the complexity of chemical processes at the surface of field emitters (Cojocaru-Mirédin et al., 2024), and are in general associated with a degradation of the spatial precision (Gault et al., 2026). Therefore, the user should always be cautious in the application of advanced data mining protocols to datasets containing large amounts of molecular ions.

### *iii. Thermal tails and background noise*

*Thermal tails*. Thermal tails are due to delayed evaporation during the phase of heat relaxation after a laser pulse. They induce broadening and skewness in mass/charge peaks, and can therefore lead to the increase of background for neighboring peaks or even to significant difficulties in identifying them and quantifying the associated signal. The impact of thermal tails can be reduced by appropriate choice of experimental parameters such as laser intensity and wavelength (Vurpillot et al., 2009; Vella, 2013; Kirchhofer et al., 2013; Caplins et al., 2023), as well as sample shape (Arnoldi et al., 2012). On the other hand, various methods are available for the optimization of the mass spectrum analysis in case of peak overlaps with thermal tails (London, 2019).

*Background noise* is generally referred to the random detection of ionized gaseous species coming from the analysis chamber or species adsorbed at the surface of the specimen. These atoms/molecules being ionized at the DC voltage, are detected at any time during the opening of the detection window and constitute a continuous signal in the time-of-flight spectra. Another small contribution may be originated from the detector itself (Pareige et al., 2016). In instrument pulsed with deep or extreme UV light, pulsing may contribute to the increase of background noise too (Miaja-Avila et al., 2021). Solute species are obviously the most exposed to loss related to background noise, as this increases the LOD.

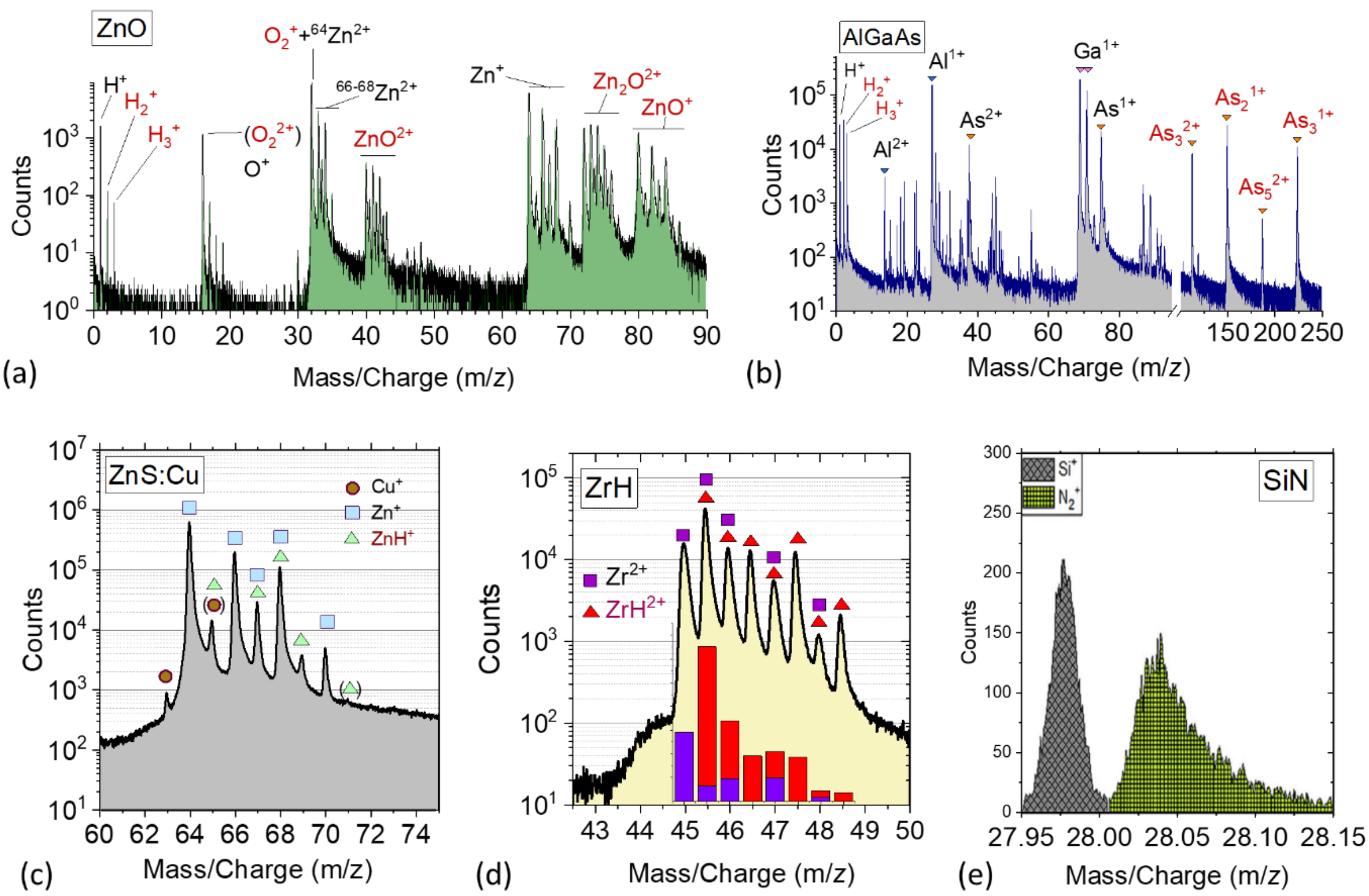

***Fig. [13].*** *Examples of mass/charge spectra potentially exhibiting problems for the attribution of ion species. (a) The shown mass/charge spectrum of ZnO is affected by (i) the significant presence of atomic and molecular hydrogen (ii) the potential ambiguity in attributing the peak at 16 m/z to* $O_+$ *or* $O_2^{2+}$*, (iii) the superposition of the peaks of* $O_2^+$ *and* $^{64}Zn^{2+}$ *at 32 m/z and (iv) the presence of a large amount of molecular ions (highlighted with the red text). (b) The spectrum of AlGaAs exhibits a large amount of* $As_n^{k+}$ *molecular clusters (red text). (c) The close-up of the mass/charge spectrum of ZnS:Cu shows the superposition of* $Zn^+$ *peaks (indicated by blue squares) with the peaks of* $ZnH^+$ *(green triangles). The* $^{63}Cu^+$ *peak can be recognized but is largely superimposed to the low-mass tail of majoritarian* $^{64}Zn^+$*. Due to the presence of* $^{64}ZnH^+$ *(and, secondarily, to the high-mass tail of* $^{64}Zn^+$*), the signal of* $^{65}Cu^+$ *is not recoverable. (d) Close up of the mass/charge spectrum of ZrH in the region containing the superimposed signals of* $Zr^{2+}$ *and* $ZrH^{2+}$ *ions. The decomposition of these peaks is shown in the histogram in the inset. (e) High-resolution mass/charge spectrum of SiN showing the possibility of resolution of* $^{28}Si^+$ *and* $^{14}N_2^+$ *peaks (Adapted with permission from (Chae et al., 2025), Springer Nature).*

### c. *Physical channels*

Physical channels of loss reflect the complexity of field ion evaporation, and the departure of its reality from the picture of ionization of atomic species. Neutral evaporation, preferential evaporation, correlated evaporation, molecular evaporation with or without dissociation can significantly complicate not only the appearance and the interpretation of mass spectra, but also the detection process itself. These phenomena can often be assessed and sometimes quantified; however, they should be managed as carefully as possible because they are not avoidable in most non-metallic materials.

#### i. *Neutral evaporation*

Neutral evaporation is the most “devious” mechanism of loss in APT. Because neutral evaporation may affect differently chemical species of the sample of interest, it can lead to a direct systematic bias in the measurement of composition. We recall that MCP detection gain is strongly dependent of the particle kinetic energy (or more rigorously particle velocity) below 1-2 keV. Neutral particles, directly emitted from the specimen surface, are not accelerated by the distribution of field close to the emitter, and are therefore completely undetectable in an APT configuration. Furthermore, the change in vacuum pressure that could be linked to an emission, in the form of an atomic gas, of neutral particles, is extremely limited due to the limited size of the probed volume in APT. A simple calculation of the increase of the gas pressure from the completed sublimation of the top apex will be significantly lower than that of the residual pressure in the analysis chamber (~$10^{-9}$ Pa in the best case). In this review we will also differentiate the case of neutral emission induced by dissociation of cationic molecules to the direct emission of atoms, or some neutral molecules from the surface. This emission will be described in the section related to molecular dissociation and Saxey’s plot representation.

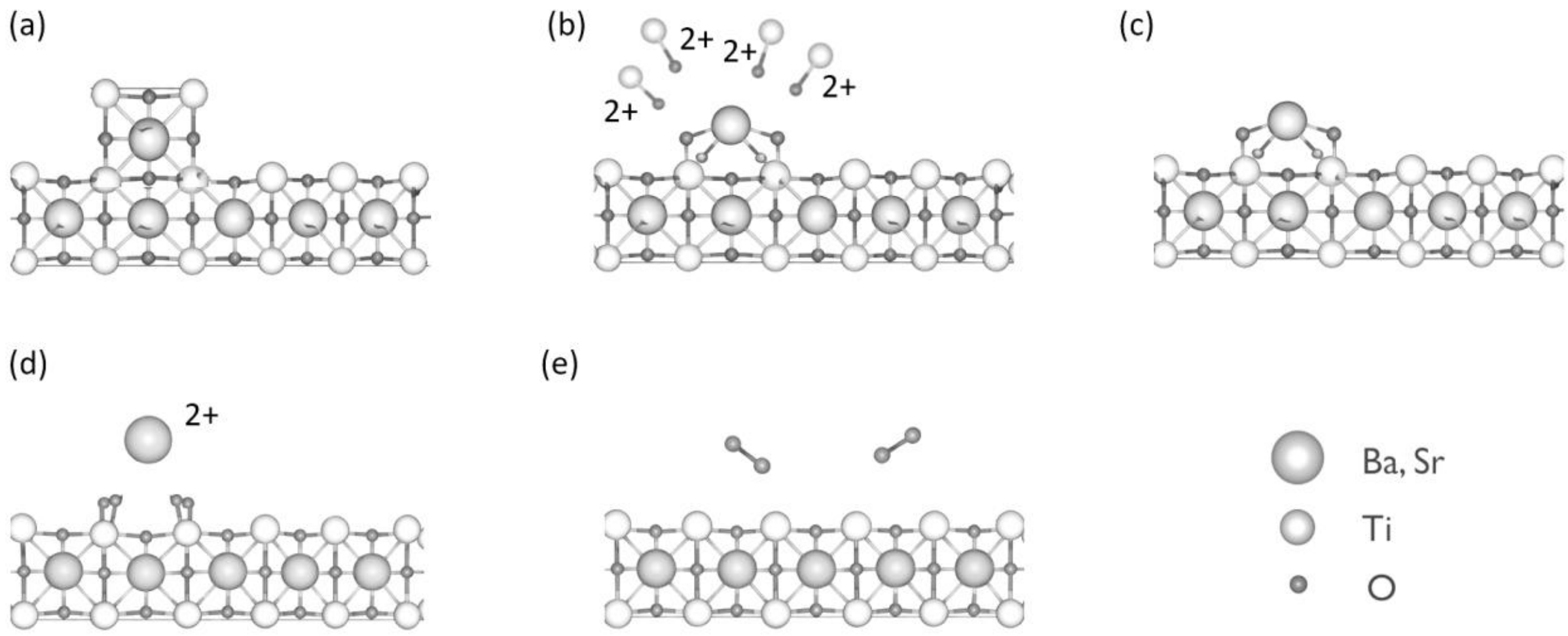

*Figure [14]. Sketch illustrating the proposed field evaporation behavior for the Ba:STO layers (a) the sample structure (assuming a perovskite structure); (b) Field evaporation of the $TiO^{2+}$ leaving behind a Ba/Sr rich surface; (c) the Ba/Sr-O bonds become stretched due to ionic polarization caused by the field; (d) $Ba^{2+}/Sr^{2+}$ are field evaporated; (e) The oxygen left behind is then thermally desorbed as neutrals. (Scheme adapted from (Morris et al., 2024))*

Because there is no direct proof of the direct emission of neutrals in the high electric field conditions of an APT, the literature of the effect is scarce. Several studies have proposed that the discrepancy between the APT measured composition of compound semiconductors and oxides and their actual compositions (which are normally stochiometric and well established) was mainly due to thermal desorption of the nonmetallic element present in the material. This was for instance proposed for oxides (Devaraj et al., 2013) (Figure [14]) or nitrides (Mancini et al., 2014a). The literature often proposed the emission of gaseous $N_2$ and $O_2$ in nitrides and oxides directly from the surface. The formation of a significant amount of these molecules which are at the end desorbed or sublimated without ionization is evoked. This hypothesis is linked to the variation of the measured composition with the surface electric field. When increasing the electric field, the non-metallic element proportion is always seen to increase progressively, indicating a strong relationship between the probability of ionization and the electric field. On a zinc oxide (ZnO) surface, Xia et al. simulated the surface stability under a positive electric field ranging from 25 to 30 V/nm (Karahka et al., 2015; Xia et al., 2015). Above 15 V/nm, the surface undergoes reorganization due to the electrostatic field effect, and beyond 25 V/nm, the geometry optimization no longer converges. The surface rapidly emits metallic species (in the form of $Zn^{2+}$), while oxygen atoms rearrange quickly to form $O_2$ molecules.

The hypothesis of neutral emission is also reinforced by the absence of evidence of strong biases that may be due to multi-hits events and detector pile-up. This last hypothesis was for instance ruled out using of a specially designed meshed electrode placed in front of the specimen to manipulate ion pile-up from co-evaporated ions (Morris et al., 2022). Pile-up is therefore not the major cause of composition inaccuracy in GaN and AlN, or TiN (Schiester et al., 2024). The preferential field evaporation (Hatzoglou et al., 2020)(described hereafter) was also ruled out, since the amount of unwindowed detection events in mass spectra is not found to drastically increase at high laser pulse energy / low field. In addition, an estimation of measurement of the detection efficiency was performed in various alloys and compounds by Diercks and Gorman (Diercks and Gorman, 2018), by using correlated electron microscopy measurement of APT specimen before and after APT measurements. Using this approach, an empirical relationship between the image compression factor and detection efficiency is found for all of the analyzed specimens. Since ICF is generally in the range [1.4-1.6] in APT, the detection efficiency is found to be lower than expected reinforcing the hypothesis of undetected neutral emission (figure [12]). The same observed low estimated value was also measured in ternary semiconductor alloys by Di Russo et al. (Di Russo et al., 2018b) A more robust similar approach, reinforced by numerical computation of the ICF in the geometry of the specimen, was also proposed by Veret et al. in the case of monazite $CePO_4$ indicating a systematic degradation of the detection efficiency, and loss of the non-metallic element probably by neutral emission (Veret et al., 2026).

The neutral emission mechanism appears to correspond to experimental observations in various oxides and nitrides. However, this mechanism relies on the fundamental hypothesis that a neutral atom or small neutral molecule may first exit the extreme field region, and second may survive the extreme electric field that exists in front of the specimen. While many authors seem to identify the direct emission of neutral molecules from the surface, one may question the likelihood of such an event occurring on the specimen, which is subjected to an electric field exceeding 20 V/nm in most of experimental cases.

Indeed, two physical phenomena oppose this mechanism. First, under these field strengths, if neutral particle emission were to occur, the particle would be strongly polarized by the electric field. Gault indicated that under the influence of the electric field, and due to the atomic polarizability of neutral molecules, the binding energy associated with surface polarization is on the order of 0.2 to 0.45 eV under a field between 20 and 30 V/nm (Gault et al., 2016) (see field polarization, section 3.b). A molecule desorbed due to thermal excitation (a few hundred K, 10 to 100 meV) has a relatively low velocity induced normally by the thermal kinetic energy due to the surface specimen temperature. We may note that the potential well induced by polarization effect is linked to the gradient of field in front of the surface (so extending on several tens of nanometer). A neutral molecule at the temperature induced by the laser pulse would normally shift only by a few nm from its initial position before being recaptured by the specimen. During the travel, density functional theory (DFT) models predict the ionization of these neutral particles (as found by Karakha for instance (Karahka and Kreuzer, 2015; Karahka et al., 2015)). Nevertheless, this is fundamentally linked to the limitations of DFT, which neglects reaction kinetics and assumes the electronic ground state at each time step during relaxation. In reality, it should be necessary to evaluate the probability of ionization throughout the entire process of molecule or atom expulsion.

This kind of calculation was done in the process of field ionization or post field ionization, in the context of a single atom or molecule escaping the surface (Haydock and Kingham, 1981; Kingham, 1982) and were recently used to calculate the probability of ionization of a neutral atom or molecule leaving the surface with an initial velocity (Veret et al., 2026). They make use of ionization theory as described in section 3.c. This approach calculates the integrated probability of ionization of a particle traveling from the specimen surface to the infinite using Schrödinger equation in simple model potential of the atom under the presence of a surface electrostatic field. For a single particle of charge leaving a planar surface submitted to a field $F$, the probability of ionization or post ionization depends essentially on the transparency of the electronic tunnel barrier existing between the center of the atom and the specimen surface. The lifetime of neutral atom or molecule is dependent on the time spent in the ionization zone which is inversely dependent of the speed of the particle. A significant shift in the ionization probability with field is predicted with kinetic energy of more than 1 eV (Veret et al., 2026). The same effect was predicted in the context of FIM by Haydock and Kingham, (Haydock and Kingham, 1981). This process is also compatible with the required energy to be liberated from the polarization effect described previously (the kinetic energy must be higher than 0.2-0.4 eV). We may note that molecules in excited states in APT were observed concurrently in some other studies during the process of field evaporation (Vurpillot et al., 2018; Zanuttini et al., 2017a). Indeed, some dissociation channels observed experimentally were associated to the presence of strongly excited internal states in $SiO_2$ or GaN, normally not existing at moderate temperature ($\ll$ 1000K). Shifts in the energy curves of electrons could reach several eV, indicating that the process of field evaporation significantly populates these states. We postulate that the atomic surface reconstruction following the emission of metallic species, as modelled by Kreuzer et al. and as proposed in Figure [14], could produce excited molecules. It could be likely for oxygen atoms of nitrogen atoms, that are not linked together in the original structures. We may note that further experimental and theoretical efforts are needed to found a consensus about this mechanism of loss in the APT community.

#### ii. *Preferential evaporation*

Historically, preferential field evaporation was identified as the most obvious physical phenomenon that induced biases in the composition measurement. The term "preferential" indicates that one or more species are easier to evaporate and may also do it in really not convenient moments, i.e. out of laser or voltage pulse, which makes them either undetectable or unidentifiable in mass spectra. Anomalous behavior and subsequent biases have been observed in the quantitative analysis of solute alloying elements, depending on the analysis conditions. The most sticking example is the case of FeCu and FeSi, where the apparent concentration of solute copper (Cu) in ferritic iron (Fe) obtained by APT tends to be much lower than the actual concentration as the specimen temperature increases, whereas the apparent concentration of solute silicon (Si) in Fe tends to be higher than the actual concentration (Miller and Smith, 1981; Worrall and Smith, 1986; Danoix et al., 2001; Miller et al., 1996; Yamaguchi et al., 2009b). Such phenomena have been qualitatively interpreted by the mechanism referred to as preferential field evaporation. A qualitative interpretation is given in both laser or voltage pulsing mode, using theoretical evaporation rate equations expressed in eq. (5).

In pulsed mode, the pulse amplitude has to be adjusted to avoid the preferential departure or retention of any given species at the surface of the specimen. This is generally explained with simple arguments using a graphical interpretation of field evaporation behavior in an alloy AB as illustrated in Figure [15]-a), which could be found in seminal atom probe books (Gault et al., 2012b, p. 138; David J. Larson et al., 2013, p. 83; Miller and Forbes, 2014c, p. 265). The black dots correspond to a setting with which species A and B will both have their field evaporation controlled by the application of the voltage pulses (scenarios 1 and 4). This is in contrast with the two scenarios represented by grey symbols (scenario 2 and 3). In scenario 2, atom A, with a higher evaporation field, will be preferentially retained on the surface because the pulsed field is not high enough to promote its field evaporation. Scenario 3 would result in the preferential evaporation of the B species because the field evaporation of the B atoms can be provoked by the combination of the temperature and the DC voltage, without the need for the pulsed electric field.

The basic theory of field evaporation can be used to explain this representation and predict how the evaporation field vary in an alloy for each species (more precisely the field at which the evaporation rate reaches the needed value for analysis, and not the theoretical ZBEF). Experimentally, this corresponds to changing both the field strength and temperature simultaneously, so as to maintain a constant rate of field evaporation. For a material A of ZBEF $F_A$ and field sensitivity $C_{A,}$ equation (5) may be rearranged to

$$\frac{F}{F_A} = 1 - \frac{k_B T}{C_A} ln\left(\frac{K_0}{K}\right) \qquad (23)$$

*K* is here the evaporation rate at the tip surface, proportional to the detection rate. $K_0$ is the evaporation rate for a zero barrier. This critical field strength is proportional to the applied voltage V, and following this simple approach, linearly dependent of applied temperature T. We may note that the slope of this linear expression is dependent to the sensitivity to the field as defined in equation (6). $C_A$ is generally linked to $F_{A,}$ ZBEF is normally stronger for element with higher *C*, since C and ZBEV are both dependent on the binding energy of atoms to the surface. Field evaporation of the surface is therefore easily tuned by either increasing the field (through the applied voltage in Voltage-pulsed APT) or the specimen temperature (through laser pulsing in LA-APT). Practically, the DC field for a standing temperature $T_0$ is set at a field *F* ensuring a very low evaporation rate $K_{DC}$ between pulses, and evaporation rate on pulse is increased to a rate $K_{pulse}$, so that the DC field evaporation is significantly lower than the detection rate on pulses generating mass peak on the mass spectrum. We may note that this difference must integrate the duty cycle of the acquisition sequence (duration of the DC evaporation is ~$10^4$ the duration of the field evaporation on pulses). Assuming classical values $F_A$ =3.5V/Å, and $C_A$ =1eV, considering that less than 1% of atoms are evaporated between pulses (DC noise), and following (23), the required evaporation field pulse (through voltage pulse $V_P$) can be estimated. $V_P$ must be larger than 4% of the DC voltage $V_{50}$ at 50K and 8% of the DC voltage $V_{100}$ at 100K in VP mode. At the same $V_{50}$ and $V_{100}$, temperature pulse of respectively ~150K and ~300K are necessary to generate the same evaporation rate in laser assisted mode (Figure [15]-b)). The critical parameter controlling the required pulse fraction (i.e. the ratio $V_P/V_{DC}$) to ensure low DC field evaporation for a given temperature is here the amplitude of the sensitivity to field $C_{A.}$ An example for a case B with $C_B$ =0.5eV and $F_B$=2.5V/Å is shown in (Figure [15]-c)). A pulse fraction of 16% is necessary in this case to ensure the good conditions of analysis.

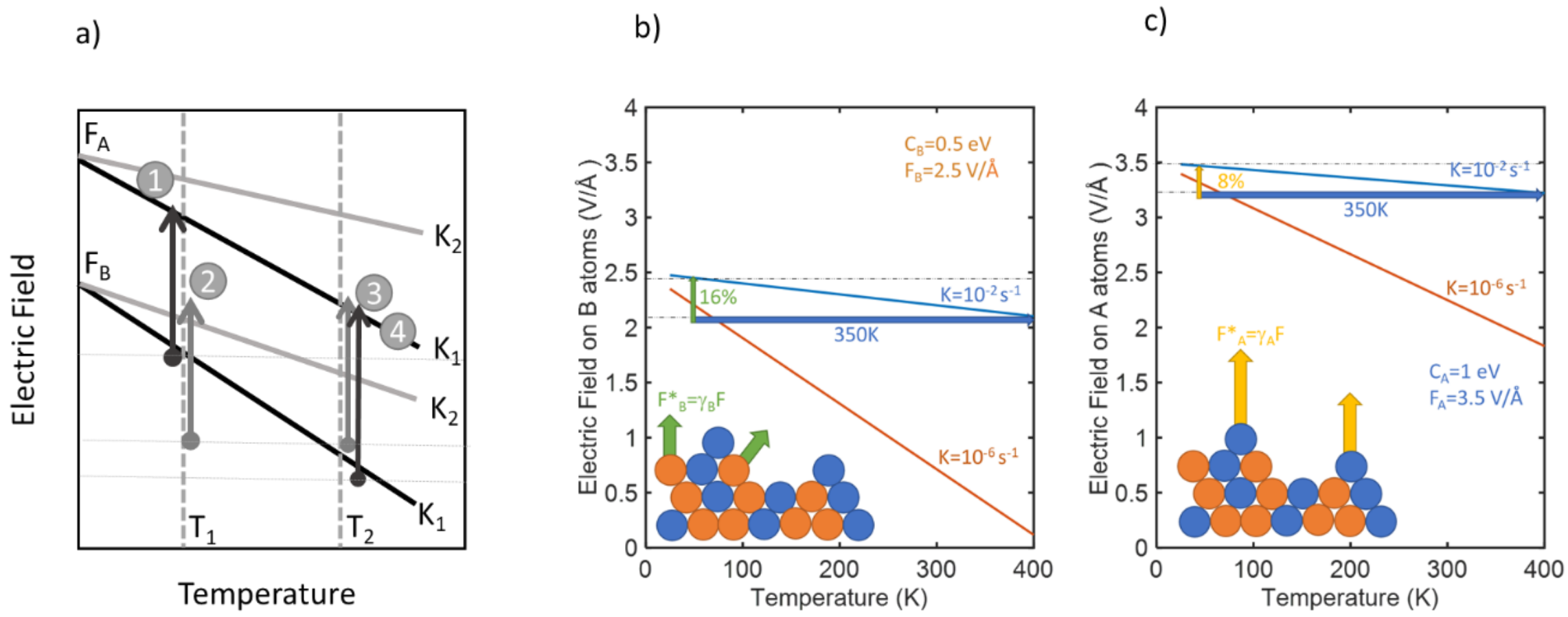

*Figure [15]. a) Classical interpretation of preferential field evaporation of a binary alloy. Electric field and base temperature calibration, illustrating various scenarios of electric field and base temperature combinations. The arrows represent the pulsed field. F1 and F2 are evaporation rates, $F_A$ and $F_B$ are the evaporation field of the A and B species, respectively and F1, F2, and F3 are different DC fields. Scenario 2 corresponds to the preferential retention of A (grey dot) b) and c) More realistic comparison between the constant evaporation probability of two species A (left) and B (right) having different ZBEF ($F_A$, $F_B$) and activation energy barrier ($C_A$,$C_B$) – a case potentially leading to the loss of B atoms by preferential evaporation. Since local field F* is different above A and B atoms for the same applied voltage, the two graphs cannot be concatenated (as done in a).*

When dealing with a compound, composed of these two elements A and B, preferential field evaporation is often interpreted by superimposing the two graphs of the two elements on the same scale of F and T. With an evaporation field difference of 40% between A and B, only with a large pulse fraction (larger that this value), good conditions of analysis may be found, by setting the DC field below the lowest curves of both species. *Nevertheless, this interpretation is over simplistic and must be avoided*, since the actual microscopic field $F_\mu$ over each atom may vary significantly due to the local depletion or retention of the neighbor atoms under field evaporation. When analyzing an alloy, the low evaporation field atoms are first removed from the surface, increasing the protrusion degree of high evaporation field neighbor atoms ($F^*=\gamma F$, with $\gamma$ a field enhancement factor). Classical electrostatics showed that the local microscopic field (see Table [II]) can be theoretically enhanced by a factor $\gamma$=2 for an atom situated on top of a planar surface for instance (Suchorski et al., 1995), with actual measured value that can reach $\gamma$=1.5. We may assume that due to this process, a quasi-equilibrium shape of the surface builds up so that the microscopic field over each atom will be close to the evaporation field of each atom. This interplay between geometry and field was demonstrated in atomistic modelling approach for instance (Vurpillot et al., 2000). In this case, a direct use of these diagrams is clearly difficult in the context of predicting composition biases.

Recently more advanced phenomenological models were developed to give a proper interpretation. These models consider the coupled evaporation rate of elements on pulse and between pulse in the analysis of binary or ternary alloy. A first analytical model was developed by Takahashi et al. to predict such phenomena in VP mode (Takahashi and Kawakami, 2014), and was improved by Hatzoglou et al. some years after (Hatzoglou et al., 2020). This model provides the more favorable analysis condition while maintaining high measurement accuracy. This model explained the temperature dependence on the apparent concentration of solute element whereas the dependence of the pulse fraction and pulse frequency is not completely explained.

An example of result is presented in Figure [16] for an FeCu alloy with 1.2% of Cu dispersed in the Fe matrix. However, the application of these models to laser assisted field evaporation is more tedious since there is no variation of the local field ($V$ is constant) and therefore no variation of the evaporation barrier on the evaporation pulse, or between pulses. Experimentally, the base temperature has also a second order effect on the composition bias, indicating that it is not possible to control the DC field evaporation of the low evaporation field species.

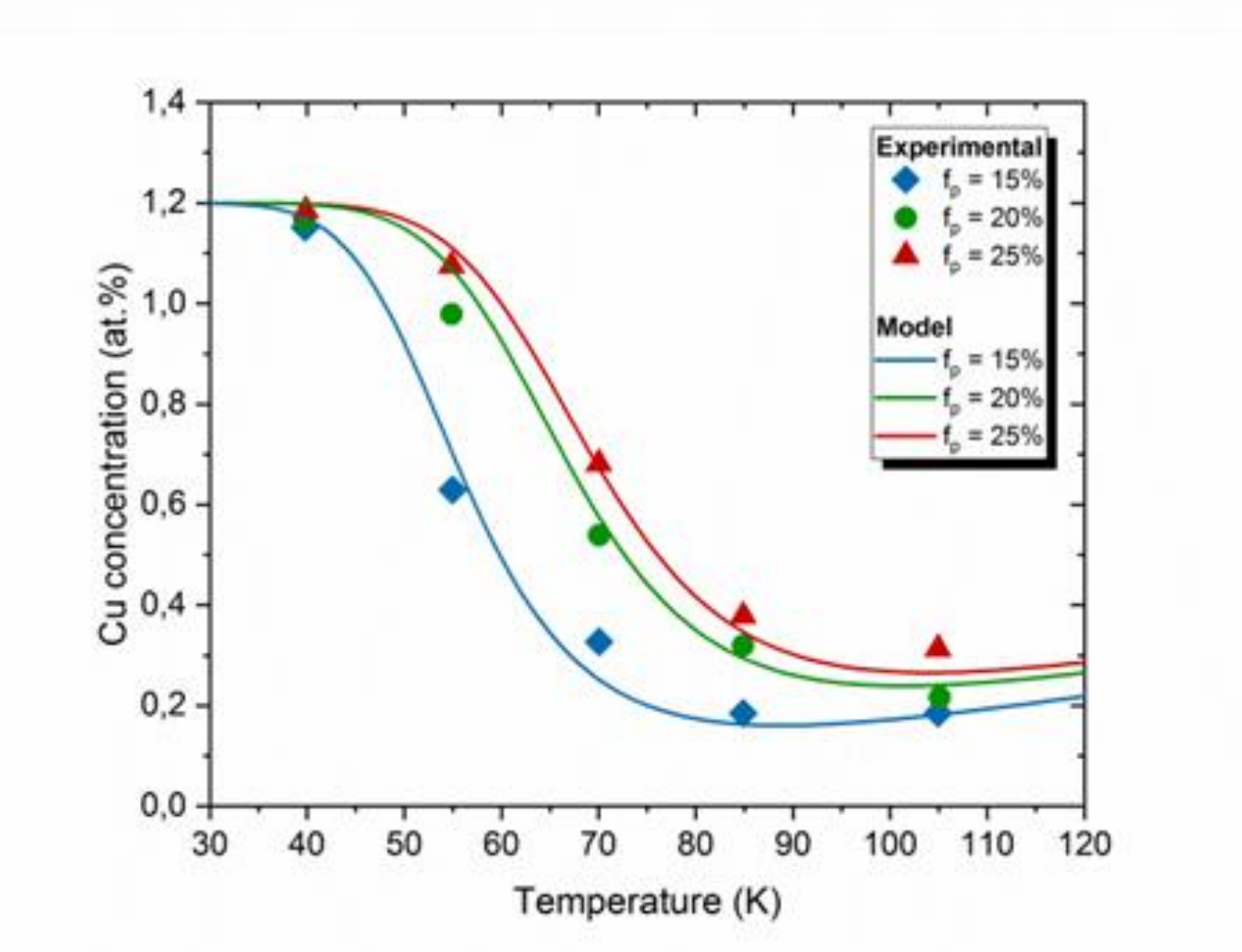

*Figure [16]. Cu compositions (at.%) as a function of the analysis temperature (K) for different pulse fractions fp=$V_P$/$V_{DC}$ (15, 20, and 25%). Experimental data (dots) have been extracted from Takahashi and Kawakami (2014). Full lines are obtained from the analytical model model proposed by Hatzoglou using variables energetic constants of Fe and Cu ($C_{Fe}$~0.5 eV; $C_{Cu}$ ~0.3 eV). Reproduced with permission from (Hatzoglou et al., 2020), Oxford University Press)*

Although preferential field evaporation is suspected in the analysis of binary or ternary semiconductor compound, a quantitative model for evaluating the difference between the real and apparent concentrations of alloying elements has not yet been proposed in laser assisted mode. Quantitative discussion concerned with measurement condition has not been carried out sufficiently (Di Russo et al., 2018b).

If the quantitative estimation of the apparent concentration can be accomplished, preferable measurement conditions can be easily suggested for different alloys. It is generally accepted that quantitative analyses are preferably realized under low specimen temperature, high-pulse fraction, and high-pulse frequency in the case of voltage pulsing. A low temperature reduces the difference in the evaporation field between the alloying element and a high-pulse frequency reduces the period of DC evaporation. In the case of laser pulsing, a high LPE is generally beneficial to reduce preferential evaporation, however it may promote diffusion and loss of light elements such as O and N in oxides and nitrides (Mancini et al., 2014a). However, it is often not convenient to apply these preferable measurement conditions because of an increased possibility of the fracture of needle-shaped specimens under such strict conditions. Therefore, it is necessary to choose measurement conditions that realize a better balance between a reduced possibility of specimen rupture and high quantitative performance.

### *iii. Correlated evaporation and multiple detection events*

Correlated evaporation occurs when two or more ions are emitted from the tip within the time window of the evaporation trigger pulse, regardless of whether the pulse is electrical or laser-based. If significant correlated evaporation occurs, the frequency of multiple detection events as a function of the number of detected ions is significantly different from the Poissonian statistics that would be observed if evaporation proceeded randomly in space and time, as illustrated in Fig. [17]-(a) (De Geuser et al., 2007). Evaporated ions may be emitted from adjacent regions or from points on the tip that are sufficiently separated so that the departure of one ion does not alter the electric field experienced by the second atom on the surface. In the first case, the departure of an ion can immediately trigger the evaporation of its nearest neighbor atoms (Müller et al., 2011). This typically occurs because these atoms are located at the top of a terrace, or at neighboring sites along terrace edges, as reproduced in the simulation data of Fig. [17]-(b). These sites correspond to regions of high local electric field due to the reduced curvature radius at the atomic scale. The departure of an ion can significantly alter the local field distribution, thereby inducing the evaporation of one or more additional atoms. Because of their close proximity on the tip, neighbouring atoms are projected onto nearby regions of the detector, leading to spatially correlated detections (Fig. [17]-(c)). In the second case, the departure of two or more ions is simply due to the temporal coincidence of these events (i.e., occurring during the same trigger pulse). However, their projection points on the detector are generally well separated, and the multi-ion detection arises solely from a temporally correlated event (Fig. [17]-(c)). Spatially correlated detection events may or may not be resolved by the detector, depending on its multi-hit capabilities, which should be sufficient to prevent the occurrence of so-called pile-up phenomena (see sections 2.c and 4.a.ii).

The field evaporation of B in Si provides a clear example of compositional bias induced by correlated evaporation effects. In 2012, m/z Costa et al. have analyzed [100]-oriented Si wafers implanted with a very high dose of $^{11}$B atoms ($10^{17}$ at./cm$^2$) (Da Costa et al., 2012). Because the evaporation field of B (64 V/nm) is higher than that of Si (33 V/nm), B is prone to significant retention effects. The use of a conventional delay-line detector (DLD) led to a selective loss of B atoms due to pronounced pile-up effects. In contrast, the use of an advanced delay-line detector (aDLD) allowed the detection of up to 30 ions from a single evaporation pulse. The capability to record multi-ion events is therefore critical for accurate compositional analysis and has been associated with an increase in the measured concentration associated to the B peak from 5 to 12 at.%. (section 2.c.iii).

Spatial correlation is not exclusively acting over the short time scale of a single pulse, but may persist over several periods. In this case spatial correlations do not yield multiple detection events, but may be assessed through the analysis of detection events occurring after one or several pulses, also referred to as *pseudo-multiples* (Müller et al., 2011; Ndiaye et al., 2024) (section 4.d.v).

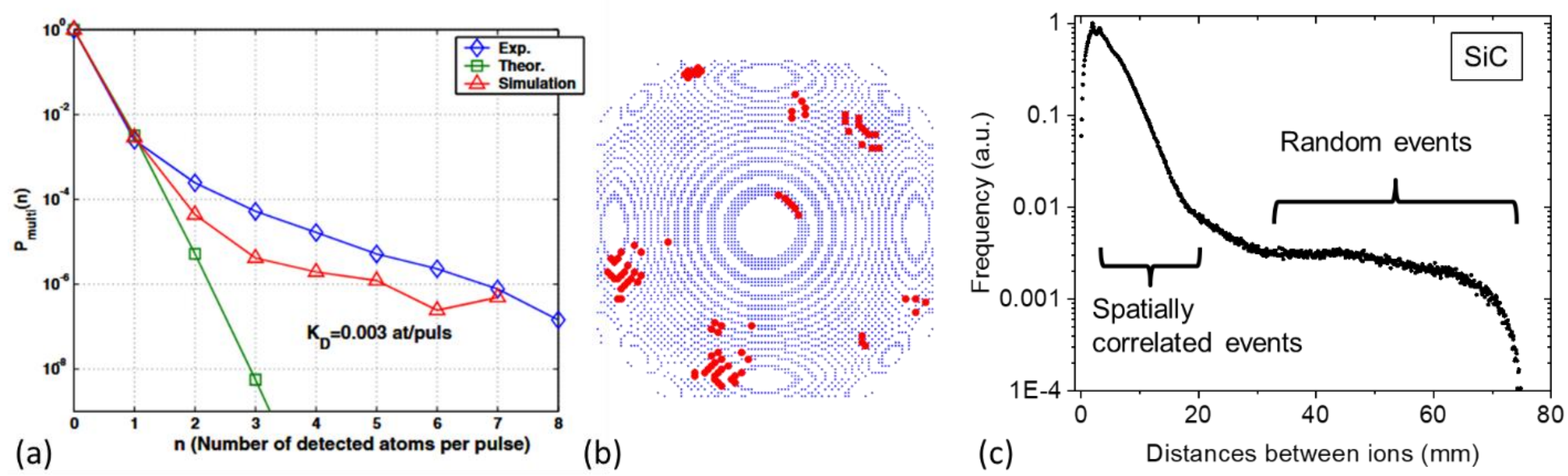

*Figure [17]. Correlated evaporation in APT. (a) Comparison of statistics of multiple detection events for purely random (uncorrelated) evaporation (green squares), experimental data (blue diamonds) and simulation accounting for spatial correlations in evaporation. (b) Surface of a simulated sample (blue dots) at a given time. The red circles are the next 100 atoms in the evaporation sequence and their positions clearly indicate spatial correlation in the sequence (Reproduced with permission from (De Geuser et al., 2007), Elsevier)). (c) Statistics of distances of ion impacts on the detector resulting from the analysis of SiC.*

### *iv. Molecular dissociation*

In APT, molecular dissociation occurs when field-evaporated molecular ions – such as metal-oxygen or nitride complexes—fragment during flight to the detector due to intrinsic metastability or bond softening in the high electric field. If this process is fast enough, i.e. if it produces neutral atoms or fragments within the first few angstroms from the surface, the neutral fragments evade detection, leading to systematic undercounting of specific elements. For example, in ZnO, dicationic species like $ZnO^{2+}$ dissociate into charged Zn ions and neutral O atoms, resulting in apparent oxygen loss and metal enrichment. Simulations indicate that stronger fields (~10 V/nm) favor neutral O emission over fully ionized products, directly biasing bulk compositions, particularly in oxides and non-metals where molecular ions dominate (see section 5.a.iii on ZnO and related reference (Zanuttini et al., 2017a)). These problems also concern dissociation reactions in which neutral fragments can still be detected, as in $SiO_2$, (Zanuttini et al., 2017b). Undetected neutral fragments may also leave a trace of their presence by influencing the position of two charged fragments in correlation histograms, as in the case of the reaction $\mathrm{GaN_3^{2+}} \rightarrow \mathrm{Ga^+} + \mathrm{N^+} + \mathrm{N_2}$ occurring during field evaporation of GaN (Gault et al., 2016).

Three types of dissociations can be distinguished base on the analysis of the correlation histograms (Section 2.f):

*(i) Heterolytic dissociations with production of two charged fragments.* If the dissociation occurs near the tip surface, both daughter ions travel through the entire potential drop and are detected with mass/charge ratios reflecting their own identities. If the dissociation happens further from the tip, the measured values are closer to the mass/charge ratio of the parent ion. When dissociation occurs mid-flight, the measured ratios are intermediate, depending on how much of the potential drop was traversed before and after dissociation. Assuming velocity conservation, the measured mass/charge ratio $m_1'$ of one fragment is given by (here and in the following, we assume that the parent ion is doubly charged and the daughter ions are simply charged):

$$m_1' = m_1 \left[1 - \frac{V_d}{V_{DC}}\left(1 - \frac{m_1}{m_p}\right)\right]^{-1} \qquad (24)$$

Here, $m_1$ is the actual mass/charge of the daughter ion, $m_p$ is the mass of the parent ion, $V_d$ is the potential at the dissociation point, $V_{DC}$ is the tip voltage (Saxey, 2011). The second ion's measured value, $m_2'$ , is then:

$$m_2' = m_2 \left[1 - \frac{m_p - m_2}{m_p - m_1}\left(1 - \frac{m_1}{m_1'}\right)\right]^{-1} \qquad (25)$$

In the histogram (Fig. [4]-(a), see also the green features in Fig. [4]-(b,c)), such dissociation tracks start at positions corresponding to the m/z values of the fragments and converge towards the point of the diagonal corresponding to the m/z value of the parent molecule. A large number of such tracks are visible in SiC. Similar tracks have been found in other materials, including GaN (Saxey, 2011; Di Russo et al., 2018a), GaAs (Cuduvally et al., 2020), $SiO_2$ (Zanuttini et al., 2017b) and $LiFePO_4$ (Santhanagopalan et al., 2015). Phosphorus molecular clusters as produced from InP are also prone to molecular dissociation (E. Di Russo et al., 2020).

*(ii) Heterolytic dissociations with production of a neutral fragment.* Even neutral atoms formed by dissociation can sometimes be detected. Though they can't be accelerated in the electric field, if a neutral fragment is produced during flight, it retains the parent ion's velocity and may reach the detector. In such cases, for a dissociation producing two fragments the following relation holds:

$$m_2' = \frac{m_2 m_1'}{m'^1 - m_p + m_2} \qquad (26)$$

Since dissociation strongly affects the apparent mass/charge ratios, accounting for it is crucial for identifying the ions involved. The track produced by neutral-producing dissociation reactions has an end on the diagonal, at the $m_p/z_p$ value when the dissociation happens close to the detector, and an asymptotic behavior approximating the mass/charge of the charged fragment when the dissociation happens at nm distance from the specimen surface (see the blue features in Fig. [4] -(b,c)). Such dissociation tracks have been clearly observed in several studies on $SiO_2$ (Zanuttini et al., 2017a) as well as in pure water specimens (Schwarz et al., 2020). In the case of reactions producing three or more fragments, the interpretation of the track is less straightforward but it is sometimes possible, as for the reaction $GaN_3^{2+} \rightarrow Ga^+ + N^+ + N_2$ in field evaporation of GaN (Di Russo et al., 2018a; Gault et al., 2016). Neutral-producing dissociations also appear to be quite frequent in the analysis of organic molecules, such as tetradecane (Meng et al., 2022a).

(iii) *Homolytic dissociations with Coulomb repulsion.* Homolytic dissociation corresponds to the reaction $A_2^{2+} \rightarrow A^+ + A^+$. If the reaction involves the same isotope, the *m/z* values of the parent and of the daughter ions coincide, so all reaction products are grouped in the correlation histogram close to the diagonal. However, due to the spatial proximity of the produced fragments undergoing similar acceleration, these reactions also produce a diagonal track with negative slope which is due to the Coulomb repulsion between the fragments (Inset of Fig. [4]-(a) and red feature in Fig. [4]-(b,c)). If the fragments are ejected with a component of their motion along the trajectory of the parent ion, they will tend to repel each other: the forward ion will be accelerated, the backward ion decelerated. This does not happen when the fragments are ejected in a direction perpendicular to the trajectory of the parent ion. In this case, Coulomb repulsion will rather tend to increase the distance of the events on the detector. This information is not retrievable from correlation histograms in m/z, but can be accessed if the correlation between time of flight difference and distance of events on the detector is considered (Blum et al., 2016b).

Correlation histograms may thus reveal systematic errors in the composition measurement, which may be more or less significant depending on the amount and on the identity of dissociation events (Saxey, 2011; Santhanagopalan et al., 2015; Blum et al., 2016b; Zanuttini et al., 2018). Dissociation fragments also project back to incorrect spatial origins, blurring 3D reconstructions. A careful analysis of correlation histograms may help quantifying the number of ions involved in dissociation reactions, although only events where both fragments are detected as ions are visible, leaving neutral fragments undetected and potentially leading to an underestimation of the total number of dissociation events. The relative importance of these events can nevertheless be estimated. In SiC, for instance, sum up to about 0.5% of the total amount of ions, and are not sufficient to account for the observed errors in composition measurements (Ndiaye et al., 2023a).

### *d. Methods for assessment and (possibly) correction of losses*

An old proverb of unknown origin cited in chapter 9 of "Atom Probe Tomography – Put theory into practice" reminds to the APT user that "*If you don't promptly identify the bias, you may feel much more uncomfortable later*" (Lefebvre-Ulrikson et al., 2016). Though the authenticity of the proverb is questionable, its message is not. The need of assessing whether a compositional bias affects an APT dataset is so common that applying a set of methods for this assessment has become just a good practice for any APT user. In most cases, an inaccuracy in compositional measurements in APT depends on some experimental parameter or is the result of detector limitations. Methods *i.* and *ii.* refers to experimental variables or surface dynamics, methods *iii.* to *v.* to detector performances. Furthermore, the user should take care about the way experimental variables influence the occurrence of multiple evaporation events (correlated evaporation, dissociation, etc.) and thus the specific detection efficiencies. The applicability of these methods clearly depends also on the material under study; needle-shaped specimens of brittle or complex materials may indeed be impossible to prepare or to sustain under the measurement conditions required for a thorough assessment of compositional bias. In these cases, the principle should be followed that a biased analysis is better than a too high risk of fracture, leading to potentially no analysis at all (Kelly and Larson, 2000).

### *i. Variation of experimental parameters*

The simplest test for ascertaining a compositional bias is varying the experimental parameters of the experiment. The most common move is varying the laser energy at constant detection rate and constant base temperature (Agrawal et al., 2011a; Mancini et al., 2014a; Morris et al., 2018). An example of this approach is illustrated in Fig. [18]-(a). In this case, the atomic fraction of Ga and N in GaN is clearly deviating from stoichiometry and is correlated with the laser energy. However, varying the laser energy at constant detection rate also implies that the DC voltage applied to the specimen adjusts accordingly. So, the same variation of measured composition can be displayed as a function of the DC voltage, as shown in Fig. [18]-(b). Thus, the variation of the laser pulse energy at constant detection rate often reveals field-driven mechanisms of loss. On the other hand, varying the laser energy allowing the detection rate to change can be a means to identify mechanisms of loss directly related to the absorption of the light, as well as other non-ideal phenomena such as the laser-induced asymmetry of the apex curvature. This effect arises because the laser light is often absorbed unevenly on the tip surface, resulting in spatially nonuniform heating (Fleischmann et al., 2018; Müller et al., 2011). Another important environmental parameter is the laser frequency. The upper limit for laser frequency is usually the inverse of the time of flight of the heaviest ions. Typically, frequencies above 500 kHz are not used in most APT systems. Low frequencies may enhance phenomena related to preferential evaporation or to the supply of foreign species (most commonly, oxygen or hydrogen (Meier et al., 2022)).

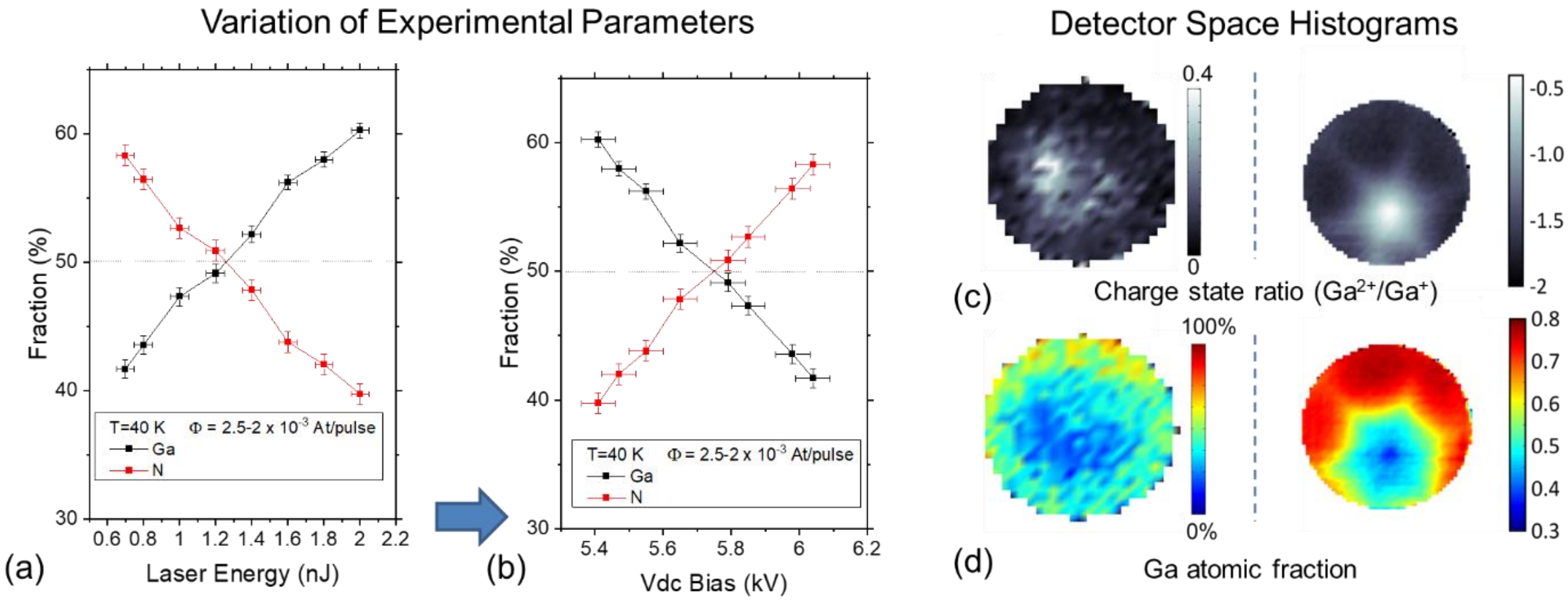

*Figure [18]. Examples of methods for the assessment of physical-chemical mechanisms leading to compositional biases. (a,b) The variation of experimental parameters. In the reported example, (a) the variation of the laser pulse energy at constant base temperature and detection rate clearly indicates a variation of the measured composition in GaN. (b) As the measurement is performed at constant detection rate, a second environmental variable varies, i.e. the DC voltage (the blue arrow indicates the dependence of $V_{DC}$ on the laser pulse energy). The composition may be thus shown to correlate also to this environmental variable. Measurements performed without the constraint of constant detection rate may lift the ambiguity on which parameter is primarily influencing the measurement. (c,d) Investigation of distributions within a given dataset. In the two reported examples, detector space histograms indicate a correlation between (c) the CSR($Ga^{2+}/Ga^{+}$) and (d) the measured composition in*

*GaN. While the diagrams on the left hand-side indicate just a sort of axial symmetry which could be attributed to the specimen shape, those on the right hand-side also contain clear information about the crystallographic symmetry of the system (Data reproduced from* (Di Russo et al., 2018a; Mancini et al., 2014a)*).*

### *ii. Variations within a given dataset*

APT datasets collected with a fixed set of experimental parameters may also offer the opportunity of assessing compositional inaccuracies. These become visible as spatial gradients of the measured composition having no relationship with the nanostructure of the sample. These spatial gradients may be assessed both in the detector space or in the reconstructed space. Such gradients are generally related to surface field non-uniformities, visualized via the CSR, which may have different origins: they may be related to the aforementioned laser-induced apex curvature asymmetry, to a non-constant apex curvature at equilibrium related to specimen preparation or to specimen crystallography (as visible in Fig. [18]-(c,d)). In specific cases, comparison between the composition pattern and the CSR may indicate different roles of field and crystallography in the determination of composition (Diagne et al., 2025). Long APT analyses at constant detection rate on specimens with a significant evolution of the apex radius may also reveal field and composition gradients when investigated in the reconstructed space. If the radius increases indeed, the imaged surface increases and the surface field must thus decrease in order to reduce the evaporation rate and keep the detection rate constant. The gradual decrease of the surface field with the analyzed depth may thus translate into a gradient of measured composition if a field-dependent bias is present (Rigutti et al., 2016b).

### *iii. Analysis of multiple events and detector pile-up*

The analysis of multiple detection events is crucial in APT composition metrology as the physical channels leading to losses in terms of molecular dissociation and correlated evaporation leave their trace in this subset of data. This part of data mining constitutes a sort of bridge between correlated evaporation and detector performance. In the framework of quantitative analyses, multiple data may also offer the possibility to quantify the loss and therefore correcting the errors in composition measurements.

Beyond providing the key to quantifying the effects of molecular dissociation (section 4.c.iv), multiple events are most useful for assessing and quantifying the possibility of pile-up and its effect on the measured composition. In most compounds, a first effect of correlated evaporation on the measured composition can be traced by looking at the statistics of multiple detection events resolved for the different chemical species present in the sample. An example of this is reported in Fig. [19]-(a) relative to the analysis of SiC. The statistics shows clearly that, while Si is mostly detected as a single event, C has the opposite behavior, i.e. high PME. This is a hint about a higher degree of correlation for C evaporation, leading to higher potential losses by detector pile-up, which is indeed the case for SiC and likely for many other carbides (Ndiaye et al., 2023a). The effect of pile-up can be estimated or quantified especially when an element has more detectable isotopes (see next section) and can be visualized as a dip in the statistics of distances for homoisotopic pairs close to zero, as reported in Fig. [19]-(b) for the frequencies of $^{71}Ga^{+}$-$^{71}Ga^{+}$ and $^{69}Ga^{+}$-$^{69}Ga^{+}$. This occurs because homoisotopic pairs corelated in space are also extremely close in time when they impinge on the detector. This is not the case for heteroisotopic pairs correlated in space, which have a sufficiently large difference in TOF to allow the heavier isotope to escape from the dead/dazzled zone (Di Russo et al., 2017).

An alternative approach was proposed in order to neutralize the impact of multiple detection events on the composition measurement, consisting in strongly reducing the detection efficiency by placing a grid on the opposite side of the local electrode aperture, thereby significantly decreasing the number of multi-hit events. This method was first proposed by Thuvander et al. in 2013 and later (Thuvander et al., 2013, 2019) and more recently applied by Morris et al. (Morris et al., 2022) to study compositional biases of both GaN and AlN. Even with a substantial reduction in multi-hits, particularly at high fields, the overall composition of GaN and AlN remained essentially unchanged, indicating that correlated evaporation phenomena have little effect on the measured composition. It should be reminded that spatial filters do not filter multiple evaporation events closely correlated in space, which would be still subjected to pile-up and be detected as single events.

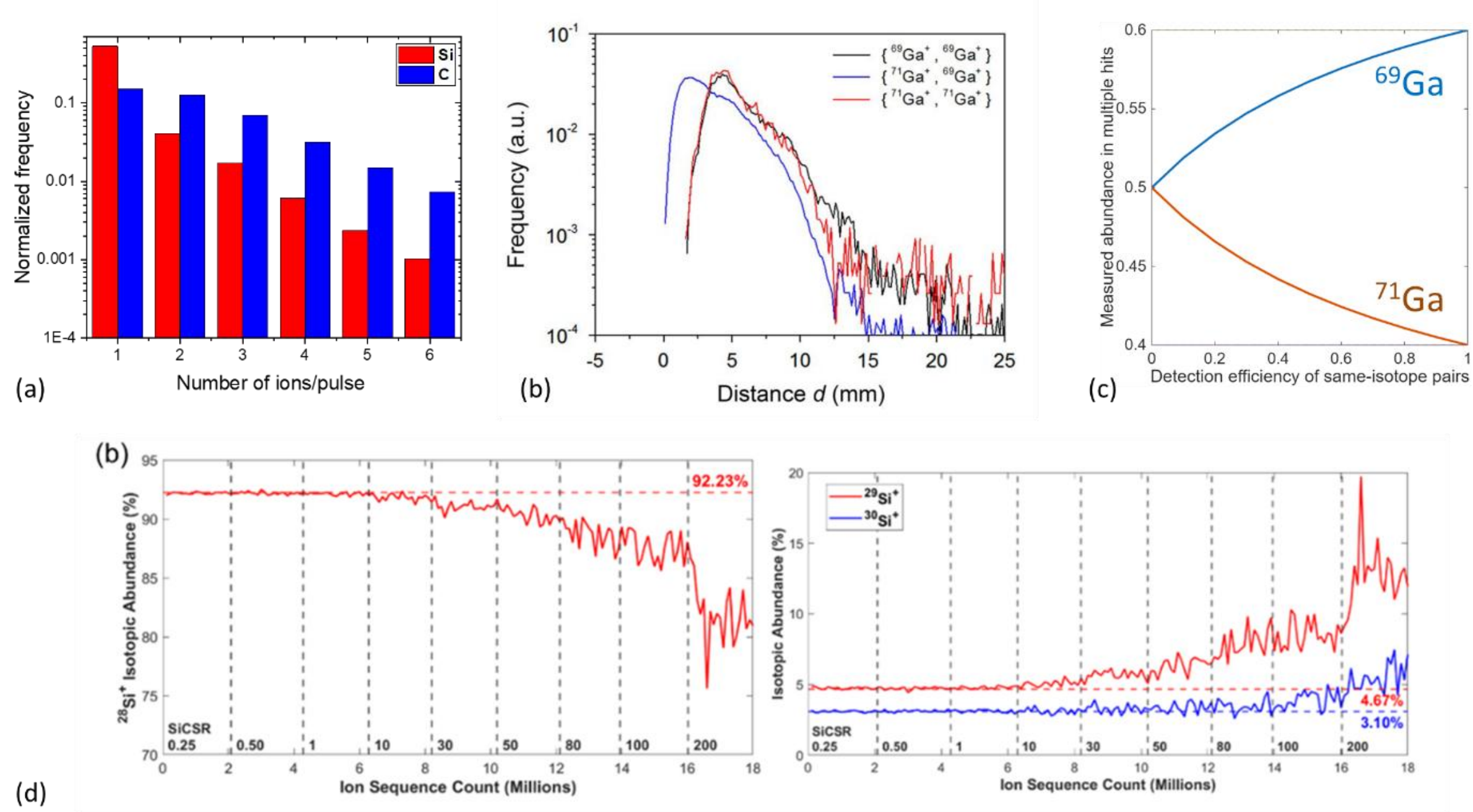

*Figure [19]. Analysis of multiple events in APT. (a) Statistics of multiple events as a function of multiplicity in SiC resolved by species. C is much more frequently found in multiple events than Si. (b) Histogram of distances on the detector of $Ga^+$ pairs from multiple detection events. The homoisotopic pairs present a dip at short distances due to detector pile-up (Reproduced with permission from* (Di Russo et al., 2018a)*, Elsevier). (c) Calculated variation of detected isotope abundances of Ga in double evaporation events as a function of the detection efficiency for same-isotope pairs. (d) Isotope abundances detected in multiple hits for $Si^+$ during the evaporation sequence of a B-doped Si sample. The data indicate a depletion of the majority isotope $^{28}Si^+$ and an enhancement of those of the minority isotopes $^{29,30}Si+$ (Reproduced with permission from (Guerguis et al., 2024), Elsevier).*

### *iv. Investigation of isotopic ratios*

A robust approach for the assessment of biases induced by correlated evaporation involves analyzing the occurrence of homo- and heteroisotopic multiple events for elements composed of two or more isotopes (Thuvander et al., 2011; Di Russo et al., 2018a; Gopon et al., 2022; Ndiaye et al., 2024). An example is Ga, which consists of two isotopes, $^{69}Ga$ and $^{71}Ga$. Hetero-isotopic multi-hits typically occur with a time delay in spatially correlated events, due to the different m/n ratios of the isotopes, which is generally sufficient to avoid pile-up effects. In contrast, homo-isotopic hits can occur within the detector's so-called "dead zone," resulting in the detection of only a single ion (Fig. [19]-(b)). In terms of effect on abundances, the process can be understood as follows. Natural gallium consists of two stable isotopes, $^{69}Ga$ and $^{71}Ga$, with (approximated) natural abundances $p_1$=60% and $p_2$=40%, respectively. Assuming random pair evaporation, the probability of evaporating an isotope pair is given by $p_1^2$=36% for $^{69}Ga$–$^{69}Ga$, $p_2^2$=16% for $^{71}Ga$–$^{71}Ga$, and $2p_1p_2$=48% for mixed pairs. Due to detector pile-up effects, same-isotope pairs are detected with a reduced efficiency factor $\eta_{\text{pile-up}}$, while mixed-isotope pairs are detected without suppression. This leads to a modified detection frequency per pair, where same-isotope contributions are scaled by $\eta_{\text{pile-up}}$. As a consequence, the measured isotopic abundance deviates from the natural abundance, as shown by the plot in Fig. [19]-(c): with decreasing detection efficiency for same-isotope pairs, the apparent abundance of $^{69}Ga$ (majority) decreases while that of $^{71}Ga$ (minority) increases. In the extreme (though hypothetical) case in which all homoisotopic pairs were undetected, only heteroisotopic pairs would be responsible for detected isotope abundances leading to 50% abundance for both isotopes. Similar considerations, though slightly more complex, apply to elements with more isotopes. A further complication to this picture is that detected isotope pairs may also results from events with multiplicity higher than 2. Applied to GaAs, this approach indicated that only ~2% of the total $Ga^+$ ions were missed due to pile-up, an amount insufficient to fully explain the observed compositional biases (Di Russo et al., 2018a, 2017).

An experimental example of biased isotopic abundances is reported in the analysis of B-doped Si (Fig. [19]-(d)). The abundance of the majority isotope $^{28}Si^+$ decreases during the analysis, while that of minority isotopes $^{29,30}Si^+$ increases. This phenomenon is related to the increase of the electric field during the evaporation sequence and to the observed increase of multiple events. Along with the analysis of $Si^{2+}$ isotopic abundance, the observation allowed the authors for an estimation of lost Si atoms and for a correction of the B doping concentration (see section 5.b.i for details, (Guerguis et al., 2024)).

### v. *Pseudo-singles and Pseudo-multiples*

The terms pseudo-singles and pseudo-multiples refer to two different categories of detection events (Müller et al., 2011; Ndiaye et al., 2024).

*Pseudo-singles* are single detection events resulting from a double event in which the second ion was undetected (detector dead zone) or was misinterpreted/unresolved as a double event (detector dazzled zone). Pseudo-singles can bias isotopic abundances of single detection events, and are undistinguishable from real single events unless the instrument allows for recording the amplitude of the MCP signal (Fig. [11]-(d,e)). In this case, it becomes possible to estimate the amount of pseudo-singles $N_{\text{pseudo singles}}$, which may help in the estimation of total losses by pile-up (Ndiaye et al., 2023a, 2024).

*Pseudo-multiples* are defined as successive detection events separated by exactly $\Delta(N_{\text{pulse}})=1$ (consecutive laser pulses), mimicking the spatial correlations of true multiples (same-pulse events) but without dead-time losses, serving as a lossless reference (Müller et al., 2011; Ndiaye et al., 2024).

By comparing the spatial distributions of isotopic pairs in true multiple events and pseudo-multiples, it becomes possible to identify and quantify the depletion of same-isotope pairs in true multiples, directly attributable to pile-up losses and to estimate the number of expected pairs $N_{\text{expected pairs}}$. Mixed-isotope pairs, which are less affected by pile-up due to their larger time-of-flight differences. Using the measured number of mixed-isotope pairs and their natural probabilities, it is possible to calculate the expected total number of pairs in the absence of loss, which is given by

$$N_{\text{corrected}} = N_{\text{expected pairs}} - \frac{N_{\text{pseudo singles}}}{2}. \qquad (27)$$

In the analysis of $LaB_6$, the deficit between expected and measured pairs reveals that about 58% of B pairs are lost in multiple events, primarily involving singly charged ions (Ndiaye et al., 2024). This approach restores the material's stoichiometry to near its nominal value, demonstrating that isotopic and pseudo-multiple event analysis can effectively quantify and correct detection biases in APT. The method provides a physically grounded framework for improving compositional accuracy in materials analysis.

### *vi. Isotopic substitution*

Isotopic substitution consists in purposedly altering the isotopic abundances of a given element, most often by replacing a majority isotope with a minority one (Kellogg, 1987; Gemma et al., 2009; Takahashi et al., 2010). This operation is meant to shift mass peaks without changing the chemistry of the compound, and to avoid mass peak superpositions or ambiguities. It can also be used in order to study the accumulation of specific isotopes at specific locations in the sample, resulting from transport, precipitation or other microscopic mechanisms. The most common application of this approach is probably deuterium charging in order to distinguish hydrogen from the sample from environmental hydrogen in APT of metals (Mouton et al., 2019; Chen et al., 2023). Isotopic substitution has been reported in studies on oxides, with the investigation of $^{18}O$-enriched hematite, revealing that the only singly charged $O^+$ monomers are detected (and not $O_2^{2+}$) (Bachhav et al., 2013). Annealing in $^{18}O$ was also applied in order to study oxygen transport along metal-perovskite heterostructures (Taylor et al., 2023). Another noteworthy case concerns the resolution of Si and N signatures via the substitution of $^{14}N$ by $^{15}N$, yielding a composition in good agreement with complementary analyses for Ti-Si-N (D. L.J. Engberg et al., 2018), as well as more reliable quantification of N in $Fe_4N$ phases in steels (Takahashi et al., 2022).

### *vii. Statistical correction of biased composition from random alloys*

Once a compositional bias has been assessed and globally corrected, a question remains whether the correction can be extended to other properties than the average composition. A correction method has been proposed for assessing the correct statistical properties of the random alloys which would be affected by compositional inaccuracies conserving their randomness. This is a quite restrictive scenario, but it is instructive on how low detection efficiencies affect the information on APT datasets. The datasets displayed in Fig. [20] are simulated, but the protocol has profitably been applied on real APT datasets (Rigutti et al., 2016a). The correction protocol aims to recover the statistical properties of a biased distribution while conserving its spatial characteristics, i.e. the positions of local minima and maxima. In principle, the statistical properties of a random alloy are known. It is thus possible to define a reference binomial function corresponding to the statistical distribution of the alloy, sampled by voxels of constant volume and number of atoms (Fig. [20]-(c)). In case depicted in Fig. [20]-(a), the bias is described by a different detection efficiency for Al ($\eta_{Al}$=0.3) and for Ga ($\eta_{Al}$=0.15) in the AlGaN alloy (the composition is determined in terms of the III-site fraction, see section 5.a.iii). The main effects of this bias are a shift of the average III-site fraction from the pristine $<y^{reference}>$=0.25 to $<y^{bias}>$=0.4, and an increase of the standard deviation of the distribution, i.e. of the amplitude of the compositional fluctuations. Based on the average site fraction $<y^{bias}>$ and the average number of atoms per voxel $<N_{bin}^{bias}>$, an associated binomial distribution and its partition function are defined.

The latter is used in the correction protocol schematized in Fig. [20]-(b). Similarly, the reference binomial distribution is used to obtain an interpolated reference partition function. The partition functions of the biased and reference datasets, interpolated to obtain a strictly monotonically increasing behavior, are shown in the diagram of Fig. [20]-(b). The interpolated partition functions are thus used to define the correction of the site fraction measured locally in any voxel (for mathematical details, see ref. (Rigutti et al., 2016a)). The application of this correction protocol makes it possible to make the biased distributions similar to the reference one, as can be seen in the diagrams in Fig. [20]-(c). However, when the efficiency becomes very low, the correction protocol cannot completely bring averages and standard deviations back to the reference values.

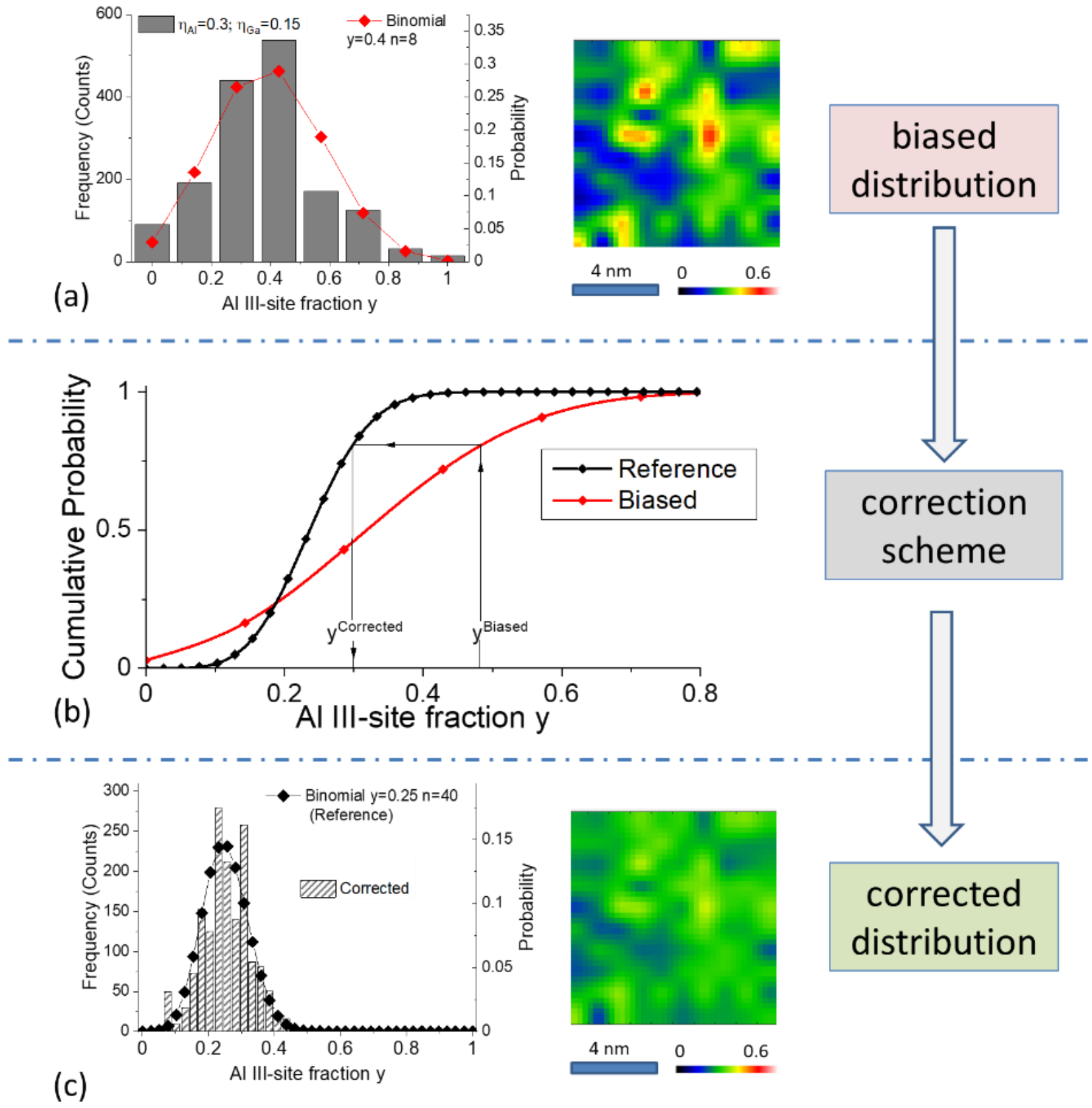

*Figure [20]. Statistical correction of random alloy distributions affected by compositional biases conserving the alloy randomness. (a) The frequency distribution analysis (left) of a simulated $Al_{0.25}Ga_{0.75}N$ dataset in which the specific detection efficiency is for $\eta_{Al}$=0.3 for Al and $\eta_{Ga}$=0.15 for Ga and the corresponding 2D III-site fraction map (right). The statistical distribution can be approximated by a binomial function. Not only the average III-site fraction is affected by an error, but also alloy fluctuations exhibit enhanced amplitudes. (b) The correction scheme is based on the knowledge of the biased partition function derived from the binomial function in (a) and on the knowledge of the reference partition function that would describe the statistical distribution of an ideally unbiased alloy. The corrected composition $y^{Corrected}$ can be evaluated for any voxel or subvolume starting from the $y^{Biased}$ composition measured. (c) The resulting corrected distribution reasonably fits the reference binomial function, has a correct average (frequency distribution analysis, left) and exhibits weaker fluctuations (right) (Adapted with permission from (Rigutti et al., 2016a), AIP).*

## 5. Review of literature data

In this part we review the main studies on compositional metrology of APT present in the literature. This part is meant to depict the state of the art, identify common problems and trends in different material systems, and provide an overview of the methods adopted to identify, describe and understand the inaccuracies in composition measurements. This part is divided into two main sections, one dealing with the measurement of the composition of the matrix, the other with the assessment of the concentration and distribution of impurities.

### *a. Composition of the matrix*

The measurement of the composition of the matrix of a compound material is obviously the first step in the understanding of its functional and chemical properties. While a relatively large number of techniques can determine global compositions and even global statistical properties with excellent precision and accuracy, APT is needed when the materials properties are determined by the distribution of chemical species at the nanoscale or at the atomic scale. For this reason, it is critical to understand whether the elements present in a compound material are detected with specific detection efficiencies, and how these may depend on the experimental parameters chosen. A representative number of studies has appeared in the recent past focusing on this problem.

#### *i. Arsenides, Antimonides, Phosphides*

Classical III-V semiconductors (i.e. III-V except III-Nitrides) were among the first compound semiconductor to be analyzed by atom probe, namely with 1D APT (either voltage- or laser-assisted) during the 1980s (Sakurai et al., 1984; Hashizume et al., 1986; Cerezo et al., 1986b). These works also provide the first examples of APT compositional metrology in semiconductors. A renewed interest in the analysis of III-V compounds occurred after the introduction of the 3D APT assisted by fs laser pulses.

Field evaporation of III-Vs is generally complicated by the occurrence of large molecular clusters of element V atoms, which can be found in the mass spectra of GaAs (Di Russo et al., 2017; Gorman et al., 2011a), InAs (Perea et al., 2006), GaSb (Müller et al., 2011), and InP (E. Di Russo et al., 2020). The presence of such clusters introduces an ambiguity in the interpretation of mass spectra, which could account for up to 10% of the element V atomic fraction (Cuduvally et al., 2020). This is not the only source of inaccuracy, as the measured composition is also field-dependent in these compounds (section 4.b).

*Binary compounds.* Figure [21]-(a) and (b) report two consistent sets of measurements performed with very different instruments in the case of GaAs, a stoichiometric compound. The measurements performed by Hashizume et al. (Figure [21]-(a)) show a significant excess of Ga at low field and a slight Ga deficit at high field (Hashizume et al., 1986). These authors do not adopt the CSR metrics, but plot their results as a function of the effective DC field fraction, i.e. the quantity $V_{DC}/(V_{DC}+V_E)$, where $V_E$ is the evaporation field as determined for a given detection rate. The abscissa is therefore a growing function of the surface field. The more recent results obtained by Di Russo et al. (Figure [21]-(b)) are reported as a function of the CSR($As^{2+}/As^{+}$) and reveal a consistent common trend for a set of measurements performed at different temperatures, laser wavelengths and pulse laser energy, with the atomic fraction of Ga varying from 60% at low field to 30% at high field. These results can be interpreted as following: the preferential evaporation of Ga atoms at high field can explain the Ga loss, while at low field, a possible mechanism accounting for the loss of As atoms is the formation of As clusters on the tip surface and their direct dissociation in smaller subunits with the formation of neutral As molecules/atoms. Furthermore, the clustering of V-element atoms exhibits only a weak dependence on the laser pulse energy at fixed DC voltage (Di Russo et al., 2017).

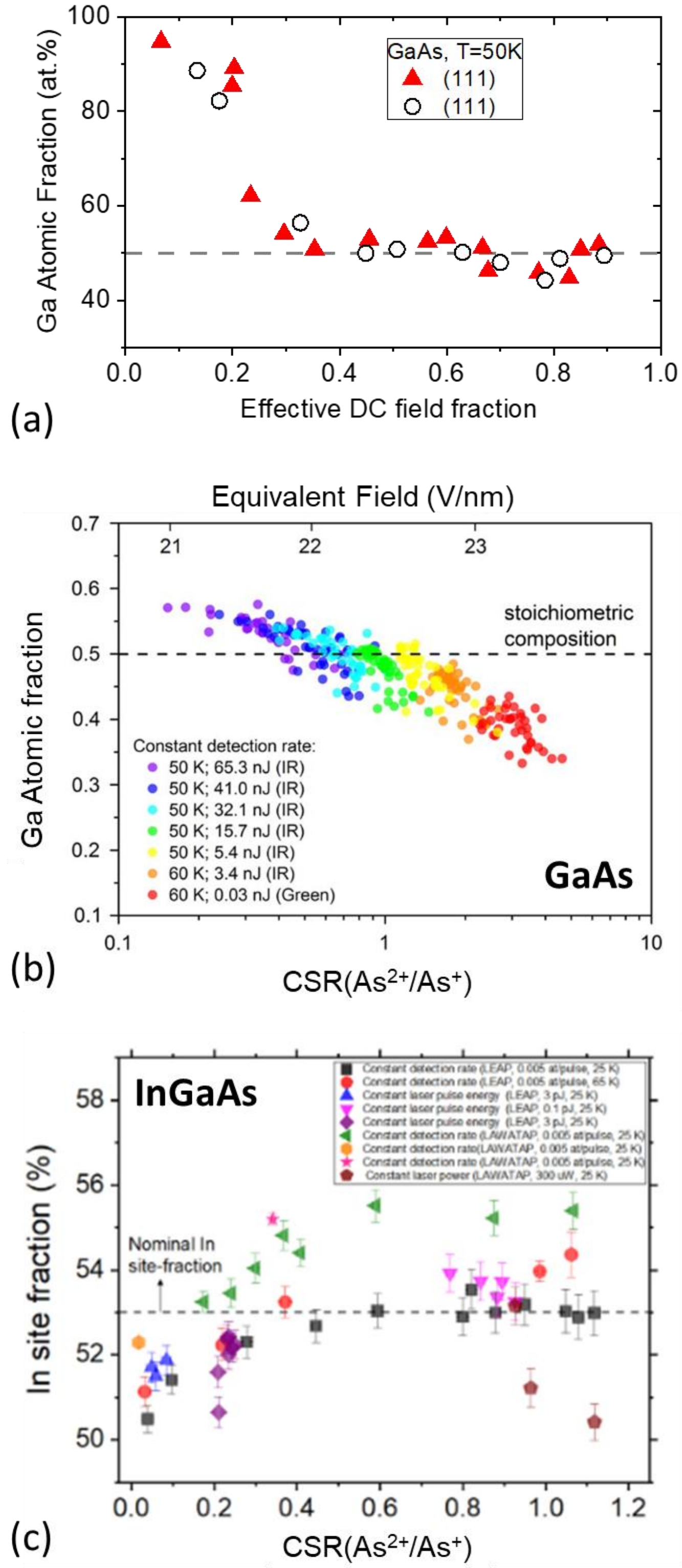

*Figure [21]. Field-dependent compositional biases in classical III-V semiconductors. (a) Atomic fraction of Ga measured in 1D-LA-APT of GaAs as a function of the effective DC field fraction (see main text, a quantity approximately proportional to the surface field) (Data adapted from (Hashizume et al., 1986), American Institute of Physics). An excess of Ga is measured at low field, while a deficit of Ga is measured at high field. A similar situation is reproduced in (b), in which measurements performed at different temperatures, laser wavelengths and intensities align on a common trend (Adapted with permission from* (Di Russo et al., 2017), *Oxford Academic,). (c) Field-dependence of the measurement of the In III-site fraction in the* $In_{0.53}Ga_{0.47}As$ *pseudobinary alloy. (Adapted with permission from (Cuduvally et al., 2020), Elsevier)*

Further works on GaSb have shown a similar complexity (Müller et al., 2011), but also that compositional gradients may occur due to the different effect of the laser light on the illuminated and on the dark side of the specimen. This effect is particularly critical for low-bandgap materials but it may be reconducted to a field effect because the steady state curvature of the field emitter apex becomes larger on the illuminated side and smaller on the dark side (Müller et al., 2012). Finally, an excess composition in As has also been reported for the magnetic compound $Ga_{0.963}Mn_{0.037}As$, but compositional biases in this systems have not been investigated systematically yet (Kodzuka et al., 2009).

*Pseudobinary alloys.* More critically than in stoichiometric compounds, field-dependent compositional biases have also been assessed in pseudobinary III-V alloys. Figure [21]-(c) displays the dependence on the CSR($As^{2+}/As^{+}$) of the measured III-site fraction of In. Differently from III-N materials (see section 5.a.ii-iii) there is no clear trend in the compositional bias. This is also due to the interplay of the possible loss mechanisms, which have been classified by relevance. Among the various mechanisms, cluster peak overlaps are identified as the most significant contributor to As loss. This effect is more pronounced at low field, where larger clusters ionize more easily, causing overlaps that obscure As detection. Multi-hit events are the second most significant loss mechanism, also more prevalent at low field—70% of multi-hits contain As versus 57% at high field. This suggests additional As loss may occur via dissociation into indistinguishable fragments. Dissociation is the third channel and is slightly more significant at high field, contributing about 0.05 at.% at low field and 0.1 at.% at high field. Neutral production through dissociation is less important and occurs mainly at high field, but experimental evidence and DFT calculations indicate such neutrals likely re-ionize, minimizing loss. DC evaporation is the least significant mechanism but increases at high field due to the higher evaporation field required for As compared to In and Ga. While this is a qualitative argument, evaporation fields defined by Müller theory are not exactly applicable to semiconductor compounds and should be used cautiously (see section 2.a). At low field, background accounts for ~2 at.%, rising to ~30 at.% at high field. Significant DC evaporation of In and Ga likely increases the measured As fraction at high field, partially offsetting losses but introducing additional compositional bias and reducing the overall accuracy of the measurement (Cuduvally et al., 2020).

### *ii. III-Nitrides : GaN*

Compositional biases have been reported in several III–nitride materials, with GaN being by far the most extensively investigated. This prominence can be attributed to three main factors. First, GaN dominated semiconductor materials research during the first two decades of the 2000s because of its exceptionally broad technological relevance, spanning power electronics and optoelectronics coupled to potential for optimization due to the persistence of defects. Second, the material is readily accessible, as it is routinely grown by epitaxial techniques in many laboratories, both as thin films and as nanowires. Finally, GaN is particularly well suited for APT: it can be analyzed with high reliability while still exhibiting pronounced compositional biases under specific experimental conditions. This behavior is especially striking given that the intrinsic composition of GaN deviates by less than 0.1 at.% from ideal 1:1 stoichiometry.

*First reports and thermal effects.* The first report of compositional biases in GaN was provided by Agrawal et al. in 2011, based on a study of nanowires grown along the <0001> direction using LEAP 3000X Si and LEAP 3000X HR instruments operated with a green laser ($\lambda$=532 nm) (Agrawal et al., 2011a). Because GaN has a wide bandgap of approximately 3.6 eV, laser absorption at 532 nm is intrinsically limited (photon energy ≈ 2.3 eV), which can hinder optimal atom probe analysis. Measurements were conducted at a fixed detection rate $\varphi$, with the laser pulse energy LPE varied between 3 and 20 pJ and the DC bias applied to the tip $V_{DC}$ was adjusted to keep $\varphi$ constant during tip evaporation. Under these conditions, the charge-state ratio CSR($Ga^{2+}/Ga^{+}$) increased from 1 to 31, accompanied by a corresponding reduction in the measured Ga fraction from 0.8 to 0.5 (Fig. [22]). The reported mass spectra reveal the presence of $N^+$, $N_2^+$, $Ga^{2+}$, and $Ga^+$ ions, together with less abundant species such as $N^{2+}$, $GaN^+$, $N_3^+$, $GaN_3^+$. However, no information is provided on how the mass spectra evolve as a function of the analysis parameters. Varying the laser repetition rate between 200 and 500 kHz did not measurably affect the composition, while a modest reduction in signal-to-noise ratio led to improved spectral quality through a decrease in uncorrelated inter-pulse events. Accordingly, a repetition rate of 200–250 kHz was selected. Analysis of two-dimensional density and composition maps extracted from the cross-section revealed the presence of a c-axis pole and a sixfold symmetry, consistent with the stereographic projection along the <0001> direction of the wurtzite hexagonal crystal structure of GaN. At the location of this pole, a reduction in both Ga and N densities was observed. This behavior was attributed to the divergence of ion trajectories caused by a local decrease in the tip curvature radius. Such a reduction also results in a local enhancement of the electric field, which in turn leads to a Ga-deficient composition at the pole. Thermal modelling indicates that the low thermal conductivity of GaN nanowires leads to non-uniform heating at high LPE, which in turn promotes preferential nitrogen loss and, consequently, a spatially non-uniform compositional distribution. This inhomogeneous heating is likely driven by surface absorption, potentially associated with defects or surface states. Similar conclusions were drawn by Dawahre et al. in 2013 (Dawahre et al., 2013) on bulk GaN grown along the c-axis, analyzed using a LEAP 3000 XS operated in laser mode ($\lambda = 532$ nm) at a repetition frequency of 250 kHz.

*The role of crystal symmetry and polarity.* In 2012, Riley et al. analysed GaN nanowires oriented along the non-polar a-axis using a LEAP 4000X Si (Riley et al., 2012). In this case, no artifacts related to non-uniform heating were observed when employing a UV laser ($\lambda$=355 nm). Analysis of hit maps showed that $Ga^+$ and $N_2^+$ follow the same trend, exhibiting low hit densities along a band connecting the $(01\bar{1}0)$, $(11\bar{2}0)$ and $(10\bar{1}0)$ poles, as well as at the Ga-polar surface. This reduced hit density is attributed to the locally enhanced electric field at the edges of these low-index planes, where the increased field promotes the preferential evaporation of doubly charged ions such as $Ga^{2+}$ and $N_2^{2+}$. It should be noted that the hexagonal wurtzite GaN crystal structure contains no N–N bonds. Consequently, the observed low abundance of $N^+$ ions, in contrast to the predominance of $N_2^+$ ions, can be explained by the relatively low diffusion barrier of N atoms (0.66 eV for diffusion on the Ga-polar surface (Fernández-Garrido et al., 2008)) compared with the desorption barrier of $N_2$. This difference facilitates the diffusion of N atoms across the surface and the subsequent formation of $N_2$ molecules. After optimizing the analysis conditions to achieve overall stoichiometry, Ga deficiency was observed on N-polar facets, whereas N deficiency occurred on non-polar and Ga-polar facets. Stoichiometric composition was detected only near the N-polar facet. These findings suggest that facet polarity critically modulates Ga and N evaporation, resulting in local compositional deviations even when the material is globally stoichiometric. This behaviour is attributed to the preferential loss of N atoms: during multiple-hit events, more $N_2$ than Ga is evaporated, leaving a higher fraction of N ions undetected. $N_2$ loss is particularly pronounced at non-polar and Ga-polar surfaces, likely due to variations in atomic bonding coordination at these facets. After optimizing the analysis conditions to achieve overall stoichiometry, Ga deficiency was observed on N-polar facets, whereas N deficiency occurred on non-polar and Ga-polar facets. Stoichiometric composition was detected only near the N-polar facet. These observations indicate that facet polarity strongly governs the evaporation rates of Ga and N, leading to local deviations from stoichiometry despite an overall balanced composition. The observed compositional biases were attributed to the preferential loss of N atoms, either between laser pulses or during multiple-hit detection events. In particular, during multiple-hit events, $N_2$ evaporates more readily than Ga, resulting in a larger fraction of undetected N ions. This $N_2$ loss mechanism is especially pronounced at non-polar and Ga-polar surfaces, likely due to variations in atomic bonding coordination at these facets. Despite local compositional variations arising from crystallography, the primary factor influencing the detection rate is the DC voltage applied to the tip. In particular, increasing $V_{DC}$ appeared to bring the Ga/N ratio closer to stoichiometry; however, this effect is largely apparent, resulting from a decrease in Ga detection. This behavior was explained by the lower evaporation field of Ga (≈15 V/nm) compared with N. This is consistent with previous observations of Ga depletion at high applied DC voltages in GaAs, but the reader should consider that the application of the values of evaporation fields (see section 2.a) does not reflect the nature of the Ga-N bonding in GaN, and even less the crystal polarity and the possible state of the surface. Both As and N, as group-V elements, display similar evaporation behavior, with thermally assisted surface migration and emission as higher-order molecular ions (Liddle et al., 1988; Gorman et al., 2007, 2011b).

Further insights into the APT analysis of GaN nanowires grown along the <0001> direction were provided by Diercks *et al.* in 2013 (Diercks, 2013). In this study, atom probe analyses were performed using both a straight-flight-path LEAP 4000X Si and a reflectron-equipped LEAP 4000X HR, the latter providing enhanced mass resolution. Both instruments operated with a UV laser (λ=355 nm) laser. Hit maps revealed that Ga and N ions exhibit sixfold symmetry around the (0001) pole. Although Riley et al. reported that nanowire polarity influences evaporation behavior and, consequently, the detected composition, Diercks et al. observed no such variations when nanowires were mounted in the opposite orientation. The specimens exhibited the same sixfold pattern, indicating that the observed evaporation behavior depends primarily on crystallography rather than bond direction. Constant detection rate measurements (0.007 ions/pulse) were then carried out while increasing LPE from 0.002 to 10 pJ. At the lowest laser energies, concentration maps revealed $N_2$ depletion and Ga enrichment at the central pole and along the six zone lines. At 0.1 pJ, this trend reverses, with $N_2$ becoming enriched at these sites and in a region opposite the incident laser direction. At 10 pJ, this effect dominates, and any correlation between local composition and crystallographic orientation disappears. A similar behavior is observed for the charge-state ratio $CSR(Ga^{2+}/Ga^{+})$, which serves as an indicator of local electric field strength. At low laser energies, a high-field region appears at the (0001) pole and along the zone lines. This feature vanishes at 0.1 pJ, and at 10 pJ only a high-field region is observed on the side opposite the laser. The high-field region at the (0001) pole coincides with the area of highest multiple-hit detection events.

*Influence of laser and temperature.* In a subsequent study published in 2015 (Diercks and Gorman, 2015). further investigated the influence of tip base temperature on the measured composition by varying LPE from 0.006 to 0.05 pJ, a range in which ion emission is not expected to be affected by the laser incidence direction. Using a LEAP 4000X Si instrument, it was observed that increasing the tip base temperature from 20 to 120 K produced no measurable change in the detected Ga fraction. The variation of the N fraction measured as a function of LPE was then examined. As LPE increased from approximately 0.001 to 10 pJ, the detected N fraction decreased from about 0.6 to 0.1, with stoichiometric composition observed at LPE ≈ 0.01 pJ (Fig.[22]). The same behavior was observed using both UV (λ=355 nm) and green (λ=532 nm) lasers, indicating comparable laser–matter absorption mechanisms despite the different excitation wavelengths. In contrast to earlier reports, the study shows that at the lowest laser energies the apparent nitrogen content exceeds 50%. These observations indicate that the apparent deviation from the 1:1 Ga:N ratio is likely due to under-detection of one or both species. To test this hypothesis, TEM images of four nanowires analyzed at LPE ranging from 0.005 to 0.1 pJ were acquired before and after atom probe analysis and used to determine the reconstruction efficiency, ensuring consistency with the known evaporation depth. The resulting detection efficiencies were 0.20 ± 0.08 for nanowires analyzed with the LEAP 4000X HR and 0.37 ± 0.08 for those analyzed with the LEAP 4000X Si. Both values are lower than the nominal instrument detection efficiency, which is approximately 0.37 for LEAP 4000X HR and 0.5 for LEAP 4000X Si. Despite the reduced detection efficiency, nominal GaN stoichiometry is recovered at intermediate laser energies (0.005–0.01 pJ), indicating that both Ga and N ions are partially lost outside the assigned mass ranges. At the lowest laser energies, where a N-rich composition is measured, the electric field required to maintain the target detection rate is sufficiently high to promote ion evaporation between laser pulses, predominantly affecting Ga and potentially N. As LPE increases, Ga losses due to between-pulse evaporation are reduced owing to the lower DC field at the specimen surface. In contrast, higher laser energies increase the probability of nitrogen loss through sublimation of neutral $N_2$ species rather than through uncorrelated evaporation of ionic nitrogen species (see section 4.c.i). Owing to the strong N≡N bond and its weak surface binding (Rapcewicz et al., 1997; Guo et al., 1998; Lymperakis and Neugebauer, 2009), nitrogen is more readily released in neutral form, whereas the lower evaporation field of Ga enables its ionization under the same conditions (Tsong, 1978b). Consequently, at low-coordination surface sites exposed to higher laser-induced heating, the available thermal energy is sufficient to release both Ga and $N_2$, while the electric field preferentially ionizes only Ga. Analysis of background counts further confirms that the apparent nitrogen deficit measured at high LPE (1 - 10 pJ) originates from the sublimation of neutral $N_2$ rather than from uncorrelated evaporation of ionic $N_2$ species. Even when the full background signal is attributed to $N_2$, the measured nitrogen counts remain insufficient to achieve a 1:1 Ga:N ratio.

*Role of electric field.* A comprehensive investigation of the field-evaporation behaviour of bulk GaN growth along the (0001) direction as a function of the analysis parameters was reported by Mancini et al. (Mancini et al., 2014a) and by Di Russo et al. (2018) (Di Russo et al., 2018a). Measurements were conducted using three complementary protocols: (i) at fixed laser pulse energy LPE while varying the DC bias applied to the tip, $V_{DC}$; (ii) at fixed $V_{DC}$ while varying LPE; and (iii) at a constant detection rate $\varphi$, by keeping LPE fixed and automatically increasing $V_{DC}$ to compensate for the increasing tip radius during field evaporation. Measurements were carried out using two tomographic atom probe instruments, LaWaTAP and FlexTAP, which provided fully consistent results. Both instruments were operated with UV femtosecond laser pulses ($\lambda$ = 343 nm, 350 fs), focused onto the specimen to a spot diameter of approximately 20 µm. The laser repetition rate was set to 100 kHz for LaWaTAP and 50 kHz for FlexTAP. In both instruments, ion detection was performed using a custom-designed detector comprising a microchannel plate coupled to an advanced delay-line detector (MCP/aDLD), with a detection efficiency $\eta_{MCP} \approx 0.6$ (Da Costa et al., 2012, 2005). Both studies indicate that the measured composition varies as a function of the charge-state ratio, CSR($Ga^{2+}/Ga^{+}$), when using protocols (i) and (iii). Specifically, a Ga fraction of 0.8 is measured at CSR ≈ 0.02, stoichiometric composition is obtained at CSR ≈ 0.08, and a Ga fraction of 0.4 is observed at CSR ≥ 0.3 Fig. [18]-(a,b). This CSR range corresponds to an equivalent field $F_{eq}$ of approximately 23–27 V/nm, with stoichiometric composition achieved at $F_{eq} \approx 24$ V/nm. Particularly noteworthy is the field-dependent evolution of the fraction of detected counts in the mass spectrum, as reported by Mancini *et al.* As the measured Ga fraction decreases from ~0.6 to ~0.4, the fraction of $Ga^{2+}$ counts increases from ~1% to ~10%, while $Ga^{+}$ counts decrease from 71% to 58%. Simultaneously, $N_2^{+}$ counts increase from 18% to 31%, and $N^{+}$ (or $N_2^{2+}$) counts rise from ~4.5% to ~9.5%. Of particular interest, given their role in molecular dissociation phenomena, is the decrease in the fraction of $GaN_3^{2+}$ counts, which drops from ~1.9% to ~0.7%, providing clear evidence that high fields do not promote the formation of heterogeneous molecular ions. Lastly, protocol (ii) involves increasing LPE while maintaining a constant field by fixing $V_{DC}$. Mancini *et al.* found that varying LPE between 0.7 and 2.4 nJ did not affect the measured Ga fraction within the error bars, which averaged ~0.56 ($F_{eq} \approx 24$ V/nm). In addition, LPE > 2.7 nJ led to asymmetric tip emission, causing severe tip shape deformations. The same experiment was repeated by Di Russo *et al.* at a much lower field ($F_{eq} \approx 22$ V/nm), where the average Ga fraction was ~0.77 while varying LPE from 3 to 40 nJ. Two-dimensional histograms of the charge-state ratio and Ga fraction in detector space reveal a clear correlation between the local electric field and the measured composition. The presence of a c-axis pole is associated with a high-field region and a Ga-poor composition. Furthermore, as the number of detected events increases from $\sim 10^5$ up to $\sim 10^7$, a sixfold pattern characteristic of the wurtzite hexagonal structure becomes evident (Fig. [18]-(c,d)). The same pattern is observed in field ion microscopy (FIM) micrographs (Di Russo et al., 2018a), which directly image the electric field distribution at the tip surface. Finally, this work examines whether the compositional biases observed in GaN originate from intrinsic detection effects that can be identified through the analysis of multiple-hit events (see section 4.d.iii). Correlation histograms indicate signatures of neutral species formation, notably $N_2$ molecules, but the inferred yield is too low to explain the compositional bias at low electric fields. In contrast, dissociation occurring near the tip surface may not appear in these histograms.

Correlation tables and detector hit distances at different electric fields indicate that strong ion-pair correlations mainly arise from correlated field evaporation. No evidence was found for dissociation producing multiple charged fragments, although events yielding one charged ion and one neutral nitrogen atom cannot be excluded.

In 2018, the possible causes of nitrogen underestimation in GaN measurements were in-depth investigated by Morris et al. (Morris et al., 2018). Experiments were conducted using a LEAP 5000 XR equipped with a UV laser (λ =355 nm) and a CAMECA LaWaTAP operated with a green laser (λ =515 nm). The measured composition as a function of the electric field was found to be highly reproducible across different instruments, laser wavelengths and excitation conditions. thereby showing a very high level of reproducibility. In contrast to earlier reports, a stoichiometric Ga:N ratio was recovered under high-field conditions, corresponding to CSR values exceeding 0.3. In 2022, the LEAP 5000 XR electrode was modified by adding a grid on its reverse side to manipulate co-evaporated ions (Morris et al., 2022). This modification reduced the detection efficiency from approximately 52% for the standard electrode to about 7.3% for the grid-equipped electrode, effectively equalizing ion pile-up across all species. However, no effect on the apparent field-dependent stoichiometry was observed, supporting the conclusion that N ion pile-up is not the primary origin of the compositional bias measured in GaN.

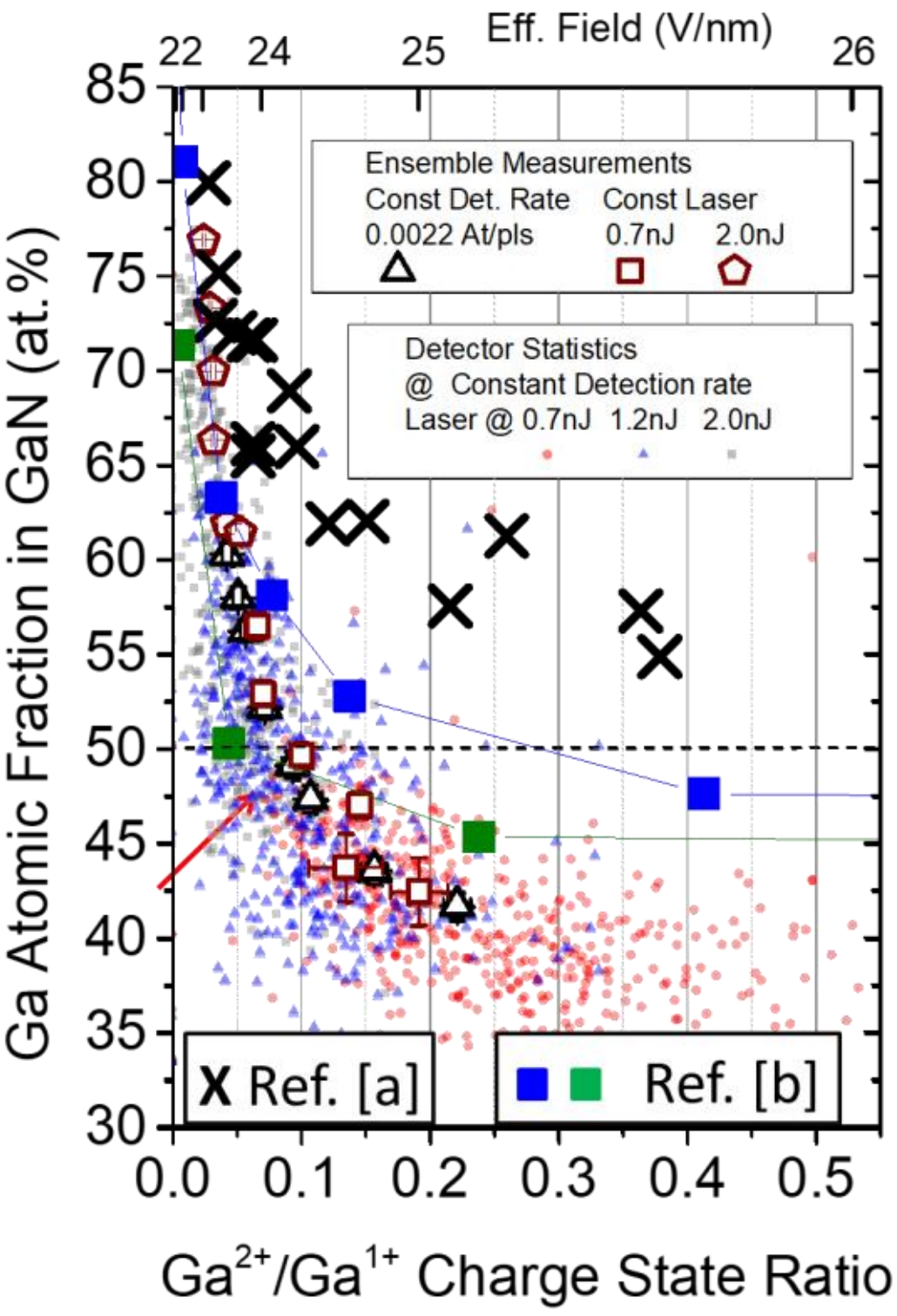

*Figure [22]. Correlation between the atomic Ga fraction in GaN and the CSR($Ga^{2+}/Ga^{+}$). (a) Atomic fraction of Ga from several constant potential measurements (black triangles) or two constant laser intensities (red square and pentagon), and comparison with the constant potential measurements contained in the references [a]=(Agrawal et al., 2011b) [b]=(Diercks and Gorman, 2015) . Adapted from reference (Mancini et al., 2014b), Copyright 2014 American Chemical Society.*

*Mechanisms for neutral N emission.* Nitrogen loss during atom probe analysis of III-nitrides continues to be debated within the community (Gault et al., 2016). Although significant $N_2$ loss via thermal decomposition has been suggested (Huang et al., 2025), it was considered unlikely in the context of APT experiments. Other processes, particularly the surface diffusion of Ga and N on the tip surface, are also likely to play a significant role, potentially affecting the quality of the data. In particular, mass spectra indicates that nitrogen predominantly recombines into $N_2$ molecules on the tip surface prior to emission. If $N_2$ molecules are desorbed from the tip during the tip evaporation, they are expected to behave similarly to the imaging gas in field ion microscopy (FIM). The applied field eventually drives gaseous species (adsorbed or desorbed) towards the highest electric field region of the tip, where a higher N-fraction is expected to be observed. This behavior could partly account for the lower specific detection efficiency for nitrogen, $\eta_N$, observed in $Al_yGa_{1-y}N$ (Di Russo et al., 2018b). However, the first ionization energy of atomic nitrogen is approximately 14.6 eV, whereas that of molecular nitrogen ($N_2$) is about 15.6 eV (Hierl and Franklin, 1967; Jaroń-Becker et al., 2004). For a specimen with a tip radius of curvature of 50 nm and a surface electric field of 20 V/nm, the field strength remains above ~17 V/nm over a distance of nearly 5 nm from the surface, which is sufficient to ionize both $N_2$ and atomic N (Gault et al., 2016). This evidence casts significant doubt on the viability of a loss mechanism involving the direct desorption of neutral molecules in the case of nitrides. An alternative hypothesis attributes the observed nitrogen loss to the emission of neutral species generated by the dissociation of molecular ions in the low-field region near the tip surface. In particular, in GaN it has been observed that low evaporation-field conditions promote the emission of molecular ions such as $N_3^+$, $GaN^{2+}$ and $GaN_3^{2+}$ (Di Russo et al., 2018a). The latter is of particular interest, as it can undergo the following dissociation reaction, leading to the emission of neutral nitrogen species: $GaN_3^{2+} \rightarrow Ga^+ + N^+ + N_2$ (Gault et al., 2016). Molecular dissociation thus represents a compelling explanation. $N_2$-bearing molecular ions may dissociate in-flight in low-field regions, producing neutral fragments that escape from the sample surface but either miss the detector or fall below the detection threshold. Only quite recently, the hypothesis of direct emission by ejection of sufficiently rapid molecules (possibly in an excited state) minimizing the probability of ionization has been given further arguments in the case of $O_2$ in oxide materials (see section 4.d.ii and (Veret et al., 2025)). This hypothesis seems to be viable for $N_2$ in nitrides.

### *iii. Other III-Nitrides: AlN, InN, AlGaN, InAlN, InGaN, ScAlN*

*AlN.* The mass spectrum of AlN closely resembles that of GaN (Mancini et al., 2014b). Nitrogen-related ions are predominantly detected at 14 and 28 m/*z*, corresponding to $N_2^{2+}$ and $N_2^+$ species, respectively. However, the signal at 14 m/*z* may also include contributions from $N^+$ ions; if the peak is entirely assigned to this species, the nitrogen fraction would be decreased by approximately 5%. Interestingly, a weak peak attributed to $N^{2+}$ is observed at 7 M/z , which appears in GaN only under high evaporation field conditions (Di Russo et al., 2018a). Aluminium-related peaks are observed at 27, 13.5, and 9 M/z , corresponding to $Al^+$, $Al^{2+}$, and $Al^{3+}$ ions, respectively. It must be noted that the simultaneous presence of these three charged states represents a deviation from the Kingham statistic, as $Al^+$ and $Al^{3+}$ ions cannot be formed simultaneously (Kingham, 1982). Measurements performed at constant detection rate and fixed laser energies (LPE = 1.0, 1.2, 4.5 and 5.3 nJ), using a LaWaTAP system ($\lambda$ = 343 nm) operated at a repetition frequency of 100 kHz, reveal that the measured composition can be expressed as a function of the charge-state ratio CSR($Al^{2+}/Al^+$) (Mancini et al., 2014b). At low field (CSR = 1), the measured Al fraction is approximately 0.55. A near-stoichiometric composition is obtained at CSR = 10, corresponding to an equivalent field $F_{eq} \approx 24.4$ V/nm. Further increasing the CSR beyond this value leads to a slight reduction in the measured Al fraction, reaching approximately 0.49. Within this CSR interval, the equivalent field $F_{eq}$ ranges from approximately 23 to 25 V/nm, comparable to the evaporation field reported for GaN. This contrasts with the simultaneous observation of $N^{2+}$ and $Al^{3+}$ ions in mass spectra, suggesting that the evaporation field of AlN is higher than that of GaN.

In 2022, Morris et al. (Morris et al., 2022) measured the field-dependent composition of AlN and provided a detailed analysis of the role of multi-hits using a LEAP 5000 XR operated in laser mode ($\lambda$ = 355 nm) using a pulse frequency of 125 kHz. As the CSR increased from 0.8 to 7, the overall detected

Al fraction decreased from approximately 0.64 to 0.58, while the fraction of multi-hits dropped from 18% to 10%. The measurements were then repeated using an electrode modified with an incorporated grid, which reduced the detection efficiency from ~52% to ~7.3%. This reduced the fraction of multi-hits by a factor of ~4. Nevertheless, the measured composition as a function of the field remained largely unchanged.

The composition of AlN thin films was further investigated by Hans et al. in 2023 (Hans et al., 2023) using two local electrode atom probe (LEAP) instruments: a LEAP 4000X HR operating with a UV laser ($\lambda = 355$ nm) and a LEAP 3000X HR equipped with a green laser ($\lambda = 532$ nm). Constant detection rate experiments were carried out by systematically varying LPE. For the UV laser, the pulse energy was adjusted between 5 and 50 pJ at a repetition rate of 200 kHz, whereas for the green laser it ranged from 0.1 to 2.0 pJ with a pulse frequency of 250 kHz. The data indicate that the detected Al-fraction decreases from approximately 0.64 to 0.51 as the equivalent field, $F_{eq}$, calculated from the CSR($Al^{2+}/Al^{+}$), rises from ~25.6 to ~29.2 V/nm, independent of the experimental conditions. This indicates that the accuracy of composition measurements in AlN is strongly dependent on the applied electric field strength.

*InN.* [0001]-oriented InN was analysed by Di Russo *et al.* (supporting information to (Di Russo et al., 2019)) using a FlexTAP instrument operated at constant detection rate with green laser pulses ($\lambda \approx 515$ nm). The resulting mass spectrum closely resembles that of GaN, with nitrogen predominantly detected as $N_2^{+}$ ions; contributions from $N^{+}$ (or $N_2^{2+}$) and $N_3^{+}$ ions are also observed. Indium is detected in both singly and doubly charged states, with $In^{+}$ peaks at 113 and 115 m/z and $In^{2+}$ peaks at 56.5 and 57.5 m/*z*. Several heteronuclear molecular ions are identified, including $InN^{2+}$, giving rise to a prominent peak at 64.5 M/z , and $InN_3^{2+}$, observed at 78.5 m/*z*. A spatially resolved analysis of the probed surface revealed pronounced local variations in the indium charge-state ratio CSR($In^{2+}/In^{+}$), exhibiting a radial symmetry centred on the [0001] crystallographic pole. These variations reflect changes in the $F_{eq}$, which was estimated to range between approximately 21 and 24 V/nm. This field range suggests that the effective evaporation field of InN may be lower than that of GaN and AlN. A strong correlation was observed between the local CSR and the measured In-fraction. At high effective fields (~24 V/nm), the detected composition was In-depleted, with indium fractions as low as ~0.45, whereas decreasing the field led to progressively In-rich compositions, reaching values up to ~0.7 at ~21 V/nm. Stoichiometric InN was recovered only within a narrow field window around ~23 V/nm, spatially confined to a thin annular region surrounding the [0001] pole. These observations were interpreted in terms of competing field-dependent loss mechanisms. Under high-field conditions, preferential evaporation of indium between laser pulses results in indium depletion. Conversely, at lower fields, nitrogen loss (most likely through the formation of neutral $N_2$ species that escape detection) dominates, leading to indium enrichment. Accurate stoichiometry is achieved only under intermediate field conditions where these two effects are balanced.

*AlGaN.* The field dependence of the detected AlGaN composition was first examined in a c-axis–oriented GaN multiple quantum-well structure presenting thick $Al_{0.25}Ga_{0.75}N$ barriers (Rigutti et al., 2016a). Constant detection rate experiments were carried out by maintaining a fixed LPE, which resulted in a progressive decrease of the surface electric field during the analysis while maintaining an inhomogeneity of surface electric field and microscopically measured composition, as traced in the detector space histograms of Fig. [23]-(a). As a consequence, each AlGaN barrier was evaporated under progressively lower electric-field conditions. For each AlGaN barrier, the average CSR and the corresponding composition were determined. The evolution of the local electric field was quantified through the charge-state ratio CSR($Ga^{2+}/Ga^{+}$), which provides an estimate of the equivalent field $F_{eq}$. An increase in the local electric field is associated with a pronounced decrease in the detected Ga fraction, which drops from approximately 0.4 to 0.3, while the detected Al fraction remains essentially unchanged. In addition, consistently with observations in GaN, increasing the electric field also leads to an enhancement of the detected nitrogen fraction, which rises from about 40% to 50%. The data show that increasing the CSR from approximately 0.03 to 0.1 results in an increase of the measured group-III site fraction $y$, defined as $y$ = Al/(Al+Ga), from 0.27 to 0.37. Notably, this range systematically exceeds the nominal Al content of the barriers ($y$ = 0.25), indicating that the measured composition consistently overestimates the true Al fraction. This work also proposed a scheme for the correction of the bias based on the statistical properties of the random AlGaN alloy, as shown in Fig. [20]. These findings are consistent with the work of Di Russo et al. (Di Russo et al., 2019), who reported a correlation between the microscopic electric-field distribution and the measured composition in an $Al_{0.15}Ga_{0.85}N$ electron-blocking layer in a c-axis–oriented LED structure. Measurements were performed at constant detection rate using FlexTAP (λ =343 nm, pulse repetition rate 50 kHz), fixing the LPE at 1.0 nJ. Results indicate that $y$ increases from 0.20 to 0.35 as CSR($Ga^{2+}$/$Ga^{+}$) rises from 0.2 to 2.0, corresponding to an equivalent electric field $F_{eq}$ ranging from approximately 25 to 27 V/nm. Interestingly, these values are again higher compared to the expected composition. Because the mass spectra typically exhibit peaks attributed to $Ga^{+}$, $Ga^{2+}$, $Al^{+}$, $Al^{2+}$, and $Al^{3+}$, the CSR($Al^{2+}$/$Al^{+}$) was also calculated. A comparison between the two metrics shows that they are nearly equivalent. In particular, CSR($Al^{2+}$/$Al^{+}$) spans from approximately 3 to 20, which corresponds to a narrower estimated field range of about 24 to 25 V/nm.

In 2018, Morris et al. have investigated bulk $Al_yGa_{1-y}N$ ($y$ = 0.08, 0.44, 0.75) using both a LEAP 5000XR (λ = 355 nm) and a CAMECA LaWaTAP (λ = 515 nm), performing measurements at a constant detection rate (Morris et al., 2018). The two instruments demonstrate nearly identical measured compositions as a function of the electric field. The evaporation field of $Al_yGa_{1-y}N$ depends on the Al fraction $x$.For sample with $y$ = 0.08, the tip field required at fixed laser energy closely matches that of GaN. In contrast, samples with higher Al content ($y$ = 0.44 and 0.75) require substantially stronger tip fields, with CSR values exceeding 5 and remaining comparable between the two compositions. This increase is consistent with the higher evaporation field of Al. Additionally, the expected 1:1 ratio between group-III elements (Al + Ga) and N is never fully achieved in $Al_yGa_{1-y}N$ samples, with the exception of the sample with $y$ = 0.44, which reaches a composition close to the expected ratio at high field (CSR = 0.3), whereas GaN exhibits a nearly stoichiometric composition at high field (CSR = 0.5). Interestingly, at the same field, the N fraction decreases as the Al content increases, suggesting that the higher LPE required to evaporate the tip are associated with a higher tip temperature, which in turn enhances $N_2$ sublimation. No variation in the site fraction $y$ with applied field is observed for $y$ = 0.08 over the CSR range 0.002 – 0.2. The sample with $y$ = 0.44 shows that $y$ decreases as CSR decreases, reaching the expected composition at CSR ≈ 0.04, before slightly decreasing again. Lastly, for $y$ = 0.75, the expected site fraction $y$ is recovered at both low and high electric fields (CSR = 0.05 and 6, respectively), whereas intermediate field conditions yield $y$ values lower than the nominal composition. This deviation originates from an overestimation of the measured Al-fraction and an underestimation of the Ga-fraction relative to the expected stoichiometry.

The accuracy of compositional measurements in c-axis–oriented $Al_yGa_{1-y}N$/Ga multilayers was investigated and modeled by Di Russo et al. in 2018, considering a series of samples with III-site fractions y ranging from 0.07 to 0.56 (Di Russo et al., 2018b). Analyses were performed with a LaWaTAP operated with a UV laser (λ = 343 nm) and equipped with a custom-designed detector system (Da Costa et al., 2005). Preliminary analyses of the CSR($Ga^{2+}$/$Ga^{+}$) and CSR($Al^{2+}$/$Al^{+}$) shows that they are effectively equivalent at low surface fields, whereas deviations from Kingham's theory emerge at higher fields. Despite this limitation, the associated uncertainty in the surface field determination under the experimental conditions considered here is confined to ~1 V/nm (~5%). The measured composition exhibits a strong dependence on the surface electric field: $y$ is accurately reproduced under low-field conditions but increases systematically with increasing surface field. This trend is general and reproducible over the entire composition range investigated. The presence of GaN layers allowed for an accurate calibration of the 3D reconstructions and for the determination of element-specific detection efficiencies $\eta_i$ (see section 4.a.iii). The Al detection efficiency, $\eta_{Al}$, remains essentially constant along the analysis direction but shows a pronounced radial variation, reaching its highest values in low-field regions, where it approaches the detector efficiency limit ($\eta_{Al} \approx \eta_{MCP} \approx 0.6$). In contrast, the Ga detection efficiency, $\eta_{Ga}$, is strongly correlated with the CSR($Ga^{2+}/Ga^{+}$) in both axial and radial directions, and approaches the detector efficiency only in low-field regions near the end of the analysis. The N detection efficiency, $\eta_N$, exhibits only a weak radial dependence and a slight increase along the analysis direction, following a trend opposite to that of $\eta_{Ga}$; however, its average value remains well below the detector efficiency limit ($\eta_N < \eta_{MCP} \approx 0.6$). These data were subsequently used to formulate a semi-quantitative model describing the preferential evaporation of Al and Ga between laser pulses, which accounts for the observed increase in the measured III-site fraction *y* with increasing surface electric field and a correct value of composition measured at low field only, as visualized in Fig. [23]-(b). The estimated energy barriers are approximately $Q_{Ga}$ = 1.0 eV and $Q_{Al}$ = 1.1 eV, with corresponding evaporation fields of $F_{ev,Ga}$ = 25 V/nm and $F_{ev,Al}$ = 27 V/nm. By contrast, nitrogen is proposed to be lost via a distinct mechanism, involving neutral evaporation and the generation of neutrals from molecular dissociation.

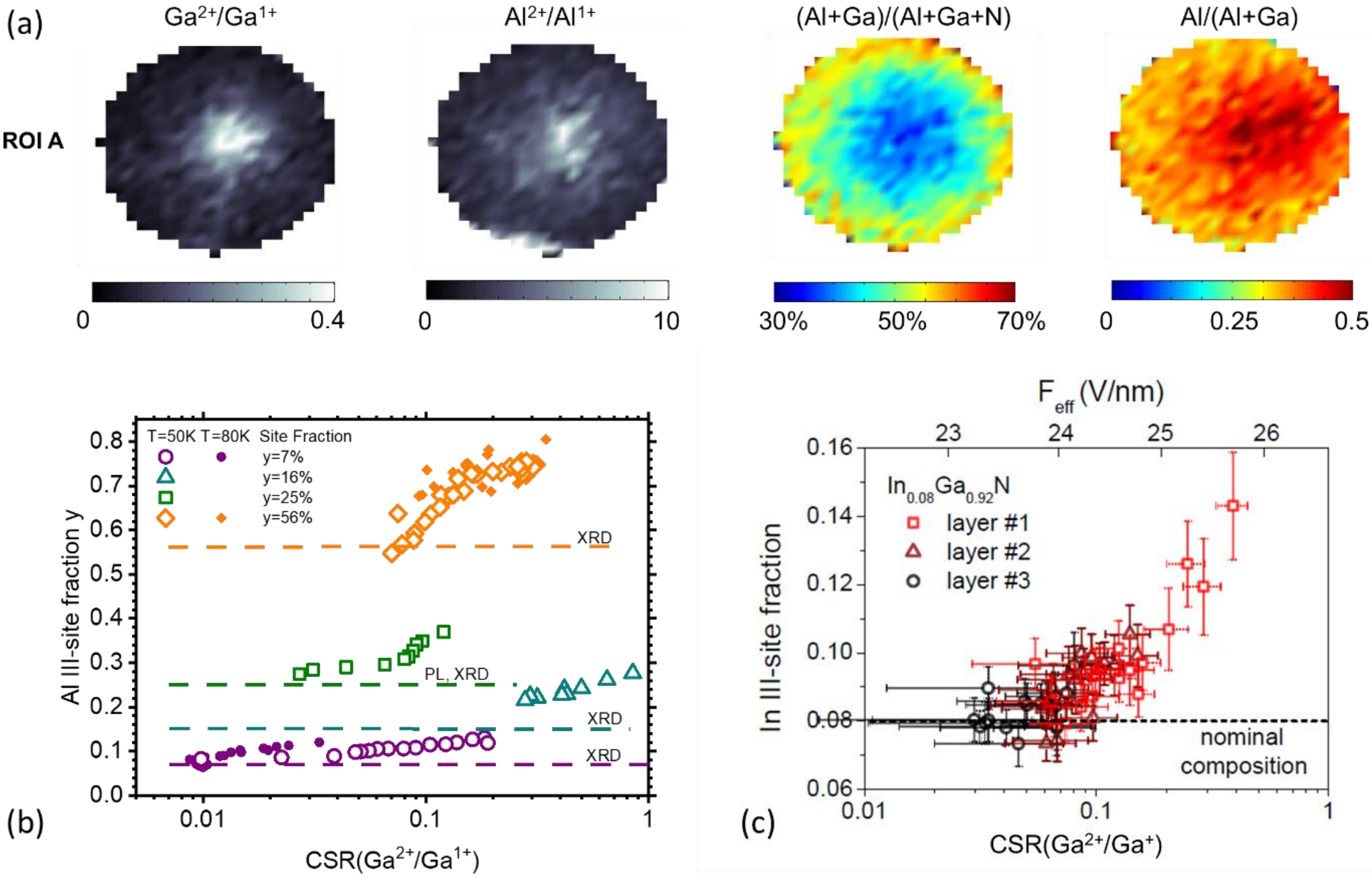

*Figure [23]. Field-dependent compositional inaccuracies in AlGaN and InGaN. (a) From left to right, detector space histograms showing the distribution of the CSR($Ga^{2+}/Ga^{+}$), CSR($Al^{2+}/Al^{+}$), atomic fraction of III-type elements and Al III-site fraction issued from an analysis of random AlGaN alloy (Adapted with permission from (Rigutti et al., 2016a), AIP). (b) Correlation between the y III-site fraction in AlGaN and the CSR($Ga^{2+}/Ga^{+}$) ratio for 6 samples with four different compositions of AlGaN (Adapted with permission from* (Di Russo et al., 2018b) *ACS). (c) In III-site fraction plotted as the function of the local CSR($Ga^{2+}/Ga^{+}$) from the analysis of quantum well layers of a LED device. The effective field $F_{eff}$ is calculated from the CSR($Ga^{2+}/Ga^{+}$) (Reproduced with permission from (Di Russo et al., 2019), AIP).*

*InAlN*. Only a limited number of studies report compositional measurements of $Al_yIn_{1-y}N$. A representative mass spectrum was reported by Dawahre et al. (Dawahre et al., 2013). The spectrum closely resembles that of AlN, with addition peaks attributed to In ions at 115 and 57.5 M/z , corresponding to $^{115}In^{+}$ and $^{115}In^{2+}$, ions respectively ($^{113}$In gives rise to a small peak, accounting for only 4.3% of the total element abundance). Nevertheless, no systematic investigation of the composition as a function of the analysis parameters has been reported. Mancini et al. measured the composition of a $Al_yIn_{1-y}N$ grown along the m-axis. Measurements were performed using a LaWaTAP system operated with a UV laser (λ = 343 nm) at LPE = 1 nJ and constant detection rate. The microscopic distributions of both CSR($Al^{2+}/Al^+$) and CSR($In^{2+}/In^+$) were reported, together with the corresponding spatial distribution of the measured In/(Al+In) fraction. However, the resulting composition maps showed no correlation with the local surface field (Mancini et al., 2014b).

*InGaN.* The correct quantification of the In III-site fraction in InGaN is technologically significant as this alloy constitutes the quantum well layers of LEDs. Riley et al. (2014) systematically investigated the reliability of indium quantification in $In_yGa_{1-y}N$ ($y \approx 0.25$) quantum wells (QWs) grown on wurtzite GaN along non-polar m-axis. APT measurements were performed using a LEAP 4000X Si instrument equipped with a UV laser (λ = 355 nm), operated at a pulse energy of 0.01 pJ and a repetition rate of 200 kHz. The evaporation behavior of In, Ga, and N atoms is strongly influenced by the structure and surface polarity of the tip facets. In particular, variations in the III/N ratio normal to the non-centrosymmetric c-plane reflect a transition from N-polar to Ga-polar surface termination. N-polar surfaces typically exhibit a balanced Ga-to-N ratio, whereas Ga-polar terminations show systematic nitrogen depletion, as previously reported for $a$-axis–oriented GaN nanowires (Riley et al., 2014). This artifact has been attributed to the preferential diffusion of nitrogen, which promotes $N_2$ formation and desorption over direct evaporation, leading to nitrogen loss either as neutral species or beyond the detection window. When moving into the QWs along the $m$-axis, Ga counts decrease more than the corresponding increase in In counts on both Ga- and N-polar facets, suggesting a reduced overall detection probability for group-III species. By contrast, Ga counts from the nonpolar $m$-plane remain relatively constant along the analysis direction, indicating enhanced detection of group-III species within the QWs, accompanied by a rise in N counts. The resulting increase in the N/III ratio within the QWs can thus be attributed to a reduced Ga detection probability, consistent with the observed elevation in the $Ga^{2+}/Ga^+$ ratio, which signals a locally increased electric field. Although $Ga^+$ and $In^+$ ions have different nominal evaporation fields (15 V/nm and 12 V/nm, respectively) (Tsong, 1978b), their field-dependent evaporation rates are sufficiently similar that variations in local field due to polarity, surface faceting, or composition do not induce measurable changes in the In mole fraction. Moreover, the evaporation characteristics of c-plane (0001) and semipolar $(20\bar{2}\bar{1})$ QWs exhibit analogous responses to surface polarity and crystallography, supporting the consistency of these conclusions across different orientations.

The composition of $In_yGa_{1-y}N$ quantum wells grown along the m-axis of a $In_yGa_{1-y}N$/GaN multi-quantum wells system was analyzed by Mancini et al. (Mancini et al., 2014b). Measurements were performed using a LaWaTAP system operated with a UV laser (λ = 343 nm) at constant detection rate, with LPE fixed at 0.3 nJ. Maps of the CSR($Ga^{2+}/Ga^+$) revealed a high-field region that was attributed to the presence of the m-pole. As observed for GaN, the measured group-III fraction deviates from the stoichiometric value of 0.5, being higher than 0.5 in high-field regions and lower in low-field regions. Within the quantum wells, In is found to form high-concentration stripes approximately aligned with the c-axis, whose position varies from well to well independently of the local surface field. Notably, the indium distribution shows no correlation with either the field distribution or the measured group-III fraction. The analysis of the average charge-state ratio and the average group-III fraction measured along the tip reveals oscillations in both quantities, with a periodicity matching the positions of the $In_yGa_{1-y}N$ quantum wells. These oscillations were interpreted as arising from local modulations of the evaporation field, which is slightly higher in the quantum wells than in the surrounding GaN matrix.

$In_yGa_{1-y}N$ layers within a c-axis–oriented LED structure, with a nominal indium fraction $y = 0.08$, were investigated by Di Russo et al. (Di Russo et al., 2019). Spatially resolved measurements of both the local field and composition were performed under constant detection detection rate using a FlexTAP instrument operated with a UV laser (λ = 343 nm, 50 kHz) and a fixed LPE = 0.3 nJ. The corresponding mass spectra exhibit all the characteristic peaks typically observed in InN and GaN. Three distinct $In_yGa_{1-y}N$ layers were analyzed, each experiencing a progressive reduction of the surface electric field due to the so-called cone-angle effect experience during constant detection rate measurements. The local field distribution was estimated using both CSR($In^{2+}/In^+$) and CSR($Ga^{2+}/Ga^+$); however, the two metrics were found to differ significantly. The indium charge-state ratio decreased from ~0.7 to ~0.1, corresponding to an equivalent field $F_{eq}$ ranging from approximately 22.5 to 21 V nm$^{-1}$. In contrast, the Ga charge-state ratio decreased from ~0.7 to ~0.03, indicating higher equivalent fields, spanning roughly 25 to 23 V nm$^{-1}$. Consequently, the Ga-based charge-state ratio was adopted as the reference metric for comparing compositional measurements in $In_yGa_{1-y}N$ with other III–N semiconductors. Field maps revealed a pronounced high-field region at the center of the tip, attributed to the alignment with the crystallographic c-axis. The local In group-III site fraction was found to be enhanced in the same regions. When plotted (Fig. [23]-(c)) as a function of the Ga-based charge-state ratio, the In fraction decreased from ~0.14 to ~0.07 (this last value approaches the nominal composition) as CSR($Ga^{2+}/Ga^+$) decreased from ~0.4 to ~0.03. As previously observed for $Al_yGa_{1-y}N$, the nominal composition in $In_yGa_{1-y}N$ is therefore recovered under lower-field evaporation conditions.

*ScAlN.* ScAlN has been the object of several studies in APT (Dzuba et al., 2022; Ndiaye et al., 2023b). The work by Ndiaye et al. has focused on its compositional metrology, monitoring the dependence of the Sc III-site fraction on the CSR(Al), indicating that this quantity is rather constant and thus pointing out for an accurate measurement of composition (Ndiaye et al., 2023b). However, these experiments were conducted on thin films and more statistical significance would be welcome for this alloy.

#### iv. *Other Nitrides*

*TiN.* Multiple detection events may represent up to 78% of detection events for TiN. These multiple hits are linked to ion dissociation and correlated evaporation, which lead to under- or over-estimation of nitrogen and titanium in the reconstructed datasets. Schiester and co-authors have studied the problem by comparing results obtained by hardware filtering (Thuvander et al., 2013), see also section 5.a.vi on carbides) or by software filtering (i.e. by filtering out the multiple detection events). Filtering obviously decreases the global amount of information, but allows for a better measurement accuracy. In voltage-mode APT, without filtering, the composition deviated by ~3.8 at.% from reference. With hardware + software filtering, deviation improved to 2.1 at.%, and software filtering alone achieved ~0.1 at.% error, within uncertainty limits. Low pulse energy (0.1 nJ) yielded ~1.4 at.% error. In LA-APT, higher pulse energy (2.0 nJ) increased error up to 8.1 at.%, corresponding with increased formation of complex molecular ions (e.g. $Ti_2N$), and weakened electric field (~36 V/nm vs ~40 V/nm). Using laser-assisted APT, the accuracy was significantly increased through software filtering the same value could be observed with 0.6 nJ LPE within the measurement uncertainties of TOF-ERDA/RBS. However, investigations based on laser-assisted APT measurements indicated that a high proportion of multiple detection events alone does not lead to significant deviations from the reference composition. (Schiester et al., 2024).

Compositional metrology of TiN has been recently studied by Schiester and co-authors not only exploring the space of analysis parameters (laser pulse energy, surface field) but also the performances of instruments differing by laser wavelength and detector technology. The results show a measured composition close to the nearly-stoichiometric one assessed by complementary techniques (Fig. [24]-(a)). For all three instruments the underestimation of the N content is reduced as the electric field strength increases. The use of a deep UV laser (λ=257 nm) is also assessed as beneficial with respect to longer wavelengths (λ=355 nm for the UV and λ=515 nm for the green). On the other hand, the impact of the detector technology (reduction of the dead zone and increase of the microchannel open area in the LEAP generations after the 3000) is not found to play a significant role (Schiester et al., 2025). Interestingly, the trends reported in this work point out a trend in the measurement which is common to the main III-N materials. However, in the case of TiN the surface field is much more intense and should be sufficient to prevent N losses via post-ionization.

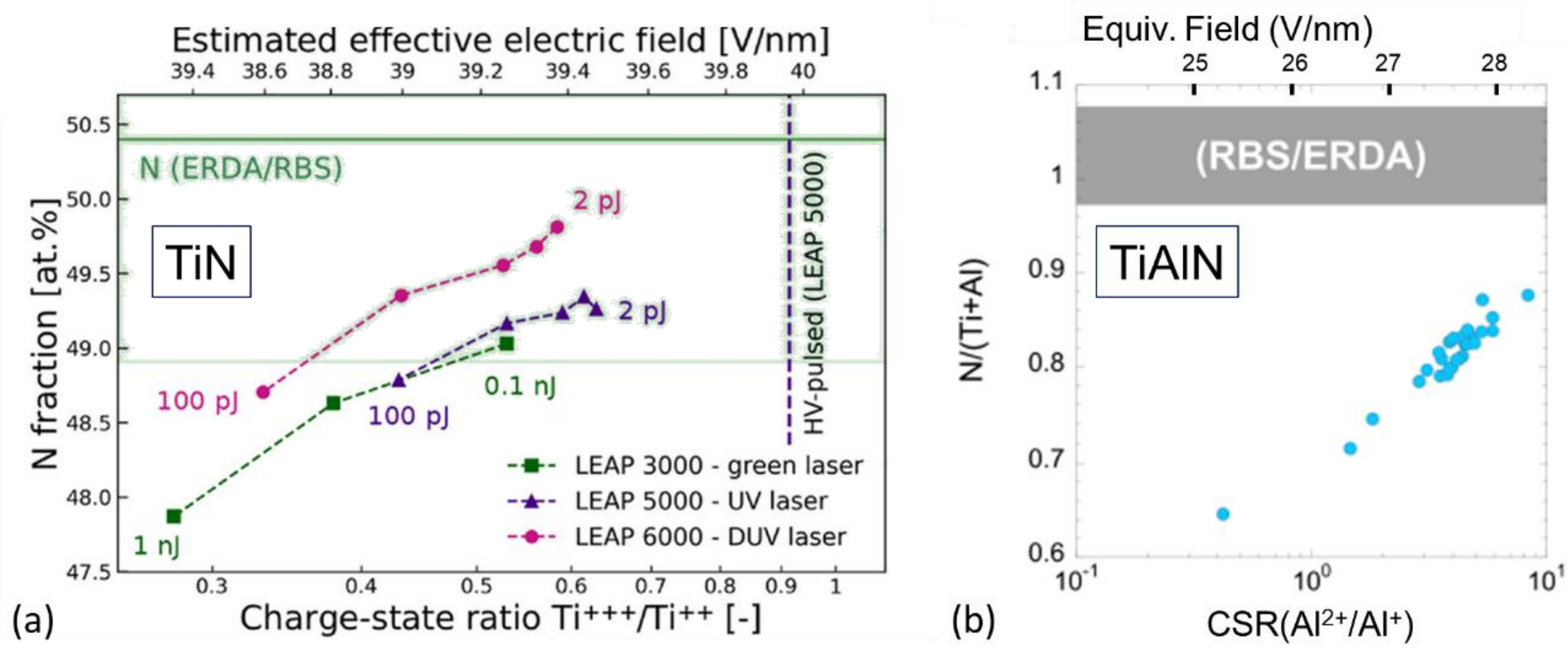

*Figure [24]. (a) Dependence of the measured atomic fraction of N in TiN as a function of the CSR($Ti^{3+}/Ti^{2+}$) and for instruments with different laser wavelength and detector technology (Reproduced from ref. (Schiester et al., 2025), Elsevier). (b) Dependence of the measured atomic ratio N/(Ti+Al) as a function of the CSR($Al^{2+}/Al^{+}$) (Adapted from ref. (Hans and Schneider, 2020), IOP Publishing).*

*TiAlN.* The accuracy of the composition measurement for the alloy TiAlN has also been investigated (Hans and Schneider, 2019, 2020). This compound evaporates at significantly lower field than TiN, as reported in ref. (Hans and Schneider, 2020) (Fig. [24]-(b)), i.e. at field intensities slightly higher than the typical values for GaN and AlN (Mancini et al., 2014a). The authors study in particular the dependence of the N/(Ti+Al) ratio on the laser pulse energy and on the CSR($Al^{2+}/Al^{+}$). The expected ratio N/(Ti+Al)~1 is gradually approximated for increasing field but not reached at the highest tested field (28 V/nm). In general, the increase of the electric field strength from ~25 V/nm to ~28 V/nm reduces the error from the independently measured composition (ERDA , RBS) from 11.4 to 4.1 at.% for N (Fig. [24]-(b)), from 8.8 to 3.0 at.% for Al and from 2.8 to 0.9 at.% for Ti. The authors interpret these phenomena as a gradually increasing ionization efficiency for possible neutral N. The authors also notice that the increase of surface field also translates into an increase of multiple detection events from 27% to 53%. They thus tend to exclude a role of the pile-up in the inaccuracies enhanced at low field (Hans and Schneider, 2020). However, this point could show that at low field the pile-up effectively translates into a larger fraction of multiple events erroneously interpreted as singles.

*TiSiN.* TiSiN is a coating material in which the main challenge for APT is the overlap between the $^{28}Si^{n+}$ and the $^{14}N_n^{n+}$ peaks (n=1,2). This problem has been circumvented by replacing $^{14}N$ with $^{15}N$, so that peaks of interest in the mass spectrum are shifted, enabling the separation of overlapping signals. This technique allowed identification and quantification of atomic fraction in reasonable agreement with other microanalysis techniques such as energy-dispersive X-ray spectroscopy (EDS) and energy elastic recoil detection analysis (ERDA) (D. L. J. Engberg et al., 2018; Naghdali et al., 2025).

### v. *Oxides*

*ZnO and MgO.* As compounds forming covalent to ionic bonds, oxides have a usually complex evaporation behavior. A first difficulty in compositional metrology of oxides is already in the mass spectrum. In ZnO, two significant peak overlap issues must be addressed. First, the peak at 32 m/z receives contributions from both $O_2^+$ and $^{64}Zn^{2+}$; separating these two contributions requires isotopic analysis by counting the heavier $Zn^{2+}$ isotopes appearing between 33 and 34 M/z , which correspond to $Zn^{2+}$ ions containing isotopes heavier than $^{64}Zn$ (Mancini et al., 2014). Second, the peak at 16 m/*z* may be attributed to $O^+$ or to $O_2^{2+}$, which is also a kinetically stable species (Nobes et al., 1991). While oxides such as $WO_3$ and NiO display a composition close to stoichiometry (Oberdorfer et al., 2007), other oxides may undergo moderate to severe compositional biases in APT measurements. Examples of measurement of composition in selected binary oxides is reported in Fig. [25]. The measurement of the atomic fraction of Zn in ZnO as a function of the CSR(Zn) is reported in Fig. [25]-(a). The composition appears to undergo important oxygen losses, which become moderate when the field increases (Mancini et al., 2014a). The main channel of loss for O is thought to be the emission of neutrals, as detailed in the following. At higher field, the composition may become O-rich as more and more Zn is lost via preferential evaporation (Amirifar et al., 2015). A similar picture can be drawn for MgO, as reported in Fig. [25]-(b). Here, however, the composition globally remains Mg-rich, as reported by different authors (Devaraj et al., 2013; Mancini et al., 2014a). In both cases, the composition is rather independent of the laser energy at constant surface field. ZnO and MgO were among the first systems for which the interpretation of experimental data was backed up by theoretical-computational efforts. These efforts developed along two main directions. The first one was the simulation of the evaporation processes from small clusters by density functional theory (DFT) (Karahka and Kreuzer, 2015; Xia et al., 2015, 2024). These calculations have shown that they can account for features such as bandgap reduction at the apex surface and evaporation of molecular ions such as $O_2^+$ and $ZnO^{2+}$, as reported in Fig. [26]-(a) (Xia et al., 2015). DFT calculations remain challenging and have significant limitations: beyond the problem of the choice of the DFT exchange potential, the simulation of realistic surfaces with a significant set of atoms is extremely demanding in computation time; furthermore, DFT describes ground states and lacks the capability of exploring excited states. In other words, DFT can provide qualitative trends, but quantitative predictions about critical fields, species formation, and compositional changes must be validated against experiment and sometimes require empirical adjustments or more advanced computational treatments (Ashton et al., 2020b). A second type of approach, tackling the shortcomings of DFT as well as the complexity of a surface process and restricting the attention to the specific problem of in-flight dissociation reactions, is based on calculations via molecular dynamics (MD). This approach has been applied to the study of ZnO, and in particular to the dissociation behavior of the $ZnO^{2+}$ molecule within the singly and doubly excited configuration interaction (MR-SDCI) model (Zanuttini et al., 2017a). According to the results of these calculations, the $ZnO^{2+}$molecular ion exhibits more field-dependent dissociation channels. As reported in Fig. [26]-(b), one channel produces a neutral O atom within a given field threshold which also depends on the state in which the molecule is evaporated. The production of neutral O close to the apex surface represents a channel of loss for O at low field, as displayed in Fig. [26]-(c). When the field increases, the O atom can more and more probably be post-ionized and thus be detected. Notice that the Fig. [26]-(c) supposes that all evaporation occurs through the channel producing a $ZnO^{2+}$ molecular ion, while the experiment clearly shows that other channels are represented. Thus, this calculation can reproduce qualitatively the O deficit at low field, but cannot account for the Zn deficit at high field (Zanuttini et al., 2017a).

*Gallium and aluminum oxides*. Gallium oxide evaporates at significantly higher CSR($Ga^{2+}/Ga^+$)~0.5-2 than GaN and III-N compounds. The composition analysis in $Ga_2O_3$ presents some common features with ZnO, as the measurement of the atomic fractions is rather independent of the laser energy at constant surface field, but presents an oxygen deficiency at low field which tends to saturate close to stoichiometry for high fields. As a peculiar feature of this compound, around 2% of the detected events are produced by the dissociation reaction $GaO^{2+} \rightarrow Ga^+ + O^+$ and are found therefore outside of the peak neighborhoods of the parent and daughter ions in the mass spectrum (Chabanais et al., 2021). In aluminum oxide ($Al_2O_3$) several authors have pointed out an excess of Al atomic fraction in the explored space of parameters (Mazumder et al., 2014; Garcia et al., 2023). Interestingly, mass spectra of $Al_2O_3$ contain significant amounts of the three charge states $A^+$, $Al^{2+}$, $Al^{3+}$, in disagreement with the post-ionization theory. Although no study has been performed on this system based on the CSR metrics, recent results showed that extreme UV (EUV)-pulsed APT yields more stable compositions and less thermal tailing in spectra compared to UV, allowing more predictable and sensitive analysis across a wider range of laser intensities (Garcia et al., 2023).

*Iron oxides.* Iron oxides represent a more complex domain as different stoichiometries should be considered (Bachhav et al., 2013, 2011; Kim et al., 2024). Figure [25]-(c) reports the O fraction measured in $Fe_2O_3$, $Fe_3O_4$ and FeO as a function of the CSR(Fe). Significantly, here the measured composition is O-poor, and, contrarily to ZnO and MgO, the oxygen loss tends to increase at higher field. The study of iron oxides points out the high degree of complexity of field evaporation in this system. Dissociation reactions appear to play here a very important role. Some of these have been identified in correlation diagrams and studied by DFT. These dissociations, especially those involving neutral molecules, show a dependence on crystallographic orientation. The significant proportion of multiple events, which exhibit considerable correlations, can be considered as the main metrological feature calling for attention for possible compositional biases (Kim et al., 2024).

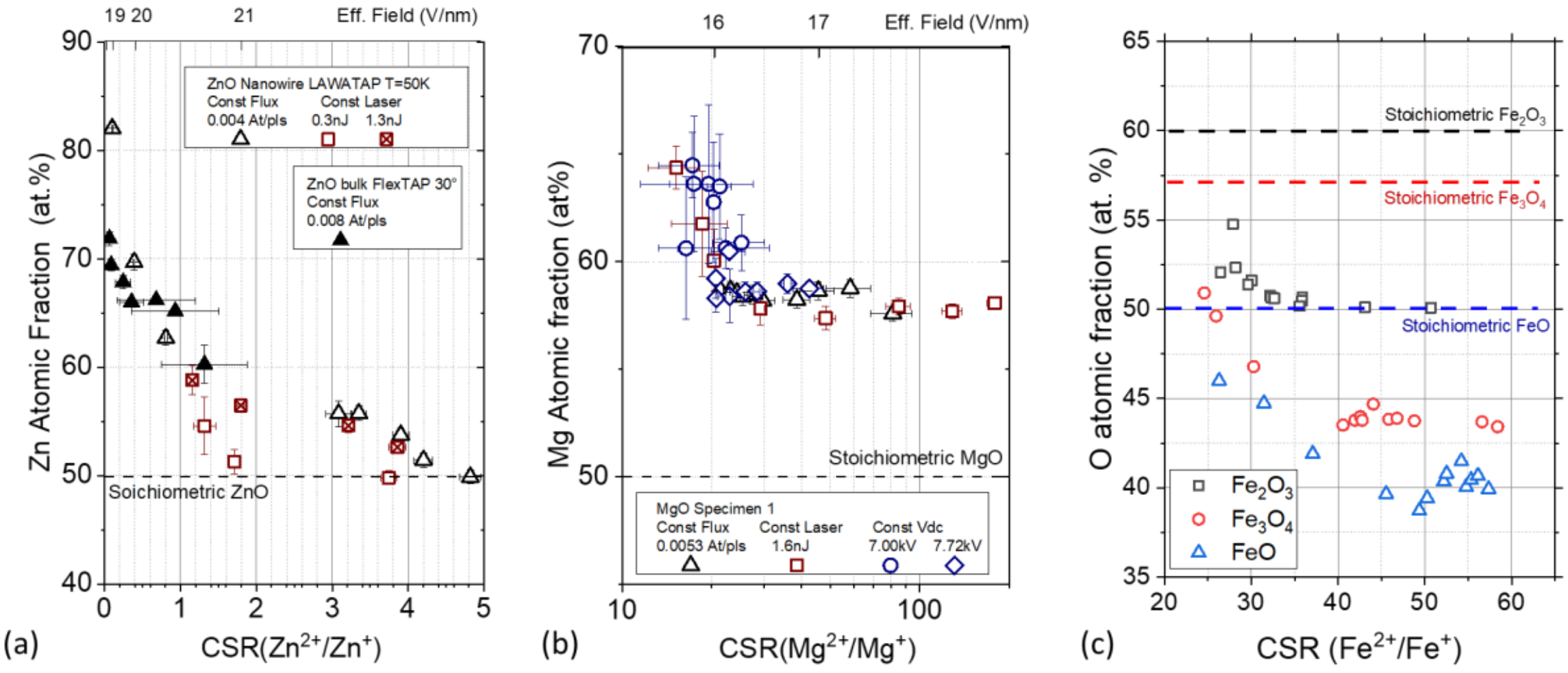

*Figure [25]. Compositional biases in binary oxides. (a) Zn atomic fraction as a function of the CSR(Zn) in ZnO and (b) Mg atomic fraction as a function of the CSR(Mg) in MgO, collected at different experimental conditions (Adapted from (Mancini et al., 2014a)); (c) O atomic fraction as a function of the CSR (Fe) from different iron oxide compounds (Data reproduced from (Kim et al., 2024)).*

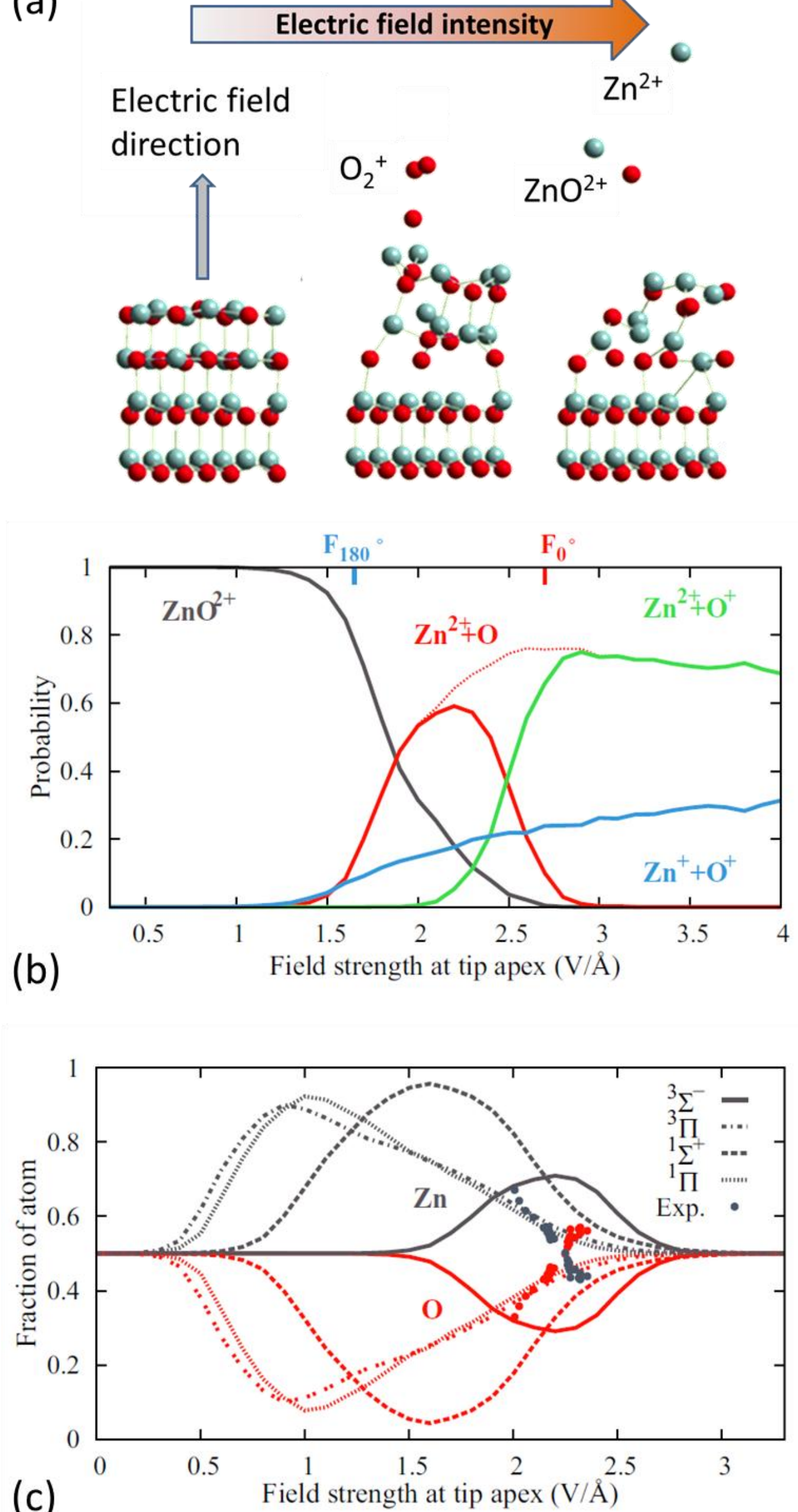

*Figure [26]. Computational approaches for the explanation of biases in binary oxides. (a) DFT calculation displaying snapshots from the evolution of the surface of a ZnO cell from field-free conditions (left) to field evaporation (middle, right). The calculations can reproduce the evaporation of $O_2^+$ and $ZnO^{2+}$ (Data adapted from (Xia et al., 2015)). (b) Abundances of species resulting from the evaporation of $ZnO^{2+}$ molecules as a function of the field in proximity of the surface. The increasing field produces a fast dissociation yielding significant amounts of a neutral fragment ($ZnO^{2+} \rightarrow Zn^{2+}+O$) between 1.5 and 2.4 V/Å. The increase of the field progressively translates into the post-ionization of the neutral O. (c) Calculated effect on the measured composition considering different initial states for the $ZnO^{2+}$ molecular ion under the hypothesis that all evaporated ions are $ZnO^{2+}$; experimental data are reported for comparison (Reproduced from from* (Zanuttini et al., 2017a)*, American Physical Society).*

*Silicon Oxides.* The case of $SiO_2$ is important due to its use in microelectronics. This system was studied varying base temperature and peak ranging and attribution within an instrument pulsed by extreme-UV laser. The measured O:Si ratios average quite close to stoichiometric $SiO_2$. Some scatter is also present depending on the experimental conditions, but no clear trend has been assessed. The standard deviation in the measured compositions suggests that each experimental variable (temperature, pulsing mode, field strength, and specimen preparation) plays a significant role (Chiaramonti et al., 2020).

*Cromium Oxides.* La Fontaine investigated thermally grown chromium oxyde ($Cr_2O_3$) using a LEAP 4000 equipped with a UV laser (λ = 355 nm), varying the LPE between 22 and 115 pJ at a base temperature of 50 K (La Fontaine et al., 2015). Under these conditions, the measured oxygen content averaged 58.4 ± 0.5 at.%, still slightly below the expected stoichiometric value of 60 at.%, with no significant dependence on laser energy. Similarly, neither varying the laser pulse frequency from 100 to 500 kHz nor increasing the base temperature from 24 to 64 K produced appreciable changes in the measured composition. Through correlation histogram analysis, several in-flight molecular ion dissociation reactions were identified, including dissociations from singly-charged molecular ions such as $CrO_2^+$ and $CrO_3^+$, each producing one charged and one neutral daughter ion (i.e., $O_2$). While the number of ions involved in such dissociative processes was relatively small compared to the thermal tail contributions, these events were shown to affect charge state ratios and to potentially generate undetected neutral species, thereby contributing to the observed oxygen deficit.

Jakob et al. (2025) performed a more extensive study on the detected composition of chromium oxyde using a LEAP 6000 XR equipped with a deep UV laser (258 nm) (Jakob et al., 2025). Measurements were carried out over a wide range of LPEs, from 0.03 to 90 pJ. The measured oxygen concentration showed a strong dependence on this parameter, dropping to 37 at% at 90 pJ and recovering to a plateau of approximately 58 at.% below 1 pJ, still slightly below the expected stoichiometric value of 60 at.%. This compositional bias was shown to be governed not by the LPE directly, but by the electrostatic field strength at the specimen apex, with higher fields yielding more accurate oxygen quantification. Additional measurements were performed using simultaneous voltage and laser pulsing (V+L), with a 20% voltage pulse fraction and LPEs between 0.05 and 50 pJ. Under these conditions, the overall trend in oxygen concentration as a function of LPE was consistent with that observed in laser-only mode, with higher LPEs leading to lower measured oxygen contents. The V+L mode offered two notable advantages over pure laser pulsing: the standing voltage was significantly reduced at equivalent LPEs, and the background noise was lowered by more than a factor of two. In terms of compositional accuracy, V+L runs performed at a detection rate of 1% yielded systematically higher oxygen concentrations than those at 0.5%, confirming that a higher detection rate improves quantification accuracy regardless of the pulsing mode. These results further support the conclusion that the electrostatic field strength at the specimen apex is the key parameter governing compositional accuracy in laser-assisted APT of chromium oxide.

*More complex oxide compounds.* Two cases of pseudobinary oxide compounds are also worth mentioning because they illustrate that the most satisfying experimental conditions can significantly differ from one compound to another. The first is (Mg,Zn)O, which has properties close to those of a random alloy such as $Al_yGa_{1-y}N$. Its evaporation behavior is also similar to that of AlGaN (see section 5.a.iii) as the measured composition in terms of the II-site fraction is more accurate at low field. The mechanism is most likely similar to that of $Al_yGa_{1-y}N$ (Di Russo et al., 2018b). A second case is the ordered compound $ErMnO_3$, in which the composition in terms of atomic fractions approaches the nominal values of 20% Er, 20% Mn, and 60% O for higher CSR($ErO^{2+}$/ $ErO^+$). The authors observe a substantial preferential retention of Er atoms, which is suppressed at higher CSRs. The high field regime also helps reducing the loss of neutral $O_2$ by increasing its post-ionization probability (Hunnestad et al., 2023).

The analysis of Barium doped SrTiO (Ba:STO) is interesting as it represents an attempt to explain the observed oxygen deficit (at least 14% of deficit within different samples and under different experimental conditions) by means of the estimation of specific evaporation fields for the main molecular ions on the basis of their binding energies within the matrix. The attempted method of estimating evaporation fields provided some insights but was not sufficient to fully explain the observed phenomena (for instance, it does not explain why certain molecular ions, such as TiO, are preferentially emitted over elemental ions). Nevertheless, the authors sketch some analogy with the main results of the DFT analyses performed on binary oxides, showing the possibility of evaporating molecular ions, and the possible emission of neutral O (Morris et al., 2024).

### *vi. Group IV compounds and carbides*

The relevance of group IV semiconductors is mainly related to their use in microelectronics. Carbides may have a large number of applications and are found not only in insulating materials, but also in metal micro and nanostructure. Their evaporation behavior is complex, as C is strongly subjected to surface reactions and yields in general a large amount of multiple hits. An early attempt to tackle the difficulties in the measurement of carbide composition have been proposed by Thuvander and co-authors for the analysis of tungsten carbide. The method consists in placing a grid a few millimeters behind the local electrode. The grid acts as a spatial filter, allowing only ions that follow particular trajectories to reach the detector. Ions produced by multiple evaporation events are more likely to be intercepted by the grid, thus reducing the likelihood they are both recorded. Thus, the overall detection efficiency decreased substantially (from around 37% to about 5%) due to the physical blocking of many ion trajectories, but more significantly the fraction of multiple hits (potentially giving rise to pile-up) was significantly reduced, by over 80% in typical analyses. In this way, a correct measurement of the C atomic fraction could be achieved, though with a global information loss (Thuvander et al., 2013).

*Metal carbides and Silicon Carbide.* The APT measurement of carbide composition in steels is generally different from the theoretical composition (Sha et al., 1992; Rementeria et al., 2017), while a strong correlation of the carbon measured fraction with the laser pulse energy is obtained during the analyses of pure cementite ($Fe_3C$) (Kitaguchi et al., 2014). Nevertheless, using peak deconvolution for the contribution at 24 m/z and the correction scheme for Fe counting loss (Miyamoto et al., 2012), nearly correct values for cementite composition can be achieved (Takahashi et al., 2020). Among the group IV semiconductors, SiC has a specific interest as a material for electronics (Nava et al., 2008; Pacchioni, 2017; She et al., 2017; Castelletto and Boretti, 2020), as an example of stoichiometric and perfectly ordered compound, and as a model system for the study of compositional metrology of carbides, making this case relevant also in metallurgy. First studies on SiC by APT evidenced a slight Si deficiency (Miller et al., 1989). Later, LA-APT studies evidenced a C deficiency and proposed an explanation of the observation through detector losses, based on the statistics of the $^{12}C$ and $^{13}C$ isotopes (Thuvander et al., 2011). The field evaporation of SiC is complex, with the occurrence of molecular ions generally composed of C, sometimes subjected to in-flight molecular dissociation. FIM analysis shows that the surface is significantly perturbed by atom motion. In SiC, the CSR(Si) expectedly decreases during the analysis due to the increase of the field of view (Fig. [27]-(a)). The measurement of composition in SiC exhibits inaccuracies, as it is shown in Fig. [27]-(b) as a function of the reconstructed depth. The exploration of experimental parameters invariably yields a Si-rich composition. This phenomenon can be consistently interpreted as due to the specific tendency of C to correlated evaporation, yielding a higher fraction of multiple evaporation and detection events for this element, which is consequently more exposed to measurement errors due to the limitations in the detector performances. This observation is corroborated by the evolution of specific quantities during the analysis. Firstly, the fraction of detected multiples decreases with depth (Fig. [27]-(c)). Secondly, the detector amplitude recorded in single events shows that the detected C has often large amplitudes, related to double or multiple neighboring impacts which could not be discriminated by the detector and which are therefore treated as a single event (*detector dazzling*, inset of Fig. [27]-(d), see also section 4.a.ii). The average amplitude of C impacts increases over time (i.e. with the analyzed depth, Fig. [27]-(d)). This observed trend of increase of C loss with increasing depth of analysis can then be explained by the decrease in geometric magnification, making it more and more likely for correlated evaporated ions to fall within the so-called "dazzled region" of the detector (Ndiaye et al., 2023a) (Fig.[11]-(f)). The measurement of composition of SiC by APT are thus subject to inaccuracies that stem from the interplay between high field surface chemistry and detector performances. We remind that the amplitude of detection events is usually not accessible in current commercial instruments, but is featured in the LaWATAP instrument used for this study.

*Silicon-Germanium and related alloys.* SiGe is an important alloy in micro and nanoelectronics as it is used for strain engineering of transistors. Compositional inaccuracies in APT analyses of SiGe can be avoided if using a CSR($Si^{2+}/Si^{+}$) in the order of 100 (Koelling et al., 2009, 2011; Dyck et al., 2017). On the other hand, SiGeC alloys display more complex mass spectra, with frequent C-containing molecular ions (occasionally formed with Si but not with Ge) and compositional inaccuracies. These have been characterized by Estivill et al., who showed that the composition is systematically poor in Ge and in C. The atomic fraction of Ge recovers to the nominal value at high CSR($Si^{2+}/Si^{+}$), while the C atomic fraction decreases further with increasing field. The loss of carbon is attributed to several factors: decrease in signal-to-noise ratio and the possibility that certain peaks are not indexed, such as hypothetical $C_3^{+}$, which would overlap with the germanium peaks. The mechanism could likely be similar to what has been ascertained in SiC (Ndiaye et al., 2023a). For germanium there is no clear answer to explain the loss at lower electric field, but a role of the laser intensity is suspected (Estivill et al., 2015).

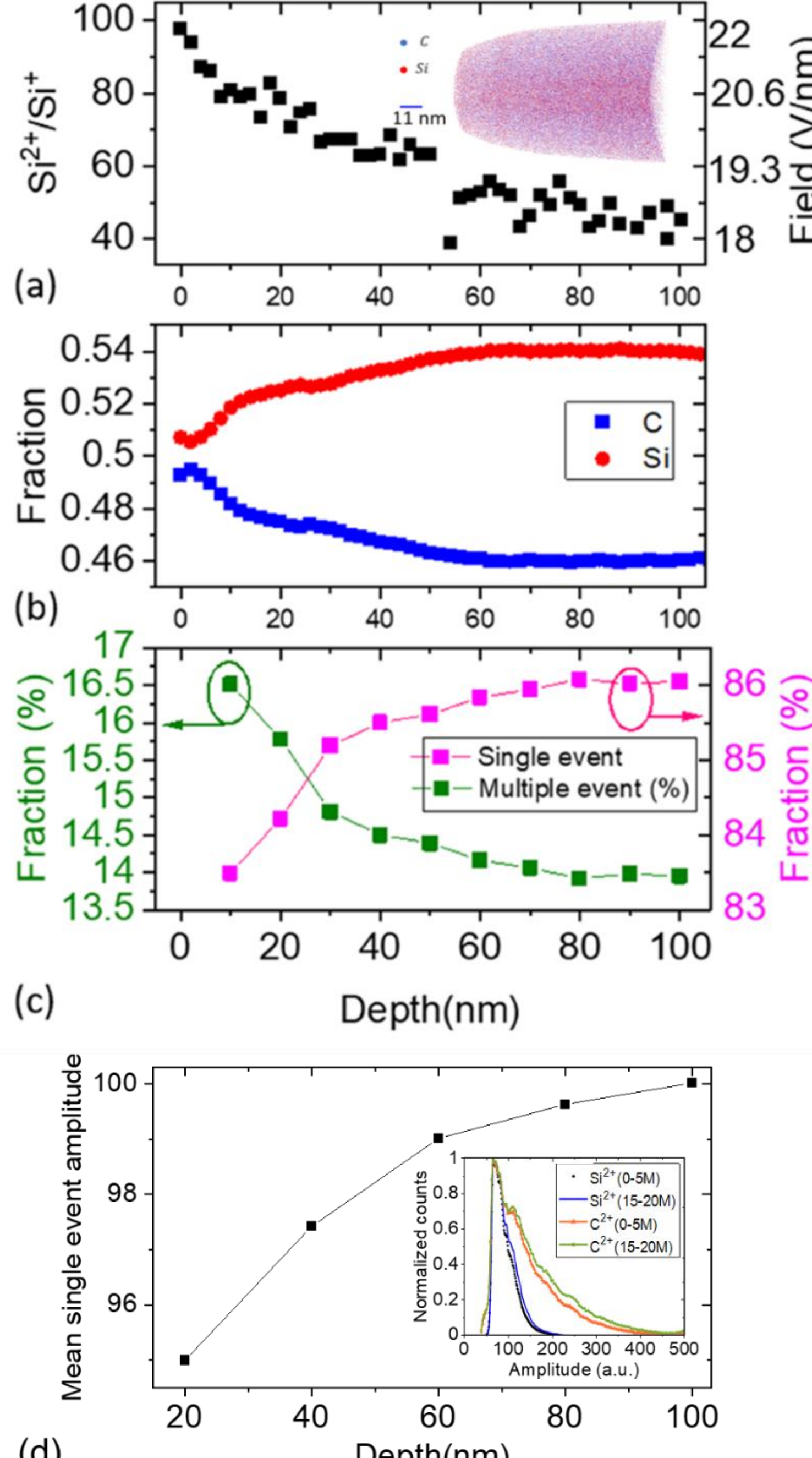

*Figure [27]. Advanced analysis of compositional biases in 4H-SiC. (a) Surface electric field around the axis [0001] pole as a function of the analysis depth; in the inset, the 3D reconstructed volume. (b) Variation of the atomic fraction of Si and C as a function of the depth coordinate of the reconstructed volume. (c) Percentage of multiple and single events as a function of the depth. (d) Mean amplitude of single detection events traced versus the depth analysis. In the inset, histograms reporting the distributions of signal amplitudes related to the impacts of $Si^{2+}$ and $C^{2+}$ at the beginning (first 5 million atoms) and the end of analysis (last 5 million atoms over 20). (Adapted with permission from* (Ndiaye et al., 2023a), *American Chemical Society)*

### b. *Impurities*

A correct measurement of dopant and impurity densities in non-metallic materials is critical because even ppm-level variations strongly influence electronic, magnetic, and catalytic properties through charge carrier control, defect formation, and strain effects (Koenraad and Flatté, 2011). Some of these species may also be characterized by statistical properties such as pairing or clustering, which can be addressed by APT and by STM only (Di Russo et al., 2025). Furthermore, the correct assessment of isotopic abundances of trace elements is paramount in geochronology (Reddy et al., 2020). APT is, in principle, a technique of choice due to its sub-nm 3D resolution and low limit of detection across all elements (Koelling et al., 2016), enabling quantification of individual atoms or clusters that determine device properties or carry other important information. However, compositional biases have been shown to occur also in this domain of application and should therefore be critically considered.

#### i. *Dopants in Si and SiGe*

*Boron in Si.* Boron is the most widely used p-type dopant in silicon and plays a central role in modern microelectronics. The accuracy of B quantification in Si by APT has shown differences across studies, possibly related to different equipment used. Several reports demonstrate good agreement with reference techniques such as SIMS, particularly in terms of concentration profiles (Kelly et al., 2007; Cojocaru-Mirédin et al., 2009b; Kambham et al., 2011; Cojocaru-Mirédin et al., 2009a). Nevertheless, other studies report notable discrepancies, including boron detection losses of 10–50 % (Da Costa et al., 2012; Kinno et al., 2017; Ngamo et al., 2009; Yeoh et al., 2020) and deviations in the implanted dopant profile compared to expected distributions (Mangelinck et al., 2014; Martin and Yatzor, 2019; Ngamo et al., 2009; Ronsheim et al., 2008; Thompson et al., 2006). These discrepancies can be largely attributed to the specific behavior of boron during field evaporation. Boron has a high evaporation threshold relative to silicon and exhibits surface migration, often co-evaporating in multi-ion bursts.

During APT, boron can remain temporarily at the specimen surface and then evaporate in these multi-ion events (Menand and Kingham, 1985; Menand et al., 1984; Da Costa et al., 2012). Detector limitations, such as pile-up effects, may prevent some ions from being correctly recorded, leading to a preferential loss of B and an underestimation of its concentration in the Si matrix. This phenomenon was clearly demonstrated in a heavily B-implanted Si ($1 \times 10^{10}$ at./cm$^2$) by (Da Costa et al., 2012). Their study showed that Fourier-domain processing of signals from a delay-line position-sensitive detector significantly improves temporal resolution, enhancing the ability to resolve multi-ion events, an approach implemented in the so-called “advanced delay-line detector” (aDLD) (see section 4.e).

Using a LEAP 3000X Si operated with a green laser ($\lambda$=532 nm), Ronsheim et al. (2010) investigate the composition of two Si:B samples presenting B concentration of $2.4 \times 10^{19}$ and $2.15 \times 10^{20}$ at./cm$^{-3}$ (Ronsheim et al., 2010). Increasing the laser pulse repetition rate modestly improves the fractional B yield, whereas changing LPE, evaporation rate and sample base temperature show no consistent change in the measured composition. The study revealed that the B collection efficiency was significantly below 100%, with the loss primarily attributed to detector limitations rather than evaporation physics. Specifically, B atoms were more likely to be lost due to multiple-event generation during evaporation. The authors suggested that enhancing multiple-event detection could mitigate this limitation and improve quantitative accuracy.

In 2015, Meisenkothen et al. (Meisenkothen et al., 2015) investigated the impact of pile-up phenomena on compositional measurement of a $^{10}$B-implanted Si sample ($1 \times 10^{15}$ at./cm) using a conventional three anode delay-line detector installed in both LEAP 4000X HR and S instruments operated with UV pulses ($\lambda$=355 nm). This detector exhibits a dead time of 5–20 ns (Jagutzki et al., 2002b), which should be compared with the signal pulse duration, typically less than 5 ns (Vallerga and McPhate, 2000).

Comparative analyses with SIMS indicate that APT measurements are affected by signal losses ranging from approximately 23% (Si configuration) to about 36% (HR configuration). In fact, B is preferentially emitted in multi-hit events, with >60% of the detected B ions belonging to multiples, despite multi-hit events accounting for <40% of total detections in pure B and <3% in B-implanted Si. Ion-correlation histograms demonstrate that same-isotope, same-charge-state B ion pairs are largely absent due to detector dead-time (~3 ns, with partial loss extending to ~6 ns), leading to severe undercounting of correlated evaporation events. This effect disproportionately suppresses the major isotope, producing apparent $^{10}B/^{11}B$ ratios up to a factor of ~2 higher than natural abundance. Probabilistic estimates indicate that dead-time-induced pile-up alone can account for ~20% B signal loss. Notably, the magnitude of the boron deficit is similar for pure B and dilute B-in-Si, indicating that the bias is largely independent of concentration. These results establish detector dead-time as a dominant, quantifiable source of B composition bias in APT and impose fundamental limits on dopant accuracy.

Further insights into the relationship between analysis parameters and the measured boron fraction in silicon were provided by Tu et al. (2017) (Tu et al., 2017a). The three-dimensional distribution of B atoms in both c-Si:B and poly-Si:B was investigated using a LEAP 4000X HR operated with a UV laser (λ=355 nm) at a repetition rate of 200 kHz. In c-Si:B, the measured B distribution is almost uniform and homogeneous at low LPE. Instead, increasing the LPE at 100 pJ leads to a B-enrichment near the (001) pole at the center of the specimen where the local density of Si atoms was lower. This accumulation was explained by electric field driven surface migration of B atoms toward high field regions prior to evaporation. This effect was observed to be enhanced by laser pulsing with a LPE fixed at 100 pJ. A similar phenomenon has been reported by Gault et al. (2012) for a wide range of solutes including P, Si, Mn, B, C, and N (Gault et al., 2012a). High LPE was also observed to lead to a segregation along the grain boundaries in poly-Si:B, which originates from a similar mechanism. In addition, due to multi-hit events for evaporating B atoms, a loss in the total count with increasing laser power was observed for poly-Si:B. In particular, the measured composition varies from 5 to $2.5 \times 10^{15}$ at./cm$^2$ with increasing LPE from 10 to 100 nJ. High laser power was also found to increase the specimen temperature, leading to a reduction of the evaporation threshold of $Si^+$ ions, this increasing their fraction among all Si ions, leading thus to a decrease of the CSR($Si^{2+}/Si^+$) from ~0.5 to ~200 (Shariq et al., 2009). Therefore, for ensuring a high-fidelity APT measurement of the B distribution in Si, high LPE is not recommended.

A systematic study of the impact of the surface electric field on the quantification accuracy of B in Si was performed by Guerguis et al. in 2024 (Guerguis et al., 2024). Si was implanted with $^{10}B$ ions at a dose of $1.0 \times 10^{15}$ at./cm$^2$. APT analyses were performed on a LEAP 5000 XS, operated using a UV laser (λ = 355 nm). During the analyses, the LPE was automatically adjusted from 41 to 9 pJ to maintain a constant detection rate, φ, while achieving CSR($Si^{2+}/Si^+$) values of 0.25, 0.5, 10, 30, 60, and 80. At the same time, the laser pulse frequency was increased from 294 to 454 kHz. Mass spectra exhibit peaks associated to $B^{2+}$ (5 m/$z$), $B^+$ (10 m/$z$), $BH^+$ (11 m/$z$) and $BH_2^+$ (12 m/$z$). The B depth profiles measured by APT were compared with SIMS measurements. Samples analysed under low-field conditions exhibit pronounced tails in the APT concentration profiles, leading to an apparent underestimation of the B concentration when compared with SIMS results (Fig. [28]-(a,c)). This is consistent with B retention at the tip surface, resulting from an insufficient field, as previously reported in other studies (Martin and Yatzor, 2019; Tu et al., 2017b). The B dose measured by APT was found to increase significantly with the tip surface field. At CSR($Si^{2+}/Si^+$) ≈ 0.25 (low field), only about half of the expected dose is detected, while increasing the field up progressively recovers the nominal B content (Fig. [28]-(b,c)). In fact, at the lowest electric field, about 69 % of B atoms are detected in multi-hit events, leading to significant B losses due to simultaneous detection of ions with the same isotope and charge state. The fraction of B multi-hit events decreases as the CSR increases, which has been interpreted as a consequence of reduced B mobility at the tip apex. Under higher laser pulse conditions, B atoms migrate (Fig. [10], top) and form clusters, which co-evaporate in bursts, leading to increased detection losses. Under high and more homogeneous field conditions, on the contrary, reduced surface migration on the specimen surface is proposed as the underlying mechanism behind this improvement (Fig. [10], bottom). With less B mobility, there is a lower likelihood of B atoms coalescing. At CSR($Si^{2+}/Si^{+}$) ≈ 80, a slight excess in the measured B dose is observed (104.7 %). This artifact arises from a loss of Si in multi-hit events, which artificially increases the apparent B dose. Under these conditions, the isotope fraction of detected $Si^{2+}$ was observed to deviate from the natural abundances (Fig. [19]-(d) and section 4.d.iv). Applying the simple double-hit correction scheme proposed by Miyamoto et al. (Miyamoto et al., 2012) the B dose estimate was improved from 104.7 % to 101.8 % (Fig. [28]-(c)).

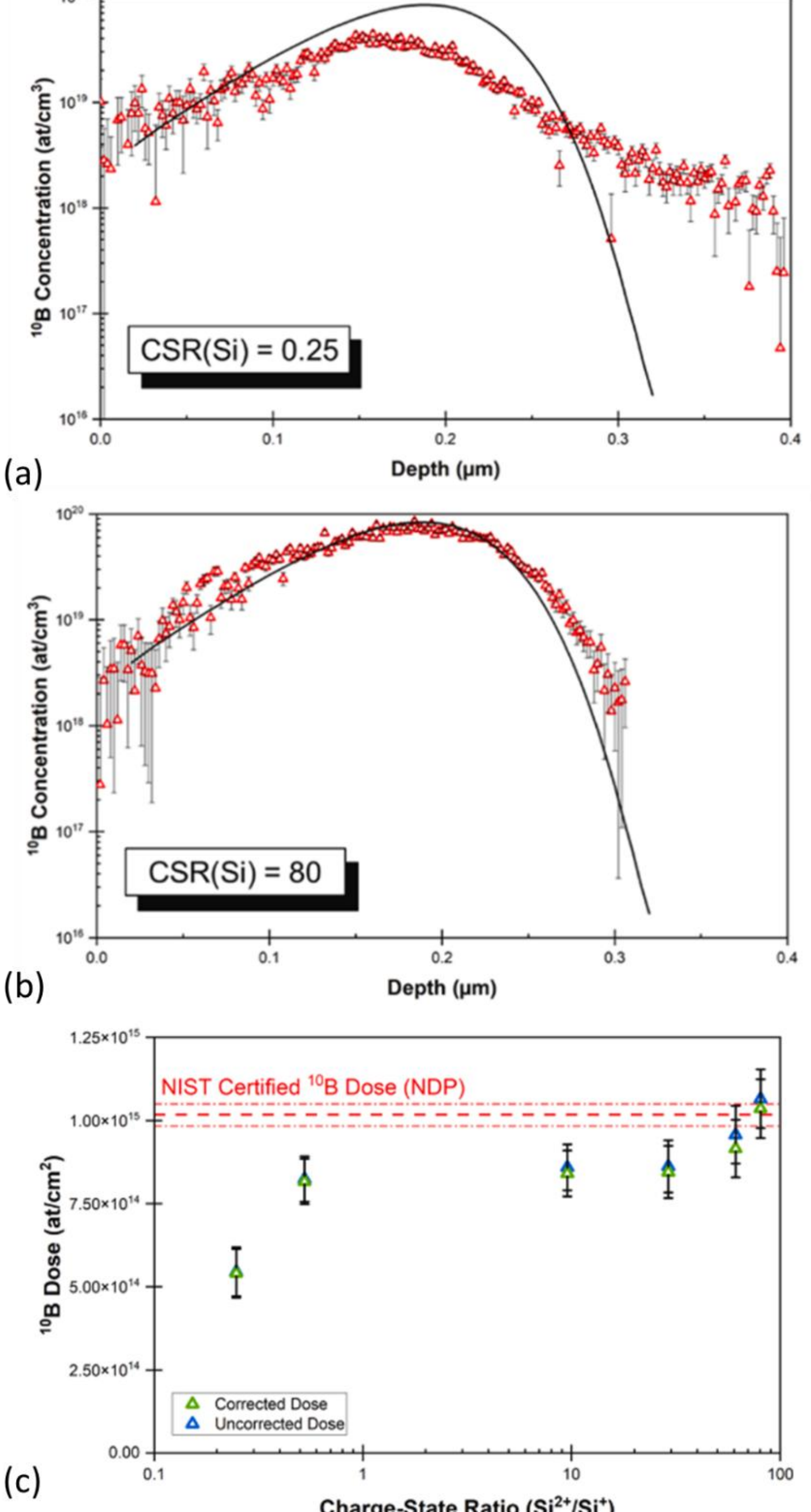

*Figure [28]. Systematic analysis of B concentration in Si by APT. (a,b) Bulk volume B concentration profiles measured at (a) low and (b) high field conditions compared with the SIMS-acquired profile shape. (c) Uncorrected and corrected B doses. The correction is performed using a factor issued from the analysis of Si multiple hits and deviations of isotopic abundances from natural abundances (Reproduced with permission from (Guerguis et al., 2024)).*

*Boron in SiGe.* Boron-doped SiGe alloys were investigated by Martin and Yatzor et al. in 2019 (Martin and Yatzor, 2019). This study focused on two 25 nm-thick SiGe layers grown on (001)-Si. The first layer contained approximately 26 % Ge, with the top half in-situ doped with ~7 × $10^{18}$ at./$cm^3$ B (as measured by SIMS), while the second layer contained ~42.7 % Ge and an in-situ B concentration of ~2 × $10^{20}$ at./$cm^3$ (as measured by SIMS). APT measurements were performed using a LEAP 4000X Si equipped with a UV laser (λ = 355 nm) operating at a repetition frequency of 125 kHz. Measurements were performed at a constant detection rate by increasing the laser energy from 1 to 21 pJ while progressively decreasing the applied DC voltage. Boron exhibits behavior similar to that observed in Si:B. Under low-field conditions (i.e., CSR($Si^{2+}$/$Si^{+}$) ≈ 0.8), boron shows resistance to field evaporation, resulting in a concentration tail extending much deeper than expected along the tip axis and in atom migration toward SiGe poles at the tip apex. Increasing the evaporation field mitigates these effects and improves the measured B profile at interfaces by promoting field evaporation of B from its original lattice sites. The measured Ge fraction agrees with SIMS within ~2 % over a CSR($Si^{2+}$/$Si^{+}$) range from 7 to 165, but decreases at very high fields. This trend suggests that $Ge^{2+}$ has a lower evaporation field than $Si^{2+}$ and is therefore more readily lost to the background. The best agreement between APT and SIMS, in terms of both profile shape and measured concentrations for Ge and B, is obtained at CSR($Si^{2+}$/$Si^{+}$) ≈ 200. Increasing the evaporation field beyond this value leads to Ge undercounting and consequently biases the measured B concentration.

*Phosphorus in Si.* In 2016, Douglas et al. (Douglas et al., 2016) investigated a 14-keV phosphorus implantation directly performed on a commercial Si micro-post array, resulting in a peak phosphorus concentration of 0.2 at.% at a depth of approximately 20 nm. To analyze the near-surface doped layer, a Ni capping layer was deposited on top of the micro-post array prior to FIB milling, enabling the preparation of needle-shaped specimens for APT analysis. In addition to facilitating specimen preparation, Ni provides good adhesion and offers an evaporation field comparable to that of Si. The specimens were subsequently analyzed using a LEAP 3000X HR equipped with a green laser (λ = 532 nm) operating at a repetition rate of 200 kHz. Measurements indicate that fixing the laser pulse energy at 0.4 nJ produces an approximately equal number of 1+ and 2+ phosphorus ions, while minimizing the tailing in silicon peaks in the mass spectra. Higher laser energies, in contrast, are reported to promote surface migration of mobile species such as phosphorus (Gault et al., 2012a). However, the presence of Ni introduces mass spectral overlaps between $Ni^{2+}$ and $^{31}P^{+}$ ions, as well as between $^{62}Ni^{+}$ and $^{31}P_2^{+}$ species, which can potentially compromise dopant quantification. Given the very low P concentration, a control sample was analysed to estimate the contribution of $^{30}SiH^{+}$ at the Si/Ni interface, which can artificially increase the apparent amount of $^{31}P^{+}$. Despite these challenges, the dopant implantation profiles measured by APT showed excellent agreement with predictions from implantation simulations performed using the SRIM software.

### *ii. Impurities in Si*

*C implantation in Si.* Dumas et al. addressed the quantification of carbon in implanted silicon, including a detailed analysis of the composition of carbon clusters (Dumas et al., 2021). Carbon was implanted into (100)-oriented Si wafers either as $^{12}C$ alone or as a mixture of $^{12}C$ and $^{13}C$, at the same dose of 1 × $10^{15}$ at./$cm^2$, featuring a C-atomic fraction of ~0.2 at.%. Importantly, the use of isotopically mixed carbon implantation provides a robust strategy to resolve mass-spectrum overlaps and to validate peak assignments based on the known isotopic ratio. Following annealing at ~750 °C to induce recrystallization and carbon clustering, APT analyses were performed using a LaWaTAP system, operated with a UV laser (λ=343 nm) at a repetition rate of 100 kHz. Measurements were first carried out under high electric field conditions and subsequently under lower field conditions by increasing the laser power. In the high-field regime, the charge-state ratio CSR($Si^{2+}$/$Si^{+}$) was approximately 50, whereas under lower field conditions it decreased substantially, ranging from ~3.6 to ~20. These results demonstrate that carbon quantification is significantly improved when operating at lower electric fields, particularly when combined with equal proportions of $^{12}C$ and $^{13}C$ implanted (see section 4.d.vi on isotope substitution). he primary reason for this improvement lies in the strong suppression of DC evaporation tails in the mass spectra at lower fields (referred to as "thermal tails" by the authors). Under these conditions, $C^+$ and $C^{2+}$ ions can be directly detected at 6 and 12 m/*z*, respectively, clearly before the onset of the $Si^+$ and $Si^{2+}$ thermal tails, which are markedly attenuated compared to high-field spectra. Lower-field conditions also enable the identification of molecular ion peaks, including $SiC_2^{2+}$, $SiC_3^{2+}$, and $SiC_2^+$. This behavior can be explained by two complementary mechanisms. First, the reduction of thermal tails reveals peaks that are partially obscured at high fields. Second, molecular ion dissociation during flight to the detector is known to be enhanced at high electric fields (Peng et al., 2019, 2018). Consequently, lowering the field reduces dissociation events, increasing the probability of detecting intact molecular ions. A pronounced attenuation of thermal tails is also observed in the mass spectra associated with cluster regions. This effect directly reflects the decrease in the local electric field, as evidenced by the strong reduction of the CSR($Si^{2+}/Si^+$) from ~50 to 0.6. In samples containing a mixture of $^{12}C$ and $^{13}C$, the presence of two carbon isotopes further enhances peak assignment by resolving otherwise ambiguous overlaps. For instance, $^{12}C^{12}C^+$ and $^{30}SiH_2O^{2+}$ coincide at 24 M/z , while $C_3^+$ and $(H_2O)_2^+$ overlap at 36 m/*z*; isotopic labelling allows these contributions to be unambiguously separated. Finally, operating at high evaporation fields is known to promote trajectory aberrations and defocusing effects, particularly for high-field species. These artefacts lead to an overestimation of interatomic distances and, consequently, to an underestimation of local carbon concentrations (Blavette et al., 2001). Consistently, lowering the electric field results in a substantial increase in the measured carbon content within the cluster core, from approximately 20 at.% (largely independent of the experimental conditions) to 46 at.%, in excellent agreement with the stoichiometric carbon concentration predicted by the SiC phase diagram.

*As implantation in Si; H, C and P impurities in Si.* Ronsheim et al. (2010) investigated the accuracy of APT measurements for quantifying an As implant (nominal dose of $2.5 \times 10^{15}$ at./cm$^2$), as well as H, C, and P impurities in a Si epitaxial layer (Ronsheim et al., 2010). Despite the lack of detailed experimental conditions and results, the APT-derived concentrations were found to be accurate within 10 % of SIMS measurements.

### *iii. Point defects in oxides*

Hunnestad et al. have recently achieved atomic-scale quantification of oxygen defects in $(LuFeO_2)_9/(LuFe_2O_4)_1$ superlattices, providing unprecedented insight into their defect structure (Hunnestad et al., 2024). 3D reconstructions shows that oxygen vacancies arranging in a layered three-dimensional structure, locally reaching ~$10^{14}$ cm$^{-2}$. However, the average oxygen concentration is found to be lower than the nominal value, a discrepancy attributed to the detection loss of neutral $O_2$ molecules. Since the CSR exhibits no significant variation in electric field strength, the formation of neutral $O_2$ is not expected to vary across the superlattice. Oxygen deficiencies arising from neutral emission were also observed in $SrTiO_3$ by Rybak et al. (Rybak et al., 2025), with the average oxygen content measured at 55 at.%, slightly lower than the expected 6 at.%. The apparent strontium deficit, with a measured composition of 17.9 at.% compared to the nominal 20.0 at.%, was recovered by extending the mass range to include the thermal tail and background beyond the peak.

### iv. *Dopants and point defects in nitrides*

*Mg and Ge dopants in GaN.* Despite Mg being the most widely studied dopant in GaN, the accuracy of Mg compositional measurements obtained by APT has not yet been conclusively validated. Amichi et al. reported that, in Mg-doped layers exhibiting Mg clustering, Mg concentrations determined by APT are in good agreement with SIMS measurements (Amichi et al., 2020). By contrast, Di Russo et al. recently applied an a posteriori correction to APT data using complementary information provided by SIMS for both GaN:Mg and GaN:Ge (E Di Russo et al., 2020). In a different approach, and building on the findings of Rigutti et al. (Rigutti et al., 2016a), Siladie et al. (Siladie et al., 2018) employed high laser energies and low DC field conditions (FlexTAP, $\lambda$ = 343 nm, LPE = 2.4 nJ) to ensure accurate determination of the III-site fractional occupancy (i.e., Mg/(Mg+Ga)). Under these conditions, the $Ga^{2+}/Ga^{+}$ ratio map was homogeneous within 0.05, meeting the criterion established by Koelling et al. for low-field regimes enabling reliable III-site occupancy measurements (Koelling et al., 2016).

*Vacancies in (Ti,Al)N.* In an early APT study of nano-lamellar cubic $(Ti_{1-x}Al_x)N_y$ thin films, Qiu et al. (Qiu et al., 2021) reported pronounced nitrogen over- and under-stoichiometry in Ti(Al)N and Al(Ti)N lamellae, respectively, which were attributed to metal and nitrogen vacancies. Vacancy concentrations of up to 12% for metals in Ti-rich regions and up to 36% for nitrogen in Al-rich volumes were inferred from APT data. However, a subsequent correlative study of Hans et al. (Hans et al., 2023) combining ion beam analysis and APT demonstrated that vacancy concentrations in nano-lamellar $(Ti_{1-x}Al_x)N_y$ cannot be reliably quantified from APT alone, as the apparent stoichiometry critically depends on the field evaporation conditions. These findings indicate that the previously reported high vacancy concentrations predominantly arise from field-induced artefacts rather than reflecting the intrinsic defect chemistry of the material.

### v. *Dopants and impurities in GaAs*

Du et al. (Du et al., 2013) investigated the spatial distributions of dopants and impurities in GaAs nanowires (NWs) grown on n-type $(\bar{1}\bar{1}\bar{1})$B substrates. APT analyses were performed using a LEAP 3000X Si instrument equipped with a green laser ($\lambda$ = 532 nm), operated at pulse energies between 10 and 100 pJ. The NWs exhibit a composition of approximately 55% Ga and 45% As, indicating a deviation from the expected 1:1 stoichiometry. This deviation is partly due to peak overlap between $As_2^{2+}$ and $As^{+}$ ions, but is primarily caused by As ions evaporating between laser pulses, generating extended high-mass tails in the As-related peaks of the mass spectrum. These ions are not correctly ranged, contributing instead to the background noise rather than to the resolved mass peaks. In addition to the matrix species, Si dopants and residual contaminants (H, C, O, etc.) were detected. Although such ions are generally expected to accumulate at the NW edges, they were also observed within the core region. To improve the reliability of the dopant analysis, multiple hit events were selectively filtered. Since instrumental noise occurs only sporadically, events involving more than one ion per pulse are rare, and the mass spectrum is therefore largely dominated by single ion detections. Applying this filtering strategy increased the signal-to-noise ratio from 100:17 to 100:2 (Yao et al., 2010). Prior to filtering, single ion events appeared relatively uniformly distributed throughout the NWs, partly reflecting background noise. After excluding multiple hits, however, the spatial distributions became more physically meaningful, with dopants concentrated in the outer regions of the nanowires, consistent with expectations.

# 6. Conclusions and perspectives

As a concluding part of this review, we will summarize some perspective directions in APT research with potential developments or consequences on compositional metrology. These research domains are categorized into two main domains, which are nevertheless strongly entangled, i.e. the implementation of instrumentation and the implementation of methods. As a conclusion, a set of recommendations will be given for present and perspective users.

## a. *Instrumental implementations*

### i. *Detector Technology*

Overcoming the current limitations of detection systems is one of the main zones of instrumental development in APT. In the near future, the development of new event encoding readouts will enable to unlock the bottleneck of DLD detectors. Indeed, the progress in microelectronics during recent decades substantially improved characteristics of various optical sensing devices, which could be combined with MCPs. Historically optical detectors were skipped due to the low achievable repetition rate (OTAP for instance was limited to the kHz regime), and the incapacity to achieve timing measurement on pixels, requiring the use of additional timing measurement. Nevertheless, the parallel multi-hit detection capabilities were without comparison to DLDs. Spots much smaller than one millimeter could be resolved (compared to the cm size for a DLD). The development of highly pixelated fast Timepix readout enabled a large number of novel hybrid detectors, where incoming particles are initially converted into a measurable charge (> 1000 e-) by MCPs (Fig. [29]). Low dark currents and low gain make MCP/Timepix hybrid devices very attractive for the high achievable spatial and temporal resolution detection of low energy ions (~200 ps in the last version Timepix4) . One of the major advantages of Timepix readout in MCP detectors is the substantial increase of counting rate capability as each pixel in the readout is an independent counter, enabling ~ GHz rates for a $28 \times 28$ mm$^2$ detector operating in event counting mode. At the same time, many simultaneous events can be encoded by the pixelated readout, which was not possible with most of the conventional high resolution MCP detectors. Another major improvement for the MCP detection technology is a substantial reduction of MCP gain, required for the detection of incoming particles. Due to the relatively low noise level in each pixel amplifier (typically 75 e− rms), the output charge value of $\sim 10^4$ electrons is sufficient for the low noise detection of individual particles, a factor of 10–1000 lower compared to charge division or time propagation readouts used in MCP detectors. The reduced gain also leads to a substantial increase of local count rate capabilities and reduction in artefacts such as Ion Feedback emission increasing the noise level (Llopart et al., 2007; Tremsin et al., 2020; Tremsin and Vallerga, 2020; Mathew et al., 2022).

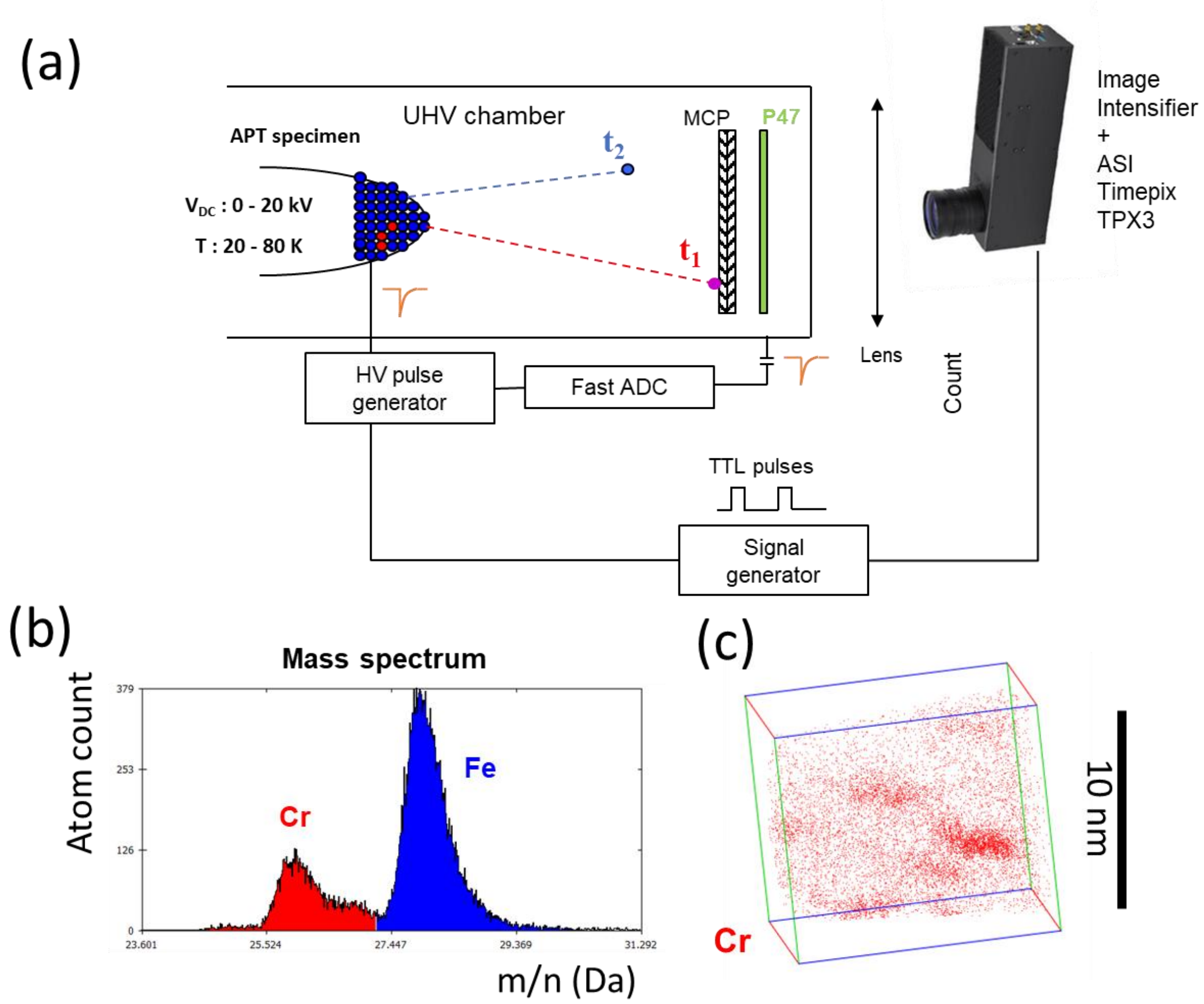

*Figure [29]. Schematic drawing of a prototype APT using a TimePix encoding anode. (a) An atom probe chamber was equiped with a detector composed of a pair of MCPs and a phosphor screen. The electrical signals are capacitively taken from the MCPs. The image on the screen is recorded through an UHV window by a TimePix detector (ASI Timepix TPX3). TPX3 instrument enables to record simultaneously the time-of flight (1.5 ns precision) and the position (X,Y with precision better than 100 microns) of every ion hit, enabling (b) mass spectrum (here an FeCr alloy) and (c) 3D reconstruction (only Cr in this case). (Unpublished results, courtesy of Gerald m/z Costa and Amsterdam Scientific Instruments)*

#### *ii. Terahertz pulsing*

Terahertz-assisted atom probe tomography (THz-APT) has recently emerged as a powerful extension of conventional laser-pulsed APT, offering a fundamentally different pathway to trigger field evaporation. Terahertz (THz) radiation occupies the spectral window between microwaves and the infrared (0.1-30 THz), corresponding to picosecond temporal scales and photon energies of only a few meV. As a consequence, THz fields are intrinsically non-ionizing and deposit negligible thermal energy compared with near-UV or visible excitation. This unique combination of low photon energy and sub- to few-picosecond duration makes THz transients particularly well matched to the intrinsic timescales of ion evaporation, enabling a regime in which emission can be driven predominantly by transient electric-field enhancement rather than by laser-induced heating. When interacting with atom probe nanotip, single-cycle THz pulses can be strongly amplified by near-field antenna effects, transiently delivering peak fields in the kV/cm range, boosting the apex field above the evaporation threshold while leaving the lattice temperature largely unchanged (Hunsche et al., 1998; Chen et al., 2000; Mair et al., 2004) .

Pump–probe experiments on Al tips have recently provided direct evidence for this athermal, field-driven mechanism, revealing a pronounced asymmetry between optical heating and THz excitation (Eriksson et al., 2025). In contrast to conventional laser pulsing, which broadens the evaporation window through thermal diffusion, THz excitation opens a sharply defined, field-limited emission gate that suppresses thermal tails and improves temporal control over ion emission. These properties position THz-APT as a compelling route toward higher-fidelity evaporation dynamics and motivate renewed interest in the role of ultrafast, non-thermal field control in atom probe tomography. However, several open challenges remain, including the origin of asymmetric time-of-flight and mass spectra and their connection to ultrafast ion dynamics (Karam et al., 2023), the quantitative accuracy of compositional measurements, and the limited efficiency of THz-triggered evaporation in wide-band-gap semiconductors and oxides and low electrical conductive materials.

### vi. *Deep and Extreme UV pulsing*

The use of Deep UV (typically, wavelength around 260-270 nm) has proven beneficial for compositional accuracy of APT measurements of insulators. Reports of this have appeared for different instruments, namely the commercial LEAP 6000 model on nitrides (Schiester et al., 2025) and the Photonic Atom Probe on mineral oxides (Veret et al., 2025). Extreme UV (EUV) pulsing in APT, developed primarily at NIST since ~2019, uses high-harmonic generation (HHG) to produce ultrafast (fs) pulses at 25-45 eV (50-28 nm), enabling precise field evaporation with minimal heat and photon fluence for semiconductors and insulators (Chiaramonti et al., 2020). This approach needs particular attention in the design of optical paths and interfaces (beam propagation occurs in vacuum only), as well as managing interferences with the detector by photoelectron generation (Caplins et al., 2023). Major achievements include stable triggering of evaporation from Si, GaN, with tunable spectra and year-long source stability, overcoming NIR-UV laser limitations like thermal tails and composition bias in III-nitrides. EUV APT yields cleaner mass spectra with sharp peaks and low background, as shown in GaN, enabling accurate 50:50 Ga:N stoichiometry without parameter-dependent bias seen in UV LAPT. For Mg-doped p-GaN, EUV clearly resolves $Mg^{2+}$ at 12 m/*z*, quantifying dopants reliably; similarly, for AlGaN alloys, it measures variable III-site fractions with minimal variation across charge-state ratios. In InGaN, EUV pulsing supports controlled evaporation, improving isotopic and alloy composition fidelity over UV methods prone to Ga or N loss (Miaja-Avila et al., 2021). Encouraging results have also been reported for oxides such as $Al_2O_3$ and MgO (Garcia et al., 2024).

### vii. *Hybrid pulsing*

Recent studies showed that combining voltage and laser pulsing in APT may deliver advantages over traditional single-mode approaches (Zhao et al., 2017; Larson et al., 2023b). By synchronizing a fast high-voltage pulse with a laser-induced thermal pulse, evaporation is confined to a narrow time window, drastically cutting background noise and sharpening mass spectra. This temporal gating effect improves signal-to-noise ratios - sometimes by an order of magnitude - especially for tricky materials like semiconductors or low-thermal-diffusivity oxides. Hybrid pulsing reduces thermal tails, enhances peak resolution, and improves the LOD for solute elements. The approach also widens the practical operating range: it lowers fracture risks compared to pure laser pulsing and avoids the limitations of voltage-only modes for fragile or poorly conductive materials. Of course, the gains depend on material-specific tuning of laser and voltage pulse fractions, but the results - t sharper peaks, cleaner spectra, and more accurate compositions - make it an interesting option for perspective instruments.

### viii. Correlative experimental setups

Correlative experiments in APT consist in coupling another technique to the atom probe. Beyond sequential correlation, that can be achieved by analyzing APT samples with other techniques before APT analysis, it is possible to design and implement in-situ correlation. Examples of these experiments are the in-situ and operando APT for the study of chemical reactions in close proximity of the APT chamber or within it (Haley et al., 2019; Lambeets et al., 2020a, 2020b), the photonic atom probe (PAP) for the study of photoluminescence from evaporating APT samples (Houard et al., 2020), and the different approaches for coupling APT with (S)TEM as in the TOMO project (Mayer et al., 2023) or in the APT-STEM instrument (Da Costa et al., 2024). Although demanding in terms of instrumental development, correlative techniques can provide important information that can be exploited for the assessment of a compositional error or its correction (as well as for the confirmation of accuracy, of course). For instance, the measurement of PL can be exploited as an indicator of the bandgap of an alloy and for its statistical distribution, which depend on the composition of the sample. In situ TEM, on the other hand, can help achieve optimal reconstructions and thus obtaining accurate quantification of losses. Scientists involved in correlative experiments know well that the global information achievable significantly overcomes the sum of the information separately issued by each technique involved (Di Russo and Rigutti, 2022).

## c. Implementation of methods

### i. Field contrast imaging of defects

Compositional biases are an issue that sometimes could be turned into an advantage. This occurs in the case in which certain structures or features contained in a given sample can be revealed just because of the effect they have on a composition measurement. The use of CSR in the study of precipitates in metals has been reported in several studies (Gault et al., 2021). In non-metallic systems, an example of this approach is given by the possibility of imaging stacking faults even in the absence of impurity segregation in them. A recent study conducted by Shu et al. reported that stacking faults in zincblende GaN have higher evaporation field than the surrounding matrix, and this translates into an apparent Ga-depletion within them, as reproduced in Fig. [30](Shu et al., 2025). The position and orientation of the defects is shown in Fig. [30]-(a), and are visualized by isoconcentration surfaces with the threshold of 80 at.% of Ga. The profiles shown in parts (b) and (c) indicate that at the defect the field increases and, consequently, the Ga fraction falls below the threshold. The reconstructed volume also shows that the defect propagates into the active region of a LED device, containing QWs. Here, it could be shown that the defect can influence the distribution of elements in the alloy. Related studies also reported that SFs can influence the distribution of dopants (Gundimeda et al., 2025; Xu et al., 2025). These studies are quite interesting in perspective as they show how CSR inhomogeneities can yield important information about features that are not associated to real compositional inhomogeneities. Density inhomogeneities could also constitute the basis for field contrast imaging, provided no effects on the CSRs or on the observed composition occur.

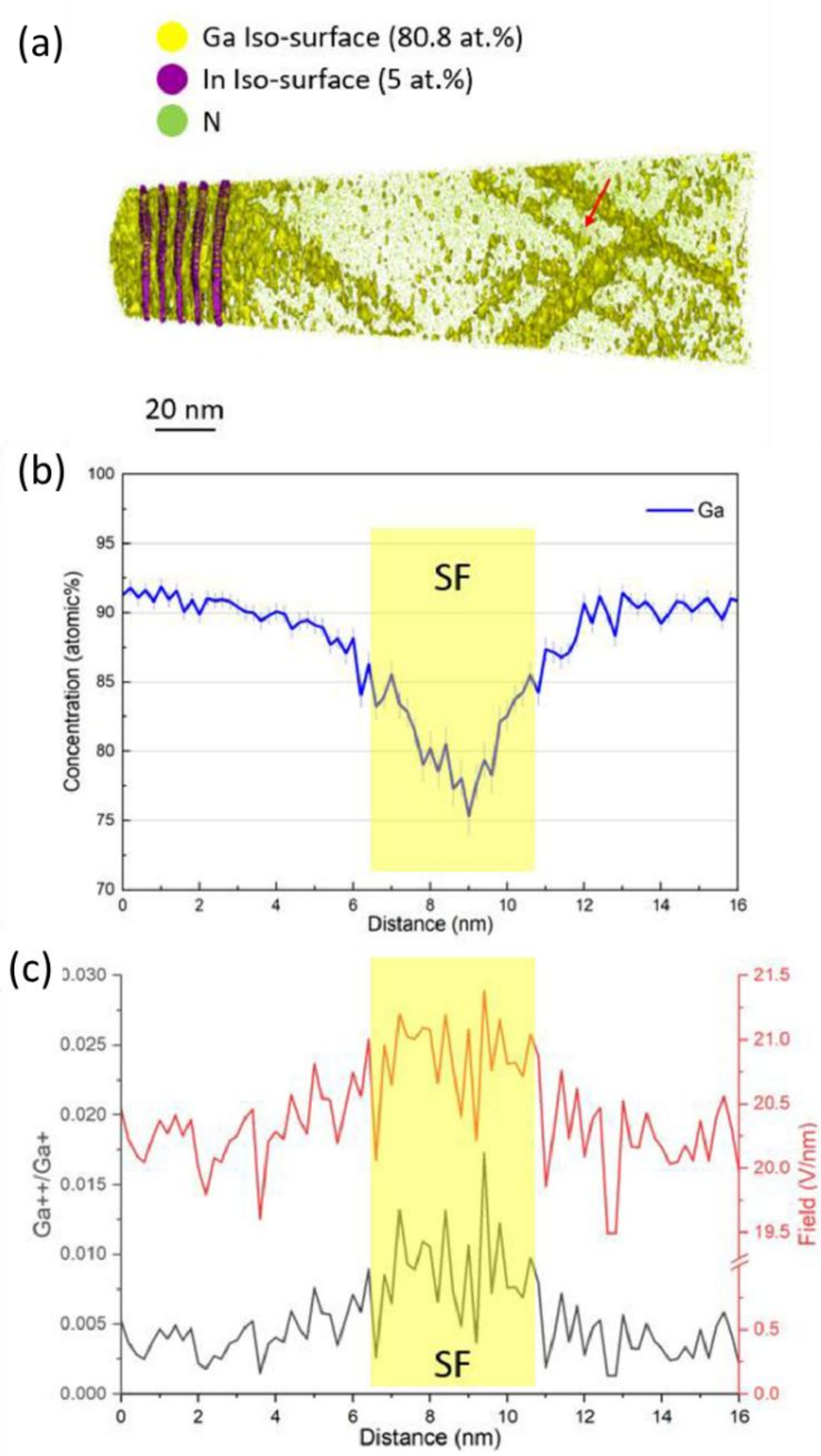

*Figure [30]. Field contrast imaging in APT revealing presence and orientation of stacking faults in ZB-GaN. (a) APT reconstruction of an InGaN/GaN MQW LED containing SFs. The yellow iso-concentration surfaces on the right indicate the SFs (red arrow). (b) One-dimensional atomic fraction and (c) CSR($Ga^{2+}$/$Ga^{+}$) profile of Ga in a volume along the red arrow in (a) (Reproduced with permission from* (Shu et al., 2025)*, Elsevier).*

### *ii. Cryo-APT and organic systems*

The analysis of organic systems can be carried out by APT provided adapted preparation methods are applied (Stoffers et al., 2012; Zhou et al., 2023). The development of cryo-preparation methods has further enabled the analysis of softer matter, such as water solutions, biological systems or organic molecules (Gerstl and Wepf, 2015). A review of APT metrology in these domains is not within the scope of this review, but several general points are worth mentioning. Recent works, demonstrate that APT can achieve sub-Dalton mass resolution and ppm-level sensitivity for small-molecule organic compounds (Proudian et al., 2019; Bingham et al., 2021). Under optimized laser conditions, intact molecular ions are preserved, enabling quantitative analysis with minimal fragmentation. This capability is critical for mapping nanoscale chemistry in organic photovoltaics and OLEDs, linking molecular segregation and interfacial reactions to device performance. DFT modeling complements experiments by predicting ion stability and fragmentation pathways (Nickerson et al., 2015; Dietrich et al., 2025).

Significant complexity arises in these systems as mass spectra are dominated by intact or partially fragmented molecular ions (often multi-charged), rather than monoatomic ions. This increases peak number, peak density and the likelihood of overlaps, complicating the interpretation. Molecular peaks and their isotopic envelopes are resolved with $m/\Delta m \sim 1000$, enabling molecular formula identification. Post-ionization fragmentation of dications into daughter ions generates multiple-hit events and additional peaks, introducing ambiguity in peak ranging and compositional quantification (Meng et al., 2022b). The influence of experimental conditions on the composition measurement has been pointed out. Laser pulse energy, specimen temperature, and surface field - significantly impact the balance between intact molecular ions, clusters, and fragments.

Some of the implications for compositional metrology are specific to these systems. On one hand, molecular-level analysis is often required: compositional metrology must operate at the molecular-ion level, matching complex isotopic patterns and charge states to candidate molecular formulas rather than atomic fractions. In this sense, a high mass resolving power is essential to separate overlapping molecular/isotopic peaks. On the other hand, fragmentation and multiple events are extremely frequent and require advanced data correction to avoid under- or over-estimation of species, as fragmentation biases compositions and varies with molecular energetics and experimental conditions (Bingham et al., 2021; Meng et al., 2022b).

#### *iv. Automated/enhanced data mining*

The development of automated or enhanced data mining and analysis methods for APT, mostly based on machine learning (ML) approaches, have been recently reviewed by Li et al. and the reader is referred to this reference for enjoying an exhaustive snapshot (Li et al., 2026). These methods are particularly adapted to the extended datasets issued from APT experiments, and may thus have a future impact on common practices. Among the main objectives of these approaches are speeding up the analysis, increasing its efficiency, reducing user and instrument dependency and enabling discoveries beyond human capability. Altogether, these methods should comply with the Findable, Accessible, Interoperable and Reusable (FAIR) principle. Within the domain of compositional metrology, these methods are expected to efficiently tackle the challenges related to user dependency and data complexity. An obvious target for their application is in mass spectrometry, with automated peak identification and decomposition, but more complex statistical analysis is well within the range of ML methods, in order to investigate short range ordering, clusters, nanoscale structures, relationship between chemistry and crystal structure, pushing resolution limits further and helping the user in visualizing and quantifying relevant features in the system. This, in turn, calls for community-driven efforts to provide open, reproducible software for APT data analysis, with possible interface to commercial solutions. This, of course, would not eliminate the need for critical attitudes by APT users, as they will still be responsible for the accuracy of data and results in order to prevent biases from being simply transferred to artificial intelligence-based methods (as it may have happened in some non-scientific domains).

### ***c. Recommendations for users***

As a distillate of this work on compositional metrology we would like to underline the awareness that compositional biases are an extremely important domain of research in the APT community, and that the users should share this awareness and agree for a minimum set of attitudes and practices concerning the occurrence of errors, the communication about them, and their management. We have therefore listed some recommendations concerning experimental practices, data analysis, standardization and research in APT compositional metrology.

### i. *Good experimental practices*

Rigor and precision form the foundation of reliable Atom Probe Tomography (APT) analysis. Absolute certainty in mass spectrum interpretation is essential, as ambiguity introduces the risk of misinterpretation and compromises result integrity. Precise terminology in quantifying composition ensures clarity and reproducibility, minimizing the potential for miscommunication in research. Rational and purposeful exploration of experimental parameters is necessary. Educated guesses in mass spectrum attributions, particularly those made to align with expected outcomes, should be avoided due to their deceptive nature and potential to undermine findings. Awareness of instrument limitations is critical, as it enables realistic interpretation of results. Field evaporation, central to APT, is an inherently complex phenomenon. Its intricacies must remain a focal point, as they directly influence observed data and experimental outcomes.

### ii. *Data mining and analysis*

Investigation into the origins of biases should extend beyond standard measurement protocols, such as constant detection rates, in order to uncover the experimental parameters responsible. They also should extend into the data mining protocols and methods. Mass spectra should be regularly reported or made available in data repositories. Portable metrics, including CSRs for the determination of the equivalent field $F_{eq}$, enhance the comparability and applicability of findings across different instruments and different studies. While global accuracy perfection may be unattainable, the pursuit of optimal conditions -for reconstruction fidelity, compositional accuracy, or specimen yield - remains a priority. Post-reconstruction corrections require careful application. Adjustments must align with the statistical properties of unbiased reference datasets in order to maintain analytical integrity. Any future application of machine learning or artificial intelligence-driven analysis should be critically scrutinized – not only by the developers, but also by any user – before it can serve as an automated or enhanced set of methods for producing results.

### iii. *Role of standards*

It is important for users and operators of APT that every step of their work is compliant with the results of the work in standardization that has been carried out over the last fifteen years within the field emission and APT community (Ulfig et al., 2009; Prosa et al., 2014). This work concerns not only terminology, which can be found on the ISO website both for general metrology concepts (“ISO/IEC Guide 99:2007(en), International vocabulary of metrology - Basic and general concepts and associated terms (VIM),” n.d.) and for specific APT concepts (“ISO 18115-1:2023(en), Surface chemical analysis - Vocabulary - Part 1: General terms and terms used in spectroscopy,” n.d.), but also the production of standard systems with known chemical characteristics, and good practices in data analysis and presentation.

### iv. *Promoting research in APT metrology*

Bias identification in material systems represents not only a setback, but also a significant opportunity. Such discoveries can yield meaningful contributions to the APT community and may define key research outcomes. No universal protocol exists for ideal APT analysis conditions. While prior studies provide useful reference points, tailored approaches are necessary to meet specific objectives, such as reconstruction fidelity, compositional accuracy, or specimen yield optimization. Measurement objectives must be clearly defined. For example, in materials like AlGaN, a stoichiometric III/V ratio may not accurately represent the Al III-site fraction. Precise goal-setting ensures meaningful and aligned measurements: the optimal conditions for the resolution of interfaces, for instance, may not be the same as those for optimizing compositional accuracy. Exact composition measurement may not always be achievable. Research should strive for the highest possible accuracy while acknowledging the inherent limitations of APT, as with any experimental technique. This pragmatic perspective fosters more robust and reliable research outcomes, which will be surely applied to the perspective work that is already ongoing. Furthermore, the work on compositional metrology is tightly bound to the efforts towards a deeper understanding of field evaporation: bond structure, bond breaking, surface reconstruction, all these aspects play a role in the determination of evaporating species and of spatial and temporal correlations (Cojocaru-Mirédin et al., 2024). This effort should be pursued in perspective within the two aforementioned entangled directions, namely the implementation of instruments and of methods.

## Acknowledgements

Financial support was provided by the French National Research Agency (ANR) via the ASCESE-3D project (ANR-21-CE50-0016), via the BE-SAFE project (ANR-24-CE08-6529), via the LABEX EMC3 QUANTIPHY project and under France 2030 investment plan (reference ANR-22-EXOR-0008). E.D.R. acknowledges the MOST - Sustainable Mobility Center and the European Union Next-GenerationEU [PIANO NAZIONALE DI RIPRESA E RESILIENZA (PNRR) - MISSIONE 4 COMPONENTE 2, INVESTIMENTO 1.4 – D.D. 1033 17/06/2022, CN00000023].

## References

Adineh, V.R., Marceau, R.K.W., Chen, Y., Si, K.J., Velkov, T., Cheng, W., Li, J., Fu, J., 2017. Pulsed-voltage atom probe tomography of low conductivity and insulator materials by application of ultrathin metallic coating on nanoscale specimen geometry. Ultramicroscopy 181, 150–159. https://doi.org/10.1016/j.ultramic.2017.05.002

Agrawal, R., Bernal, R.A., Isheim, D., Espinosa, H.D., 2011a. Characterizing Atomic Composition and Dopant Distribution in Wide Band Gap Semiconductor Nanowires Using Laser-Assisted Atom Probe Tomography. J. Phys. Chem. C 115, 17688–17694. https://doi.org/10.1021/jp2047823

Agrawal, R., Bernal, R.A., Isheim, D., Espinosa, H.D., 2011b. Characterizing Atomic Composition and Dopant Distribution in Wide Band Gap Semiconductor Nanowires Using Laser-Assisted Atom Probe Tomography. J. Phys. Chem. C 115, 17688–17694. https://doi.org/10.1021/jp2047823

Amichi, L., Mouton, I., Di Russo, E., Boureau, V., Barbier, F., Dussaigne, A., Grenier, A., Jouneau, P.-H., Bougerol, C., Cooper, D., 2020. Three-dimensional measurement of Mg dopant distribution and electrical activity in GaN by correlative atom probe tomography and off-axis electron holography. J. Appl. Phys. 127, 065702. https://doi.org/10.1063/1.5125188

Amirifar, N., Lardé, R., Talbot, E., Pareige, P., Rigutti, L., Mancini, L., Houard, J., Castro, C., Sallet, V., Zehani, E., Hassani, S., Sartel, C., Ziani, A., Portier, X., 2015. Quantitative analysis of doped/undoped ZnO nanomaterials using laser assisted atom probe tomography: Influence of the analysis parameters. J. Appl. Phys. 118, 215703. https://doi.org/10.1063/1.4936167

Arnoldi, L., Vella, A., Houard, J., Deconihout, B., 2012. Antenna effect in laser assisted atom probe tomography: How the field emitter aspect ratio can enhance atomic scale imaging. Appl. Phys. Lett. 101, 153101. https://doi.org/10.1063/1.4757884

Ashton, M., Mishra, A., Neugebauer, J., Freysoldt, C., 2020a. Ab initio Description of Bond Breaking in Large Electric Fields. Phys. Rev. Lett. 124, 176801. https://doi.org/10.1103/PhysRevLett.124.176801

Ashton, M., Mishra, A., Neugebauer, J., Freysoldt, C., 2020b. Ab initio Description of Bond Breaking in Large Electric Fields. Phys. Rev. Lett. 124, 176801. https://doi.org/10.1103/PhysRevLett.124.176801

Aziz, M.J., 2001. Stress effects on defects and dopant diffusion in Si. Mater. Sci. Semicond. Process. 4, 397–403. https://doi.org/10.1016/S1369-8001(01)00014-2

Bacchi, C., Da Costa, G., Cadel, E., Cuvilly, F., Houard, J., Vaudolon, C., Normand, A., Vurpillot, F., 2022. Development of an Energy-Sensitive Detector for the Atom Probe Tomography. Microsc. Microanal. 28, 1076–1091. https://doi.org/10.1017/S1431927621012708

Bacchi, C., Da Costa, G., Vurpillot, F., 2019. Spatial and Compositional Biases Introduced by Position Sensitive Detection Systems in APT: A Simulation Approach. Microsc. Microanal. 25, 418–424. https://doi.org/10.1017/S143192761801629X

Bachhav, M., Danoix, F., Hannoyer, B., Bassat, J.M., Danoix, R., 2013. Investigation of O-18 enriched hematite ($\alpha$-$Fe_2O_3$) by laser assisted atom probe tomography. Int. J. Mass Spectrom. 335, 57–60. https://doi.org/10.1016/j.ijms.2012.10.012

Bachhav, M., Danoix, R., Danoix, F., Hannoyer, B., Ogale, S., Vurpillot, F., 2011. Investigation of wüstite ($Fe_{1-x}O$) by femtosecond laser assisted atom probe tomography. Ultramicroscopy, Special Issue: 52nd International Field Emission Symposium 111, 584–588. https://doi.org/10.1016/j.ultramic.2010.11.023

Bas, P., Bostel, A., Deconihout, B., Blavette, D., 1995. A general protocol for the reconstruction of 3D atom probe data. Appl. Surf. Sci., Proceedings of the 41st International Field Emission Symposium 87–88, 298–304. https://doi.org/10.1016/0169-4332(94)00561-3

Beavis, R.C., Chait, B.T., Standing, K.G., 1989. Factors affecting the ultraviolet laser desorption of proteins. Rapid Commun. Mass Spectrom. 3, 233–237. https://doi.org/10.1002/rcm.1290030708

Beinke, D., Oberdorfer, C., Schmitz, G., 2016. Towards an accurate volume reconstruction in atom probe tomography. Ultramicroscopy 165, 34–41. https://doi.org/10.1016/j.ultramic.2016.03.008

Bhatt, S., Katnagallu, S., Neugebauer, J., Freysoldt, C., 2023. Accurate computation of chemical contrast in field ion microscopy. Phys. Rev. B 107, 235413. https://doi.org/10.1103/PhysRevB.107.235413

Bingham, J.T., Proudian, A.P., Vyas, S., Zimmerman, J.D., 2021. Understanding Fragmentation of Organic Small Molecules in Atom Probe Tomography. J. Phys. Chem. Lett. 12, 10437–10443. https://doi.org/10.1021/acs.jpclett.1c02277

Blavette, D., Vurpillot, F., Pareige, P., Menand, A., 2001. A model accounting for spatial overlaps in 3D atom-probe microscopy. Ultramicroscopy 89, 145–153. https://doi.org/10.1016/S0304-3991(01)00120-6

Blum, I., Cuvilly, F., Lefebvre-Ulrikson, W., 2016a. Chapter Four - Atom Probe Sample Preparation, in: Atom Probe Tomography. Academic Press, pp. 97–121. https://doi.org/10.1016/B978-0-12-804647-0.00004-8

Blum, I., Rigutti, L., Vurpillot, F., Vella, A., Gaillard, A., Deconihout, B., 2016b. Dissociation Dynamics of Molecular Ions in High dc Electric Field. J. Phys. Chem. A 120, 3654–3662. https://doi.org/10.1021/acs.jpca.6b01791

Bostel, A., Blavette, D., Menand, A., Sarrau, J.M., 1989. TOWARD A TOMOGRAPHIC ATOM-PROBE. J. Phys. Colloq. 50, C8-506. https://doi.org/10.1051/jphyscol:1989886

Brandon, D.G., 1966. On field evaporation. Philos. Mag. J. Theor. Exp. Appl. Phys. 14, 803–820. https://doi.org/10.1080/14786436608211973

Breen, A.J., Babinsky, K., Day, A.C., Eder, K., Oakman, C.J., Trimby, P.W., Primig, S., Cairney, J.M., Ringer, S.P., 2017. Correlating Atom Probe Crystallographic Measurements with Transmission Kikuchi Diffraction Data. Microsc. Microanal. 23, 279–290. https://doi.org/10.1017/S1431927616012605

Bunton, J.H., Olson, J.D., Lenz, D.R., Kelly, T.F., 2007. Advances in Pulsed-Laser Atom Probe: Instrument and Specimen Design for Optimum Performance. Microsc. Microanal. 13, 418–427. https://doi.org/10.1017/S1431927607070869

Caplins, B.W., Chiaramonti, A.N., Garcia, J.M., Sanford, N.A., Miaja-Avila, L., 2023. Atom probe tomography using an extreme ultraviolet trigger pulse. Rev. Sci. Instrum. 94, 093704. https://doi.org/10.1063/5.0160797

Cappelli, C., Smart, S., Stowell, H., Pérez-Huerta, A., 2021. Exploring Biases in Atom Probe Tomography Compositional Analysis of Minerals. Geostand. Geoanalytical Res. 45, 457–476. https://doi.org/10.1111/ggr.12395

Castelletto, S., Boretti, A., 2020. Silicon carbide color centers for quantum applications. J. Phys. Photonics 2, 022001. https://doi.org/10.1088/2515-7647/ab77a2

Cerezo, A., Clifton, P.H., Gomberg, A., Smith, G.D.W., 2007. Aspects of the performance of a femtosecond laser-pulsed 3-dimensional atom probe. Ultramicroscopy 107, 720–725. https://doi.org/10.1016/j.ultramic.2007.02.025

Cerezo, A., Gibuoin, D., Kim, S., Sijbrandij, S.J., Venker, F.M., Warren, P.J., Wilde, J., Smith, G.D.W., 1996. Materials Applications of an Advanced 3-Dimensional Atom Probe. J. Phys. IV 06, C5-210. https://doi.org/10.1051/jp4:1996533

Cerezo, A., Godfrey, T.J., Grovenor, C.R.M., Hetherington, M.G., Hoyley, R.M., Jakubovics, J.P., Liddle, J.A., Smith, G.D.W., Worrall, G.M., 1989. Materials analysis with a position-sensitive atom probe. J. Microsc. 154, 215–225. https://doi.org/10.1111/j.1365-2818.1989.tb00584.x

Cerezo, A., Grovenor, C.R.M., Smith, G.D.W., 1986a. Pulsed laser atom probe analysis of semiconductor materials. J. Microsc. 141, 155–170. https://doi.org/10.1111/j.1365-2818.1986.tb02712.x

Cerezo, A., Grovenor, C.R.M., Smith, G.D.W., 1986b. Pulsed laser atom probe analysis of semiconductor materials. J. Microsc. 141, 155–170. https://doi.org/10.1111/j.1365-2818.1986.tb02712.x

Cerezo, A., Grovenor, C.R.M., Smith, G.D.W., 1985. Pulsed laser atom probe analysis of GaAs and InAs. Appl. Phys. Lett. 46, 567–569. https://doi.org/10.1063/1.95541

Cerezo, A., Smith, G.D.W., Waugh, A.R., 1984. THE FIM100 - PERFORMANCE OF A COMMERCIAL ATOM PROBE SYSTEM. J. Phys. Colloq. 45, C9-335. https://doi.org/10.1051/jphyscol:1984955

Cerezo, A., Warren, P.J., Smith, G.D.W., 1999. Some aspects of image projection in the "eld-ion microscope. Ultramicroscopy 79, 251–257.

Chabanais, F., Russo, E.D., Karg, A., Eickhoff, M., Lefebvre, W., Rigutti, L., 2021. Behavior of the ε-Ga2O3:Sn Evaporation During Laser-Assisted Atom Probe Tomography. Microsc. Microanal. 27, 687–695. https://doi.org/10.1017/S1431927621000544

Chae, B.-G., Won, J.Y., Shin, Y.S., Yun, D.J., Ahn, J. min, Park, S.T., Lee, K., An, H., Seol, M., Ro, I.-J., Kim, S.-H., Chung, C., Lee, E., 2025. Direct observation of 3D nitrogen distribution in silicon-based dielectrics using atom probe tomography. Nat. Commun. 16, 5612. https://doi.org/10.1038/s41467-025-60732-2

Chakravarthy, S.S., Curtin, W.A., 2011. Stress-gradient plasticity. Proc. Natl. Acad. Sci. 108, 15716–15720. https://doi.org/10.1073/pnas.1107035108

Chen, M., Wehrs, J., Sologubenko, A.S., Rabier, J., Michler, J., Wheeler, J.M., 2020. Size-dependent plasticity and activation parameters of lithographically-produced silicon micropillars. Mater. Des. 189, 108506. https://doi.org/10.1016/j.matdes.2020.108506

Chen, Q., Jiang, Z., Xu, G.X., Zhang, X.-C., 2000. Near-field terahertz imaging with a dynamic aperture. Opt. Lett. 25, 1122–1124. https://doi.org/10.1364/OL.25.001122

Chen, Y.-S., Liu, P.-Y., Niu, R., Devaraj, A., Yen, H.-W., Marceau, R.K.W., Cairney, J.M., 2023. Atom Probe Tomography for the Observation of Hydrogen in Materials: A Review. Microsc. Microanal. 29, 1–15. https://doi.org/10.1093/micmic/ozac005

Chiaramonti, A.N., Miaja-Avila, L., Caplins, B.W., Blanchard, P.T., Diercks, D.R., Gorman, B.P., Sanford, N.A., 2020. Field Ion Emission in an Atom Probe Microscope Triggered by Femtosecond-Pulsed Coherent Extreme Ultraviolet Light. Microsc. Microanal. 26, 258–266. https://doi.org/10.1017/S1431927620000203
Coakley, K.J., Sanford, N.A., 2022. Learning Atom Probe Tomography time-of-flight peaks for mass-to-charge ratio spectrometry. Ultramicroscopy 237, 113521. https://doi.org/10.1016/j.ultramic.2022.113521
Cojocaru-Mirédin, O., Cadel, E., Vurpillot, F., Mangelinck, D., Blavette, D., 2009a. Three-dimensional atomic-scale imaging of boron clusters in implanted silicon. Scr. Mater. 60, 285–288. https://doi.org/10.1016/j.scriptamat.2008.10.008
Cojocaru-Mirédin, O., Mangelinck, D., Blavette, D., 2009b. Nucleation of boron clusters in implanted silicon. J. Appl. Phys. 106, 113525. https://doi.org/10.1063/1.3265998
Cojocaru-Mirédin, O., Yu, Y., Köttgen, J., Ghosh, T., Schön, C.-F., Han, S., Zhou, C., Zhu, M., Wuttig, M., 2024. Atom Probe Tomography: a Local Probe for Chemical Bonds in Solids. Adv. Mater. 36, 2403046. https://doi.org/10.1002/adma.202403046
Connétable, D., Maugis, P., 2020. Effect of stress on vacancy formation and diffusion in fcc systems: Comparison between DFT calculations and elasticity theory. Acta Mater. 200, 869–882. https://doi.org/10.1016/j.actamat.2020.09.053
Cuduvally, R., Morris, R.J.H., Ferrari, P., Bogdanowicz, J., Fleischmann, C., Melkonyan, D., Vandervorst, W., 2020. Potential sources of compositional inaccuracy in the atom probe tomography of InxGa1-xAs. Ultramicroscopy 210, 112918. https://doi.org/10.1016/j.ultramic.2019.112918
Currie, L.A., 1968. LIMITS FOR QUALITATIVE DETECTION AND QUANTITATIVE DETERMINATION. APPLICATION TO RADIOCHEMISTRY. Anal Chem 40 586-93 Mar 1968. https://doi.org/10.1021/ac60259a007
Da Costa, G., 2016. Chapter Six - Atom Probe Tomography: Detector Issues and Technology, in: Lefebvre-Ulrikson, W., Vurpillot, F., Sauvage, X. (Eds.), Atom Probe Tomography. Academic Press, pp. 155–181. https://doi.org/10.1016/B978-0-12-804647-0.00006-1
Da Costa, G., Castro, C., Normand, A., Vaudolon, C., Zakirov, A., Macchi, J., Ilhami, M., Edalati, K., Vurpillot, F., Lefebvre, W., 2024. Bringing atom probe tomography to transmission electron microscopes. Nat. Commun. 15, 9870. https://doi.org/10.1038/s41467-024-54169-2
Da Costa, G., Vurpillot, F., Bostel, A., Bouet, M., Deconihout, B., 2005. Design of a delay-line position-sensitive detector with improved performance. Rev. Sci. Instrum. 76, 013304–013304. https://doi.org/10.1063/1.1829975
Da Costa, G., Wang, H., Duguay, S., Bostel, A., Blavette, D., Deconihout, B., 2012. Advance in multi-hit detection and quantization in atom probe tomography. Rev. Sci. Instrum. 83, 123709–123709. https://doi.org/10.1063/1.4770120
Dalapati, P., Beainy, G., Di Russo, E., Blum, I., Houard, J., Moldovan, S., Vella, A., Vurpillot, F., Le Biavan, N., Hugues, M., Chauveau, J.M., Rigutti, L., 2021. In Situ Spectroscopic Study of the Optomechanical Properties of Evaporating Field Ion Emitters. Phys. Rev. Appl. 15, 024014. https://doi.org/10.1103/PhysRevApplied.15.024014
Danoix, F., Miller, M.K., Bigot, A., 2001. Analysis conditions of an industrial Al–Mg–Si alloy by conventional and 3D atom probes. Ultramicroscopy 89, 177–188. https://doi.org/10.1016/S0304-3991(01)00098-5
Dawahre, N., Shen, G., Renfrow, S.N., Kim, S.M., Kung, P., 2013. Atom probe tomography of AlInN/GaN HEMT structures. J. Vac. Sci. Technol. B 31, 041802. https://doi.org/10.1116/1.4807321
De Geuser, F., Gault, B., 2020. Metrology of small particles and solute clusters by atom probe tomography. Acta Mater. 188, 406–415. https://doi.org/10.1016/j.actamat.2020.02.023
De Geuser, F., Gault, B., 2017. Reflections on the Projection of Ions in Atom Probe Tomography. Microsc. Microanal. 23, 238–246. https://doi.org/10.1017/S1431927616012721
De Geuser, F., Gault, B., Bostel, A., Vurpillot, F., 2007. Correlated field evaporation as seen by atom probe tomography. Surf. Sci. 601, 536–543. https://doi.org/10.1016/j.susc.2006.10.019

Deconihout, B., Renaud, L., Da Costa, G., Bouet, M., Bostel, A., Blavette, D., 1998. Implementation of an optical TAP: preliminary results. Ultramicroscopy 73, 253–260. https://doi.org/10.1016/S0304-3991(97)00164-2

Deconihout, B., Vurpillot, F., Bouet, M., Renaud, L., 2002. Improved ion detection efficiency of microchannel plate detectors. Rev. Sci. Instrum. 73, 1734–1740. https://doi.org/10.1063/1.1461882

Devaraj, A., Colby, R., Hess, W.P., Perea, D.E., Thevuthasan, S., 2013. Role of Photoexcitation and Field Ionization in the Measurement of Accurate Oxide Stoichiometry by Laser-Assisted Atom Probe Tomography. J. Phys. Chem. Lett. 4, 993–998. https://doi.org/10.1021/jz400015h

Di Russo, E., Blum, I., Houard, J., Costa, G.D., Blavette, D., Rigutti, L., 2017. Field-Dependent Measurement of GaAs Composition by Atom Probe Tomography. Microsc. Microanal. 1–9. https://doi.org/10.1017/S1431927617012582

Di Russo, E., Blum, I., Houard, J., Gilbert, M., Da Costa, G., Blavette, D., Rigutti, L., 2018a. Compositional accuracy of atom probe tomography measurements in GaN: Impact of experimental parameters and multiple evaporation events. Ultramicroscopy 187, 126–134. https://doi.org/10.1016/j.ultramic.2018.02.001

Di Russo, E., Blum, I., Rivalta, I., Houard, J., Da Costa, G., Vurpillot, F., Blavette, D., Rigutti, L., 2020. Detecting Dissociation Dynamics of Phosphorus Molecular Ions by Atom Probe Tomography. J. Phys. Chem. A 124, 10977–10988. https://doi.org/10.1021/acs.jpca.0c09259

Di Russo, E., Cherkashin, N., Korytov, M., Nikolaev, A.E., Sakharov, A.V., Tsatsulnikov, A.F., Bonef, B., Blum, I., Houard, J., Da Costa, G., Blavette, D., Rigutti, L., 2019. Compositional accuracy in atom probe tomography analyses performed on III-N light emitting diodes. J. Appl. Phys. 126, 124307. https://doi.org/10.1063/1.5113799

Di Russo, E., et al., 2020. Phosphorus cation dissociation in laser-assisted atom probe tomography of indium phosphide. J Phys Chem C Press.

Di Russo, E, Mavel, A., Fan Arcara, V., Damilano, B., Dimkou, I., Vézian, S., Grenier, A., Veillerot, M., Rochat, N., Feuillet, G., Bonef, B., Rigutti, L., Duboz, J.-Y., Monroy, E., Cooper, D., 2020. Multi-microscopy nanoscale characterization of the doping profile in a hybrid Mg/Ge-doped tunnel junction. Nanotechnology 31, 465706. https://doi.org/10.1088/1361-6528/ab996c

Di Russo, E., Moyon, F., Gogneau, N., Largeau, L., Giraud, E., Carlin, J.-F., Grandjean, N., Chauveau, J.M., Hugues, M., Blum, I., Lefebvre, W., Vurpillot, F., Blavette, D., Rigutti, L., 2018b. Composition Metrology of Ternary Semiconductor Alloys Analyzed by Atom Probe Tomography. J. Phys. Chem. C 122, 16704–16714. https://doi.org/10.1021/acs.jpcc.8b03223

Di Russo, E., Rigutti, L., 2022. Correlative atom probe tomography and optical spectroscopy: An original gateway to materials science and nanoscale physics. MRS Bull. 47, 727–735. https://doi.org/10.1557/s43577-022-00367-6

Di Russo, E., Verstijnen, T., Koenraad, P., Pantzas, K., Patriarche, G., Rigutti, L., 2025. Order and disorder at the atomic scale: Microscopy applied to semiconductors. Rev. Mod. Phys. 97, 025006. https://doi.org/10.1103/RevModPhys.97.025006

Diagne, A., Anglade, P.-M., Blum, I., Rigutti, L., 2025. Microscopic Behavior of Hydrogen and Hydride Molecules in Atom Probe Tomography of Zirconium. J. Phys. Chem. C 129, 3084–3095. https://doi.org/10.1021/acs.jpcc.4c08027

Diercks, D.R., 2013. Atom probe tomography evaporation behavior of C-axis GaN nanowires: Crystallographic, stoichiometric, and detection efficiency aspects. J. Appl. Phys. 114, 184903. https://doi.org/10.1063/1.4830023

Diercks, D.R., Gorman, B.P., 2018. Self-consistent atom probe tomography reconstructions utilizing electron microscopy. Ultramicroscopy 195, 32–46. https://doi.org/10.1016/j.ultramic.2018.08.019

Diercks, D.R., Gorman, B.P., 2015. Nanoscale Measurement of Laser-Induced Temperature Rise and Field Evaporation Effects in CdTe and GaN. J. Phys. Chem. C 119, 20623–20631. https://doi.org/10.1021/acs.jpcc.5b02126

Dietrich, C.A., Schwarz, T.M., Meng, K., Stender, P., Schmitz, G., Kästner, J., 2025. Postdesorption Fragmentation of Tetradecane in Atom-Probe Tomography Simulated by DFT. J. Phys. Chem. A 129, 3963–3968. https://doi.org/10.1021/acs.jpca.5c01332

Douglas, J.O., Bagot, P.A.J., Johnson, B.C., Jamieson, D.N., Moody, M.P., 2016. Optimisation of sample preparation and analysis conditions for atom probe tomography characterisation of low concentration surface species. Semicond. Sci. Technol. 31, 084004. https://doi.org/10.1088/0268-1242/31/8/084004

Du, S., Burgess, T., Gault, B., Gao, Q., Bao, P., Li, L., Cui, X., Kong Yeoh, W., Liu, H., Yao, L., Ceguerra, A.V., Hoe Tan, H., Jagadish, C., Ringer, S.P., Zheng, R., 2013. Quantitative dopant distributions in GaAs nanowires using atom probe tomography. Ultramicroscopy 132, 186–192. https://doi.org/10.1016/j.ultramic.2013.02.012

Dumas, P., Duguay, S., Borrel, J., Hilario, F., Blavette, D., 2021. Atom probe tomography quantification of carbon in silicon. Ultramicroscopy 220, 113153. https://doi.org/10.1016/j.ultramic.2020.113153

Dyck, O., Leonard, D.N., Edge, L.F., Jackson, C.A., Pritchett, E.J., Deelman, P.W., Poplawsky, J.D., 2017. Accurate Quantification of Si/SiGe Interface Profiles via Atom Probe Tomography. Adv. Mater. Interfaces 4, 1700622. https://doi.org/10.1002/admi.201700622

Dzuba, B., Nguyen, T., Sen, A., Diaz, R.E., Dubey, M., Bachhav, M., Wharry, J.P., Manfra, M.J., Malis, O., 2022. Elimination of remnant phases in low-temperature growth of wurtzite ScAlN by molecular-beam epitaxy. J. Appl. Phys. 132, 175701. https://doi.org/10.1063/5.0118075

Engberg, D. LJ., Johnson, L.J.S., Jensen, J., Thuvander, M., Hultman, L., 2018. Resolving mass spectral overlaps in atom probe tomography by isotopic substitutions – case of TiSi15N. Ultramicroscopy 184, 51–60. https://doi.org/10.1016/j.ultramic.2017.08.004

Engberg, D. L. J., Johnson, L.J.S., Jensen, J., Thuvander, M., Hultman, L., 2018. Resolving mass spectral overlaps in atom probe tomography by isotopic substitutions – case of TiSi15N. Ultramicroscopy 184, 51–60. https://doi.org/10.1016/j.ultramic.2017.08.004

Eriksson, G., De Tullio, M., Carnovale, F., Inverardi, G.N., Morresi, T., Houard, J., Ropitaux, M., Blum, I., Cadel, E., Lattanzi, G., Thuvander, M., Andersson, M., Hulander, M., Taioli, S., Vella, A., 2025. Role of Defects in Atom Probe Analysis of Sol−Gel Silica. ACS Omega 10, 33741–33754. https://doi.org/10.1021/acsomega.5c04733

Ernst, N., 1979. Experimental investigation on field evaporation of singly and doubly charged rhodium. Surf. Sci. 87, 469–482. https://doi.org/10.1016/0039-6028(79)90542-9

Ernst, N., Jentsch, Th., 1981. Post-field ionization of singly charged rhodium: An experimental and theoretical study. Phys. Rev. B 24, 6234–6241. https://doi.org/10.1103/PhysRevB.24.6234

Estivill, R., Grenier, A., Duguay, S., Vurpillot, F., Terlier, T., Barnes, J.-P., Hartmann, J.-M., Blavette, D., 2015. Quantitative investigation of SiGeC layers using atom probe tomography. Ultramicroscopy 150, 23–29. https://doi.org/10.1016/j.ultramic.2014.11.020

Eswara, S., Pshenova, A., Yedra, L., Hoang, Q.H., Lovric, J., Philipp, P., Wirtz, T., 2019. Correlative microscopy combining transmission electron microscopy and secondary ion mass spectrometry: A general review on the state-of-the-art, recent developments, and prospects. Appl. Phys. Rev. 6, 021312. https://doi.org/10.1063/1.5064768

Fehre, K., Trojanowskaja, D., Gatzke, J., Kunitski, M., Trinter, F., Zeller, S., Schmidt, L.Ph.H., Stohner, J., Berger, R., Czasch, A., Jagutzki, O., Jahnke, T., Dörner, R., Schöffler, M.S., 2018. Absolute ion detection efficiencies of microchannel plates and funnel microchannel plates for multi-coincidence detection. Rev. Sci. Instrum. 89, 045112. https://doi.org/10.1063/1.5022564

Fernández-Garrido, S., Koblmüller, G., Calleja, E., Speck, J.S., 2008. In situ GaN decomposition analysis by quadrupole mass spectrometry and reflection high-energy electron diffraction. J. Appl. Phys. 104, 033541. https://doi.org/10.1063/1.2968442

Fleischmann, C., Paredis, K., Melkonyan, D., Vandervorst, W., 2018. Revealing the 3-dimensional shape of atom probe tips by atomic force microscopy. Ultramicroscopy 194, 221–226. https://doi.org/10.1016/j.ultramic.2018.08.010

Fletcher, C., Moody, M.P., Fleischmann, C., Dialameh, M., Porret, C., Geiser, B., Haley, D., 2022. Automated calibration of model-driven reconstructions in atom probe tomography. J. Phys. Appl. Phys. 55, 375301. https://doi.org/10.1088/1361-6463/ac7986

Forbes, R.G., 1995. Field evaporation theory: a review of basic ideas. Appl. Surf. Sci., Proceedings of the 41st International Field Emission Symposium 87–88, 1–11. https://doi.org/10.1016/0169-4332(94)00526-5

Forbes, R.G., Biswas, R.K., Chibane, K., 1982. FIELD EVAPORATION THEORY: A REANALYSIS OF PUBLISHED FIELD SENSITIVITY DATA. Surf. Sci. 114, 498–514.

Fowler, R.H., Nordheim, L., 1928. Electron emission in intense electric fields. Proc. R. Soc. Lond. Ser. Contain. Pap. Math. Phys. Character 119, 173–181. https://doi.org/10.1098/rspa.1928.0091

Fraser, G.W., 2002. The ion detection efficiency of microchannel plates (MCPs). Int. J. Mass Spectrom., Detectors and the Measurement of Mass Spectra 215, 13–30. https://doi.org/10.1016/S1387-3806(01)00553-X

Frasinski, L.J., Codling, K., Hatherly, P.A., 1989. Covariance Mapping: A Correlation Method Applied to Multiphoton Multiple Ionization. Science 246, 1029–1031. https://doi.org/10.1126/science.246.4933.1029

Garcia, J.M., Caplins, B.W., Chiaramonti, A.N., Miaja-Avila, L., Sanford, N.A., 2023. A Comprehensive Examination of Aluminum Oxide (Al2O3) Using Extreme and Near Ultraviolet Laser-Assisted Atom Probe Tomography. Microsc. Microanal. 29, 83–84. https://doi.org/10.1093/micmic/ozad067.033

Garcia, J.M., Chiaramonti, A.N., Caplins, B.W., Miaja-Avila, L., Sanford, N.A., 2024. Comprehensive Experimental Study of Insulating Aluminum Oxide (α-Al2O3) Using NUV- and EUV-Pulsed Atom Probe Tomography. Microsc. Microanal. 30, ozae044.051. https://doi.org/10.1093/mam/ozae044.051

Gault, B., Chiaramonti, A., Cojocaru-Mirédin, O., Stender, P., Dubosq, R., Freysoldt, C., Makineni, S.K., Li, T., Moody, M., Cairney, J.M., 2021. Atom probe tomography. Nat. Rev. Methods Primer 1, 1–30. https://doi.org/10.1038/s43586-021-00047-w

Gault, B., Danoix, F., Hoummam/z , K., Mangelinck, D., Leitner, H., 2012a. Impact of directional walk on atom probe microanalysis. Ultramicroscopy 113, 182–191. https://doi.org/10.1016/j.ultramic.2011.06.005

Gault, B., Geuser, F.D., Freysoldt, C., Klaes, B., Vurpillot, F., 2026. Spatial resolution(s) in atom probe tomography. https://doi.org/10.48550/arXiv.2601.04586

Gault, B., Haley, D., De Geuser, F., Moody, M.P., Marquis, E.A., Larson, D.J., Geiser, B.P., 2011. Advances in the reconstruction of atom probe tomography data. Ultramicroscopy 111, 448–457. https://doi.org/10.1016/j.ultramic.2010.11.016

Gault, Baptiste, Loi, S.T., Araullo-Peters, V.J., Stephenson, L.T., Moody, M.P., Shrestha, S.L., Marceau, R.K.W., Yao, L., Cairney, J.M., Ringer, S.P., 2011. Dynamic reconstruction for atom probe tomography. Ultramicroscopy 111, 1619–1624. https://doi.org/10.1016/j.ultramic.2011.08.005

Gault, B., Moody, M.P., Cairney, J.M., Ringer, S.P., 2012b. Atom Probe Microscopy. Springer Science & Business Media.

Gault, B., Saxey, D.W., Ashton, M.W., Sinnott, S.B., Chiaramonti, A.N., Moody, M.P., Schreiber, D.K., 2016. Behavior of molecules and molecular ions near a field emitter. New J. Phys. 18, 033031. https://doi.org/10.1088/1367-2630/18/3/033031

Gault, B., Vurpillot, F., Vella, A., Gilbert, M., Menand, A., Blavette, D., Deconihout, B., 2006. Design of a femtosecond laser assisted tomographic atom probe. Rev. Sci. Instrum. 77, 043705. https://doi.org/10.1063/1.2194089

Gautam, S.K., Ndiaye, S., Houard, J., Lefebvre, D., Chauveau, J.-M., Hugues, M., Vella, A., Rigutti, L., 2025. A Photonic Atom Probe Study of Thermal Effects at the Nanosecond and Nanometer scale. Nano Lett. 25, 8589–8595. https://doi.org/10.1021/acs.nanolett.5c01289

Geiser, B., Larson, D., Oltman, E., Gerstl, S., Reinhard, D., Kelly, T., Prosa, T., 2009. Wide-Field-of-View Atom Probe Reconstruction. Microsc. Microanal. 15, 292–293. https://doi.org/10.1017/S1431927609098249

Gemma, R., Al-Kassab, T., Kirchheim, R., Pundt, A., 2009. APT analyses of deuterium-loaded Fe/V multi-layered films. Ultramicroscopy, IFES 2008 109, 631–636. https://doi.org/10.1016/j.ultramic.2008.11.005

Gerstl, S.S.A., Wepf, R., 2015. Methods in Creating, Transferring, & Measuring Cryogenic Samples for APT. Microsc. Microanal. 21, 517–518. https://doi.org/10.1017/S1431927615003384

Giddings, A.D., Koelling, S., Shimizu, Y., Estivill, R., Inoue, K., Vandervorst, W., Yeoh, W.K., 2018. Industrial application of atom probe tomography to semiconductor devices. Scr. Mater. 148, 82–90. https://doi.org/10.1016/j.scriptamat.2017.09.004

Gomer, R., 1994. Field emission, field ionization, and field desorption. Surf. Sci. 299–300, 129–152. https://doi.org/10.1016/0039-6028(94)90651-3

Gomer, R., 1959. Field Desorption. J. Chem. Phys. 31, 341–345. https://doi.org/10.1063/1.1730354

Gopon, P., Douglas, J.O., Meisenkothen, F., Singh, J., London, A.J., Moody, M.P., 2022. Atom Probe Tomography for Isotopic Analysis: Development of the 34S/32S System in Sulfides. Microsc. Microanal. 28, 1127–1140. https://doi.org/10.1017/S1431927621013568

Gorman, B.P., Norman, A.G., Lawrence, D., Prosa, T., Guthrey, H., Al-Jassim, M., 2011a. Atomic scale characterization of compound semiconductors using atom probe tomography, in: 2011 37th IEEE Photovoltaic Specialists Conference. Presented at the 2011 37th IEEE Photovoltaic Specialists Conference, pp. 003357–003359. https://doi.org/10.1109/PVSC.2011.6186667

Gorman, B.P., Norman, A.G., Lawrence, D., Prosa, T., Guthrey, H., Al-Jassim, M., 2011b. Atomic scale characterization of compound semiconductors using atom probe tomography, in: 2011 37th IEEE Photovoltaic Specialists Conference. Presented at the 2011 37th IEEE Photovoltaic Specialists Conference, pp. 003357–003359. https://doi.org/10.1109/PVSC.2011.6186667

Gorman, B.P., Norman, A.G., Yan, Y., 2007. Atom Probe Analysis of III–V and Si-Based Semiconductor Photovoltaic Structures. Microsc. Microanal. 13, 493–502. https://doi.org/10.1017/S1431927607070894

Greiwe, G.-H., Balogh, Z., Schmitz, G., 2014. Atom probe tomography of lithium-doped network glasses. Ultramicroscopy 141, 51–55. https://doi.org/10.1016/j.ultramic.2014.03.007

Gribb, T.T., Olson, J.D., Martens, R.L., Shepard, J.D., Wiener, S.A., Kunicki, T.C., Ulfig, R.M., Lenz, D.R., Strennen, E.M., Oltman, E.X., Bunton, J.H., Strait, D.R., Kelly, T.F., 2002. First Data from a Commercial Local Electrode Atom Probe. Microsc. Microanal. 8, 1094–1095. https://doi.org/10.1017/S1431927602103540

Guerguis, B., Cuduvally, R., Morris, R.J.H., Arcuri, G., Langelier, B., Bassim, N., 2024. The impact of electric field strength on the accuracy of boron dopant quantification in silicon using atom probe tomography. Ultramicroscopy 266, 114034. https://doi.org/10.1016/j.ultramic.2024.114034

Guest, A.J., 1971. Un modele math ematique pour l'etude par ordinateur du fonctionnement d'une galette de microcanaux. Acta Electron. 14, 79–97.

Gundimem/z , A., Kusch, G., Frentrup, M., Xiu, H., Shu, R., Hofer, C., Bagot, P.A.J., Moody, M.P., Kappers, M.J., Wallis, D.J., Oliver, R.A., 2025. Impact of stacking faults on the luminescence of a zincblende InGaN/GaN single quantum well. J. Phys. Appl. Phys. 58, 025112. https://doi.org/10.1088/1361-6463/ad8662

Guo, C., Li, M., Nibarger, J.P., Gibson, G.N., 1998. Single and double ionization of diatomic molecules in strong laser fields. Phys. Rev. A 58, R4271–R4274. https://doi.org/10.1103/PhysRevA.58.R4271

Haley, D., McCarroll, I., Bagot, P.A.J., Cairney, J.M., Moody, M.P., 2019. A Gas-Phase Reaction Cell for Modern Atom Probe Systems. Microsc. Microanal. 25, 410–417. https://doi.org/10.1017/S1431927618016240

Hans, M., Schneider, J.M., 2020. Electric field strength-dependent accuracy of TiAlN thin film composition measurements by laser-assisted atom probe tomography. New J. Phys. 22, 033036. https://doi.org/10.1088/1367-2630/ab7770
Hans, M., Schneider, J.M., 2019. On the chemical composition of TiAlN thin films - Comparison of ion beam analysis and laser-assisted atom probe tomography with varying laser pulse energy. Thin Solid Films 688, 137251. https://doi.org/10.1016/j.tsf.2019.04.026
Hans, M., Tkadletz, M., Primetzhofer, D., Waldl, H., Schiester, M., Bartosik, M., Czettl, C., Schalk, N., Mitterer, C., Schneider, J.M., 2023. Is it meaningful to quantify vacancy concentrations of nanolamellar (Ti,Al)N thin films based on laser-assisted atom probe data? Surf. Coat. Technol. 473, 130020. https://doi.org/10.1016/j.surfcoat.2023.130020
Hashizume, T., Hasegawa, Y., Kobayashi, A., Sakurai, T., 1986. Atom-probe investigation of III–V semiconductors: Comparison of voltage-pulse and laser-pulse modes. Rev. Sci. Instrum. 57, 1378–1380. https://doi.org/10.1063/1.1138604
Hatzoglou, C., Da Costa, G., Wells, P., Ren, X., Geiser, B.P., Larson, D.J., Demoulin, R., Hunnestad, K., Talbot, E., Mazumder, B., Meier, D., Vurpillot, F., 2023a. Introducing a Dynamic Reconstruction Methodology for Multilayered Structures in Atom Probe Tomography. Microsc. Microanal. 29, 1124–1136. https://doi.org/10.1093/micmic/ozad054
Hatzoglou, C., Klaes, B., Delaroche, F., Costa, G.D., Geiser, B., Kühbach, M., Wells, P.B., Vurpillot, F., 2023b. Mesoscopic modeling of field evaporation on atom probe tomography. J. Phys. Appl. Phys. 56, 375301. https://doi.org/10.1088/1361-6463/acd649
Hatzoglou, C., Rouland, S., Radiguet, B., Etienne, A., Costa, G.D., Sauvage, X., Pareige, P., Vurpillot, F., 2020. Preferential Evaporation in Atom Probe Tomography: An Analytical Approach. Microsc. Microanal. 26, 689–698. https://doi.org/10.1017/S1431927620001749
Haydock, R., Kingham, D.R., 1981. FIELD IONIZATION THEORY: A NEW, ANALYTIC, FORMALISM. Surf. Sci. 103, 239–247.
Heller, M., Ott, B., Felfer, P., 2024. Compensating Image Distortions in a Commercial Reflectron-Type Atom Probe. Microsc. Microanal. 30, 1152–1162. https://doi.org/10.1093/mam/ozae052
Hierl, P.M., Franklin, J.L., 1967. Appearance Potentials and Kinetic Energies of Ions from N2, CO, and NO. J. Chem. Phys. 47, 3154–3161. https://doi.org/10.1063/1.1712367
Houard, J., Normand, A., Di Russo, E., Bacchi, C., Dalapati, P., Beainy, G., Moldovan, S., Da Costa, G., Delaroche, F., Vaudolon, C., Chauveau, J.M., Hugues, M., Blavette, D., Deconihout, B., Vella, A., Vurpillot, F., Rigutti, L., 2020. A photonic atom probe coupling 3D atomic scale analysis with in situ photoluminescence spectroscopy. Rev. Sci. Instrum. 91, 083704. https://doi.org/10.1063/5.0012359
Huang, C.-Y., Chao, Y.-C., Yen, H.-W., 2025. Atomic insights into strain-induced nanoscopic compositional fluctuation in AlGaN quantum well epitaxy. Scr. Mater. 266, 116788. https://doi.org/10.1016/j.scriptamat.2025.116788
Hunnestad, K.A., Das, H., Hatzoglou, C., Holtz, M., Brooks, C.M., van Helvoort, A.T.J., Muller, D.A., Schlom, D.G., Mundy, J.A., Meier, D., 2024. 3D oxygen vacancy distribution and defect-property relations in an oxide heterostructure. Nat. Commun. 15, 5400. https://doi.org/10.1038/s41467-024-49437-0
Hunnestad, K.A., Hatzoglou, C., Vurpillot, F., Nylund, I.-E., Yan, Z., Bourret, E., van Helvoort, A.T.J., Meier, D., 2023. Correlating laser energy with compositional and atomic-level information of oxides in atom probe tomography. Mater. Charact. 203, 113085. https://doi.org/10.1016/j.matchar.2023.113085
Hunsche, Koch, Brener, Nuss, 1998. THz near-field imaging. Opt. Commun. 150, 22–26. https://doi.org/10.1016/S0030-4018(98)00044-3
Hyde, J.M., Cerezo, A., Setna, R.P., Warren, P.J., Smith, G.D.W., 1994. Lateral and depth scale calibration of the position sensitive atom probe. Appl. Surf. Sci. 76–77, 382–391. https://doi.org/10.1016/0169-4332(94)90371-9

ISO 18115-1:2023(en), Surface chemical analysis — Vocabulary — Part 1: General terms and terms used in spectroscopy [WWW Document], n.d. URL https://www.iso.org/obp/ui#iso:std:iso:18115:-1:ed-3:v1:en:term:21.1 (accessed 8.16.24).

ISO/IEC Guide 99:2007(en), International vocabulary of metrology — Basic and general concepts and associated terms (VIM) [WWW Document], n.d. URL https://www.iso.org/obp/ui/es/#iso:std:iso-iec:guide:99:ed-1:v2:en (accessed 8.7.25).

Jagutzki, O., Cerezo, A., Czasch, A., Dorner, R., Hattas, M., Huang, M., Mergel, V., Spillmann, U., Ullmann-Pfleger, K., Weber, T., Schmidt-Bocking, H., Smith, G.D.W., 2002a. Multiple hit readout of a microchannel plate detector with a three-layer delay-line anode. IEEE Trans. Nucl. Sci. 49, 2477–2483. https://doi.org/10.1109/TNS.2002.803889

Jagutzki, O., Cerezo, A., Czasch, A., Dorner, R., Hattas, M., Huang, M., Mergel, V., Spillmann, U., Ullmann-Pfleger, K., Weber, T., Schmidt-Bocking, H., Smith, G.D.W., 2002b. Multiple hit readout of a microchannel plate detector with a three-layer delay-line anode. IEEE Trans. Nucl. Sci. 49, 2477–2483. https://doi.org/10.1109/TNS.2002.803889

Jakob, S., Fazi, A., Thuvander, M., 2025. Laser-Assisted Field Evaporation of Chromia with Deep Ultraviolet Laser Light. Microsc. Microanal. 31, ozae111. https://doi.org/10.1093/mam/ozae111

Jakob, S., Thuvander, M., 2024. Revisiting Compositional Accuracy of Carbides Using a Decreased Detector Efficiency in a LEAP 6000 XR Atom Probe Instrument. Microsc. Microanal. 30, 1163–1171. https://doi.org/10.1093/mam/ozae069

Jaroń-Becker, A., Becker, A., Faisal, F.H.M., 2004. Ionization of ${\mathrm{N}}_{2},$ ${\mathrm{O}}_{2},$ and linear carbon clusters in a strong laser pulse. Phys. Rev. A 69, 023410. https://doi.org/10.1103/PhysRevA.69.023410

Johnson, L.J.S., Thuvander, M., Stiller, K., Odén, M., Hultman, L., 2013. Blind deconvolution of time-of-flight mass spectra from atom probe tomography. Ultramicroscopy 132, 60–64. https://doi.org/10.1016/j.ultramic.2013.03.015

Kambham, A.K., Mody, J., Gilbert, M., Koelling, S., Vandervorst, W., 2011. Atom-probe for FinFET dopant characterization. Ultramicroscopy 111, 535–539. https://doi.org/10.1016/j.ultramic.2011.01.017

Karahka, M., Kreuzer, H.J., 2015. Field evaporation of insulators and semiconductors: Theoretical insights for ZnO. Ultramicroscopy, 1st International Conference on Atom Probe Tomography & Microscopy 159, 156–161. https://doi.org/10.1016/j.ultramic.2015.03.011

Karahka, M., Xia, Y., Kreuzer, H.J., 2015. The mystery of missing species in atom probe tomography of composite materials. Appl. Phys. Lett. 107, 062105. https://doi.org/10.1063/1.4928625

Karam, M., Houard, J., Damarla, G., Rousseau, L., Bhorade, O., Vella, A., 2023. THz driven field emission: energy and time-of-flight spectra of ions. New J. Phys. 25, 113017. https://doi.org/10.1088/1367-2630/ad0855

Katnagallu, S., Dagan, M., Parviainen, S., Nematollahi, A., Grabowski, B., Bagot, P.A.J., Rolland, N., Neugebauer, J., Raabe, D., Vurpillot, F., Moody, M.P., Gault, B., 2018. Impact of local electrostatic field rearrangement on field ionization. J. Phys. Appl. Phys. 51, 105601. https://doi.org/10.1088/1361-6463/aaaba6

Kellogg, G.L., 1987. Ion signal calibration in the imaging atom-probe with an external, time-gated image intensifier. Rev. Sci. Instrum. 58, 38–42. https://doi.org/10.1063/1.1139563

Kellogg, G.L., 1984. Measurement of activation energies for field evaporation of tungsten ions as a function of electric field. Phys. Rev. B 29, 4304–4312. https://doi.org/10.1103/PhysRevB.29.4304

Kellogg, G.L., Tsong, T.T., 1980. Pulsed-laser atom-probe field-ion microscopy. J. Appl. Phys. 51, 1184–1193. https://doi.org/10.1063/1.327686

Kelly, T.F., Gorman, B.P., Ringer, S.P., 2022. Atomic-Scale Analytical Tomography: Concepts and Implications, 1st ed. Cambridge University Press. https://doi.org/10.1017/9781316677292

Kelly, T.F., Gribb, T.T., Olson, J.D., Oltman, E., Wiener, S.A., Lenz, D.R., Shepard, J.D., Martens, R.L., Ulfig, R.M., Strennen, E.M., Bunton, J.H., Strait, D.R., Kunicki, T.C., Payne, T., Watson, J., 2003.

Configuration and Performance of a Local Electrode Atom Probe. Microsc. Microanal. 9, 564–565. https://doi.org/10.1017/S1431927603442827
Kelly, T.F., Larson, D.J., 2000. Local Electrode Atom Probes. Mater. Charact. 44, 59–85. https://doi.org/10.1016/S1044-5803(99)00055-8
Kelly, T.F., Larson, D.J., Thompson, K., Alvis, R.L., Bunton, J.H., Olson, J.D., Gorman, B.P., 2007. Atom Probe Tomography of Electronic Materials. Annu. Rev. Mater. Res. 37, 681–727. https://doi.org/10.1146/annurev.matsci.37.052506.084239
Kelly, T.F., Vella, A., Bunton, J.H., Houard, J., Silaeva, E.P., Bogdanowicz, J., Vandervorst, W., 2014. Laser pulsing of field evaporation in atom probe tomography. Curr. Opin. Solid State Mater. Sci. 18, 81–89. https://doi.org/10.1016/j.cossms.2013.11.001
Kim, S.-H., Antonov, S., Zhou, X., T. Stephenson, L., Jung, C., A. El-Zoka, A., K. Schreiber, D., Conroy, M., Gault, B., 2022. Atom probe analysis of electrode materials for Li-ion batteries: challenges and ways forward. J. Mater. Chem. A 10, 4926–4935. https://doi.org/10.1039/D1TA10050E
Kim, S.-H., Bhatt, S., Schreiber, D.K., Neugebauer, J., Freysoldt, C., Gault, B., Katnagallu, S., 2024. Understanding atom probe's analytical performance for iron oxides using correlation histograms and ab initio calculations. New J. Phys. 26, 033021. https://doi.org/10.1088/1367-2630/ad309e
Kingham, D.R., 1982. The post-ionization of field evaporated ions: A theoretical explanation of multiple charge states. Surf. Sci. 116, 273–301. https://doi.org/10.1016/0039-6028(82)90434-4
Kinno, T., Sasaki, T., Tomita, M., Ohkubo, T., 2017. Quantitativeness in laser-assisted atom probe analysis of boron and carbon codoped in silicon. Jpn. J. Appl. Phys. 56, 116601. https://doi.org/10.7567/JJAP.56.116601
Kirchhofer, R., Teague, M.C., Gorman, B.P., 2013. Thermal effects on mass and spatial resolution during laser pulse atom probe tomography of cerium oxide. J. Nucl. Mater. 436, 23–28. https://doi.org/10.1016/j.jnucmat.2012.12.052
Kitaguchi, H.S., Lozano-Perez, S., Moody, M.P., 2014. Quantitative analysis of carbon in cementite using pulsed laser atom probe. Ultramicroscopy 147, 51–60. https://doi.org/10.1016/j.ultramic.2014.06.004
Kobayashi, Y., Takahashi, J., Kawakami, K., 2011. Anomalous distribution in atom map of solute carbon in steel. Ultramicroscopy, Special Issue: 52nd International Field Emission Symposium 111, 600–603. https://doi.org/10.1016/j.ultramic.2011.01.016
Kodzuka, M., Ohkubo, T., Hono, K., Matsukura, F., Ohno, H., 2009. 3DAP analysis of (Ga,Mn)As diluted magnetic semiconductor thin film. Ultramicroscopy, IFES 2008 109, 644–648. https://doi.org/10.1016/j.ultramic.2008.11.011
Koelling, S., Gilbert, M., Goossens, J., Hikavyy, A., Richard, O., Vandervorst, W., 2011. Quantitative depth profiling of SiGe-multilayers with the Atom Probe. Surf. Interface Anal. 43, 163–166. https://doi.org/10.1002/sia.3544
Koelling, S., Gilbert, M., Goossens, J., Hikavyy, A., Richard, O., Vandervorst, W., 2009. High depth resolution analysis of Si/SiGe multilayers with the atom probe. Appl. Phys. Lett. 95, 144106. https://doi.org/10.1063/1.3243461
Koelling, S., Li, A., Cavalli, A., Assali, S., Car, D., Gazibegovic, S., Bakkers, E.P.A.M., Koenraad, P.M., 2016. Atom-by-Atom Analysis of Semiconductor Nanowires with Parts Per Million Sensitivity. Nano Lett. https://doi.org/10.1021/acs.nanolett.6b03109
Koenraad, P.M., Flatté, M.E., 2011. Single dopants in semiconductors. Nat. Mater. 10, 91–100. https://doi.org/10.1038/nmat2940
Krishnaswamy, S.V., Messier, R., Wu, C.S., McLane, S.B., Tsong, T.T., 1981. Atom probe analysis of rf-sputtered a-Si:H films. J. Vac. Sci. Technol. 18, 309–312. https://doi.org/10.1116/1.570748
La Fontaine, A., Gault, B., Breen, A., Stephenson, L., Ceguerra, A.V., Yang, L., Dinh Nguyen, T., Zhang, J., Young, D.J., Cairney, J.M., 2015. Interpreting atom probe data from chromium oxide scales. Ultramicroscopy, 1st International Conference on Atom Probe Tomography & Microscopy 159, 354–359. https://doi.org/10.1016/j.ultramic.2015.02.005

Lambeets, S.V., Cardwell, N., Onyango, I., Wirth, M.G., Vo, E., Wang, Y., Gaspard, P., Ivory, C.F., Perea, D.E., Visart de Bocarmé, T., McEwen, J.-S., 2025. Elucidating the Role of Electric Fields in Fe Oxidation via an Environmental Atom Probe. Angew. Chem. 137, e202423434. https://doi.org/10.1002/ange.202423434

Lambeets, S.V., Kautz, E.J., Wirth, M.G., Orren, G.J., Devaraj, A., Perea, D.E., 2020a. Nanoscale Perspectives of Metal Degradation via In Situ Atom Probe Tomography. Top. Catal. 63, 1606–1622. https://doi.org/10.1007/s11244-020-01367-z

Lambeets, S.V., Visart de Bocarmé, T., Perea, D.E., Kruse, N., 2020b. Directional Gateway to Metal Oxidation: 3D Chemical Mapping Unfolds Oxygen Diffusional Pathways in Rhodium Nanoparticles. J. Phys. Chem. Lett. 11, 3144–3151. https://doi.org/10.1021/acs.jpclett.0c00321

Larson, D. J., Gault, B., Geiser, B.P., De Geuser, F., Vurpillot, F., 2013. Atom probe tomography spatial reconstruction: Status and directions. Curr. Opin. Solid State Mater. Sci., Atom Probe Tomography 17, 236–247. https://doi.org/10.1016/j.cossms.2013.09.002

Larson, D.J., Prosa, T.J., Chen, Y., Reinhard, D.A., Martin, I., Ulfig, R.M., Holman, M., Robinson, J., Lenz, D., 2023a. Improving Analytical Capability via Simultaneous Voltage and Laser Pulsing in Atom Probe Tomography. Microsc. Microanal. 29, 609–610. https://doi.org/10.1093/micmic/ozad067.295

Larson, D.J., Prosa, T.J., Chen, Y., Reinhard, D.A., Martin, I., Ulfig, R.M., Holman, M., Robinson, J., Lenz, D., 2023b. Improving Analytical Capability via Simultaneous Voltage and Laser Pulsing in Atom Probe Tomography. Microsc. Microanal. 29, 609–610. https://doi.org/10.1093/micmic/ozad067.295

Larson, D.J., Prosa, T.J., Oltman, E., Reinhard, D.A., Geiser, B.P., Ulfig, R.M., Merkulov, A., 2018. A High Multiple Hits Correction Factor for Atom Probe Tomography. Microsc. Microanal. 24, 1084–1085. https://doi.org/10.1017/S1431927618005901

Larson, David J., Prosa, T.J., Ulfig, R.M., Geiser, B.P., Kelly, T.F., 2013. Applications of the Local Electrode Atom Probe, in: Local Electrode Atom Probe Tomography. Springer, New York, NY, pp. 201–247. https://doi.org/10.1007/978-1-4614-8721-0_7

Lauhon, L.J., Adusumilli, P., Ronsheim, P., Flaitz, P.L., Lawrence, D., 2009. Atom-Probe Tomography of Semiconductor Materials and Device Structures. MRS Bull. 34, 738–743. https://doi.org/10.1557/mrs2009.248

Lawitzki, R., Stender, P., Schmitz, G., 2021. Compensating Local Magnifications in Atom Probe Tomography for Accurate Analysis of Nano-Sized Precipitates. Microsc. Microanal. 27, 499–510. https://doi.org/10.1017/S1431927621000180

Lefebvre-Ulrikson, W., Da Costa, G., Rigutti, L., Blum, I., 2016. Chapter Nine - Data Mining, in: Atom Probe Tomography. Academic Press, pp. 279–317. https://doi.org/10.1016/B978-0-12-804647-0.00009-7

Li, Y., Wei, Y., Saxena, A., Kühbach, M., Freysoldt, C., Gault, B., 2026. Machine learning enhanced atom probe tomography analysis. Prog. Mater. Sci. 156, 101561. https://doi.org/10.1016/j.pmatsci.2025.101561

Liddle, J.A., Norman, A., Cerezo, A., Grovenor, C.R.M., 1988. PULSED LASER ATOM PROBE ANALYSIS OF TERNARY AND QUATERNARY III-V EPITAXIAL LAYERS. J. Phys. Colloq. 49, C6-514. https://doi.org/10.1051/jphyscol:1988686

Llopart, X., Ballabriga, R., Campbell, M., Tlustos, L., Wong, W., 2007. Timepix, a 65k programmable pixel readout chip for arrival time, energy and/or photon counting measurements. Nucl. Instrum. Methods Phys. Res. Sect. Accel. Spectrometers Detect. Assoc. Equip., VCI 2007 581, 485–494. https://doi.org/10.1016/j.nima.2007.08.079

Loi, S.T., Gault, B., Ringer, S.P., Larson, D.J., Geiser, B.P., 2013. Electrostatic simulations of a local electrode atom probe: The dependence of tomographic reconstruction parameters on specimen and microscope geometry. Ultramicroscopy 132, 107–113. https://doi.org/10.1016/j.ultramic.2012.12.012
London, A.J., 2019. Quantifying Uncertainty from Mass-Peak Overlaps in Atom Probe Microscopy. Microsc. Microanal. 25, 378–388. https://doi.org/10.1017/S1431927618016276
London, A.J., Haley, D., Moody, M.P., 2017. Single-Ion Deconvolution of Mass Peak Overlaps for Atom Probe Microscopy. Microsc. Microanal. 23, 300–306. https://doi.org/10.1017/S1431927616012782
Lüken, J., Fleischmann, C., Sijbers, J., De Beenhouwer, J., 2024. AdAPTS: An Adaptive Atom Probe Tomography Simulation Library. Microsc. Microanal. 30, ozae044.017. https://doi.org/10.1093/mam/ozae044.017
Lymperakis, L., Neugebauer, J., 2009. Large anisotropic adatom kinetics on nonpolar GaN surfaces: Consequences for surface morphologies and nanowire growth. Phys. Rev. B 79, 241308. https://doi.org/10.1103/PhysRevB.79.241308
Mair, S., Gompf, B., Dressel, M., 2004. Spatial and spectral behavior of the optical near field studied by a terahertz near-field spectrometer. Appl. Phys. Lett. 84, 1219–1221. https://doi.org/10.1063/1.1647707
Mancini, L., Amirifar, N., Shinde, D., Blum, I., Gilbert, M., Vella, A., Vurpillot, F., Lefebvre, W., Lardé, R., Talbot, E., Pareige, P., Portier, X., Ziani, A., Davesnne, C., Durand, C., Eymery, J., Butté, R., Carlin, J.-F., Grandjean, N., Rigutti, L., 2014a. Composition of Wide Bandgap Semiconductor Materials and Nanostructures Measured by Atom Probe Tomography and Its Dependence on the Surface Electric Field. J. Phys. Chem. C 118, 24136–24151. https://doi.org/10.1021/jp5071264
Mancini, L., Amirifar, N., Shinde, D., Blum, I., Gilbert, M., Vella, A., Vurpillot, F., Lefebvre, W., Lardé, R., Talbot, E., Pareige, P., Portier, X., Ziani, A., Davesnne, C., Durand, C., Eymery, J., Butté, R., Carlin, J.-F., Grandjean, N., Rigutti, L., 2014b. Composition of Wide Bandgap Semiconductor Materials and Nanostructures Measured by Atom Probe Tomography and Its Dependence on the Surface Electric Field. J. Phys. Chem. C 118, 24136–24151. https://doi.org/10.1021/jp5071264
Mangelinck, D., Panciera, F., Hoummam/z , K., El Kousseifi, M., Perrin, C., Descoins, M., Portavoce, A., 2014. Atom probe tomography for advanced metallization. Microelectron. Eng. 120, 19–33. https://doi.org/10.1016/j.mee.2013.12.018
Martin, A.J., Yatzor, B., 2019. Examining the Effect of Evaporation Field on Boron Measurements in SiGe: Insights into Improving the Relationship Between APT and SIMS Measurements of Boron. Microsc. Microanal. 25, 617–624. https://doi.org/10.1017/S1431927619000291
Mathew, A., Eijkel, G.B., Anthony, I.G.M., Ellis, S.R., Heeren, R.M.A., 2022. Characterization of microchannel plate detector response for the detection of native multiply charged high mass single ions in orthogonal-time-of-flight mass spectrometry using a Timepix detector. J. Mass Spectrom. 57, e4820. https://doi.org/10.1002/jms.4820
Matoba, S., Takahashi, R., Io, C., Koizumi, T., Shiromaru, H., 2011. Absolute Detection Efficiency of a High-Sensitivity Microchannel Plate with Tapered Pores. Jpn. J. Appl. Phys. 50, 112201. https://doi.org/10.1143/JJAP.50.112201
Mayer, J., Barthel, J., Vayyala, A., Dunin-Borkowski, R., Bischoff, M., van Leeuwen, H., Kujawa, S., Bunton, J., Lenz, D., Kelly, T.F., 2023. The TOMO Project – Integrating a Fully Functional Atom Probe in an Aberration-Corrected TEM. Microsc. Microanal. 29, 593–594. https://doi.org/10.1093/micmic/ozad067.286
Mazumder, B., Liu, X., Yeluri, R., Wu, F., Mishra, U.K., Speck, J.S., 2014. Atom probe tomography studies of Al2O3 gate dielectrics on GaN. J. Appl. Phys. 116, 134101. https://doi.org/10.1063/1.4896498
McCarroll, I.E., Bagot, P.A.J., Devaraj, A., Perea, D.E., Cairney, J.M., 2020. New frontiers in atom probe tomography: a review of research enabled by cryo and/or vacuum transfer systems. Mater. Today Adv. 7, 100090. https://doi.org/10.1016/j.mtadv.2020.100090

McPhail, D.S., 2006. Applications of Secondary Ion Mass Spectrometry (SIMS) in Materials Science. J. Mater. Sci. 41, 873–903. https://doi.org/10.1007/s10853-006-6568-x
Meier, M.S., Jones, M.E., Felfer, P.J., Moody, M.P., Haley, D., 2022. Extending Estimating Hydrogen Content in Atom Probe Tomography Experiments Where H2 Molecule Formation Occurs. Microsc. Microanal. 28, 1231–1244. https://doi.org/10.1017/S1431927621012332
Meisenkothen, F., Steel, E.B., Prosa, T.J., Henry, K.T., Prakash Kolli, R., 2015. Effects of detector dead-time on quantitative analyses involving boron and multi-hit detection events in atom probe tomography. Ultramicroscopy 159, 101–111. https://doi.org/10.1016/j.ultramic.2015.07.009
Menand, A., Kingham, D.R., 1985. Evidence for the quantum mechanical tunnelling of boron ions. J. Phys. C Solid State Phys. 18, 4539. https://doi.org/10.1088/0022-3719/18/23/015
Menand, A., Martin, C., Sarrau, J.M., 1984. FIELD EVAPORATION CHARGE STATE OF BORON IONS : A TEMPERATURE EFFECT STUDY. J. Phys. Colloq. 45, C9-98. https://doi.org/10.1051/jphyscol:1984917
Meng, K., Schwarz, T.M., Weikum, E.M., Stender, P., Schmitz, G., 2022a. Frozen n-Tetradecane Investigated by Cryo-Atom Probe Tomography. Microsc. Microanal. 28, 1289–1299. https://doi.org/10.1017/S143192762101254X
Meng, K., Schwarz, T.M., Weikum, E.M., Stender, P., Schmitz, G., 2022b. Frozen n-Tetradecane Investigated by Cryo-Atom Probe Tomography. Microsc. Microanal. 28, 1289–1299. https://doi.org/10.1017/S143192762101254X
Miaja-Avila, L., Caplins, B.W., Chiaramonti, A.N., Blanchard, P.T., Brubaker, M.D., Davydov, A.V., Diercks, D.R., Gorman, B.P., Rishinaramangalam, A., Feezell, D.F., Bertness, K.A., Sanford, N.A., 2021. Extreme Ultraviolet Radiation Pulsed Atom Probe Tomography of III-Nitride Semiconductor Materials. J. Phys. Chem. C 125, 2626–2635. https://doi.org/10.1021/acs.jpcc.0c08753
Miller, M.K., Angelini, P., Cerezo, A., More, K.L., 1989. PULSED LASER ATOM PROBE CHARACTERIZATION OF SILICON CARBIDE. J. Phys. Colloq. 50, C8-464. https://doi.org/10.1051/jphyscol:1989878
Miller, M.K., Cerezo, A., Hetherington, M.G., Smith, G.D.W., 1996. Atom Probe Field Ion Microscopy. Oxford University Press. https://doi.org/10.1093/oso/9780198513872.001.0001
Miller, M.K., Forbes, R.G., 2014a. Introduction to the Physics of Field Ion Emitters, in: Atom-Probe Tomography. Springer, Boston, MA, pp. 51–109. https://doi.org/10.1007/978-1-4899-7430-3_2
Miller, M.K., Forbes, R.G., 2014b. Data Reconstruction, in: Atom-Probe Tomography. Springer, Boston, MA, pp. 259–302. https://doi.org/10.1007/978-1-4899-7430-3_6
Miller, M.K., Forbes, R.G., 2014c. Field Evaporation and Related Topics, in: Atom-Probe Tomography. Springer, Boston, MA, pp. 111–187. https://doi.org/10.1007/978-1-4899-7430-3_3
Miller, M.K., Smith, G.D.W., 1981. An atom probe study of the anomalous field evaporation of alloys containing silicon. J. Vac. Sci. Technol. 19, 57–62. https://doi.org/10.1116/1.571017
Miyamoto, G., Shinbo, K., Furuhara, T., 2012. Quantitative measurement of carbon content in Fe–C binary alloys by atom probe tomography. Scr. Mater. 67, 999–1002. https://doi.org/10.1016/j.scriptamat.2012.09.007
Monajem, M., Ott, B., Heimerl, J., Meier, S., Hommelhoff, P., Felfer, P., 2025. PyCCAPT: A Python Package for Open-Source Atom Probe Instrument Control and Data Calibration. Microsc. Res. Tech. n/a. https://doi.org/10.1002/jemt.70011
Morris, Richard.J.H., Cuduvally, R., Melkonyan, D., Fleischmann, C., Zhao, M., Arnoldi, L., van der Heide, P., Vandervorst, W., 2018. Toward accurate composition analysis of GaN and AlGaN using atom probe tomography. J. Vac. Sci. Technol. B 36, 03F130. https://doi.org/10.1116/1.5019693
Morris, R.J.H., Cuduvally, R., Lin, J.-R., Zhao, M., Vandervorst, W., Thuvander, M., Fleischmann, C., 2022. Field dependent study on the impact of co-evaporated multihits and ion pile-up for the apparent stoichiometric quantification of GaN and AlN. Ultramicroscopy 241, 113592. https://doi.org/10.1016/j.ultramic.2022.113592

Morris, R.J.H., Lin, J.-R., Scheerder, J.E., Popovici, M.I., Meersschaut, J., Goux, L., Kar, G.S., van der Heide, P., Fleischmann, C., 2024. Significant Oxygen Underestimation When Quantifying Barium-Doped SrTiO Layers by Atom Probe Tomography. Microsc. Microanal. 30, 49–58. https://doi.org/10.1093/micmic/ozad144

Mouton, I., Breen, A.J., Wang, S., Chang, Y., Szczepaniak, A., Kontis, P., Stephenson, L.T., Raabe, D., Herbig, M., Britton, T.B., Gault, B., 2019. Quantification Challenges for Atom Probe Tomography of Hydrogen and Deuterium in Zircaloy-4. Microsc. Microanal. 25, 481–488. https://doi.org/10.1017/S143192761801615X

Müller, E.W., 1960. Field Ionization and Field Ion Microscopy, in: Marton, L., Marton, C. (Eds.), Advances in Electronics and Electron Physics. Academic Press, pp. 83–179. https://doi.org/10.1016/S0065-2539(08)60210-3

Müller, E.W., 1956. Field Desorption. Phys. Rev. 102, 618–624. https://doi.org/10.1103/PhysRev.102.618

Müller, E.W., 1941. Abreißen adsorbierter Ionen durch hohe elektrische Feldstärken. Naturwissenschaften 29, 533–534. https://doi.org/10.1007/BF01481175

Müller, E.W., Bahadur, K., 1956. Field Ionization of Gases at a Metal Surface and the Resolution of the Field Ion Microscope. Phys. Rev. 102, 624–631. https://doi.org/10.1103/PhysRev.102.624

Müller, E.W., Panitz, J.A., McLane, S.B., 1968. The Atom-Probe Field Ion Microscope. Rev. Sci. Instrum. 39, 83–86. https://doi.org/10.1063/1.1683116

Müller, M., Saxey, D.W., Smith, G.D.W., Gault, B., 2011. Some aspects of the field evaporation behaviour of GaSb. Ultramicroscopy, Special Issue: 52nd International Field Emission Symposium 111, 487–492. https://doi.org/10.1016/j.ultramic.2010.11.019

Müller, M., Smith, G.D.W., Gault, B., Grovenor, C.R.M., 2012. Compositional nonuniformities in pulsed laser atom probe tomography analysis of compound semiconductors. J. Appl. Phys. 111, 064908. https://doi.org/10.1063/1.3695461

Naghdali, S., Schiester, M., Waldl, H., Terziyska, V., Hans, M., Primetzhofer, D., Schalk, N., Tkadletz, M., 2025. Improving the elemental and imaging accuracy in atom probe tomography of (Ti,Si)N single and multilayer coatings using isotopic substitution of N. Ultramicroscopy 276, 114200. https://doi.org/10.1016/j.ultramic.2025.114200

Nava, F., Bertuccio, G., Cavallini, A., Vittone, E., 2008. Silicon carbide and its use as a radiation detector material. Meas. Sci. Technol. 19, 102001. https://doi.org/10.1088/0957-0233/19/10/102001

Ndiaye, S., Bacchi, C., Klaes, B., Canino, M., Vurpillot, F., Rigutti, L., 2023a. Surface Dynamics of Field Evaporation in Silicon Carbide. J. Phys. Chem. C 127, 5467–5478. https://doi.org/10.1021/acs.jpcc.2c08908

Ndiaye, S., Bhorade, O., Blum, I., Klaes, B., Bacchi, C., Houard, J., Vella, A., Vurpillot, F., Rigutti, L., 2024. Isotopic Correction of Compositional Inaccuracies in the Atom Probe Analysis of LaB6. J. Phys. Chem. C 128, 2937–2947. https://doi.org/10.1021/acs.jpcc.3c07595

Ndiaye, S., Elias, C., Diagne, A., Rotella, H., Georgi, F., Hugues, M., Cordier, Y., Vurpillot, F., Rigutti, L., 2023b. Alloy distribution and compositional metrology of epitaxial ScAlN by atom probe tomography. Appl. Phys. Lett. 123, 162102. https://doi.org/10.1063/5.0167855

Ngamo, M., Duguay, S., Cristiano, F., Daoud-Ketata, K., Pareige, P., 2009. Atomic scale study of boron interstitial clusters in ion-implanted silicon. J. Appl. Phys. 105, 104904. https://doi.org/10.1063/1.3126498

Nickerson, B.S., Karahka, M., Kreuzer, H.J., 2015. Disintegration and field evaporation of thiolate polymers in high electric fields. Ultramicroscopy 159, 173–177. https://doi.org/10.1016/j.ultramic.2015.03.013

Nilsen, J.S., van Helvoort, A.T.J., 2022. Composition Analysis by STEM-EDX of Ternary Semiconductors by Internal References. Microsc. Microanal. 28, 61–69. https://doi.org/10.1017/S1431927621013672

Nobes, R.H., Moncrieff, D., Wong, M.W., Radom, L., Gill, P.M.W., Pople, J.A., 1991. The structure and stability of the O2+2 dication: a dramatic failure of Møller—Plesset perturbation theory. Chem. Phys. Lett. 182, 216–224. https://doi.org/10.1016/0009-2614(91)80204-B

Oberdorfer, C., Eich, S.M., Schmitz, G., 2013. A full-scale simulation approach for atom probe tomography. Ultramicroscopy 128, 55–67. https://doi.org/10.1016/j.ultramic.2013.01.005

Oberdorfer, C., Stender, P., Reinke, C., Schmitz, G., 2007. Laser-Assisted Atom Probe Tomography of Oxide Materials. Microsc. Microanal. 13, 342–346. https://doi.org/10.1017/S1431927607070274

Ono, T., Sasaki, T., Otsuka, J., Hirose, K., 2005. First-principles study on field evaporation of surface atoms from W(0 1 1) and Mo(0 1 1) surfaces. Surf. Sci. 577, 42–46. https://doi.org/10.1016/j.susc.2004.12.024

Pacchioni, G., 2017. Spin qubits: Useful defects in silicon carbide. Nat. Rev. Mater. 2, 17052. https://doi.org/10.1038/natrevmats.2017.52

Panayi, P., Clifton, P.H., Lloyd, G., Shellswell, G., Cerezo, A., 2006. A Wide Angle Achromatic Reflectron for the Atom Pprobe, in: 2006 19th International Vacuum Nanoelectronics Conference. Presented at the 2006 19th International Vacuum Nanoelectronics Conference, pp. 63–63. https://doi.org/10.1109/IVNC.2006.335353

Pareige, C., Lefebvre-Ulrikson, W., Vurpillot, F., Sauvage, X., 2016. Chapter Five - Time-of-Flight Mass Spectrometry and Composition Measurements, in: Atom Probe Tomography. Academic Press, pp. 123–154. https://doi.org/10.1016/B978-0-12-804647-0.00005-X

Pavlina, E.J., Van Tyne, C.J., 2008. Correlation of Yield Strength and Tensile Strength with Hardness for Steels. J. Mater. Eng. Perform. 17, 888–893. https://doi.org/10.1007/s11665-008-9225-5

Peng, Z., Vurpillot, F., Choi, P.-P., Li, Y., Raabe, D., Gault, B., 2018. On the detection of multiple events in atom probe tomography. Ultramicroscopy 189, 54–60. https://doi.org/10.1016/j.ultramic.2018.03.018

Peng, Z., Zanuttini, D., Gervais, B., Jacquet, E., Blum, I., Choi, P.-P., Raabe, D., Vurpillot, F., Gault, B., 2019. Unraveling the Metastability of Cn2+ (n = 2–4) Clusters. J. Phys. Chem. Lett. 10, 581–588. https://doi.org/10.1021/acs.jpclett.8b03449

Peralta, J., Broderick, S.R., Rajan, K., 2013. Mapping energetics of atom probe evaporation events through first principles calculations. Ultramicroscopy 132, 143–151. https://doi.org/10.1016/j.ultramic.2013.02.007

Perea, D.E., Allen, J.E., May, S.J., Wessels, B.W., Seidman, D.N., Lauhon, L.J., 2006. Three-Dimensional Nanoscale Composition Mapping of Semiconductor Nanowires. Nano Lett. 6, 181–185. https://doi.org/10.1021/nl051602p

Prosa, T.J., Geiser, B.P., Lawrence, D., Olson, D., Larson, D.J., 2014. Developing detection efficiency standards for atom probe tomography, in: Instrumentation, Metrology, and Standards for Nanomanufacturing, Optics, and Semiconductors VIII. Presented at the Instrumentation, Metrology, and Standards for Nanomanufacturing, Optics, and Semiconductors VIII, SPIE, pp. 38–45. https://doi.org/10.1117/12.2062211

Prosa, T.J., Oltman, E., 2022. Study of LEAP® 5000 Deadtime and Precision via Silicon Pre-Sharpened-Microtip$^{TM}$ Standard Specimens. Microsc. Microanal. 28, 1019–1037. https://doi.org/10.1017/S143192762101206X

Proudian, A.P., Jaskot, M.B., Diercks, D.R., Gorman, B.P., Zimmerman, J.D., 2019. Atom Probe Tomography of Molecular Organic Materials: Sub-Dalton Nanometer-Scale Quantification. Chem. Mater. 31, 2241–2247. https://doi.org/10.1021/acs.chemmater.8b04476

Qiu, R., Aboulfadl, H., Bäcke, O., Stiens, D., Andrén, H.-O., Halvarsson, M., 2021. Atom probe tomography investigation of 3D nanoscale compositional variations in CVD TiAlN nanolamella coatings. Surf. Coat. Technol. 426, 127741. https://doi.org/10.1016/j.surfcoat.2021.127741

Rapcewicz, K., Buongiorno Nardelli, M., Bernholc, J., 1997. Theory of surface morphology of wurtzite GaN (0001) surfaces. Phys. Rev. B 56, R12725–R12728. https://doi.org/10.1103/PhysRevB.56.R12725

Reddy, S.M., Saxey, D.W., Rickard, W.D.A., Fougerouse, D., Montalvo, S.D., Verberne, R., van Riessen, A., 2020. Atom Probe Tomography: Development and Application to the Geosciences. Geostand. Geoanalytical Res. 44, 5–50. https://doi.org/10.1111/ggr.12313

Rementeria, R., Jimenez, J.A., Allain, S.Y.P., Geandier, G., Poplawsky, J.D., Guo, W., Urones-Garrote, E., Garcia-Mateo, C., Caballero, F.G., 2017. Quantitative assessment of carbon allocation anomalies in low temperature bainite. Acta Mater. 133, 333–345. https://doi.org/10.1016/j.actamat.2017.05.048

Rigutti, L., Di Russo, E., Chabanais, F., Blum, I., Houard, J., Gogneau, N., Largeau, L., Karg, A., Eickhoff, M., Lefebvre, W., Vurpillot, F., 2021. Surface Microscopy of Atomic and Molecular Hydrogen from Field-Evaporating Semiconductors. J. Phys. Chem. C 125, 17078–17087. https://doi.org/10.1021/acs.jpcc.1c04778

Rigutti, L., Mancini, L., Hernández-Maldonado, D., Lefebvre, W., Giraud, E., Butté, R., Carlin, J.F., Grandjean, N., Blavette, D., Vurpillot, F., 2016a. Statistical correction of atom probe tomography data of semiconductor alloys combined with optical spectroscopy: The case of Al0.25Ga0.75N. J. Appl. Phys. 119, 105704. https://doi.org/10.1063/1.4943612

Rigutti, L., Mancini, L., Lefebvre, W., Houard, J., Hernàndez-Maldonado, D., Russo, E.D., Giraud, E., Butté, R., J-F Carlin, Grandjean, N., Blavette, D., Vurpillot, F., 2016b. Statistical nanoscale study of localised radiative transitions in GaN/AlGaN quantum wells and AlGaN epitaxial layers. Semicond. Sci. Technol. 31, 095009. https://doi.org/10.1088/0268-1242/31/9/095009

Rigutti, L., Venturi, L., Houard, J., Normand, A., Silaeva, E.P., Borz, M., Malykhin, S.A., Obraztsov, A.N., Vella, A., 2017. Optical Contactless Measurement of Electric Field-Induced Tensile Stress in Diamond Nanoscale Needles. Nano Lett. 17, 7401–7409. https://doi.org/10.1021/acs.nanolett.7b03222

Riley, J.R., Bernal, R.A., Li, Q., Espinosa, H.D., Wang, G.T., Lauhon, L.J., 2012. Atom Probe Tomography of a-Axis GaN Nanowires: Analysis of Nonstoichiometric Evaporation Behavior. ACS Nano 6, 3898–3906. https://doi.org/10.1021/nn2050517

Riley, J.R., Detchprohm, T., Wetzel, C., Lauhon, L.J., 2014. On the reliable analysis of indium mole fraction within InxGa1−xN quantum wells using atom probe tomography. Appl. Phys. Lett. 104, 152102. https://doi.org/10.1063/1.4871510

Rolander, U., Andrén, H.-O., 1989. STATISTICAL CORRECTION FOR PILE-UP IN THE ATOM-PROBE DETECTOR SYSTEM. J. Phys. Colloq. 50, C8-534. https://doi.org/10.1051/jphyscol:1989891

Rolland, N., Larson, D.J., Geiser, B.P., Duguay, S., Vurpillot, F., Blavette, D., 2015a. An analytical model accounting for tip shape evolution during atom probe analysis of heterogeneous materials. Ultramicroscopy, 1st International Conference on Atom Probe Tomography & Microscopy 159, Part 2, 195–201. https://doi.org/10.1016/j.ultramic.2015.03.010

Rolland, Nicolas, Vurpillot, F., Duguay, S., Blavette, D., 2015. A Meshless Algorithm to Model Field Evaporation in Atom Probe Tomography. Microsc. Microanal. 21, 1649–1656. https://doi.org/10.1017/S1431927615015184

Rolland, N., Vurpillot, F., Duguay, S., Blavette, D., 2015b. Dynamic evolution and fracture of multilayer field emitters in atom probe tomography: a new interpretation. Eur. Phys. J. Appl. Phys. 72, 21001. https://doi.org/10.1051/epjap/2015150233

Ronsheim, P., Flaitz, P., Hatzistergos, M., Molella, C., Thompson, K., Alvis, R., 2008. Impurity measurements in silicon with D-SIMS and atom probe tomography. Appl. Surf. Sci. 255, 1547–1550. https://doi.org/10.1016/j.apsusc.2008.05.247

Ronsheim, P.A., Hatzistergos, M., Jin, S., 2010. Dopant measurements in semiconductors with atom probe tomography. J. Vac. Sci. Technol. B 28, C1E1-C1E4. https://doi.org/10.1116/1.3242422

Rousseau, L., Normand, A., Morgado, F.F., Marie Scisly Søreide, H.-S., Stephenson, L.T., Hatzoglou, C., Da Costa, G., Tehrani, K., Freysoldt, C., Gault, B., Vurpillot, F., 2023. Introducing field evaporation energy loss spectroscopy. Commun. Phys. 6, 100. https://doi.org/10.1038/s42005-023-01203-2

Rousseau, L., Normand, A., Morgado, F.F., Stephenson, L., Gault, B., Tehrani, K., Vurpillot, F., 2020. Dynamic Effects in Voltage Pulsed Atom Probe. Microsc. Microanal. 26, 1133–1146. https://doi.org/10.1017/S1431927620024587

Rybak, J.M., Arlt, J., Gault, B., Volkert, C.A., 2025. Advancing Atom Probe Tomography of SrTiO3: Measurement Methodology and Impurity Detection Limits. Microsc. Microanal. 31, ozaf051. https://doi.org/10.1093/mam/ozaf051

Sakurai, T., Hashizume, T., Jimbo, A., Sakata, T., 1984. AN ATOM-PROBE STUDY OF III-V COMPOUND SEMICONDUCTORS. J. Phys. Colloq. 45, C9-458. https://doi.org/10.1051/jphyscol:1984975

Santhanagopalan, D., Schreiber, D.K., Perea, D.E., Martens, R.L., Janssen, Y., Khalifah, P., Meng, Y.S., 2015. Effects of laser energy and wavelength on the analysis of LiFePO4 using laser assisted atom probe tomography. Ultramicroscopy 148, 57–66. https://doi.org/10.1016/j.ultramic.2014.09.004

Saxey, D.W., 2011. Correlated ion analysis and the interpretation of atom probe mass spectra. Ultramicroscopy 111, 473–479. https://doi.org/10.1016/j.ultramic.2010.11.021

Saxey, D.W., Cairney, J.M., McGrouther, D., Honma, T., Ringer, S.P., 2007. Atom probe specimen fabrication methods using a dual FIB/SEM. Ultramicroscopy 107, 756–760. https://doi.org/10.1016/j.ultramic.2007.02.024

Schiester, M., Waldl, H., Hans, M., Thuvander, M., Primetzhofer, D., Schalk, N., Tkadletz, M., 2024. Influence of multiple detection events on compositional accuracy of TiN coatings in atom probe tomography. Surf. Coat. Technol. 477, 130318. https://doi.org/10.1016/j.surfcoat.2023.130318

Schiester, M., Waldl, H., Rice, K.P., Hans, M., Primetzhofer, D., Schalk, N., Tkadletz, M., 2025. Effects of laser wavelength and pulse energy on the evaporation behavior of TiN coatings in atom probe tomography: A multi-instrument study. Ultramicroscopy 270, 114105. https://doi.org/10.1016/j.ultramic.2025.114105

Schwarz, T.M., Dumont, M., Garcia-Giner, V., Jung, C., Porter, A.E., Gault, B., 2025. Advancing atom probe tomography capabilities to understand bone microstructures at near-atomic scale. Acta Biomater. 198, 319–333. https://doi.org/10.1016/j.actbio.2025.03.051

Schwarz, T.M., Weikum, E.M., Meng, K., Hadjixenophontos, E., Dietrich, C.A., Kästner, J., Stender, P., Schmitz, G., 2020. Field evaporation and atom probe tomography of pure water tips. Sci. Rep. 10, 20271. https://doi.org/10.1038/s41598-020-77130-x

Sha, W., Chang, L., Smith, G.D.W., Liu Cheng, Mittemeijer, E.J., 1992. Some aspects of atom-probe analysis of Fe�C and Fe�N systems. Surf. Sci. 266, 416–423. https://doi.org/10.1016/0039-6028(92)91055-G

Shariq, A., Mutas, S., Wedderhoff, K., Klein, C., Hortenbach, H., Teichert, S., Kücher, P., Gerstl, S.S.A., 2009. Investigations of field-evaporated end forms in voltage- and laser-pulsed atom probe tomography. Ultramicroscopy 109, 472–479. https://doi.org/10.1016/j.ultramic.2008.10.001

She, X., Huang, A.Q., Lucía, Ó., Ozpineci, B., 2017. Review of Silicon Carbide Power Devices and Their Applications. IEEE Trans. Ind. Electron. 64, 8193–8205. https://doi.org/10.1109/TIE.2017.2652401

Shu, R., Oliver, R.A., Frentrup, M., Kappers, M.J., Xiu, H., Kusch, G., Wallis, D.J., Hofer, C., Bagot, P.A.J., Moody, M.P., 2025. Beyond transmission electron microscopy imaging: Atom probe tomography reveals chemical inhomogeneity at stacking fault interfaces in InGaN/GaN light-emitting diodes. Materialia 40, 102417. https://doi.org/10.1016/j.mtla.2025.102417

Sijbrandij, S.J., Cerezo, A., Deconihout, B., Godfrey, T.J., Smith, G.D.W., 1996. Characterization of Efficiency Enhancement in Microchannel Plate Detectors. J. Phys. IV 06, C5-302. https://doi.org/10.1051/jp4:1996548

Siladie, A.-M., Amichi, L., Mollard, N., Mouton, I., Bonef, B., Bougerol, C., Grenier, A., Robin, E., Jouneau, P.-H., Garro, N., Cros, A., Daudin, B., 2018. Dopant radial inhomogeneity in Mg-doped GaN nanowires. Nanotechnology 29, 255706. https://doi.org/10.1088/1361-6528/aabbd6

Silaeva, E.P., Arnoldi, L., Karahka, M.L., Deconihout, B., Menand, A., Kreuzer, H.J., Vella, A., 2014. Do Dielectric Nanostructures Turn Metallic in High-Electric dc Fields? Nano Lett. 14, 6066–6072. https://doi.org/10.1021/nl502715s

Silaeva, E.P., Karahka, M., Kreuzer, H.J., 2013. Atom Probe Tomography and field evaporation of insulators and semiconductors: Theoretical issues. Curr. Opin. Solid State Mater. Sci. 17, 211–216. https://doi.org/10.1016/j.cossms.2013.08.001

Smith, R., Walls, J.M., 1978. Ion trajectories in the field-ion microscope. J. Phys. Appl. Phys. 11, 409. https://doi.org/10.1088/0022-3727/11/4/005

Stoffers, A., Oberdorfer, C., Schmitz, G., 2012. Controlled Field Evaporation of Fluorinated Self-Assembled Monolayers. Langmuir 28, 56–59. https://doi.org/10.1021/la204126x

Suchorski, Yu., Schmidt, W.A., Ernst, N., Block, J.H., Kreuzer, H.J., 1995. Electrostatic fields above individual atoms. Prog. Surf. Sci. 48, 121–134. https://doi.org/10.1016/0079-6816(95)93420-C

Takahashi, J., Kawakami, K., 2014. A quantitative model of preferential evaporation and retention for atom probe tomography. Surf. Interface Anal. 46, 535–543. https://doi.org/10.1002/sia.5555

Takahashi, J., Kawakami, K., Kobayashi, Y., 2020. Study on Quantitative Analysis of Carbon and Nitrogen in Stoichiometric θ-Fe3C and γ′-Fe4N by Atom Probe Tomography. Microsc. Microanal. 26, 185–193. https://doi.org/10.1017/S1431927620000045

Takahashi, J., Kawakami, K., Kobayashi, Y., Tarui, T., 2010. The first direct observation of hydrogen trapping sites in TiC precipitation-hardening steel through atom probe tomography. Scr. Mater. 63, 261–264. https://doi.org/10.1016/j.scriptamat.2010.03.012

Takahashi, J., Kawakami, K., Miura, K., Hirano, M., Ohtsu, N., 2022. Quantitative Analysis of Nitrogen by Atom Probe Tomography Using Stoichiometric γ′-Fe4N Consisting of 15N Isotope. Microsc. Microanal. 28, 42–52. https://doi.org/10.1017/S1431927621013623

Taylor, S.D., Yano, K.H., Sassi, M., Matthews, B.E., Kautz, E.J., Lambeets, S.V., Neuman, S., Schreiber, D.K., Wang, L., Du, Y., Spurgeon, S.R., 2023. Resolving Diverse Oxygen Transport Pathways Across Sr-Doped Lanthanum Ferrite and Metal-Perovskite Heterostructures. Adv. Mater. Interfaces 10, 2202276. https://doi.org/10.1002/admi.202202276

Tegg, L., Breen, A.J., Huang, S., Sato, T., Ringer, S.P., Cairney, J.M., 2023. Characterising the performance of an ultrawide field-of-view 3D atom probe. Ultramicroscopy 253, 113826. https://doi.org/10.1016/j.ultramic.2023.113826

Tegg, L., Stephenson, L.T., Cairney, J.M., 2024. Estimation of the Electric Field in Atom Probe Tomography Experiments Using Charge State Ratios. Microsc. Microanal. 30, 466–475. https://doi.org/10.1093/mam/ozae047

Thompson, K., Bunton, J.H., Kelly, T.F., Larson, D.J., 2006. Characterization of ultralow-energy implants and towards the analysis of three-dimensional dopant distributions using three-dimensional atom-probe tomography. J. Vac. Sci. Technol. B Microelectron. Nanometer Struct. Process. Meas. Phenom. 24, 421–427. https://doi.org/10.1116/1.2141621

Thuvander, M., Kvist, A., Johnson, L.J.S., Weidow, J., Andrén, H.-O., 2013. Reduction of multiple hits in atom probe tomography. Ultramicroscopy, IFES 2012 132, 81–85. https://doi.org/10.1016/j.ultramic.2012.12.005

Thuvander, M., Shinde, D., Rehan, A., Ejnermark, S., Stiller, K., 2019. Improving Compositional Accuracy in APT Analysis of Carbides Using a Decreased Detection Efficiency. Microsc. Microanal. 25, 454–461. https://doi.org/10.1017/S1431927619000424

Thuvander, M., Weidow, J., Angseryd, J., Falk, L.K.L., Liu, F., Sonestedt, M., Stiller, K., Andrén, H.-O., 2011. Quantitative atom probe analysis of carbides. Ultramicroscopy, Special Issue: 52nd International Field Emission Symposium 111, 604–608. https://doi.org/10.1016/j.ultramic.2010.12.024

Tremsin, A.S., Vallerga, J.V., 2020. Unique capabilities and applications of Microchannel Plate (MCP) detectors with Medipix/Timepix readout. Radiat. Meas. 130, 106228. https://doi.org/10.1016/j.radmeas.2019.106228

Tremsin, A.S., Vallerga, J.V., Siegmund, O.H.W., 2020. Overview of spatial and timing resolution of event counting detectors with Microchannel Plates. Nucl. Instrum. Methods Phys. Res. Sect. Accel. Spectrometers Detect. Assoc. Equip. 949, 162768. https://doi.org/10.1016/j.nima.2019.162768

Tsong, T.T., 2005. Atom-Probe Field Ion Microscopy: Field Ion Emission, and Surfaces and Interfaces at Atomic Resolution. Cambridge University Press.

Tsong, T.T., 1978a. Measurement of the field evaporation rate of several transition metals. J. Phys. F Met. Phys. 8, 1349. https://doi.org/10.1088/0305-4608/8/7/008

Tsong, T.T., 1978b. Field ion image formation. Surf. Sci. 70, 211–233. https://doi.org/10.1016/0039-6028(78)90410-7

Tsong, T.T., 1971. Measurement of the Polarizabilities and Field Evaporation Rates of Individual Tungsten Atoms. J. Chem. Phys. 54, 4205–4216. https://doi.org/10.1063/1.1674660

Tsong, T.T., Cole, M.W., 1987. Dissociation of compound ions in a high electric field: Atomic tunneling, orientational, and isotope effects. Phys. Rev. B 35, 66–73. https://doi.org/10.1103/PhysRevB.35.66

Tsong, T.T., Kellogg, G., 1975. Direct observation of the directional walk of single adatoms and the adatom polarizability. Phys. Rev. B 12, 1343–1353. https://doi.org/10.1103/PhysRevB.12.1343

Tsong, T.T., Liou, Y., 1985. Time-of-Flight Energy and Mass Analysis of Metal-Helide Ions and Their Formation and Dissociation. Phys. Rev. Lett. 55, 2180–2183. https://doi.org/10.1103/PhysRevLett.55.2180

Tsong, T.T., Müller, E.W., 1969. Effects of Static-Field Penetration and Atomic Polarization on the Capacity of a Capacitor, Field Evaporation, and Field Ionization Processes. Phys. Rev. 181, 530–534. https://doi.org/10.1103/PhysRev.181.530

Tsong, T.T., Ng, Y.S., Krishnaswamy, S.V., 1978. Quantification of atom-probe FIM data and an application to the investigation of surface segregation of alloys. Appl. Phys. Lett. 32, 778–780. https://doi.org/10.1063/1.89892

Tu, Y., Takamizawa, H., Han, B., Shimizu, Y., Inoue, K., Toyama, T., Yano, F., Nishim/z , A., Nagai, Y., 2017a. Influence of laser power on atom probe tomographic analysis of boron distribution in silicon. Ultramicroscopy 173, 58–63. https://doi.org/10.1016/j.ultramic.2016.11.023

Tu, Y., Takamizawa, H., Han, B., Shimizu, Y., Inoue, K., Toyama, T., Yano, F., Nishim/z , A., Nagai, Y., 2017b. Influence of laser power on atom probe tomographic analysis of boron distribution in silicon. Ultramicroscopy 173, 58–63. https://doi.org/10.1016/j.ultramic.2016.11.023

Ulfig, R., Kelly, T., Gault, B., 2009. Promoting Standards in Quantitative Atom Probe Tomography Analysis. Microsc. Microanal. 15, 260–261. https://doi.org/10.1017/S143192760909881X

Vallerga, J.V., McPhate, J.B., 2000. Optimization of the readout electronics for microchannel plate delay line anodes, in: Instrumentation for UV/EUV Astronomy and Solar Missions. Presented at the Instrumentation for UV/EUV Astronomy and Solar Missions, SPIE, pp. 34–42. https://doi.org/10.1117/12.410543

Vella, A., 2013. On the interaction of an ultra-fast laser with a nanometric tip by laser assisted atom probe tomography: A review. Ultramicroscopy, IFES 2012 132, 5–18. https://doi.org/10.1016/j.ultramic.2013.05.016

Vella, A., Houard, J., 2016. Chapter Eight - Laser-Assisted Field Evaporation, in: Lefebvre-Ulrikson, W., Vurpillot, F., Sauvage, X. (Eds.), Atom Probe Tomography. Academic Press, pp. 251–278. https://doi.org/10.1016/B978-0-12-804647-0.00008-5

Vella, A., Shinde, D., Houard, J., Silaeva, E., Arnoldi, L., Blum, I., Rigutti, L., Pertreux, E., Maioli, P., Crut, A., Del Fatti, N., 2018. Optothermal response of a single silicon nanotip. Phys. Rev. B 97, 075409. https://doi.org/10.1103/PhysRevB.97.075409

Veret, T., Delaroche, F., Blum, I., Houard, J., Klaes, B., Mouton, I., De-Geuser, F., Seydoux-Guillaume, A.-M., Vurpillot, F., 2026. Exploring Mechanisms Leading to Composition Errors in Monazite (CePO4) Analyzed With Atom Probe Tomography. Microsc. Microanal. 32, ozag009. https://doi.org/10.1093/mam/ozag009

Veret, T., Delaroche, F., Blum, I., Houard, J., Klaes, B., Mouton, I., De-Geuser, F., Seydoux-Guillaume, A.-M., Vurpillot, F., 2025. Exploring mechanisms leading to composition errors in monazite (CePO4) analysed with atom probe tomography. https://doi.org/10.48550/arXiv.2511.02617

Vurpillot, F., 2016. Chapter Seven - Three-Dimensional Reconstruction in Atom Probe Tomography: Basics and Advanced Approaches, in: Atom Probe Tomography. Academic Press, pp. 183–249. https://doi.org/10.1016/B978-0-12-804647-0.00007-3

Vurpillot, F., Bostel, A., Cadel, E., Blavette, D., 2000. The spatial resolution of 3D atom probe in the investigation of single-phase materials. Ultramicroscopy 84, 213–224. https://doi.org/10.1016/S0304-3991(00)00035-8

Vurpillot, F., Gaillard, A., Da Costa, G., Deconihout, B., 2013. A model to predict image formation in Atom probeTomography. Ultramicroscopy, IFES 2012 132, 152–157. https://doi.org/10.1016/j.ultramic.2012.12.007

Vurpillot, F., Hatzoglou, C., Klaes, B., Rousseau, L., Maillet, J.-B., Blum, I., Gault, B., Cerezo, A., 2024. Crystallographic Dependence of Field Evaporation Energy Barrier in Metals Using Field Evaporation Energy Loss Spectroscopy Mapping. Microsc. Microanal. 30, 1091–1099. https://doi.org/10.1093/mam/ozae083

Vurpillot, F., Houard, J., Vella, A., Deconihout, B., 2009. Thermal response of a field emitter subjected to ultra-fast laser illumination. J. Phys. Appl. Phys. 42, 125502. https://doi.org/10.1088/0022-3727/42/12/125502

Vurpillot, F., Oberdorfer, C., 2015. Modeling Atom Probe Tomography: A review. Ultramicroscopy, 1st International Conference on Atom Probe Tomography & Microscopy 159, 202–216. https://doi.org/10.1016/j.ultramic.2014.12.013

Vurpillot, F., Parviainen, S., Djurabekova, F., Zanuttini, D., Gervais, B., 2018. Simulation tools for atom probe tomography: A path for diagnosis and treatment of image degradation. Mater. Charact. 146, 336–346. https://doi.org/10.1016/j.matchar.2018.04.024

Vurpillot, F., Rolland, N., Estivill, R., Duguay, S., Blavette, D., 2016. Accuracy of analyses of microelectronics nanostructures in atom probe tomography. Semicond. Sci. Technol. 31, 074002. https://doi.org/10.1088/0268-1242/31/7/074002

Wam/z , M., 1984. ON THE THERMALLY ACTIVATED FIELD EVAPORATION OF SURFACE ATOMS. Surf. Sci. 145, 451–465.

Waugh, A.R., Boyes, E.D., Southon, M.J., 1976. Investigations of field evaporation with a field-desorption microscope. Surf. Sci. 61, 109–142. https://doi.org/10.1016/0039-6028(76)90411-8

Westman, A., Brinkmalm, G., Barofsky, D.F., 1997. MALDI induced saturation effects in chevron microchannel plate detectors. Int. J. Mass Spectrom. Ion Process. 169–170, 79–87. https://doi.org/10.1016/S0168-1176(97)00205-X

Wilkes, T.J., Smith, G.D.W., Smith, D.A., 1974. On the quantitative analysis of field-ion micrographs. Metallography 7, 403–430. https://doi.org/10.1016/0026-0800(74)90041-X

Woodhead, A.W., Ward, R., 1977. The channel electron multiplier and its use in image intensifiers. Radio Electron. Eng. 47, 545–553. https://doi.org/10.1049/ree.1977.0079

Woods, E.V., 2025. Mapping the Path to Cryogenic Atom Probe Tomography Analysis of Biomolecules. Microsc. Microanal. 31.

Worrall, G.M., Smith, G.D.W., 1986. THE QUANTITATIVE ANALYSIS OF COPPER IN IRON BASED ALLOYS. J. Phys. Colloq. 47, C2-250. https://doi.org/10.1051/jphyscol:1986237

Xia, Y.), Karahka, M., Kreuzer, H.J., 2015. Field evaporation of ZnO: A first-principles study. J. Appl. Phys. 118, 025901. https://doi.org/10.1063/1.4926489

Xia, Y., Tang, L., Lu, X., Zhu, S., 2024. Theoretical insights into laser-assisted field evaporation of ionic compounds. J. Appl. Phys. 136, 134304. https://doi.org/10.1063/5.0231078

Xu, X., Frentrup, M., Kusch, G., Shu, R., Hofer, C., Bagot, P.A.J., Moody, M.P., Kappers, M.J., Wallis, D.J., Oliver, R.A., 2025. Point defect luminescence associated with stacking faults in magnesium doped zincblende GaN. J. Appl. Phys. 137, 235301. https://doi.org/10.1063/5.0274599

Yamaguchi, Y., Takahashi, J., Kawakami, K., 2009a. The study of quantitativeness in atom probe analysis of alloying elements in steel. Ultramicroscopy, IFES 2008 109, 541–544. https://doi.org/10.1016/j.ultramic.2008.11.017

Yamaguchi, Y., Takahashi, J., Kawakami, K., 2009b. The study of quantitativeness in atom probe analysis of alloying elements in steel. Ultramicroscopy, IFES 2008 109, 541–544. https://doi.org/10.1016/j.ultramic.2008.11.017

Yang, L., Chen, E.Y.-S., Qu, J., Garbrecht, M., McCarroll, I.E., Mosiman, D.S., Saha, B., Cairney, J.M., 2025. Improved atom probe specimen preparation by focused ion beam with the aid of multi-dimensional specimen control. Microstructures 5, N/A-N/A. https://doi.org/10.20517/microstructures.2024.53

Yao, L., Gault, B., Cairney, J.M., Ringer, S.P., 2010. On the multiplicity of field evaporation events in atom probe: A new dimension to the analysis of mass spectra. Philos. Mag. Lett. 90, 121–129. https://doi.org/10.1080/09500830903472997

Yeoh, W.K., Hung, S.-W., Chen, S.-C., Lin, Y.-H., Lee, J.J., 2020. Quantification of dopant species using atom probe tomography for semiconductor application. Surf. Interface Anal. 52, 318–323. https://doi.org/10.1002/sia.6706

Zanuttini, D., Blum, I., di Russo, E., Rigutti, L., Vurpillot, F., Douady, J., Jacquet, E., Anglade, P.-M., Gervais, B., 2018. Dissociation of GaN2+ and AlN2+ in APT: Analysis of experimental measurements. J. Chem. Phys. 149, 134311. https://doi.org/10.1063/1.5037010

Zanuttini, D., Blum, I., Rigutti, L., Vurpillot, F., Douady, J., Jacquet, E., Anglade, P.-M., Gervais, B., 2017a. Simulation of field-induced molecular dissociation in atom-probe tomography: Identification of a neutral emission channel. Phys. Rev. A 95, 061401. https://doi.org/10.1103/PhysRevA.95.061401

Zanuttini, D., Blum, I., Rigutti, L., Vurpillot, F., Douady, J., Jacquet, E., Anglade, P.-M., Gervais, B., 2017b. Electronic structure and stability of the SiO2+ dications produced in tomographic atom probe experiments. J. Chem. Phys. 147, 164301. https://doi.org/10.1063/1.5001113

Zhao, L., Normand, A., Houard, J., Blum, I., Delaroche, F., Latry, O., Ravelo, B., Vurpillot, F., 2017. Optimizing Atom Probe Analysis with Synchronous Laser Pulsing and Voltage Pulsing. Microsc. Microanal. Off. J. Microsc. Soc. Am. Microbeam Anal. Soc. Microsc. Soc. Can. 23, 221–226. https://doi.org/10.1017/S1431927616012666

Zhou, Z., Wang, Z., Niu, R., Liu, P.-Y., Huang, C., Sun, Y.-H., Wang, X., Yen, H.-W., Cairney, J.M., Chen, Y.-S., 2023. Cryogenic atom probe tomography and its applications: a review. Microstructures 3, N/A-N/A. https://doi.org/10.20517/microstructures.2023.38